\pdfoutput=1

\documentclass[12pt,twoside]{myreport}
\usepackage{amsmath}
\usepackage{amssymb}
\usepackage{natbib}
\usepackage{graphics}
\usepackage{deluxetable}
\usepackage{lscape}
\usepackage{float}
\usepackage{rotating}
\usepackage{xspace}
\usepackage{mydefs}
\usepackage{indentfirst}
\usepackage{setspace}
\renewcommand{\baselinestretch}{2}
\numberwithin{equation}{chapter}

\title{THE THREE-DIMENSIONAL BEHAVIOR OF SPIRAL SHOCKS IN PROTOPLANETARY DISKS}

\author{Aaron Christopher Boley}

\date{September 2007}

\begin{document}
\maketitle

\newpage

%
%

\pagestyle{plain}
\pagenumbering{roman}
\setcounter{page}{2}

%
%
\signaturepage{Richard H.~Durisen, PhD}{Andrew Bacher, PhD}
       {Haldan Cohn, PhD}{Megan K.~Pickett, PhD}
       {Liese van Zee, PhD}

\clearpage

%
%
%




%
%
%

\vspace*{2.5in}
\begin{center}
To Karen,\\
Marilyn,\\
Meredith,\\
and Monte
\end{center}
\clearpage
%
%
%
\begin{acknowledgments}

ACKNOWLEDGMENTS: I would like to thank my Dissertation Committee and the faculty of the Indiana University
Astronomy Department for helping me to cultivate my writing, research, and teaching 
abilities.  Inside and outside of the department, I have benefited from conversations with many scientists. 
In particular I would like to thank \AA ke Nordlund for hosting me at NBIfA and for initiating much of the radiation transfer work contained in this dissertation.  Fred Ciesla, Jeff Cuzzi, Steve Desch, and Edward Scott, I thank you for many useful comments regarding solids in disks.  I thank Tom Hartquist for stimulating discussions about disk chemistry. Nuria Calvet and Lee Hartmann, I am grateful for your elucidations on the observations of protoplanetary disks.  As regards disk dynamics, I owe my gratitude to Alan Boss, Charles Gammie, Giueseppi Lodato, Artur Gawryszczak, Lucio Mayer (thank you for hosting me at ITP), Andy Nelson, Shangli Ou, Megan Pickett, Tom Quinn, Ken Rice, Dimitris Stamatellos, and Joel Tohline.  I am indebted to Johnny Chang, Bob Hood, and Steve Johnson for their work on the optimization of the parallel version of CHYMERA.  Scott Michael, thank you for managing the IU hydro research group's dedicated workstations. I thank Annie Mej\'ia and Kai Cai for laying the ground work for several of the studies presented here, and I thank Kevin Croxall for our many chats.  

\newpage

\noindent Liese van Zee:  You have helped me throughout my entire tenure as a graduate student.  You have taught me many valuable lessons, and have pushed me to be aggressive but professional in pursuing career, grant, and research opportunities.  Thank you.
\vspace*{1.5cm}

\noindent Richard H. Durisen:  You have acted as my teacher, my mentor, my colleague, and my friend.  You have allowed me to engage in my own research interests without ever being too far away.  You have aided me greatly in seeking career opportunities.  You are largely responsible for any success I have achieved as a researcher and for any future success I may be so fortunate to earn.  I cherish your lessons.  Thank you.
\vspace*{1.5cm}

\noindent To my family:  Mom, you are the one who truly taught me the basics on which all of this work is built.  Thank you.  Meredith, you have provided me with many delightful diversions.  Thank you.  Karen, you have endured all of graduate school with me.  You were there to celebrate achievements, and you were beside me during my dark hours.  Any fruit that this work produces is as much borne out of your labors as is borne out of mine. Thank you.

\clearpage

Finally, I would like to acknowledge the generous financial support given to me as a NASA Graduate Student Researchers Program fellow by the Science Mission Directorate.  This work was supported in part by the IU Astronomy Department IT facilities, by systems made available by the NASA Advanced Supercomputing Division at NASA Ames, by systems obtained by Indiana University through Shared University Research grants through IBM, Inc., to Indiana University, and by dedicated workstations provided to my research group by IU's University Information Technology Services.
\end{acknowledgments}

%
%

\begin{abstract}

ABSTRACT: In this dissertation, I describe theoretical and numerical studies that address the three-dimensional behavior of spiral shocks in protoplanetary disks and the controversial topic of gas giant formation by disk instability.  For this work, I discuss characteristics of gravitational instabilities (GIs) in bursting and asymptotic phase disks; outline a theory for the three-dimensional structure of spiral shocks, called shock bores, for isothermal and adiabatic gases;  consider convection as a source of cooling for protoplanetary disks; investigate the effects of opacity on disk cooling; use multiple analyses to test for disk stability against fragmentation; test the sensitivity of GI behavior to radiation boundary conditions; measure shock strengths and frequencies in GI-bursting disks; evaluate temperature fluctuations in unstable disks; and investigate whether spiral shocks can form chondrules when GIs activate.  The numerical methods developed for these studies are discussed, including a radiation transport routine that explicitly couples the low and high optical depth regimes and a routine that models ortho and parahydrogen.  Finally, I explore the hypothesis that chondrule formation and the FU Ori phenomenon are driven by GI activation in dead zones. 
\end{abstract}



\tableofcontents
\listoffigures
\listoftables

\pagestyle{headings}
\pagenumbering{arabic}
\setcounter{chapter}{0}
\chapter{INTRODUCTION}

The formation of planetary systems in disks was anticipated as early as 1755 by Kant.  Since that time, analytic and numerical work has demonstrated that disks should form during protostellar core collapse (e.g., Ulrich 1976; Cassen \& Moosman 1981; Durisen et al.~1989; Yorke et al.~1993; Pickett et al.~1997;  Vorobyov \& Basu 2006; Krumholz et al.~2007), and direct imaging and spectral energy distribution (SED) fitting have verified that disks do surround young stars (e.g., Beckwith et al. 1990; Strom et al.~1993; Padgett et al.~1999; Calvet et al.~2005; Andrews \& Williams 2005; D'Alessio et al.~2006; Eisner \& Carpenter 2006).   It is in these disks where grain growth and annealing occur (Natta et al.~2007),  where planetesimals form from the processed dust, and where planets are built.    Therefore, these {\it protoplanetary disks} do not simply act as a mass reservoir for accretion onto the star, but also serve as an astrophysical factory for chemical and dynamical processes leading to planet formation.   For this discussion, I will focus on T Tauri-mass stars and their disks, although Herbig Ae/Be stellar disks are likely to be susceptible to similar phenomena.  \newpage

T Tauri stars are pre-main-sequence (PMS) stars that lie between the birthline,  where PMS stars first begin quasi-static gravitational contraction (Stahler 1983, 1988), and the zero-age main sequence (ZAMS), and have masses $\lesssim 2.5 M_{\odot}$ (Lawson et al.~1996). These objects are typically divided into two groups: classical T Tauri stars (CTTS) and weak-line (weak-emission) T Tauri stars (WTTS).  CTTSs have excess infrared emission, excess blue continuum emission, and strong line emission.  These data are interpreted to indicate the presence of a disk and active accretion onto the star.  WTTSs have some infrared excess, but much weaker line and blue continuum emission; the gaseous inner disk surrounding the star is believed to be mostly dissipated.  The distinction between a CTTS and a WTTS is quantified by the equivalent width (EW) of the H${\alpha}$ line, where WTTSs have an EW(H${\alpha}) \le 10$~\AA~(see Andrews \& Carpenter 2005 for a summary). The higher mass analogs of T Tauri stars are the Herbig Ae/Be stars (Herbig 1960).  These lie between the intermediate-mass birthline (Palla \& Stahler 1990) and the ZAMS, exhibit infrared excess, and have strong emission lines  (Th\'e et al.~1994; van den Ancker et al.~1997).

T Tauri stars fit into a more general classification of PMS objects, the Young Stellar Objects (YSOs).  The term YSO should be understood to include the forming star, disk, and envelope, when present. Typically, YSOs are divided into three general classifications (Lada \& Wilking 1984; Adams et al.~1987; Kenyon \& Hartmann 1987; Lada 1987; Greene et al.~1994; Andrews \& Williams 2005): Class I objects are embedded protostars with envelope material that is accreting onto a surrounding disk.  For these sources, energy dissipation by disk accretion is an important component of the total disk emission.  In Class II objects, most of the disk emission is reprocessed star light, and envelope accretion has essentially ceased.  Finally, Class III objects have very little disk emission.  In addition to Classes I through III, Andr\'e et al.~(1993) proposed the categorization Class 0, with VLA 1623 as the prototype.  They suggest that Class 0 sources are the precursors to the Class I stage and that most of the circumstellar material is distributed in an envelope.  In contrast, Jayawardhana et al.~(2001) argue that Class 0 and Class I YSOs may be at a similar stage in evolution, with Class 0 YSOs forming in a denser environment than Class I YSOs.

Classes I, II, and III can be specified quantitatively by calculating the spectral index, $\alpha=d \ln \nu F_{\nu}/d\ln\nu$, from the object's SED.  Following Greene et al.~(1994), Class I objects have a negative $\alpha$ as measured between 2.2 and 10 $\mu$m, Class II objects have a flat or rising spectrum with $0\lesssim\alpha \le 1.6$, and Classes III have an $\alpha>1.6$.  CTTSs span Classes I and II, and WTTSs are Class III objects.  The classification is interpreted as an evolutionary sequence, but absolute time intervals are not necessarily associated with a given phase. Recent observations by Eisner \& Carpenter (2006) seem to indicate that a Class I object evolves to a Class II object in about 1 Myr. 

It is generally accepted that during the Class I stage of T Tauri star evolution, disks are massive enough for gravitational instabilities (GIs) to develop (see below). For example, recent observations of FU Ori, which is transitioning from the Class I to II stage, indicate that the inner 1 AU may be gravitationally unstable (Zhu et al.~2007).   Moreover, detailed measurements of L1551 IRS 5, the prototypical Class I object (Adams et al.~1987), indicate that the circumstellar disk mass around the northern source is comparable with the star's mass and that it too is gravitationally unstable (Osorio et al.~2003).  Three-dimensional hydrodynamics simulations by Cai et al.~(2007) of L1551 based on the disk parameters reported by Osorio et al.~confirm that GIs develop in the disk.  

A {\it gravitational instability} is a dynamic instability driven by self-gravity. 
Toomre (1964) demonstrated that an infinitely thin gaseous disk is unstable to self-gravitating ring instabilities when the Toomre $Q$ parameter,
\begin{equation}Q=\frac{c_s\kappa}{\pi G\Sigma},\end{equation}
approaches unity. The stabilizing quantities are the sound speed $c_s$ and the epicyclic frequency of the gas $\kappa$, and the destabilizing quantity is the surface density $\Sigma$.  When $Q\lesssim 1.7$, a disk with finite thickness is susceptible to nonaxisymmetric instabilities (see Durisen et al.~2007a for a review), and spiral waves develop.  For a GI-active disk, the spiral waves driven by self-gravity may be the dominate way angular momentum is transfered outward and mass transfered inward (Lynden-Bell \& Kalnajs 1972;  Boss 1984b; Larson 1984; Durisen et al.~1986; Laughlin \& Bodenheimer 1994).  However, the role that GIs play in T Tauri disks of ages $>$ 1 Myr is uncertain, especially their role in planet formation.  Are disks that are older than 1 Myr cold and/or massive enough for GIs to activate?

Even in relatively low-mass disks, GIs may activate for large $r$. A back-of-the-envelope calculation shows that for a Keplerian disk, i.e, $\kappa=\Omega=\left(GM_{\rm star}/r^3\right)^{1/2}$, $Q\sim r^{-3/2-q/2+p}$, where $T\sim r^{-q}$ and $\Sigma\sim r^{-p}$ represent the power laws for the effective temperature and surface density profiles, respectively. The typical disk effective temperature profile falls off as roughly $r^{-1/2}$, with a large amount of scatter (Beckwith et al.~1990; Kitamura et al.~2002), and disk surface density profiles are believed to lie between about $p\approx1$ to 1.5 (Beckwith et al.~1990; Dullemond et al.~2007).  With $q=1/2$ and $p=3/2$, one expects $Q\sim r^{-1/4}$, and at least the outer disk will be unstable against GIs.  What about GIs in the planet-formation region of the disk, presumably $r\lesssim 40$ AU?  

Over the past several decades, the typical picture of a T Tauri disk has been the {\it Minimum Mass Solar Nebula} model (Weidenschilling 1977; Hayashi et al.~1985), which is based on the known mass distribution of solids in the Solar System and which represents the minimum mass, $\sim10^{-2}~M_{\odot}$, required to form these bodies.  Beckwith et al.~(1990) found that for a sample of 86 T Tauri stars in the Taurus-Auriga star forming region, 42\% of the systems have detectable protoplanetary disks, the disk masses $M_d$ range between 0.001 $M_{\odot}$ and about 1 $M_{\odot}$, and the average disk mass is about 0.02 $M_{\odot} $, with massive disks being rare.  Andrews \& Williams (2005) also studied the Taurus-Auriga star formation region and found that 61\% of the 153 targets have detectable disks,  the median disk mass is $ M_d\sim0.005~M_{\odot}$, and less than a few percent of disk masses are likely to be gravitationally unstable based on their mass estimates and model assumptions.  Eisner \& Carpenter (2006) found a similar median mass for a sample of 336 disks in the Orion Nebular Cluster, and they found a fraction of high mass disks consistent with the fraction in Taurus-Auriga.  

With these disk masses, one might expect GIs to play a minor role in T Tauri disk evolution, except for possibly $r\gtrsim100$ AU, and expect the MMSN to be a fairly accurate description of T Tauri disks.  However, there are several potential problems with these observations.  Grain growth beyond 1 mm will likely result in underestimating disk masses because the surface area of the emitting dust grains will have changed (Andrews \& Williams 2005; Hartmann et al.~2006).  In addition, as argued by Hartmann et al.~(2006), inferred mass fluxes from excess UV and blue-optical continuum measurements suggest that mass accretion rates are incompatible with disk masses and lifetimes, e.g., 0.005 $M_{\odot}$ with $\dot{M}\sim10^{-8}~M_{\odot}\rm~yr^{-1}$ in a 1-2 Myr old disk.  With the uncertainty in measured accretion rates, it is unclear how much the $M_d$ or $\dot{M}$ measurements are off, but it does indicate that the measurements may only be accurate to within a factor of four (Andrews \& Williams 2005).  Finally, FU Orionis events (see below) accrete 0.01 $M_{\odot}$ onto the star during their hundred-year-long outbursts.  Although these disks are somewhere between Class I and II objects, one should keep in mind that any one outburst accretes a MMSN, and a typical T Tauri disk may go through approximately 10 FU Ori-like events in less than about 1 Myr (Hartmann \& Kenyon 1996).  

As suggested by Eisner \& Carpenter (2006), the average low disk masses that are observed may be more indicative of fast disk evolution than T Tauri disks never having a high-mass disk phase.  Based on massive disk estimates of several clusters at different ages, Eisner \& Carpenter argue that disk masses may change by a factor of about a few between 0.3 and 2 Myr.  In addition, Andrews \& Williams (2005)
argue that the ubiquity of planets, where 10\% of low-mass stars have {\it detectable} planets, may indicate disk measurements are systematically underestimated.  The majority of T Tauri disks with ages of 1 to 2 Myr may very well be too low mass for GIs to activate, but GIs should not be discounted during the first Myr of evolution for most T Tauri objects and even in the late stages of disk evolution around the high mass T Tauri stars (about 1 to 2.5 $M_{\odot}$).  

\section{The $\alpha$ Disk}

Shakura \& Sunyaev (1973) made the ansatz that the turbulent viscosity in an accretion disk 
responsible for carrying angular momentum outward can be approximated
by
\begin{equation}\nu=\alpha c_s h,\end{equation}
where $h$ is the disk vertical scale height and $\alpha \le 1$ is a parameter that sets the magnitude of the turbulent viscosity.
Shakura \& Sunyaev did not identify the source of the turbulence, but their heuristic approach allows for 
analytic solutions to the accretion disk problem and relatively easy numerical modeling.  In a razor thin Keplerian disk, mass transport is described by (e.g., Hartmann 1998; Balbus \& Papaloizou 1999)
\begin{equation}\frac{\partial \Sigma}{\partial t} = \frac{3}{r}\frac{\partial}{\partial r}\left(r^{1/2}
\frac{\partial}{\partial r}\left(\nu r^{1/2}\Sigma\right)\right).\end{equation}
In an isothermal Keplerian disk with negligible self-gravity, one can show that $\rho(z)=\rho_0\exp
(-\Omega^2z^2/2c_i^2)$, where $c_i$ is the isothermal sound speed.  By defining the vertical scale height $h=\Sigma/2\rho_0$, one finds that to within a factor of order unity, $h\approx c_i/\Omega$.  Varying the gas equation of state does not change the factor of order unity significantly, and so the appropriate sound speed can be used in place of $c_i$.  Using this relation, equation (1.3) can be rewritten according to $\alpha$ disk theory:
\begin{equation}\frac{\partial \Sigma}{\partial t} = \frac{3}{r}\frac{\partial}{\partial r}\left(r^{1/2}
\frac{\partial}{\partial r}\left(\alpha c_s^2 r^{1/2}\Sigma\Omega^{-1}\right)\right),\end{equation}
and so by specifying $\alpha$, mass transport can be completely described for a given initial surface density profile.  In a steady state accretion disk, $\nu\Sigma=\rm constant$ (Hartmann 1998), and equations (1.3) and (1.4) can be reduced to 
\begin{equation}\dot{M}=3\pi\nu\Sigma\approx 3\pi\alpha  c_s^2\Sigma\Omega^{-1}.\end{equation}
Note that equation (1.4) indicates that an $\alpha$ disk is an inherently local description for mass transport.  Because the energy dissipation per unit area in an accretion disk can be described by (Pringle 1981)
\begin{equation}Q_e\approx\frac{GM\dot{M}}{8\pi r^3},\end{equation}
energy dissipation in an $\alpha$ disk is local as well (see Balbus \& Papaloizou 1999).  If long-range energy transport and/or angular momentum transport are negligible, then an $\alpha$ prescription for disk evolution would be fairly accurate and advantageous, because local simulations, e.g., shearing sheets (Balbus \& Hawley 1998), would capture disk evolution well.  Otherwise, the disk behavior will be dissimilar to what $\alpha$ disk theory predicts due to long-range angular momentum and energy fluxes (Balbus \& Papaloizou 1999).  Identifying the principal angular momentum transport mechanism and how it behaves is critical to understanding whether disks can be described by an $\alpha$ disk model.

\section{GIs, MRI, and Dead Zones}

There are two mechanisms that have been demonstrated to work efficiently at transporting mass inward and angular momentum outward: GIs and the magnetorotational instability (MRI; see Balbus \& Hawley 1991; Desch 2004).  As discussed above, GIs require a cold, massive environment to activate.  In contrast, the MRI in principle only needs a weak magnetic field coupled to ionized species in the gas.  To see how the MRI produces angular momentum transfer, consider two parcels of gas at slightly different radii through which a magnetic field is threaded.  As the gas orbits, the parcels of gas are separated due to the shear in the disk.  Assuming that the magnetic flux cannot diffuse, i.e., $\partial {\bf B}/ \partial t =  \nabla\times{\bf v} \times {\bf B}$ (the {\it flux freezing} approximation), the magnetic field remains entrained with the gas. As the gas shears, the magnetic fields act as a spring between the two gas parcels.  This produces tension and an exchange of angular momentum.  Because the inner parcel leads in a Keplerian disk, the angular momentum transfer is outward, and the mass elements drift further apart.  This leads to a greater force mediated by the magnetic field lines, and even more angular momentum is transfered; an instability ensues (Balbus \& Hawley 1998).  

In order for the MRI to activate, ionized species must be present in the gas phase.   Ionization may be from a thermal source, e.g., collisional ionization of alkalis, or a nonthermal source, e.g., cosmic rays, energetic particles from the star, and X-ray irradiation. I refer to these nonthermal sources simply as energetic particles (EPs) for this discussion.  Thermal ionization of alkalis only occurs for $T\gtrsim1000$ K, and so only the innermost portion of a T Tauri disk is expected to be thermally ionized.  If most of a T Tauri disk is MRI active, it must be due to nonthermal sources.  Gammie (1996) had the insight that because one expects EPs to be attenuated by the gas, there may be regions where the MRI is active and other areas where it is absent.  In the inner regions of a disk where the column densities are large, MRI may only be active in a thin layer, resulting in {\it layered accretion}.  As one moves outward in the disk and the column density decreases, the entire disk can become MRI active.  The region where MRI is mostly absent, except for a thin layer at high altitude, is called the {\it dead zone}.  This dead zone is of particular interest to GI studies because mass may pile up in dead zones due to the sudden drop in accretion rate as the disk transitions from a fully active MRI $\alpha$ disk to a thin, layered accretion flow.  

EPs are attenuated with a scale length of about 100 g cm$^{-2}$ (Stepinski 1992), and so even for a MMSN, the disk will likely exhibit layered accretion (Desch 2004) and a dead zone can form.  However, a dead zone may not be tranquil due to Reynolds stresses from the thin, MRI-active upper layers (Fleming \& Stone 2003).  Nonetheless, even if mass accretion is only reduced and not altogether halted, mass may still pile up in the dead zone (Oishi et al.~2007).  If enough mass accumulates, then even for an otherwise low-mass disk, GIs can activate.  Such mass concentrations may play an important role in the FU Ori phenomenon.

 \section{FU Orionis Phenomenon}

The FU Ori phenomenon is
characterized by a rapid (1-10s yr) increase in optical brightness of a young T Tauri
object, typically by 5 magnitudes.   Emission line spectra and strong near and mid infrared excess indicate that the event is driven by
sudden mass accretion of the order $10^{-4} M_{\odot}~\rm yr^{-1}$ from the inner disk
onto the star (Hartmann \& Kenyon 1996). 
Because FU Ori objects appear to have
decay timescales of about 100 yr, 0.01 $M_{\odot}$ can be accreted onto the
star.  Note that this is the entire mass of the MMSN.

FU Ori objects are very
rare (5-10 known objects, with some objects uncertain; Green et al.~2006), but when compared with local star formation rates, it is plausible
for most T Tauri stars to have approximately 10 FU Ori outbursts with a low state lasting $10^{4}$ to $10^5$ yr between outbursts (Hartmann \& Kenyon 1996). Some of these systems are still surrounded by a substantial remnant envelope (e.g., V1057 Cyg), with continued infall onto the disks.  Vorobyov \& Basu (2005, 2006) find that in their 2D magnetohydrodynamics simulations with self-gravity, mass accretion onto a disk from an envelope results in episodic bursts of GI activity with about the same frequency as expected for FU Ori objects.  They suggest that the FU Ori phenomenon may be driven by fragmented clumps accreting onto the star.  However, Zhu et al.~(2007) find that this scenario may provide too much disk emission for $r>10$ AU to explain  the FU Ori observations.  Moreover,  it is unclear whether all FU Ori objects have significant envelopes (e.g., FU Ori; Herbig 1977; Kenyon \& Hartmann 1991; Green et al.~2006; Zhu et al.~2007); the FU Ori phenomenon may not require an infalling envelope to activate.  Another external source for driving the FU Ori phenomenon is close encounters with a companion.   Given the assumption that most T Tauri stars go through an FU Ori phase, this appears to be a reasonable idea.   Even though several FU Orionis objects are confirmed binaries (e.g., L1551, RNO 1B and 1C), some show no signs of a nearby companion as determined from the absence of spectral line drift (Hartmann \& Kenyon 1996; but see also arguments by Reipurth 2005 in support of a binary-driven mechanism). 

To date, the best explanation for the
optical outburst is a thermal instability. Models of this mechanism 
produce timescales and observational features that are similar to the objects that are observed (e.g., Bell \& Lin 1994). For a simple
$\alpha$ disk model, $\dot{M}\approx 3\pi \nu \Sigma \sim c^2\Sigma/\Omega$, and so an increase in the sound speed increases mass accretion.
If dissipation from mass accretion begins to occur faster than radiative
cooling, the disk could heat up until hydrogen thermally ionizes.  This
ionization creates an H$^-$ opacity front that strongly reduces the efficiency
of radiative cooling. The disk continues to heat, and a thermal runaway begins until the 
unstable region is completely ionized. The MRI would likely be
active in such a hot disk due to thermal ionization, which can lead to $\alpha$
disk-like accretion (cf.~Fromang et al.~2004).  Despite the success of the thermal
instability in explaining the FU Ori phenomenon, there is still an unanswered
question: What caused the dissipation to become greater than radiative
cooling?  

Armitage et al.~(2001) suggested that GIs in a bursting dead zone 
might be able to trigger an
FU Ori outburst by rapidly increasing the accretion into the inner disk and initiating a 
thermal MRI.
Likewise, Hartmann (2007, private communication)  and Zhu et al.~(2007) suggest that gravitational torques might drive the temperature near
1 AU above 1000 K and thermally ionize alkalis.  The temperatures may need to be greater than 1400 K to prevent depletion of ions by dust grains (Sano et al.~2000; Desch 2004), but
this does not change the general picture. Once the alkalis are ionized, a thermal MRI could operate and feed mass inside 0.1 AU until a thermal instability activates.  The FU Ori phenomenon may be a result of a cascade of instabilities, starting with a burst of GI activity in a dead zone, followed by accretion due to a thermal MRI, followed finally by a thermal instability.  Indeed, recent observations of FU Ori indicate that very large mass fluxes are present out to at least $r=0.5$ AU (Zhu et al.~2007).  

\section{Chondrules}

 Chondrules are small, 0.1-1.0 mm in diameter, igneous globules that were flash melted during their formation. They are a fundamental primitive solid inasmuch as they can account for over 80\% 
of some chondritic meteorite masses (Hewins et al.~2005). Although details are still being debated, most chondrules formed in the first 1 to 3 Myr of the Solar Nebula's evolution (Bizzarro et al.~2004; Russell et al.~2005).  Furthermore, about half of all material accreted onto Earth each year is in the form of chondritic meteorites (Hewins et al.~1996), and therefore chondrules are arguably the building blocks of the terrestrial worlds.  A theoretical description of the environments in which chondrules can form is a crucial step in understanding the origin of the Solar System. 

Meteoritics combines astrophysics with geology, and unlike many areas of astrophysics, specimens almost literally fall into our hands.  As a result, formation constraints for chondrules can be tested through
laboratory experiments.  Chondrule precursors were flash melted from solidus to liquidus, where high temperatures \linebreak $T\sim 1700$ K were experienced by the precursors for a few minutes. The melts then cooled over hours, with the actual cooling time depending on chondrule type (Scott \& Krot 2005).  Chondrules have diverse petrologies, varying in volatile abundances, mineral composition, oxygen isotope ratios, and textures (Jones et al.~2005).  Many chondrules have fine-grained and igneous coarse-grained rims, and many chondrules indicate multiple collisions, as inferred from chondrule fragments and compound chondrules, i.e., chondrules inside chondrules (Scott \& Krot 2005).  In order to remain liquid and preserve volatiles, chondrules are believed to have formed in regions of high pressure ($10^{-4}$ to $10^{-3}$ bar) or  in a dusty environment, with a dust to gas ratio 10 to 100 times greater than the typically 1/100 value assumed for the Solar Nebula (Wood 1963; Scott \& Krot 2005; Hewins et al.~2005).  


Ca-Al-rich Inclusions (CAIs) and Amoeboid Olivine Aggregates (AOAs) (e.g., MacPherson et al.~2005) also provide constraints on the environment of the early Solar Nebula.  These refractory inclusions are thought to be among the first solids processed in the Solar Nebula, and the formation of CAIs is often used to set the age $t=0$.  CAI and possibly AOA formation occurred in less than 0.3 Myr, and chondrule formation began 0-2 Myr later and lasted for $<$ 3 Myr for normal chondrules (Amelin et al.~2002; Itoh et al.~2002; Bizzarro et al.~2004; see Scott \& Krott 2005 for a review).  CAIs formed at pressures near $10^{-3}$ atm and at temperatures $T\sim1400$ K. Some CAIs were completely melted and others were not (MacPherson et al.~2005).  Moreover, some CAIs show signs of reprocessing, and so they likely experienced chondrule-forming events.  I do not dismiss the importance of CAI and AOA formation in understanding the early conditions of the Solar Nebula. Indeed, they are a critical component, but for this dissertation, I focus on chondrule formation.

Chondritic parent bodies show a range in composition, and this variation has been interpreted to be due to temporal and spatial formation differences (Wood 2005).   There are three basic types of chondrites in which almost all chondrules can be classified: carbonaceous, ordinary, and enstatite chondrites. For an in-depth discussion of chondrite classification and petrology, I refer the reader to the recent review articles  by Hewins et al.~(2005), Jones et al.~(2005), and Scott \& Krot (2005).

Chondrules are separated in their parent bodies by a material called the {\it matrix}.  This material consists of fine-grained minerals and amorphous material, and may be related to, if not the same as, the material in fine-grained rims surrounding some chondrules.  Generally, chondrules and the matrix show deviations from solar abundance, but together are closer to solar values than any component alone, with variation in fractionation between chondrule types (Cuzzi et al.~2005; Huss et al.~2005).  This abundance correspondence is referred to as {\it chondrule-matrix complementarity}.  

Chondrule collisional histories, isotopic fractionation data (see below), chondrule-matrix complementarity, fine-grained rim accumulation (Morfill et al.~1998), and petrological and parent body location arguments  (e.g., Wood 2005) suggest that chondrules formed in the Solar Nebula, as suggested by Wood (1963), in strong, localized, repeatable heating events.  For the discussions in this dissertation, I will focus on the shock wave model for chondrule formation, because it is the most well-developed and viable formation hypothesis (Iida et al.~2001; Desch \& Connolly 2002; Cielsa \& Hood 2002; Boss \& Durisen 2005a,b; Hartmann 2005; Miura \& Nakamoto 2006).  However, other formation mechanisms may play a role in processing some solids, e.g., the X-Wind model (Shu et al.~2001) and lightning (Whipple 1966; Pilipp et al.~1998; Desch \& Cuzzi 2000).  It should also be noted that the shock wave model does not necessarily identify the shock-driving mechanism, and identifying this mechanism is an ongoing topic for investigation.  

Shock waves can form chondrules through gas-drag heating.  The stopping time for a particle with radius $a<$ 1 cm 
\begin{equation}t_s=\frac{\rho_a a}{\rho c_s},\end{equation}  
where $\rho_a$ is the particle density and $\rho$ the gas density (Cuzzi et al.~2001).  As the stopping time becomes very short, it is possible to melt chondrule precursors for a range of pre-shock conditions through a combination of friction and  interaction with the hot, post-shock gas (Desch \& Connolly 2002; Ciesla \& Hood 2002). The melts' interaction with each other and the gas prevents the newly-forming chondrules from cooling too quickly.  

A critical chondrule-formation constraint is that chondrules are 
depleted in some elements, e.g., K, Fe, Si, Mg,
presumably from devolatization during the precusor melting, but the corresponding isotopic
fractionation of species in chondrules, e.g, sulfur in troilite, is mostly
absent (Tachibana \& Huss 2005). There are two favored explanations for 
this problem: (1) Chondrule heating is so rapid that
isotopic fractionation is suppressed.  (2) Chondrules were embedded within the
evaporated gas of other chondrules, and came to equilibrium with that gas. 
Miura \& Nakamoto (2006) demonstrate through 1D radiation-shock calculations that the
optical depth as measured between the shock front to about $10^{10}$ cm
upstream cannot exceed $\tau\sim$ 1-10, depending on the density and the
pre-shock velocity. They found that unless this criterion is met, chondrule
precursors spend too much time above $T\sim 1400$ K to avoid isotopic
fractionation of sulfur. However, laboratory experiments indicate that the
rapid heating and cooling needed to avoid isotopic fractionation is incompatible with observed chondrule textures (Hewins et al.~2005), and
so the criterion may not be valid. In contrast, Cuzzi \& Alexander (2006) argue that chondrules can equilibrate with
their surroundings if the chondrule-forming region is larger than a few thousand
kilometers because the evaporated gas cannot diffuse in the time it takes for the
chondrule to be processed. 

Another point to consider is that chondrule formation may not be limited to asteroid belt distances.  {\it Stardust} results (e.g, McKeegan et al.~2006) as well as comet observations (Wooden et al.~2005) indicate that high-temperature thermally processed solids are present in planetesimals from the outer Solar System (comets), and so either large-scale radial transport occurred in the Solar Nebula or some high-temperature processing occurred in the outer disk.  Understanding how chondrules form and how processed solids are mixed in the Solar Nebula is critical to describing its evolution, and ultimately, planet formation.

One plausible shock-driving mechanism is a global spiral wave (Wood 1996).   Harker \& Desch (2002) suggest that spiral waves could also explain thermal processing at distances as large as 10 AU, and Boss \& Durisen (2005a,b) suggest that GIs may be able to produce the required shock strengths to form chondrules.  In addition, Boley et al.~(2005) suggest that spiral waves could be a source of turbulence as well as a shock mechanism.  Global spiral shocks are appealing because they fit many of the constraints above.  They may be repeatable, depending on the formation mechanism for the spiral waves; they are global but produce fairly local heating; they can form chondrules in the disk; and they can work in the inner disk as well as the outer disk.

\renewcommand{\baselinestretch}{1}
\section{Disk Fragmentation and the Planet Formation Debate}
\renewcommand{\baselinestretch}{2}

Knowing under what conditions protoplanetary disks can fragment is crucial to understanding disk evolution inasmuch as a fragmented disk may produce gravitationally bound clumps. This is the {\it disk instability}  hypothesis  for the formation of gas giant planets (Kuiper 1951; Cameron 1978; Boss 1997, 1998).   The strength of GIs is regulated by the cooling rate in disks (Tomley et al.~1991, 1994; Pickett et al.~1998, 2000a, 2003), and if the cooling rate is high enough in a low-$Q$ disk, a disk can fragment (Gammie 2001).  Consider the two-dimensional adiabatic index $\Gamma$, such that  $\int p dz = P = K\Sigma^{\Gamma}$, where $K$ is the polytropic constant. Gammie quantified that for $\Gamma=2$, a disk will fragment when $t_{\rm cool}\Omega\lesssim 3$. Here,  $t_{\rm cool}$ is the local cooling time and $\Omega$ is the angular speed of the gas.  This criterion was approximately confirmed in 3D disk simulations by Rice et al.~(2003) and Mej\'ia et al.~(2005).  Rice et al.~(2005) showed through 3D disk simulations that this fragmentation criterion depends on the adiabatic index and, for $\gamma=5/3$ or 7/5, the fragmentation limit occurs when $t_{\rm cool}\Omega\lesssim6$ or 12, respectively.  These results show that a change by a factor of about 1.2 in $\gamma$ has a factor of two effect on the critical cooling time.  In addition, these results indicate that the cooling time must be roughly equal to the dynamical time of the gas for the disk to be unstable against fragmentation when $\gamma=5/3$.  Do these prodigious cooling rates occur in disks when realistic opacities are used with self-consistent radiation physics?

Nelson et al.~(2000) used 2D SPH simulations with radiation physics to study protoplanetary disk evolution.  Because their simulations were evolved in 2D, they assumed that the disk at any given moment was in vertical hydrostatic equilibrium.  Using a polytropic vertical density structure and Pollack et al.~(1994) opacities, they cooled each particle according to an appropriate effective temperature.  In their simulations, the cooling rates are too low for fragmentation.   In contrast, Boss (2001, 2005) employed radiative diffusion in his 3D grid-based code, and fragmentation occurs in his simulated disks.  Besides the difference in dimensionality of the simulations, Boss assumed a fixed temperature structure for Rosseland mean optical depths less than 10, as measured along the radial coordinate. Boss (2002) found that the fragmentation in his disks is insensitive to the metallicity of the gas and attributed this independence to fast cooling by convection (Boss 2004a).  However, it must be noted that Nelson et al.~(2000) assumed a vertically polytropic density structure.  Because the entropy $S\sim \ln K$, where $p=K\rho^{\gamma}$ is the polytropic equation of state, the Nelson et al.~approximation assumes efficient convection.  One of the principal objectives in this dissertation is to understand and resolve these discrepant results.

\section{About This Dissertation}

In this dissertation, I explore the three-dimensional behavior of spiral waves in gravitationally unstable disks, and discuss their effects on vertical disk structure, mass transport, the FU Orionis phenomenon, and chondrule formation.  In addition, I investigate under what conditions, if any, unstable disks fragment, whether sudden vertical motions in these unstable disks can be better explained by shock bores than by convection, whether the details of the treatment for radiation transport are important to disk evolution, and whether dust settling affects the strength of spiral shocks.  The culmination of this work is an effort to understand to what degree chondrule formation, planet formation, the FU Ori phenomenon, dead zones, and GI activity are interdependent. 

In Chapter 2, I discuss numerical methods that are employed to assess shock strengths in these disks, to approximate radiative cooling, and to account for the rotational and vibrational states of molecular hydrogen.  I also outline the hydrodynamics codes used for these studies.  In order to test the radiation physics described in Chapter 2, several radiation hydrodynamics tests along with their results for the radiation algorithms implemented here are discussed in Chapter 3.  The initial models for these studies are detailed in Chapter 4, and in Chapter 5, I outline shock bore theory.  The role of convection, the locality of mass transport, the importance of radiation boundary conditions, and radiative cooling rates
in GI-active disks are examined in Chapter 6.  In Chapter 7, I explore the connection between massive, bursting dead zones, the FU Ori phenomenon, and chondrule formation.   Finally in Chapter 8, I summarize the principal conclusions for the work presented in this dissertation.

 \chapter{NUMERICAL METHODS}

There are two versions of the IU hydrodynamics code that have been used for the
studies described in this dissertation. I outline both versions in this Chapter, and
layout algorithms and analysis routines that I have developed for these studies.

\section{The Standard IU Code}

The standard version of the IU code (SV) is an explicit, Eulerian code that
solves the equations of hydrodynamics in conservative form on an evenly spaced,
cylindrical grid.  The code is second-order accurate in space and in time, and
it solves for self-gravity after the first source half-step (see below).  The SV
is based on the Tohline (1980) code.
Yang (1992) made the code second order in space by including a van Leer
advection scheme (van Albada et al.~1982), made the code second order in time,
and modified the mesh staggering (see below) to its current form.
Pickett (1995) included an energy equation, artificial
viscosity (AV),  and {\it ad hoc} cooling prescriptions. Mej\'ia (2004) added
radiation physics, and Cai (2006) made several improvements to the Mej\'ia radiation
algorithm.  

The equations of hydrodynamics with self-gravity are
\begin{eqnarray}\frac{\partial\rho}{\partial t} + \nabla\cdot \rho{\bf v} &=& 0{\rm,}\\
\frac{\partial\rho {\bf v}}{\partial t}+ \nabla\cdot\rho {\bf v v}&=&-\nabla p -\rho\nabla \Phi-\nabla\cdot
\left(\rho {\bf Q}\right){\rm,}\\
\frac{\partial \epsilon}{\partial t} + \nabla\cdot {\bf v}\epsilon &=& 
- p \nabla\cdot {\bf v} +\Gamma +\Lambda\rm,~{\rm and}\\
\nabla^2\Phi&=&4\pi G\rho\rm,\end{eqnarray}
i.e., the mass continuity equation, the equation of motion, the energy equation,
and Poisson's equation,
respectively.  Here, $\rho$ is the mass density, ${\bf v}$ is the velocity of
the fluid, $p$ is the gas pressure, $\Phi$ is the total gravitational potential,
and $\epsilon=\rho e$, where $e$ is the specific internal energy of the gas.
The last term in the
equation of motion is the artificial viscosity term. The tensor $Q_{ij}=C_Q(\Delta v_i)^2$ 
for $j=i$ and $\Delta v_i < 0$, but 0 otherwise, as in the von Neumann \& Richtmeyer 
scheme (Norman \& Winkler 1986).
The constant $C_Q$ is an
experimentally determined parameter that spreads shocks over several cells to include
 the appropriate amount of entropy generation and to stabilize shocks against ringing
  (see Pickett 1995).  The 
AV contribution to the energy equation is given by $\Gamma = \rho
\left(Q_{rr}\partial_r v_r + Q_{\phi\phi}r^{-1}\partial_{\phi}v_{\phi}+Q_{zz}\partial_z v_z\right)$,
 and radiation cooling and heating
are described by the $\Lambda$ term (see \S 2.4).   If an equation of state is chosen such
that $p=f\left(\rho\right)$, e.g., barotropic gases, only the mass continuity
equation and the equation of motion are required to describe the flow. However,
if $p=f\left(\rho,e\right)$, the energy equation is also required.  The SV
version of the code employs an ideal gas equation of state with a constant ratio
of specific heats $\gamma$ such that 
\begin{equation}p = \left(\gamma-1\right)\epsilon\rm .\end{equation}
The simple form for $p$ in equation (2.5) allows the energy equation to be recast
in conservative form (Williams 1988), namely
\begin{equation}\frac{\partial\epsilon^{1/\gamma}}{\partial t} + \nabla\cdot{\bf
v}\epsilon^{1/\gamma} =
\frac{\epsilon^{1/\gamma-1}}{\gamma}\left(\Gamma+\Lambda\right)\rm.\end{equation}
Pickett (1995) implemented equation (2.6) in the SV and found that it produced
slightly better results than using equation (2.3).  

In addition to the mass density and the internal energy density of the fluid,
the SV uses the radial momentum density $S$, the vertical momentum density $T$,
and the angular momentum density $A$ to evolve the flow. The array values for
$\rho$, $\epsilon$, and $A$ are set at cell centers, while the $S$ and $T$ array
values are set on the center of cell faces (see Fig.~2.1).  Yang (1992) found
that this {\it staggered grid} formalism is more stable than a complete
cell-centered formalism. 

Because a cylindrical grid is used, equations (2.1-2.4) are solved in cylindrical
coordinates, and so equation (2.2), in particular, needs to be rewritten in a more 
useful form.  Let $\rho{\bf v}=S\hat{e}_r+A/r\hat{e}_{\phi}+T\hat{e}_z$.  The divergence
term in equation (2.2) can be written as
\begin{equation}\nabla\cdot\rho{\bf vv}=\rho{\bf v}\nabla\cdot{\bf v}+{\bf v}\cdot \nabla\rho{\bf v},\end{equation}
for position vector ${\bf r}=r\hat{e}_r +z \hat{e}_z$. The radial component of equation (2.7) is
\begin{eqnarray}\left(\nabla\cdot\rho{\bf vv}\right)\cdot\hat{e}_r&=&r^{-1} \partial_r \left( r S v_r\right)  + 
r^{-1}\partial_{\phi}\left(Sv_{\phi}\right)+\partial_z\left(Sv_z\right)-r^{-2}v_{\phi}A\nonumber\\
& =&\nabla\cdot S{\bf v}-\frac{A^2}{\rho r^3}.
\end{eqnarray}
Likewise, the azimuthal component can be written
\begin{eqnarray}r\left(\nabla\cdot\rho{\bf vv}\right)\cdot\hat{e}_{\phi}&=& r^{-1}\partial_r\left(r A v_r\right) +
r^{-1}\partial_{\phi}\left(Av_{\phi}\right)+\partial_z\left(Av_z\right)\nonumber \\
& = & \nabla\cdot A{\bf v}.
\end{eqnarray}
The vertical component is straightforward, and the equation of motion can now be written as
\begin{eqnarray}
\frac{\partial S}{\partial t} +\nabla \cdot S{\bf v} &=& -\rho\frac{\partial p}{\partial r} - \rho \frac{\partial \Phi}{\partial r} 
- \frac{\partial r  \rho Q_{rr}}{r\partial r} + \frac{A^2}{\rho r^3},\\
\frac{\partial A}{\partial t} +\nabla \cdot A{\bf v}& = &-\rho\frac{\partial p}{\partial \phi} -\rho \frac{\partial \Phi}{\partial \phi}
- \frac{\partial  \rho Q_{\phi\phi}}{\partial \phi},{\rm~and}\\
\frac{\partial T}{\partial t} +\nabla \cdot T{\bf v} &=& -\rho\frac{\partial p}{\partial z} - \rho \frac{\partial \Phi}{\partial z}
- \frac{\partial  \rho Q_{zz}}{\partial z}.
\end{eqnarray}

\clearpage
\begin{figure}[h]
\begin{center}
\includegraphics[width=8cm]{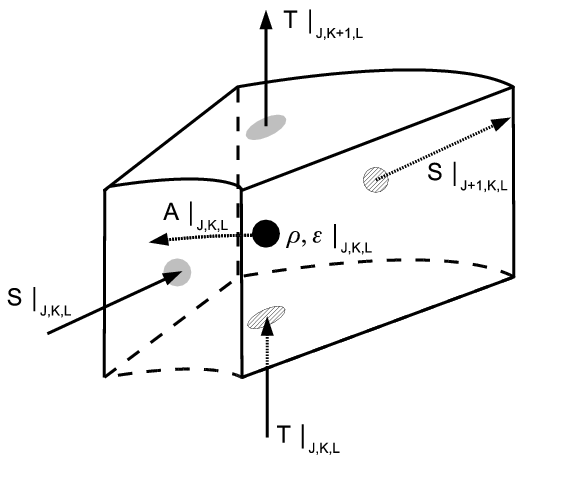}
\caption[Location of values for the five principal hydrodynamics arrays]
{Location of values for the five principal hydrodynamics arrays.  The indexing for the cylindrical grid $(r, z, \phi)$ = (J,K,L). }
\label{2.1}
\end{center}
\end{figure}
\clearpage

The SV uses an operator splitting method to evolve the hydrodynamics.  
The two operations are called {\it sourcing}, $\mathcal{S}$, which advances the right hand side
in equations (2.10-2.12), and {\it fluxing}, $\mathcal{F}$, which advances
the advection terms in equations  (2.1), (2.3), and (2.10-2.12).  The
solution for a hydrodynamic variable $X$ for any one time step is $\partial
X/\partial t = \mathcal{S}+\mathcal{F}$.  As described in Yang (1992), the
fluxing is determined through a van Leer advection scheme (van Albada et
al.~1982), with the fluxed quantities being $S$, $T$, $A$, $\rho$, and
$\epsilon^{1/\gamma}$. 
For sourcing the arrays, three separate calculations are
made.  First, the forces due to pressure gradients and potential gradients are
determined by center-differencing schemes and used to update the momentum
densities.  Second, the effects of AV on the momentum densities are included.
Finally,
heating due to AV and cooling due to either {\it ad hoc} prescriptions or
radiation transport are used to update the internal energy densities.   It is
possible to run the simulation with cold AV (Pickett \& Durisen 2007), i.e., AV
only affects the equation of motion.  However, this is only done under special
circumstances.  To illustrate how the fluxing and sourcing is used to achieve
second-order time integration, I reproduce the flow chart from Mej\'ia (2004,
Fig.~2.4) in Figure 2.2.  Note that the pressure is updated before each sourcing
and the potential is calculated only once before the second sourcing. 

\clearpage
\begin{figure}[h]
\begin{center}
\includegraphics[width=15cm]{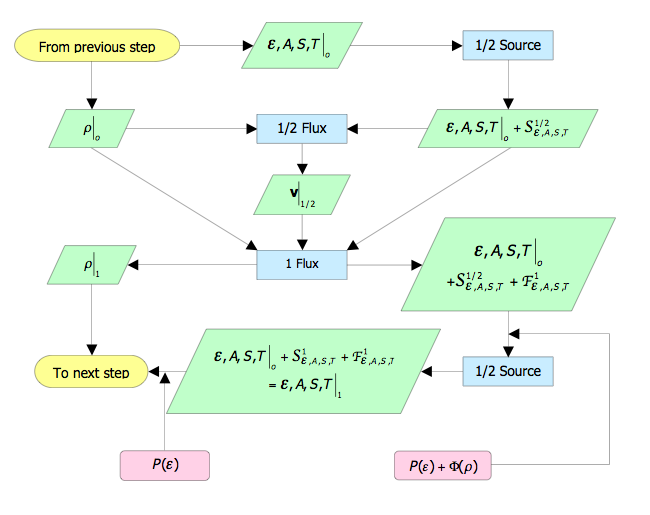}
\caption[Hydrodynamics code flow chart]
{A flow chart illustrating the second-order time integration scheme.  Reproduced from Mej\'ia (2004), Figure 2.4. }
\label{2.2}
\end{center}
\end{figure}
\clearpage


The total potential $\Phi=\Phi_b+\Phi_{\rho}$, where $\Phi_{b}$ is some
background potential and $\Phi_{\rho}$ is the potential due to the mass on the
grid.  The background potential can be set to zero, as is done for protostar+disk 
models (e.g., Yang 1992; Pickett 1995; Pickett et al.~1996, 1998). For the
models presented here and for a multitude of previous studies (Pickett et
al.~2001, 2003; Mej\'ia 2004; Mej\'ia et al.~2005; Cai et al.~2006, 2007;
Pickett \& Durisen 2007), only the disk is evolved, and so the background
potential is a point mass or fixed potential due to the star that is otherwise excluded from the
simulations.  This background potential is held fixed at the center of the grid.  Although
keeping the star rigid may create some spurious one-arm behavior in the
simulations, simply moving the star to locations that stringently keep the
center of mass at the center of the grid (see Boss 1998) may be problematic as
well.  Routines are being developed to explicitly model the motion of the star, but have not been employed for the studies discussed here. 

The potential due to the mass is found in two parts.  First, the potential on
the grid {\it boundary} is calculated through a spherical harmonics expansion
with $l=|m|=10$.  The boundary solution then provides the Dirichlet boundary
conditions for a direct Poisson solver (Tohline 1980).  The Poisson solver uses
a discrete Fourier transform to decompose the 3D potential problem into LMAX 2D
problems, for grid dimensions $(r, z, \phi)$ = (JMAX, KMAX, LMAX).  The
series of 2D problems can then be solved in parallel with a cyclic reduction
method (Tohline 1980; Pickett et al.~1996).  Pickett et al.~(2003) demonstrated
that when a constant density blob was loaded onto the grid, the $l=|m|=10$
expansion is sufficient to describe the boundary potential and that the
solution's accuracy is determined by grid resolution, not the boundary potential expansion.  Moreover, I have used the Cohl \&
Tohline (1999) Bessel expansion to test the accuracy of the SV
boundary potential solver for the Wengen test 4, which is a set of disk simulations by a wide variety
of codes for the same initial conditions (see Mayer et al.~2007, in preparation).  I find that the
outcome is unchanged when the Legendre half-integer polynomials are carried
out to $m=30$, where the Bessel expansion showed convergence to 1 part in 10 million when $m$ was
increased to 40.  The boundary potential solution reaches the required
accuracy with $l=|m|=10$.  I note that the Wengen test 4 is challenging for a potential solver because the disk 
is very flat, asymmetric, and develops dense clumps.  The potential solver is sufficient for these tests.

\section{CHYMERA}

The hydrodynamics routines in CHYMERA, Computational HYdrodynamics with MultiplE Radiation Algorithms, are
different from those in the SV in one significant way: the rotational and vibrational
states of molecular hydrogen are explicitly calculated (see \S 2.3), and so a
$\gamma=\rm constant$ approximation is no longer valid.  This, in turn, means that
the energy equation cannot be recast into its conservative form (equation [2.3]),
and so $\epsilon$ must be fluxed and $-p\nabla\cdot{\bf v}$ must be
explicitly calculated during sourcing.  For the sourcing of the energy equation,
the work term is calculated by
\begin{eqnarray}-p\nabla\cdot{\bf v} &=&
 -\tilde{p}\left(\frac{r_{J+1} v_{r}(J+1) - r_J
v_{r}(J)}{r_{J+1/2}\Delta r}\right.\nonumber \\
&+&\frac{v_{\phi}(L+1)-v_{\phi}(L)}{r_{J+1/2} \Delta\phi} \\
 &+&\left.\frac{v_z(K+1)-v_z(K)}{\Delta z}\right),\nonumber\end{eqnarray}
where only the relevant indices are shown, all velocities are face-centered, and
$\tilde{p}$ is the average of the current half step's pressure and a provisional
pressure (Black \& Bodenheimer 1975), i.e., the pressure if the internal energy
were updated according to $\tilde{p}=p$.  Without the inclusion of the
provisional pressure, CHYMERA cannot match the Sod shock tube (Sod 1978; Hawley
et al.~1984) results of the SV when the shock is forced to propagate in the
vertical direction.  However, when the provisional pressure is included, the
solutions agree to within a few hundredths of a percent, and energy loss is kept
to about a tenth of a percent (Figure 2.3; see Pickett 1995, Appendix B for test
details).  

%
%

In addition to changing the energy equation, the equation of state must also be
altered.  Instead of $p=(\gamma-1)\epsilon$, $p=k\rho T/\mu m_p$, where $T$ is
interpolated from a table based on a given cell's $e$ and where $\mu$ is the mean
molecular weight specified for calculating the $e-T$ table.  

\clearpage
\begin{figure}
\begin{center}
\includegraphics[width=7cm]{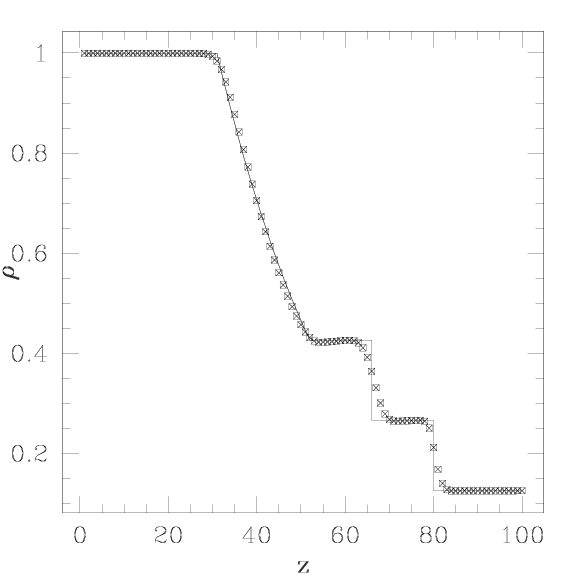}\includegraphics[width=7cm]{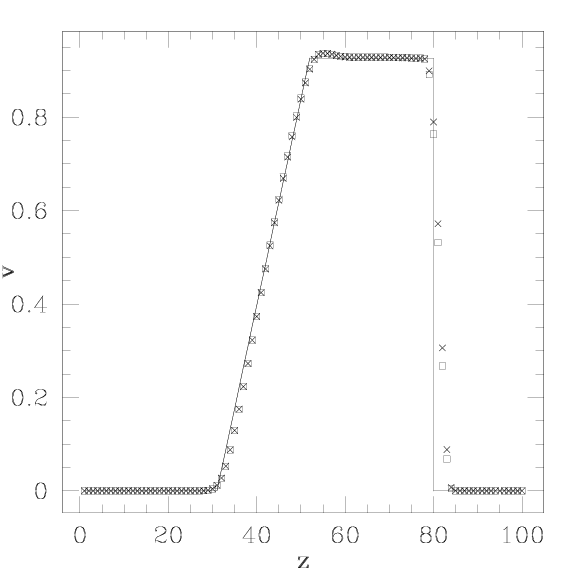}
\includegraphics[width=7cm]{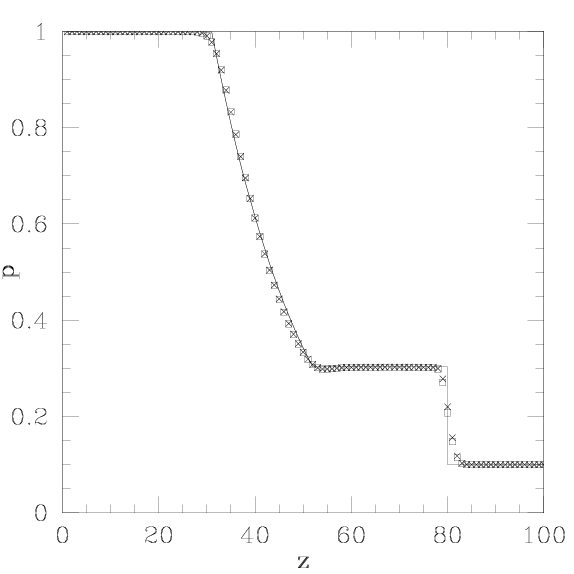}\includegraphics[width=7cm]{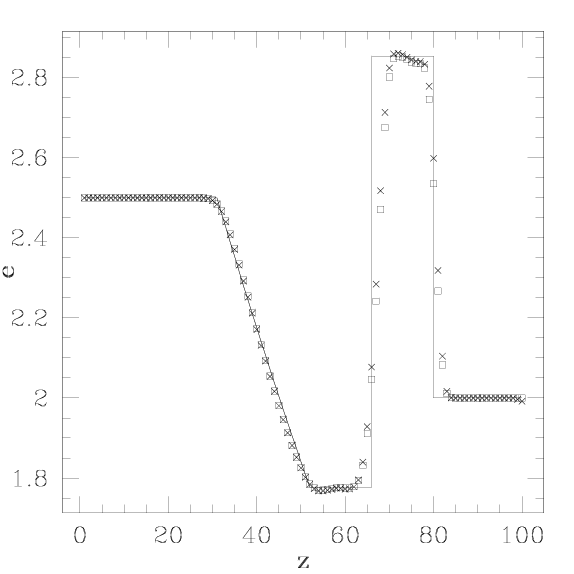}
\caption[Sod shock tube]{Results for the Sod shock
tube test for the SV (squares) and CHYMERA (Xs).  The setup is two uniform regions of $\gamma=7/5$ gas in contact with each other and initially at rest.  One region is a $\rho=1$ g cm$^{-3}$, $p=1$ dyne cm$^{-2}$ gas, and the other a $\rho=0.125$ g cm$^{-3}$, $p=0.1$ dyne cm$^{-2}$ gas. The plots for density, velocity, pressure, and specific energy, in cgs units, are compared with analytic values (solid curves).  The results are, for the most part,
indistinguishable when the shock propagates along the vertical direciton. }
\label{2.3}
\end{center}
\end{figure}
\clearpage

\section{H$_2$ Thermodynamics}

As discussed in the Introduction, the disparities between disk evolutions
reported by various hydrodynamics groups are likely due to a combination between
different treatments of radiation transfer and the rotational states of
molecular hydrogen.  Because the thermodynamics can strongly change
the outcome of a simulation (Pickett  et al.~1998, 2000), I have implemented in
CHYMERA an internal energy that takes into account the translational,
rotational, and vibrational states of H$_2$.   During its development, Boley et
al.~(2007a) noticed that in all treatments to date of planet formation by disk
instability, the effects of the rotational states of H$_2$ have been, at best,
only poorly approximated, and they drew attention to possible consequences of
various approximations for the internal energy of H$_2$ that are in the
literature.  I outline the algorithm employed in CHYMERA, and recapitulate
several of the cautions presented by Boley et al.~(2007a).

For this discussion, I refer the reader to Pathria (1996).  Consider
the following thermodynamic properties of an ideal gas: Let $E$ be the internal
energy for $N$ particles, $e$ the specific internal energy, $\epsilon$ the
internal energy density, $p$ the pressure, $T$ the gas temperature, $\rho$ the
gas density, $\mu$ the mean molecular weight in proton masses, $c_v$ the
specific heat capacity at constant volume, $Z$ the partition function for the
ensemble, $z$ the partition function for a single particle, and $R = k/m_p$,
where $k$ is Boltzmann's constant and $m_p$ is the proton mass.  I only consider
independent contributions to the partition function from translation, rotation,
and vibration represented by $ Z = Z_{\rm tran}Z_{\rm rot}Z_{\rm vib}= z^N=
\left( z_{\rm tran} z_{\rm rot} z_{\rm vib}\right)^N$.  The internal energy $E$
and the specific internal energy $e$ can be calculated by
\begin{equation}E=NkT^2\frac{\partial \ln z}{\partial
T};~e=\frac{R}{\mu}T^2\frac{\partial \ln z}{\partial T}\label{eq1}\end{equation}
for constant $\rho$. Because the gas is ideal, $c_v = d e/d T$.

\subsection{Molecular Hydrogen}

Molecular hydrogen exists as parahydrogen and as orthohydrogen where the proton
spins are antiparallel and parallel, respectively.  The partition function for
parahydrogen is
\begin{equation}z_p=\sum_{j_{\rm even}}\left(2j + 1\right) \exp\left(
-j\left(j+1\right)\theta_{\rm rot}/T\right),\end{equation}
and the partition function for orthohydrogen is
\begin{equation}z_o=\sum_{j_{\rm odd}}3\left(2j + 1\right) \exp\left(
-j\left(j+1\right)\theta_{\rm rot}/T\right),\end{equation}
where $\theta_{\rm rot}=85.4$ K  (Black \& Bodenheimer 1975).  When the two
species are in equilibrium, $z_{\rm rot} = z_p + z_o$.   However, the ortho/para
ratio (b:a) could also be frozen if no efficient mechanism for converting
between the species is available. This leads to $z_{\rm rot} =
z_p^{\left(a/\left(a+b\right)\right)} z_o'^{\left(b/\left( a +
b\right)\right)}$, where $z_o' =z_o \exp\left( 2\theta_{\rm rot}/T \right)$.
The additional exponential is required in the orthohydrogen partition function
when the ortho and para species are at some fixed ratio to ensure that rotation
only contributes to the internal energy once the rotational states are excited,
i.e., $z_o'\rightarrow \rm constant$ as $T \rightarrow 0$.

To consider the vibrational states, I approximate the molecule as an infinitely
deep harmonic oscillator, where
\begin{equation} z_{\rm vib} = \frac{1}{1-\exp\left(-\theta_{\rm
vib}/T\right)};\end{equation}
$\theta_{\rm vib} = 5987$ K  (Draine et al.~1983).  At this time,
CHYMERA is only designed to investigate temperatures $T\lesssim1500$ K where
dissociation of H$_2$ is insignificant. For this reason, I choose to ignore differences between
equation (2.17) and a proper $z_{\rm vib}$, which would take into account the
anharmonicity of the molecule and that the molecule has a finite number of
vibrationally excited states. 

I can use equation (2.14) to write the specific internal energy for H$_2$
\begin{equation} e\left({\rm H}_2\right) = \frac{R}{2}\left( \frac{3}{2}T +
\frac{T^2}{z_{\rm rot}}\frac{\partial z_{\rm rot}}{\partial T} + \theta_{\rm
vib}\frac{\exp\left( -\theta_{\rm vib}/T\right)}{1-\exp\left( -\theta_{\rm
vib}/T\right)}\right).  \end{equation}
The specific internal energy profiles as calculated by equation (2.18) are shown in
Figure 2.4 (left panel) for an equilibrium mixture (solid black), pure
parahydrogen (solid red), and a 3:1 ortho/para mixture (solid blue). The offset of the
correct 3:1 mix profile is due to the energy stored in the parallel spins of the
protons.  When the gas is ideal and dissociation and ionization can be ignored,
the first adiabatic exponent $\Gamma_1 = 1+ R/\mu c_v = c_p/c_v = \gamma$ (Cox
\& Giuli 1968).  Solid curves in Figure 2.4 (right panel) indicate the $\Gamma_1$ profiles for
the corresponding solid curves in the left panel, which is consistent with Figure 2 of Decampli et al.~(1978), as it should be because my derivation of $c_v$ is
equivalent to theirs.  All dashed curves and the curves in the center panel result
from various approximations for $e$, as discussed below.

\subsection{Approximations for $e$}

There are several approximations for $e$ that are employed in the literature,
and several of these approximations can have strong consequences for disk
dynamics, as argued by Boley et al.~(2007a). If $c_v$ is constant, then $e = c_v
T$.  Boley et al.~(2007a) pointed out that Black \& Bodenheimer (1975) calculate
$c_v$ from the Helmholtz free energy, which is valid, but then assume $e=c_v T$,
which is invalid, because $c_v$ is dependent on $T$.  Other authors have
followed suit (e.g., Whitehouse \& Bate 2006; Stamatellos et al.~2007), and this
assumption could be troublesome for gas dynamics in a hydrodynamics simulation.
Curves for  $c_v T$ are also shown in Figure 2.4 (left panel) by dashed lines.

\begin{figure}
\begin{center}
\includegraphics[width=7.cm]{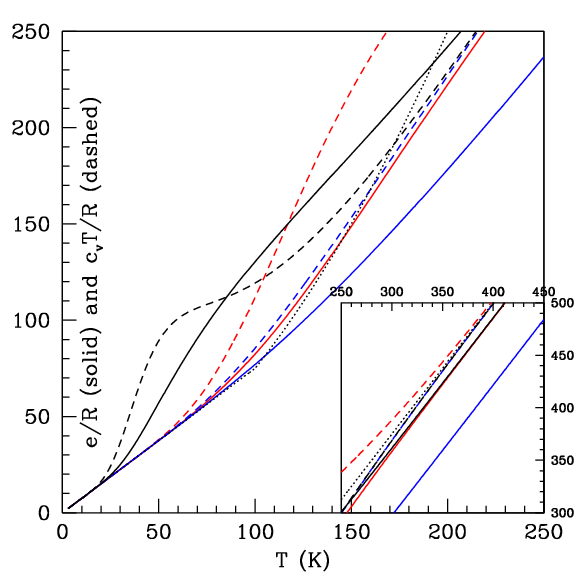}
\includegraphics[width=7.cm]{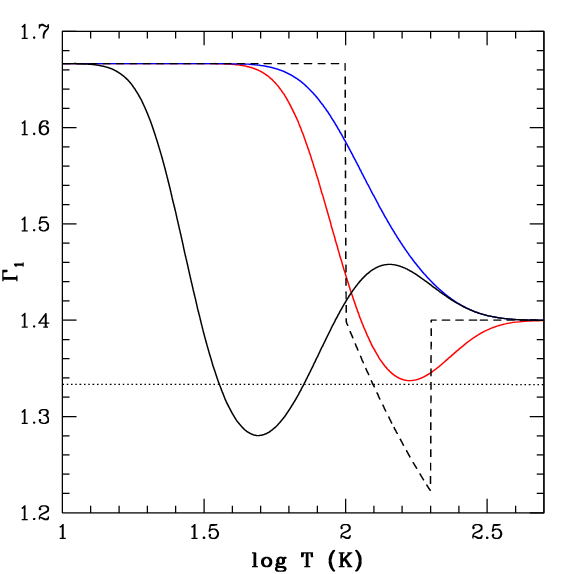}
\includegraphics[width=7.cm]{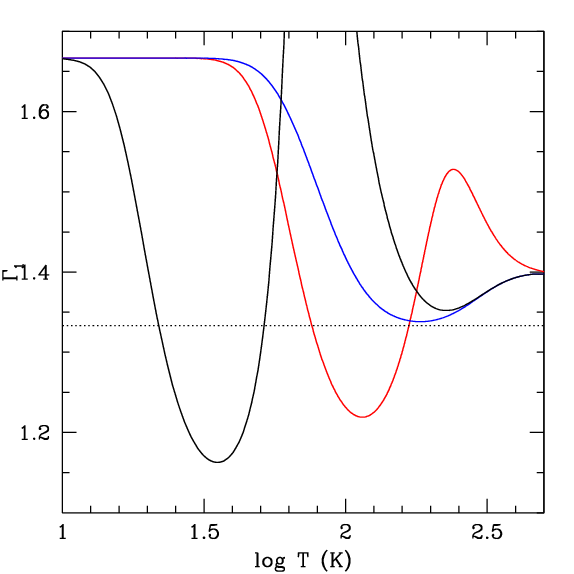}
\caption[Internal energy and $\Gamma_1$ curves for H$_2$]
{Left panel: The solid curves show the proper specific internal energy
for an equilibrium mixture (black), a pure parahydrogen gas (red), and a
constant 3:1 ortho/parahydrogen mixture (blue).  The dashed curves indicate
$e=c_v T$, which can deviate strongly from the correct curves.  Right panel:
$\Gamma_1$ curves for the corresponding solid curves in the left panel.  The
dashed curve demonstrates the consequences for $\Gamma_1$ when a seemingly
innocent interpolation is used for $e$ (see Boley et al.~2007b).  Center panel:
Corresponding $\Gamma_1$ curves for the dashed curves in the left panel.  The
deviations from the correct $e$ have strong consequences for the $\Gamma_1$
profiles.   } \label{2.4}
\end{center}
\end{figure}

The approximation $e=c_v T$ gives quite different behavior from the correct $e$,
e.g., the incorrect 3:1 curve most closely follows the correct pure parahydrogen
curve.  The dynamical effects that could result from assuming $e=c_v T$ are
evaluated by taking the temperature derivative of the dashed curves in
Figure 2.4 (left panel).   The resulting $\Gamma_1$ profiles are shown in the
center panel of Figure 2.4, and these curves are very different from the
$\Gamma_1$ profiles that they are meant to approximate (right panel).  As
mentioned above, $c_v$ can be calculated from the Helmholtz free energy as is
done by Black \& Bodenheimer (1975).  This makes it possible to compute
$\Gamma_1$ from $c_v$ correctly but then evolve the gas with an erroneous
effective specific heat because many hydrodynamics codes evolve $\epsilon=\rho
e$ (e.g., Black \& Bodenheimer 1975; Boss 1984a, 2001; Monaghan 1992; Stone \&
Norman 1992; Pickett 1995; Wadsley et al.~2004).  If the ideal gas law $p=\rho R
T/\mu$ is assumed as well, then effective $\Gamma_1$ profiles like those shown
in Figure 2.4 (center panel) seem to be unavoidable when $e=c_v T$ is assumed for a temperature
dependent $c_v$.  Because fragmentation becomes more likely as $\Gamma_1$
becomes smaller (Rice et al.~2005), the $e=c_v T$
assumption should artificially make fragmentation more likely in some
temperature regimes and less likely in others.   Boley et al.~do note, however,
that the severity of this error may depend on the state variables evolved
in a given code, and only the authors who employ $e=c_v T$ will be able to say
in detail how it affects their simulations.   

Boley et al.~(2007a,b) also caution against simple approximations to $e$
because they too may have strong consequences for $\Gamma_1$. For example,
Boss (2007) uses a quadratic interpolation for $e$
between 100 K and 200 K.\footnote{Boley et al.~(2007a) stated that Boss uses a discontinuous
internal energy for H$_2$, where H$_2$ contributes $3kT/4 m_p$ to the specific
internal energy $e$ for $T\le100$ K and $5kT/4 m_p$ for  $T>100$ K, based on his
own citations to Boss (1984a). Through private communication, Boss indicated that
he now uses a quadratic interpolation for $e$. This change is not mentioned in
the literature, but according to Boss, the interpolation has been used in all
his simulations since Boss (1989). Boley et al.~(2007b) clarified this
discrepancy.}  In Figure 2.4 (right panel, dashed curve), I illustrate the
consequences for $\Gamma_1$ when using this $e$ approximation by calculating
the specific heat at constant volume $c_v = de/dT$ and by relating that to
$\Gamma_1$ (see Cox \& Giuli 1968).   Because fragmentation becomes more likely for lower $\Gamma_1$s (Tomley et al.~1991; Boss 1997,
2000; Pickett 1998; Rice et al.~2005), an approximation for $e$ like that used by Boss between 100 and 200 K is likely to make the disk susceptible to fragmentation for that temperature
range. 
Finally, a constant $\Gamma_1$ approximation (Pickett et al.~2003; Rice et
al.~2003; Lodato \& Rice 2004; Mayer et al.~2004, 2007; Rice et al.~2005; Mej\'ia et
al.~2005; Cai et al.~2006; Boley et al.~2006) poorly
represents $e$ between about 60 and 300 K,  a plausible temperature regime for the formation of Jupiter; neither
$\Gamma_1=5/3$ nor 7/5 can be assumed confidently.

Preliminary simulations of a disk with solar composition indicate that when GIs
activate between 30 and 50 K for an equilibrium ortho/para mixture, the $e=c_v T$
simulation evolves more rapidly and has a more flocculent spiral structure than
the correct $e$ simulation for the same cooling rates.  In addition, denser
substructures form in some spiral arms of the $e=c_v T$ simulation throughout
the simulation, while dense substructures only form during the burst of the
correct $e$ simulation\footnote{The evolution of these simulations can be viewed
at http:/hydro.astro.indiana.edu/westworld.  Click on the link titled ``H2
ortho-para equilibrium tests'' under the ``Movies'' tab.}. When the
instabilities occur outside this temperature regime, the differences are
diminished.  

\subsection{The Ortho/Para Ratio}

As indicated in the previous section, the dynamical behavior of the gas is dependent on the ortho/para ratio and whether the species are in equilibrium for all $T$.  This ratio for various astrophysical conditions has been addressed by several authors (e.g., Osterbrock 1962; Dalgarno et al.~1973; Decampli et al.~1978; Flower \& Watt 1984; Sternberg \& Neufeld 1999; Fuente et al.~1999; Rodr\'iguez-Fern\'andez et al.~2000; Flower et al.~2006),  typically in the context of interstellar clouds or photodissociation regions.  However, for plausible Solar Nebula conditions, the ortho/para ratio has been inadequately addressed. For example, Decampli et al.~(1978) used an estimate for the H$^+$ number density that was derived originally to give the total gas phase ion number density in gas for which dissociative recombination dominates the removal of ions.  At protoplanetary disk number densities, however, ion removal should be primarily on grain surfaces if the ratio of grain surface area to hydrogen nucleon number density is the same as it is in diffuse interstellar clouds.

For protoplanetary disk conditions, the conversion between ortho and parahydrogen is principally due to  protonated ions such as H$^+_3$.
 Consequently, I will assume that all ionizations lead to H$^+_3$ formation.
Another possible conversion mechanism is through interactions between H$_2$ and grains.  However, this conversion might only be significant when the temperature drops below about 30 K (Le Bourlot 2000).

Consider the balance between $\rm H_3^+$ production by cosmic rays (CR) and H$_3^+$ depletion by dust grains:
\begin{equation} \zeta n\left({\rm H}_2\right)= n_g\pi a^2 v n_i, \end{equation}  
 where $n\left({\rm H_2}\right)$, $n_i$, and $n_g$ are the $\rm H_2$, $\rm H_3^+$, and grain number densities, respectively, $\zeta$ is the ionization rate by CRs and other energetic particles (EP), $a$ is the average radius of the grains, and $v$ is the thermal velocity of $\rm H_3^+$.  If standard interstellar extinction is assumed, then $\sigma=n\left({\rm H_2}\right)/n_g \pi a^2 \approx 10^{21}$ cm$^{-2}$, but as discussed below, this number is ambiguous.  The $\zeta$ appropriate for a protoplanetary disk is also ambiguous.  Cosmic rays and stellar EPs are important in ionizing the disk surface (Desch 2004; Dullemond et al.~2007), but because these particles are attenuated exponentially with a scale length of about 100 g cm$^{-2}$ (Umebayashi \& Nakano 1981), stellar EPs probably do not contribute to $n_i$.  Moreover, protostellar winds could lead to a significant reduction of $\zeta$ in analogy to CR modulation by 
the solar wind (Webber 1998). However, at a surface density of roughly 380 g cm$^{-2}$, EP production by $^{26}$Al decay is as important as CRs, with $\zeta\sim10^{-19}~\rm s^{-1}$ (Stepinski 1992). It is likely that $10^{-19}~{\rm s}^{-1}< \zeta<10^{-17} {\rm\ s}^{-1}$.   For the following estimate, I adopt the interstellar rate $\zeta = 10^{-17} {\rm\ s}^{-1}$ (Spitzer \& Tomasko 1968).  Using these numbers in equation (2.19) and adopting a thermal velocity of 1 km s$^{-1}$, $n_i\approx 0.1\rm\ cm^{-3}$.  By adopting a collisional rate coefficient $\alpha=1\times10^{-9}\rm\ cm^3\ s^{-1}$ for the H$_3^+$ interaction with H$_2$ (Walmsley et al.~2004), the lower limit timescale for ortho and parahydrogen to reach equilibrium is $t_e=\left( \alpha n_i\right)^{-1} = 300$ yr. 

The equilibrium timescale is short enough that the ortho/para ratio can thermalize in the lifetime of a disk, but the equilibrium timescale is longer than the dynamical timescale inside about 40 AU: ortho and parahydrogen should be treated as independent species for hydrodynamical simulations of young protoplanetary disks. 

What ortho/para ratio should a dynamicist assume for gravitationally unstable protoplanetary disk simulations? The answer is uncertain. Vertical and radial stirring induced by shock bores (Boley \& Durisen 2006, see Chapter 5), which could possibly lead to mixing of the low altitude disk interior with the high-altitude photodissociation region in the disk atmosphere (Dullemond et a.~2007), will transport gas through different temperature regimes on dynamic timescales.  This could lead to nonthermalized ortho/para ratios like those that are measured from H$_2$ rotational transition lines in some photodissociation regions (Fuente et al.~1999; Rodr\'iguez-Fern\'andez et al.~2000) and in Neptune's stratosphere (Fouchet et al.~2003).  Moreover, accretion of the outer disk will bring material with a cold history into warmer regions of the disk.  It is unclear whether the ortho/para ratio of, say, 15 K gas will be thermalized with $z_o/z_p\approx 0$ or 
whether the ortho/para ratio will be 3:1, which is the expected ratio for H$_2$ formation on cold grains (Flower et al.~2006). Unfortunately, the ortho/para ratio may be critical to the evolution of a protoplanetary disk.  As can be seen in Figure 2.4, the pure parahydrogen mix has a $\Gamma_1$ that approaches 4/3 for $T\approx160$ K. 
This could make the 160 K regime the most likely region of the disk to fragment because, as $\Gamma_1$ decreases, it becomes harder for the gas to support itself against local gravitational and hydrodynamic stresses (Rice et al.~2005).
   Hydrodynamicists need to consider ortho/para ratios between pure parahydrogen and 3:1 because of our ignorance of this ratio in protoplanetary disks. 

 The above discussion is based on the assumption that $\sigma\approx 10^{21}~\rm cm^{-2}$, which is probably reasonable for very young protoplanetary disks but may not be reasonable for disks with ages of about 1 Myr.  Grain growth and dust settling may significantly lower the value of $\sigma$ by depleting the total grain area (e.g., Sano et al.~2000).  Because models of T Tauri disks must take into account the effects of grain growth in order to match observed spectral energy distributions (D'Alessio et al.~2001, 2006; Furlan et al.~2006), there may be a period in a disk's evolution when the ortho and parahydrogen change from dynamically independent species to species in statistical equilibrium.  Such a transition may also take place at certain radii in a disk, e.g., near edges of a dead zone (Gammie 1996).  As indicated by Figure 2.4, a transition to statistical equilibrium could have significant dynamical consequences for disk evolution and may induce clump formation by GIs.

\section{Radiation Algorithms}

\subsection{The M2004 and C2006 Schemes}

In this section I briefly describe the radiation algorithms developed by Mej\'ia
(2004; hereafter M2004) and Cai (2006; hereafter C2006), which are also
described briefly in Boley et al.~(2006, 2007c).  In the M2004 scheme,
flux-limited diffusion is used in the $r$, $\phi$, and $z$ directions on the
cylindrical grid everywhere that the vertically integrated Rosseland optical
depth $\tau > 2/3$, which defines the disk's interior.  For mass at lower
optical depths, which defines the disk's atmosphere, the gas is allowed to
radiate as much as its emissivity allows, with the Planck mean opacity used
instead of the Rosseland mean opacity.  The disk interior and atmosphere are
coupled with an Eddington-like boundary condition over one cell.  This boundary
condition defines the flux leaving the interior, which can be partly absorbed by
the overlaying atmosphere.  Likewise, feedback from the atmosphere is explicitly
used when solving for the boundary flux.  However, cell-to-cell radiative
coupling is not explicitly modeled in the disk's atmosphere.  This method allows
for a self-consistent boundary condition that can evolve with the rest of the
disk.  The C2006 routine (see also Cai et al.~2006) improves the stability of
the routine, as described in Chapter 3, by extending the interior/atmosphere fit
over two cells.

A problem with the M2004 and C2006 routines (see Chapter 3) is a sudden drop in the temperature
profile where $\tau=2/3$.  The drop is due to the omission of complete
cell-to-cell coupling in the optically thin regime $(\tau < 2/3)$.  However, as
shown in Boley et al.~(2006; see Appendix B), the boundary does permit the
correct flux through the disk's interior.  Because the flux through the disk is
correct, the temperature drop is mainly a dynamic concern inasmuch as it might
seed convection (Boley et al.~2006).  In order to obtain the correct flux and
temperature profiles, a method for calculating fluxes that takes into account
the long-range effects of radiative transfer is required. 

\subsection{The BDNL Scheme}

In order to account for the long-range effects of radiative transfer, at least
in part, I have developed a radiation algorithm that couples the
flux-limited diffusion used in the M2004 and C2006 routines with vertical rays.
The method was first described by Boley et al.~(2007c), and so I refer to it as
the BDNL scheme.  Consider some column in a disk with fixed $r$ and $\phi$.
Take that column out of context, and imagine that it is part of a plane-parallel
atmosphere.  In this case, heating and cooling by radiation can easily be
described with the method of discrete ordinates (see, e.g., Chandrasekhar 1960;
Mihalas \& Weibel-Mihalas 1986).  This method uses discrete angles that best
sample the solid angle, as determined by Gaussian quadrature. In a
plane-parallel atmosphere, a single ray can provide decent accuracy if the
cosine of the angle measured downward from the vertical to the ray is $ \mu
=1/\sqrt 3$.  I use this approach to approximate radiative transfer in the
vertical direction, and include flux-limited diffusion  (Bodenheimer et
al.~1990)  in the $r$ and $\phi$ directions everywhere that $\tau \ge 1/\sqrt
3$. Naturally, this is only a crude approximation when one places the column
back into context.  However, Boley et al.~(2007c) argue that this method
represents the best implementation of radiative physics for simulating
protoplanetary disks with three-dimensional hydrodynamics thus far, because it
captures the long-range effects of radiative transfer that are excluded in pure
flux-limited diffusion routines and handles optically thick and thin regions. 
 As demonstrated by Boley et al.~(2007c),
such coupling can affect disk evolution.  In addition, the emphasis on the 
vertical direction is well-justified for these vertically thin systems, and capturing
the vertical transport well should guarantee that the algorithm calculates 
reasonable cooling rates.

Consider now some incoming intensity $I_-$ and some outgoing intensity $I_+$. In
the context of the approximation outlined above, the vertical flux at any cell
face can be evaluated by computing the outgoing and incoming rays for a given
column and by relating them to the flux with
\begin{equation} F=2\pi\mu\left(I_+-I_-\right).\label{eq1} \end{equation}
Once the vertical fluxes at cell faces are known, the vertical component of the
divergence of the flux can be computed for the cell center by differencing
fluxes at cell faces.

The outgoing ray by is computed
\begin{equation} I_+ = I_+(t_d)\exp (-\Delta t) + \int_{t_u}^{t_d} S(t')
\exp(t'-t_d)d t'\rm, \end{equation}
where $\Delta t=t_d-t_u$, $t_d$ is the optical depth at the base of the cell
measured {\it along the ray}, $t_u$ is the optical depth at the top of the cell,
and $I_+(t_d)$ is the upward intensity at the base of the cell.  Because I have
assumed that each column in the disk is part of a plane-parallel atmosphere, the
optical depth along the ray can be computed by $t=\tau/\mu$.

Similar to $I_+$, the incoming ray solution across one cell is defined as
\begin{equation} I_- = I_-(t_u)\exp (-\Delta t ) + \int_{t_d}^{t_u} S(t')
\exp(t_u-t')d t'\rm, \end{equation}
where $I_-(t_u)$ is the incoming intensity at the top of the cell.

The 0th approximation for $S(t)$ is that it is constant over the entire cell.
This approximation leads to 
\begin{eqnarray} I_+=I_+(t_d)\exp\left(-\Delta t\right)+ S_0(1-\exp(-\Delta
t))\\ I_-=I_-(t_u)\exp\left(-\Delta t\right)+ S_0(1-\exp(-\Delta t))\rm,
\end{eqnarray}
and $S_0=\sigma T_0^4/\pi$, where $T_0$ is the temperature at the cell center. 

Because the source function is in principle a function of optical depth, additional complexity
is necessary to obtain good accuracy.  Consider a source function that may be represented by the
quadratic
\begin{equation} S(t)=c + bt + at^2.  \end{equation}
To find the constants $c$, $b$, and $a$, Taylor expand the source function
about the optical depth defined at the cell center $t_0$:
\begin{equation} S(t)\approx \bigg\{ S_0-\frac{dS}{d t}\bigg
|_{t_0}t_0+\frac{d^2S}{2dt^2}\bigg |_{t_0}t_0^2 \bigg \} +  \bigg \{
\frac{dS}{dt}\bigg |_{t_0}-\frac{d^2S}{dt^2}\bigg |_{t_0}t_0 \bigg \}t +  \bigg
\{\frac{d^2S}{2dt^2}\bigg |_{t_0} \bigg \}t^2.  \end{equation}
The first term in curly brackets in equation (2.26) is $c$, the second is $b$, and
the third is $a$.  Using equation  (2.26), I can find solutions for equations (2.21)
and (2.22) across any given cell (see also Heinemann et al.~2006).
However, in order to use equation (2.26), the source function's derivatives must be
evaluated:
\begin{eqnarray} \frac{dS}{dt}\bigg |_{t_0}=S'_0&=&2\mu \sigma
T_0^3\frac{(T_{-1}-T_{+1})}{\pi\rho_0\kappa_0\Delta z}{\rm,~and}\\
\frac{dS'}{dt}\bigg |_{t_0}&=&\mu\frac{(S'_{-1}-S'_{+1})}{2\rho_0\kappa_0\Delta
z}.  \end{eqnarray}
Here, the 0 denotes the {\it center} of the cell of interest, the -1 denotes the
cell center below the cell of interest, and the +1 denotes the cell center above
the cell of interest. This difference scheme is used unless the following
conditions are met: (A) If the +1 cell's density is below the cutoff value,
i.e., the minimum density at which radiative physics is still computed, or the
-1 cell's density is below the cutoff value, the derivatives are set to zero,
which reduces the solutions for $I_+$ and $I_-$ to equations (2.23) and (2.24).  (B)
If cell 0 is the midplane cell, i.e., the first cell in the upper plane, a
five-point center derivative is used for $S'$, i.e., 
\begin{equation} \frac{dS}{dt}\bigg |_{t_0}=\mu\sigma
T_0^3\frac{8T_0-7T_{+1}-T_{+2}}{3\pi\rho_0\kappa_0\Delta z}\rm, \end{equation}
unless exception (A) is met.  The simple form of equation (2.29) is due to the
reflection symmetry about the midplane that is built into the grid, which means
that the -1 cell's values are equal to the midplane cell's values and that the
-2 cell's values are equal to the +1 cell's values.  In addition, the second
derivative of the source function at the midplane is taken to be the average of
the three-point centered difference method and a forward difference method,
i.e., equation (2.28) is used as one would normally use it to compute the second
derivative but that answer is averaged with the derivative obtained by
differencing $S'_0$ and $S'_{+1}$.  Various differencing schemes have been
tested, and this differencing scheme yields the best results for the widest
range of optical depths and cell resolution. (C) If $| \Delta t dS'/dt |_0 > | 2
dS/dt|_0$, then the second derivative is set to zero.  Condition C may appear to
be  strange, because it throws away the second derivative as soon as it becomes
as important as the first derivative.  However, this contradicts the assumption
that a second-order expansion can describe the source function. When the
second derivative is important, including only the first and second derivatives
is inaccurate and the routine may become numerically unstable.  I find that including the second
derivative in highly
optically thick disks along strong shocks can
lead to unphysically large radiative heating, which results in rapid gas
expansion and disk destruction (see Figure 2.5 in \S 2.4.3).  Condition C is not used
 for the simulation presented in Boley et al.~(2007c)
because the temperatures in the disk and the midplane optical depths were low
enough that the second derivative always served as a correction term.  

Now that  a solution for the source function integral is known, the incoming and
outgoing intensities can be computed.  The incoming ray is computed first by
summing the solutions to the source function integral as one moves down into the
disk along the ray with the previous sum serving as $I_-(t_u)$.  If desired, an
incident intensity at $t=0$, as in Cai et al.~(2006), can be added to the
solution by extincting the intensity according to the optical depth.  Because
reflection symmetry is assumed about the midplane, the incoming intensity
solution at the midplane serves as the $I_+(t_d)$ for the outgoing intensity at
the midplane.  

For the $r$ and $\phi$ directions, the flux-limited diffusion scheme described
by Mej\'ia (2004) and Boley et al.~(2007c) is employed when the following
conditions are met: (A) The vertical Rosseland mean optical depth at the center
of the cell of interest is greater than or equal to $1/\sqrt 3$.  This condition
ensures that we only compute flux-limited diffusion where photons moving
vertically have less than about a 50\% chance of escaping.  (B) The cells
neighboring the cell of interest  also have a $\tau \ge 1/\sqrt3$.  This should
ensure that the code only calculates temperature gradients between relevant
cells; the flux at this cell face is accounted for in the total energy loss
(gain) of the system.  If a neighboring cell has a $\tau < 1/\sqrt3$, then the
flux at that face is taken to be the vertical flux through the first cell that
is below $\tau < 1/\sqrt3$ in the column of interest.  These conditions are
similar to those employed by Mej\'ia (2004), Cai et al.~(2006), and Boley et al.~(2007c).
Once fluxes have been determined for all cell faces, the divergence of the flux
can be calculated with
\begin{equation}\nabla\cdot {\bf F}=\frac{\partial \left(r F_r\right)}{r\partial
r} + \frac{\partial F_{\phi}}{r \partial \phi} + \frac{\partial F_z}{\partial
z}.\end{equation}

\subsection{Limiters}

There is a considerable difficulty with combining hydrodynamics,
shock heating, and radiative heating and cooling together in an explicit scheme, namely the disparate
timescales in concert with coarse resolution and numerical derivatives.
Consider the hydrodynamics timescale $\Delta t_h$, the shock heating timescale
$\Delta t_s$, and the radiation timescale $\Delta t_r$.  The hydrodynamics timescale
 is the well-known Courant condition: $t_h \le {\rm min_{grid}}\{\Delta x_i/\left(v_i+c_s+c^{AV}_i
 \right)\}$, where $\Delta x_i$ is the cell size in direction $i$, $v_i$ is $i$th velocity component of the gas, $c_s$ is the sound speed, and $c^{AV}_i=4\left(Q_{ii}\right)^{1/2}$ is the AV diffusion speed (Pickett
 1995). 
  I define the shock heating and
radiation timescales as $\Delta t_s = f e/\Gamma$ and $\Delta t_r = f
e/\mid\Lambda\mid$, where $f$ is some number $< 1$, $\Gamma$ is the shock heating rate,
and $\Lambda$ is the radiation cooling/heating rate.  Even though the AV timescale is effectively 
accounted for in the definition of $t_h$, strictly adhering to this definition can result in time steps that
become extremely small and computationally inhibitive.  As described below, a heating limiter is employed to avoid this behavior. Likewise, a time step determined by $t_r$ can become too small to explicitly evolve the simulation, and so a radiation cooling/heating limiter is enforced.  In
the SV, the limiters are set such that $| \Lambda_{\rm max}| = |\Gamma_{\rm max}|
= f\epsilon/1~\rm orp $.   For the simulations presented by Mej\'ia (2004) and Boley et
al.~(2006, 2007c), $f=0.1$, and for the simulations presented by Cai et al.~(2006,
2007) and Cai (2006), $f=0.03$.  Boley et al.~(2007c) monitored the number of
cells affected by these limiters during the calculation, and found that during
the asymptotic phase, less than a few percent of the relevant  AV heated cells
were limited and less than a percent of the relevant radiatively cooling cells
were limited.  

Although limiting the heating and cooling according to an absolute timescale
worked for simulations presented by Boley et al.~(2006) and Cai et al.~(2006),
which used the same initial model, my highly optically thick disk simulations (see Chapters 4 and 7)
often resulted in a numerical runaway.  Consider the following situation.  Over one step, too much energy, relative to
an appropriate $\Delta t_s$, may be deposited into a shock and create an
unrealistic temperature gradient.  The following hydrodynamic step may be much
longer than the corresponding radiation timescale, and the very hot cell will deposit large
amounts of energy into the surrounding medium.  If the timescales are largely
disparate, this could quickly lead to a numerical runaway and result in the
rapid expansion of gas and the destruction of the disk if an appropriate limiter
is not chosen (see Figure 2.5).   
I found that this numerical catastrophe was largely due to the absolute limiting timescale employed in
previous simulations, which was typically too aggressive in some parts of the
simulation and not aggressive enough in others.  As  a result, I employ
limiters that are more commensurate with my definitions for $\Delta t_s$ and
$\Delta t_r$ described above.  In CHYMERA, the limiters are now set so that
heating, whether shock or radiation, can only change the internal energy by a
small percentage, typically a percent, for every iteration on $\epsilon$.
However, the radiative cooling limiter is adjusted to allow a cell to lose half
of its energy for any one iteration.  This limiting scheme usually leads to better
stability, and it is biased toward faster cooling.   Unfortunately, this too does not always stabilize a simulation against numerical runaways. I propose several possible code improvements in
Chapter 8 to address this issue.

\begin{figure}
\begin{center}
\includegraphics[width=4in]{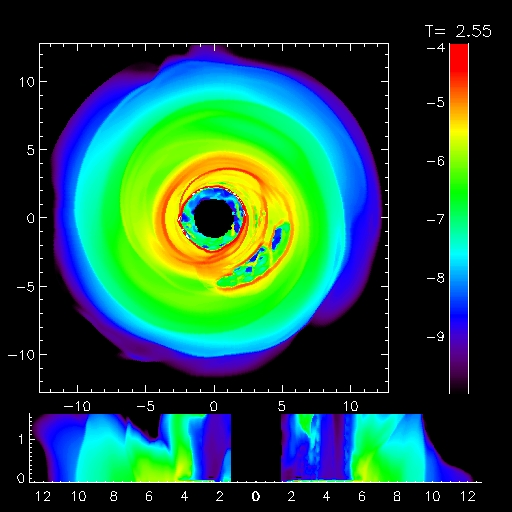}
\caption[Result of a numerical, radiation runaway] {Result of a numerical
radiation runaway, where radiative heating resulted in an unphysical hot bubble
that rapidly expanded and destroyed the disk.  Such a runaway typically occurs
near strong shocks, and is believed to be caused by highly disparate
hydrodynamics, shock heating, and radiative transfer timescales. The axes are
in AU and the logarithmic color scale shows the midplane density and a
meridional density slice in code units. }
\label{2.5}
\end{center}
\end{figure}

\subsection{Opacities}

The opacities that are used in the SV and in CHYMERA are the D'Alessio et
al.~(2001) opacities.  The details of the opacity tables are also discussed in
Mej\'ia (2004) and in Boley et al.~(2006; see Appendix A).  In addition to the
opacity tables, the SV uses the corresponding D'Alessio et al.~(2001) mean
molecular weight ($\mu$) tables.  Due to an
error with the inclusion of He, the typical $\mu$ in the simulation presented by Mej\'ia (2004) and Boley et al.~(2006) is 2.7  instead of the standard
$\mu=2.3$ for solar metallicity.  Inasmuch as the simulations of Cai et
al.~(2006, 2007), Cai (2006), and Boley et al.~(2007c) were, in part, companion
studies for the simulation presented by Mej\'ia (2004) and Boley et al.~(2006),
the incorrect $\mu$ tables were used.  CHYMERA, in contrast, does not use the
D'Alessio $\mu$ tables because the hydrodynamics is tied to the choice of $\mu$.
In this respect, the hydrodynamics and the opacities are slightly inconsistent,
but because the typical $\mu=2.3$ for the opacities, the problem is very minor.  

In the study presented by Boley et al.~(2007c), which employed the BDNL
algorithm as part of  the SV, only the Rosseland optical depths were used, with
the opacity evaluated at the cell's local temperature. This was a step backwards
from the M2004 scheme, which employs Planck means for regions where the
Rosseland $\tau\lesssim 2/3$.  Simply switching opacities for different regions
of the disk can lead to erroneous physics when tracing rays in the BDNL scheme, e.g.,
changing the location of the photosphere of the disk.  Regardless, as
demonstrated in \S 4 of Boley et al.~(2007c), the BDNL scheme performs better
overall.  For CHYMERA, the mean opacity issue is obviated by using a midplane-weighted
average between Rosseland and Planck mean opacities.  Consider the vertically
integrated midplane Rosseland optical depth $\tau_m$.  For any column of the
disk, the weighted optical depth, $\tau_w$, can be found by 
\begin{equation}\tau_w(z) =\frac{ \int_\infty^z dz \left(\kappa_R \tau_m +
\kappa_P/\tau_m\right)}{\tau_m+1/\tau_m},\end{equation}
where $\kappa_R$ and $\kappa_P$ are the Rosseland and Planck mean opacities,
respectively.  This method smoothly interpolates between the Rosseland and
Planck regimes and ensures that the photosphere remains at $\tau_R=1/\sqrt 3$
when the midplane optical depth is large.  However, other techniques for 
smoothly switching
between the Planck mean and the Rosseland mean opacities during the optical
integration are being explored.

\section{Analysis Tools}

\subsection{Fluid Element Tracer}

The hydrodynamics schemes in the SV and CHYMERA are Eulerian, and do not give direct information on the histories of fluid elements.  In order to derive detailed and statistical shock information and to capture the
complex gas motions in unstable disk simulations, I have combined a tri-Akima spline interpolation algorithm
with a fourth-order Runge-Kutta integrator (e.g.,
Press 1986) 
for tracing a large sample of fluid elements.   During the integration,
the Runge-Kutta scheme calls
the interpolator each time updated velocities are required, and thermodynamic quantities,
e.g., temperature and density,
are found through interpolation at the beginning of each time step. Because the hydrodynamics code 
explicitly solves the equation of motion for the gas, the algorithm only needs 
time and velocity information to advance the fluid elements.

An Akima spline is similar to a natural cubic spline, but typically yields better results
for curves with sudden changes (see Akima 1970), as are expected in shock profiles.  
Although the spline fits a curve to a one-dimensional set of data points,
the interpolation can be extended to data in three dimensions.  First, consider a
cubic volume with data at the vertices. A value anywhere in the volume can be
approximated by seven linear interpolations: four to calculate values between each vertex in 
a particular direction, two to calculate the values along the projections of the desired point onto the
interpolated lines, and a final interpolation through the point of
interest.  This scheme is depicted graphically in Figure 2.6. Extending this
from a simple tri-linear fit to a tri-Akima spline is relatively
straightforward.  Instead of using the eight nearest points that enclose a volume, 
one uses the 125
closest points, with five data points used for every Akima spline fit. The central
data point is the closest data value to the point of interest. I use the 
GNU Scientific Library Akima spline algorithm for performing fits.  

\clearpage
\begin{figure}
\begin{center}
\includegraphics[width=7.3cm]{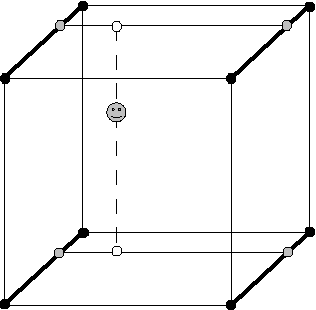}
\caption[Graphical representation of a tri-linear fit]
{Graphical representation of a tri-linear fit.  The smiley face
indicates the point for which the value is desired.  The black dots indicate the
vertices, the points for which the values are known.  The values along lines
adjoining the vertices are calculated first.  The values for the projection of
the smiley face onto the dark lines (gray circles) are then used to find the values along the
light line.  Finally, the values for the projection of the smiley face onto the
light line (white circles) are used to interpolate a value at the position of
the smiley face. The same principle is used for a tri-Akima spline
interpolation, but with 125 points instead of eight.}
\label{2.6}
\end{center}
\end{figure}
\clearpage

The integration of the fluid elements can be done by two
methods: a post-analysis evolution or a real time integration, i.e., during the
hydrodynamics evolution. For the post-analysis version, restart files from
either CHYMERA or the SV are read in, typically with 1/100-1/20 of an outer
orbit resolution, and the velocity field and thermodynamic properties are
linearly interpolated between the data snapshots.  This method is typically
dominated by read-in time, and interpolation between data snapshots results in
smeared thermodynamic quantites and ragged profiles (e.g., density vs.~time).
The data storage demands for having sufficient data for interpolation are quite
restrictive.  However, the scheme's significant advantage is that it is
post-analysis; different fluid element sampling with varying numbers of
fluid elements can be used.  The real time integration results in very smooth
profiles, captures shocks well, and is not subject to inhibitive storage
demands.  The disadvantage is that the entire simulation must be rerun
if a different fluid element sampling is desired.

\clearpage
\begin{figure}
\begin{center}
\includegraphics[width=10cm]{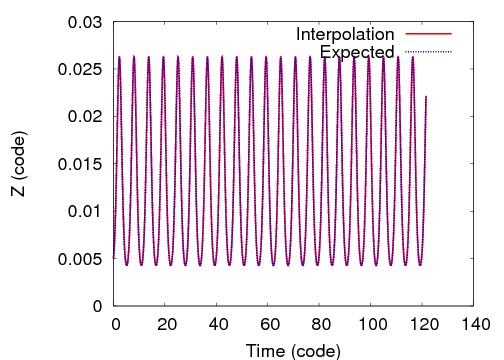}
\includegraphics[width=10cm]{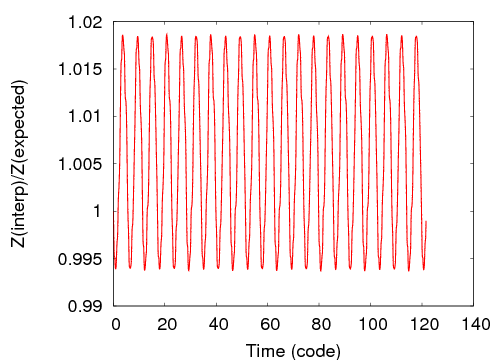}
\caption[Fluid element tracer test]{Top panel: The trajectory of a fluid element in a rapidly varying
velocity field.  For comparison, the expected trajectory is also shown.  Bottom
panel: The fractional difference between the integrated trajectory and the
expected trajectory is shown.}
\label{2.7}
\end{center}
\end{figure}
\clearpage

The accuracy of the interpolator and the integrator has been assessed through a
series of tests, and is found to be satisfactory for the purpose of the fluid
element tracer, which is to capture shock strengths and thermodynamic
fluctuations. For example, one test is to place a fluid element in a prescribed velocity 
field: $v_r=0$, $\Omega =(GM_{\rm star}/r^3)^{1/2}$, and $v_z=z\sin\left(12\phi\right)$. 
The vertical motion of a fluid element is then
$z\left(t\right)=z_0\exp\left(-\cos\left(12 \left[\Omega t+\phi_0\right]\right)/12\Omega\right)$.
 Figure 2.7 shows the calculated and expected trajectories. 
The agreement between the known and interpolated values is generally within two percent. 
Other interpolation methods were tested, including a tri-cubic interpolation, a
tri-cubic spline interpolation, and an inverse distance$^n$ interpolation, where
$n=1$ and 2 were tested.  Generally, none of these performs better than the
Akima spline method, although differences between the Akima spline, the
cubic spline, and the cubic methods are slight for smoothly varying functions. 
In addition, all methods seem to be able to capture
the motion of a fluid element in a simple velocity field well.  Because I
am interested in capturing a rapid change in the velocity field, I choose to use
the Akima spline.

\subsection{Torque and Effective $\alpha$ Analysis}

The integrated gravitational torque on the inner disk due to the outer disk can be calculated directly
from the mass distribution by
\begin{equation}
{\bf C}=\int_{V(R)}\rho{\bf x}\times\nabla\Phi dV
\end{equation}
for gravitational potential $\Phi$, position vector ${\bf x}$, and volume $V$ inside
$r=R$. When the integration is carried over the entire volume containing the
disk, ${\bf C}=0$ if angular momentum is conserved. As shown by Lynden-Bell \& Kalnajs
(1972), a symmetric stress tensor can be defined, where $ T_{ij}=(4\pi G)^{-1} g_ig_j - g^2I(8\pi G)^{-1}$
and
 $\partial T_{ij}/\partial x_j = -\rho \partial_i \Phi=-\rho g_i$. Note that I use Cartesian coordinates to
avoid the need to make distinctions between covariant and contravariant
tensors. The stress tensor is related to
the total torque on the inner disk due to the outer disk by
\begin{equation}
C_i=-\int_S \epsilon_{ijk}x_jT_{kl} dS_l,
\end{equation}
where $S$ is some surface containing the inner disk.  It can be shown, as follows, 
that equations (2.32)
and (2.33) are equivalent.  The divergence theorem states that for some rank 2
tensor,
\begin{equation}
\int\frac{\partial A_{ij}}{\partial x_j}dV = \int A_{ij}dS_j.
\end{equation}
Because $\epsilon_{ijk}x_jT_{kl}$ is a rank 2 tensor,
\begin{equation}
C_i=-\int_S \epsilon_{ijk}x_j T_{kl} dS_l =- \int_V \frac{\partial}{\partial
x_l}\left(\epsilon_{ijk}x_j T_{kl}\right)dV.
\end{equation}
Recall that $\partial x_j / \partial x_l =\delta_{jl}$ by definition, and so 
\begin{equation}
C_i=-\int_V\epsilon_{ijk}\left(T_{kj}+x_j\frac{\partial T_{kl}}{\partial
x_l}\right) dV.
\end{equation}
Inasmuch as $T_{jk}$ is a symmetric tensor, $\epsilon_{ijk}T_{kj}=0$, and by
definition of $T_{jk}$,
\begin{equation}
C_i=\int_V\rho\epsilon_{ijk}x_j g_k dV,
\end{equation}
which is equation (2.32).

In the analysis of these three-dimensional disk simulations, I use equation
(2.32) to calculate the gravitational torque because the volume intergral can be neglected in
regions where the density goes to zero.  Moreover, I am only interested in the
torque in the $\hat{e}_z$ direction, and so the torque intergral can be rewritten as
\begin{equation}
C_z=\int_V \rho \frac{\partial \Phi}{\partial \phi} dV.
\end{equation}
Using this description for the torque at any given radius, one expects for
extrema in the torque profile to indicate transitions between mass inflow and
outflow. 

In addition to the gravitational torque, velocity fluctuations in a mean flow can create Reynolds stresses (Landau \& Lifshitz 1987).  
The torque in the $\hat{e}_z$ direction due to these fluctuations for a given $r$ is given by
\begin{equation}
C_z^{\rm Reyn}=-\int_0^{2\pi}\int_{-\infty}^{\infty} T_{r\phi}r^2d\phi dz,
\end{equation}
where the Reynolds stress tensor for the $r$-$\phi$ component is 
$T_{r\phi}=\rho\delta v_r\delta v_{\phi}$.  The fluctuating components of the
velocity field $\delta v_i$ are found by subtracting the mean flow from the
velocity at a given cell.  Calculating an accurate Reynolds stress is therefore
 dependent
on determining a meaningful average flow for the gas.  In the analysis
presented here, I find the mean flow by calculating density weighted
velocity averages for each $r$ in the disk.  The Reynolds stresses determined this way are still extremely noisy, and it is unclear whether they represent the actual Reynolds stress contribution to the total torque well.  This difficulty is due to the relatively coarse
resolution used in these global simulations.  As will be shown in Chapter 6, the gravitational torque alone
represents the mass transport in these disk simulations fairly well, and so I typically ignore the Reynolds
stress. 

Once the torque on the disk is determined, an
effective $\alpha$ can be calculated, i.e., the Shakura \& Sunyaev $\alpha$ parameter
needed to describe mass transport if the disk were an $\alpha$ disk, where (Gammie 2001)
\begin{equation}
\alpha \equiv \bigg |\frac{d\ln\Omega}{d\ln r}\bigg |^{-1}\frac{\mathcal{T}_z}{c_s^2\Sigma}.
\end{equation}
Here, $\mathcal{T}_z$ is the vertically integrated, azimuthally averaged total (gravitational+Reynolds) stress
tensor and  $c_s$ is the sound speed defined for a razor-thin disk.  Gammie (2001) showed that if the disk evolved toward a state where local disk cooling was balanced by local shock heating and gravitational work, the effective $\alpha$ can be described by
\begin{equation}
      \alpha = \left(\frac{9}{4} \gamma_{2D}\left(\gamma_{2D}-1\right) t_{\rm cool}\Omega \right)^{-1}
\end{equation}
for a Keplerian disk, where
$t_{\rm cool}$ is the local cooling time and $\gamma_{\rm 2D}$ is the two-dimensional
adiabatic index. The two- and three-dimensional $\gamma$s are related by
$\gamma_{\rm 2D}=3-2/\gamma$ and $\gamma_{\rm 2D}=(3\gamma-1)/(\gamma+1)$ for the strong and negligible
self-gravitating limits, respectively. Because the torque in
the vertical direction is desired, the surface on which to evaluate $\mathcal{T}_z$  is 
an infinite cylinder  centered along the $z$ axis.  If only the
gravitational stress tensor needs to be considered, then the azimuthally averaged, 
vertically integrated
stress can be related to equation (2.32) by
\begin{equation}
\mathcal{T}_z(R)=\frac{1}{2\pi}\int_0^{2\pi}\int_{-\infty}^{\infty} T_{r\phi} dz =
-\frac{1}{2\pi R^2} \int_{V(R)}\rho\frac{\partial \Phi}{\partial \phi} dV.
\end{equation}
In order to calculate effective $\alpha$ profiles for the simulations presented
in this dissertation, I use midplane sound speeds and equation (2.41) to
evaluate equation (2.40).  Because the vertically integrated $c_s^2$ in equation (2.40)
is probably less than the midplane $c_s^2$, this approach likely understimates $\alpha$. 

\chapter{RADIATION TESTS}

A crucial demand on any radiation algorithm designed for the protoplanetary disk
problem is the necessity to handle both the high and low optical depth regimes.  
An analytic solution to a relevant test problem must be found in order to
evaluate the accuracy of radiative transport algorithms in disks.  In this
Chapter, I describe a toy problem based on Hubeny (1990) that can be used to
test the accuracy of a radiation routine for a disk geometry. 

Consider a plane-parallel slab with constant vertical gravitational acceleration
$g$ but with a midplane about which reflection symmetry is assumed.  Suppose there is some heating mechanism that produces a known
distribution of astrophysical flux once the system reaches hydrostatic and
radiative equilibrium; in equilibrium, energy transport is only vertical. Make
the ansatz that the vertical astrophysical flux has the form
\begin{equation} \mathcal{F}_z \left( \tau \right) =\mathcal{F}_0\left(
1-\frac{\tau}{\tau_m}\right), \label{eq1_02}\end{equation}
where $\mathcal{F}_0=\sigma T_e^4/\pi$ is the astrophysical flux from the
atmosphere with effective temperature $T_e$, $\tau$ is the Rosseland mean
optical depth measured vertically downward, and $\tau_m$ is the optical depth at
the midplane.  This function ensures that the flux goes to zero at the midplane
and that $\mathcal{F}_z=\mathcal{F}_0$ at  $\tau=0$.  The heating term required to achieve this flux
distribution is then
\begin{equation} \Gamma = -\pi \frac{\partial\mathcal{ F}\left(\tau\right)}{\partial z}
                           =\pi \mathcal{F}_0 \frac{\rho\kappa}{\tau_m},
\label{eq2}\end{equation}
where $\rho$ is the density at the point of interest and $\kappa$ is the
Rosseland mean mass absorption coefficient.

If $\tau_m \gtrsim 10$, the temperature structure may be derived from the flux
by using the standard Eddington approximation, which relates the mean intensity
$J$ to the astrophysical flux by
\begin{equation}\frac{4}{3}\frac{d J\left(\tau\right)}{d\tau} =
\mathcal{F}_z\left(\tau\right). \label{eq3}\end{equation}
Integrating equation (3.3) yields
\begin{equation}
T^4=\frac{3}{4}T_e^4\left(\tau\left[1-\frac{\tau}{2\tau_m}\right] +
q\right).\label{eq4}
\end{equation}
The constant $q$ can be determined by considering the low optical depth limit.
In that limit, the atmosphere reduces to a standard stellar atmosphere;
therefore $q=1/\sqrt3$ (Mihalas \& Weibel-Mihalas 1986).

In the limit that $\tau_m \lesssim 0.5$, the atmosphere approaches an isothermal
structure.  Because the source function becomes constant, the observed flux can
be found from $F=2\pi \mu I_+$ where $\mu=1/\sqrt 3$ for the single ray scheme (Chapter 2), 
which gives the temperature of the isothermal
atmosphere:

\begin{equation} T^4=\frac{T_e^4}{2\mu
\left(1-\exp\left[-2\tau_m/\mu\right]\right)}\rm ,\end{equation}
where the factor of 2 in the exponential accounts for both sides of the
atmosphere.  In addition, I explicitly include $\mu$ because the vertically
integrated $\tau$ is used.  As $\tau_m\ll 1$, $T^4=T_e^4/4\tau_m$. 

An additional assumption about the form of the opacity law permits analytic
evaluation of the hydrostatic structure of the disk and allows one to control
whether the atmosphere will be convective.  For example, I assume that $\kappa$
is constant throughout the disk for two of the tests presented below.  This
assumption should make the disk convectively stable (Lin \& Papaloizou 1980;
Ruden \& Pollack 1991).  It also makes the relation between pressure and optical
depth simple:
\begin{equation}p=\frac{g}{\kappa}\tau. \label{eq5}\end{equation}
Moreover, the midplane pressure is related to the full disk surface density
$\Sigma$ by
\begin{equation}p_m=g\frac{\Sigma}{2}.\end{equation}

\section{Relaxation Test}

For the relaxation test, I heat the gas slab in accordance with equation (3.2)
to test whether the radiation routine in question allows it to relax to the
correct hydrostatic and radiative equilibrium configuration.  By selecting
different midplane optical depths, the effects of resolution on the routine can
be tested. 

Following Boley et al.~(2007c), I test the radiation routines used by  M2004, 
C2006, and BDNL.   The
initial condition for the atmosphere is a constant density structure with a
$\tau_m = 0.05$, 0.5,  5, 10, and 100.  The target effective temperature of the
atmosphere, which controls the magnitude of the heating term, is set to
$T_e=100$ K, and $\kappa=\kappa_0=\rm~constant$.  For the discussion that
follows, I take the high optical depth regime to indicate where $\tau\ge
1/\sqrt 3$ and the low optical depth regime to indicate where $\tau < 1/\sqrt
3$.   Note that this boundary is slightly different from the M2004 and C2006
schemes; these schemes use $\tau = 2/3 $ to set the boundary between the low and
high optical depth regimes, as is done in the standard Eddington solution.

Figure 3.1 compares the M2004, C2006, and BDNL solutions to the analytic
temperature profile and to the flux profile for $\tau_m=100$.  The BDNL routine
matches the analytic curves very well. The M2004 routine does well in matching
the temperature curve for most of the high optical depth regime, which is
resolved by 12 cells, but the low optical depth regime has a sudden temperature
drop.  As reported by Boley et al.~(2006), this temperature drop is an artifact of the lack
of complete cell-to-cell coupling in the optically thin region.  Despite the
temperature drop, the boundary condition between the two regimes typically
yields the correct boundary flux, i.e., the flux leaving the high optical depth
regime.  An unfortunate result of this boundary condition is that it produces
oscillations in the flux profile and in the temperature profile when the
low/high optical depth boundary is near a cell boundary or when the entire disk
height is only resolved by a few vertical cells.  These numerical oscillations
produce artificial heating, and it is this behavior that may contribute to
keeping the inner 7 AU of the disk presented in Boley et al.~(2006) hot.   
Although this is
problematic, even when oscillations occur, the time-averaged cooling time is
within 10\% of the expected value for the test shown in Figure 3.1.  The C2006
improvement lessens this problem. 

Figure 3.2 compares the results of the M2004 and BDNL routines with each other for
$\tau_m=10$.  The high optical depth regime is now resolved by only six cells.
Both methods compute the correct flux through the slab. The temperature profiles
are both skewed more than in the $\tau_m=100$ case  mainly because the solution
deviates from the Eddington approximation, which is used to derive the analytic
temperature profile, as $\tau_m$ becomes small.   Figure 3.3 shows the same
comparison as done in Figures 3.1 and 3.2, but for $\tau_m=5$.  The high optical
depth regime is now resolved by only four cells.  Again, both methods allow for
the correct flux through the slab, and the BDNL temperature distribution is
close to the analytic value, with slight departures again due mainly to the
inaccuracy of the analytic curve.

To demonstrate the accuracy of each algorithm in the low optical depth limit, I
show the temperature profiles in Figure 3.4 for the M2004 routine and the BDNL
routine for $\tau_m=0.05$ and 0.5, and compare each curve with the
temperature estimate calculated from equation (3.5).  The M2004 routine yields
temperature profiles that are colder than the expected temperature profiles, and
the departure is more severe as $\tau_m$ increases.  This is a result of the
lack of complete cell-to-cell coupling in the atmosphere.  The BDNL routine is
in excellent agreement with the predicted temperature.  The departure from the
analytic estimate observed in the $\tau_m=0.5$ case is a result of the
inaccurate assumption in the analytic curve that the source function is constant for $\tau\sim1$. 

\begin{figure}
   \centering
   \includegraphics[width=6cm]{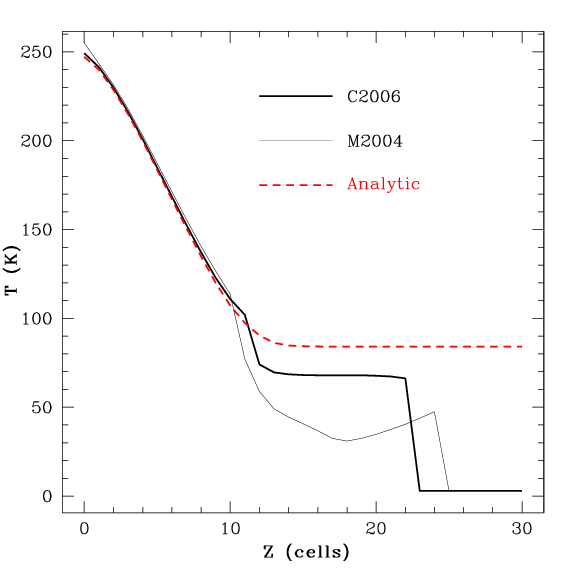}\includegraphics[width=6cm]{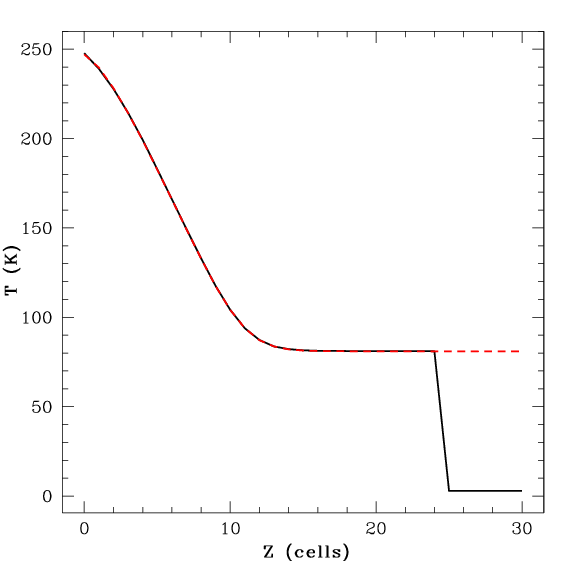} 
   \includegraphics[width=6cm]{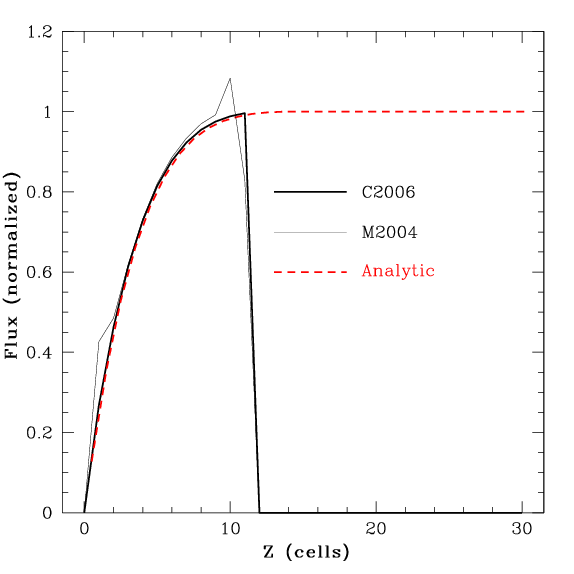}\includegraphics[width=6cm]{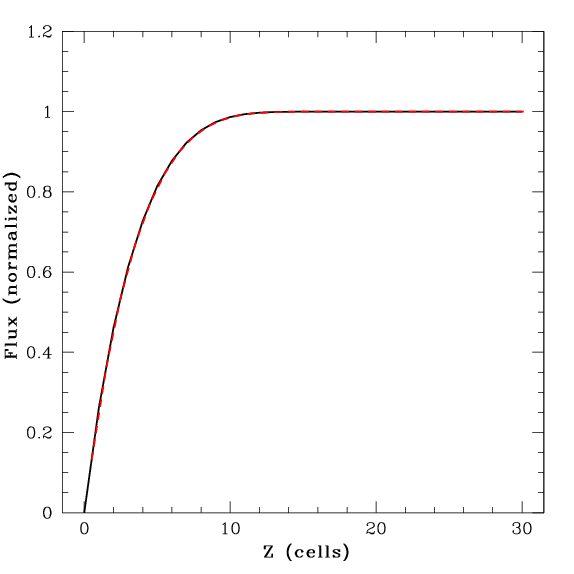}
\clearpage
   \caption[Results of the relaxation test for the M2004, C2006, and BDNL
schemes with $\tau_m=100$]
{Results of the relaxation test for the M2004, C2006, and BDNL
schemes with $\tau_m=100$. Top: The left panel shows the relaxed temperature
profile for the M2004 and C2006 routines, while the right panel shows the
relaxed temperature profile for the BDNL routine.  In both panels, the analytic
curve is represented by the red, dashed curve.  The first sudden drop in the M2004
and C2006 schemes is due to the lack of complete cell-to-cell coupling required
by radiative transfer.  The second drop, which is also in the BDNL profile, is
where the density drops to background and where the algorithm stops following radiation
physics.  Bottom: Similar to the top panels but for the flux profile.  The
undulations in M2004's flux profile are believed to be due to the low/high
optical depth boundary lying near a cell face intersection.  The C2006
modification helps avoid this problem. Finally, the sudden drop in the M2004 and
C2006 flux profile occurs because these schemes only explicitly track the flux
through the optically thick disk.   The right panels are the same as Figure 1 in
Cai et al.~(2007).}
   \label{fig1}
\end{figure}

\begin{figure}
\centering
\includegraphics[width=7cm]{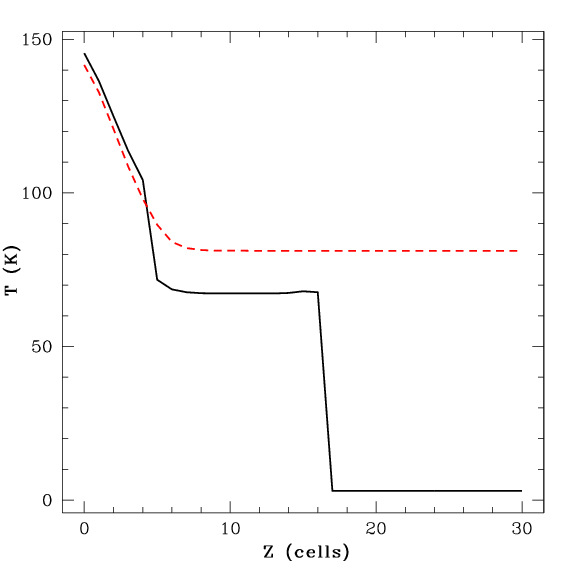}\includegraphics[width=7cm]{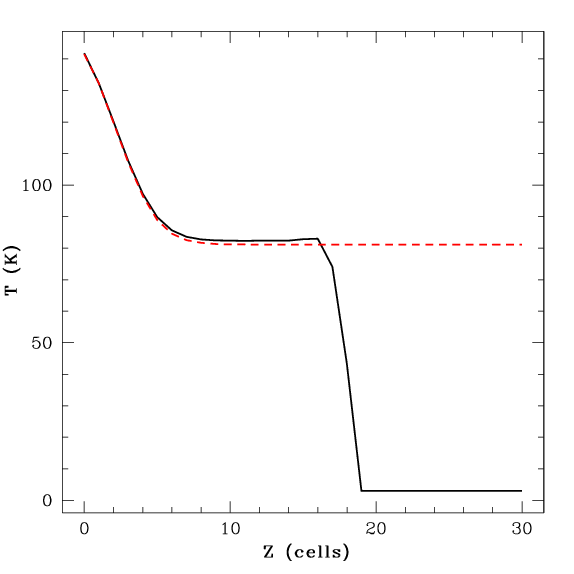} 
\includegraphics[width=7cm]{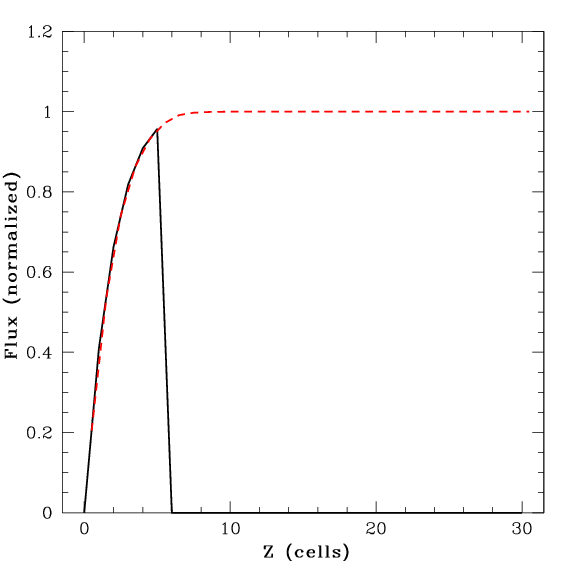}\includegraphics[width=7cm]{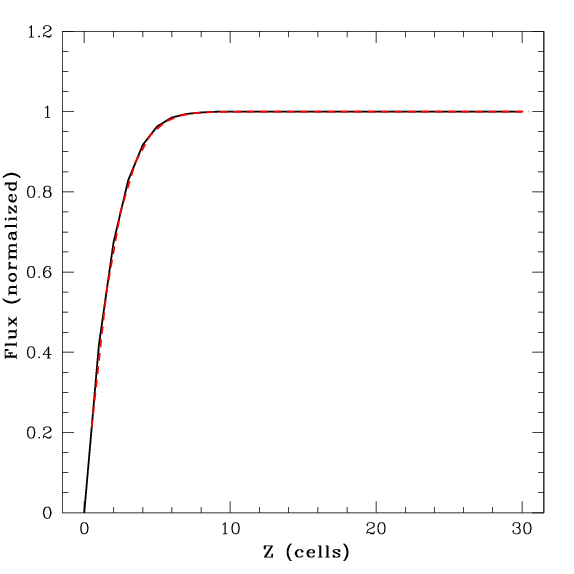}
\caption[Same as Figure 3.1, but with $\tau_m=10$ and only for the M2004 and
BDNL schemes]
{Same as Figure 3.1, but with $\tau_m=10$ and only for the M2004 and
BDNL schemes.  The undulations in the M2004 flux profile are no longer present.
The slight departure of the BDNL solution from the analytic temperature curve is
mainly due to the breakdown in the Eddington approximation, which is assumed in
the analytic temperature profile, as $\tau_m$ becomes small. }
   \label{fig2}
\end{figure}

\begin{figure}
\centering
\includegraphics[width=7cm]{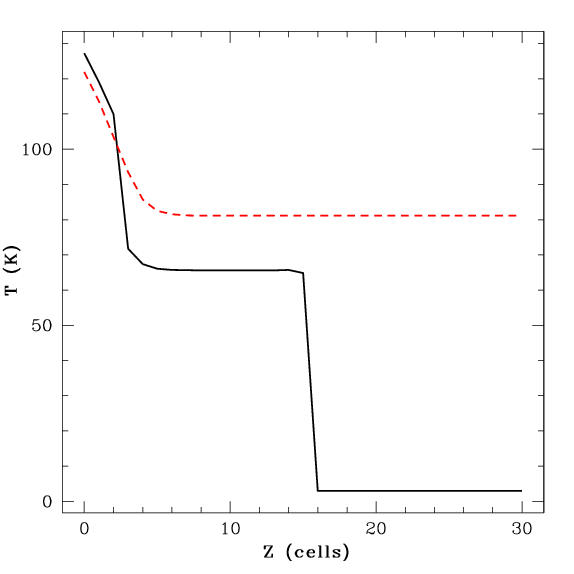}\includegraphics[width=7cm]{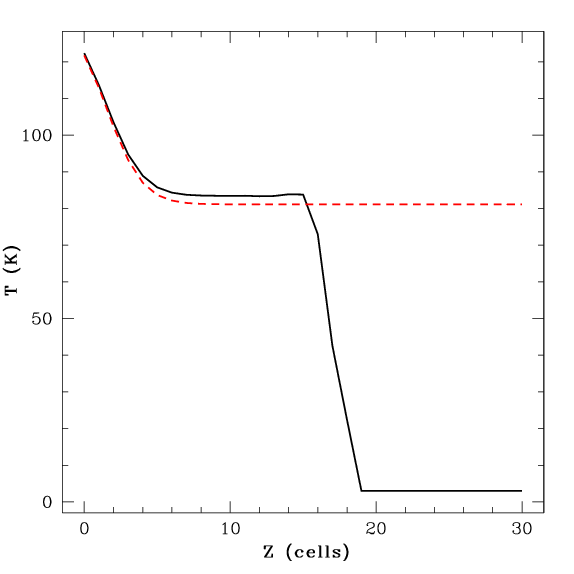} 
\includegraphics[width=7cm]{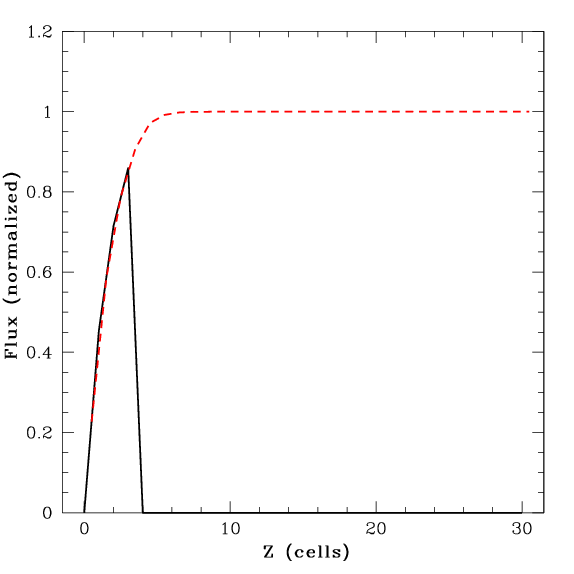}\includegraphics[width=7cm]{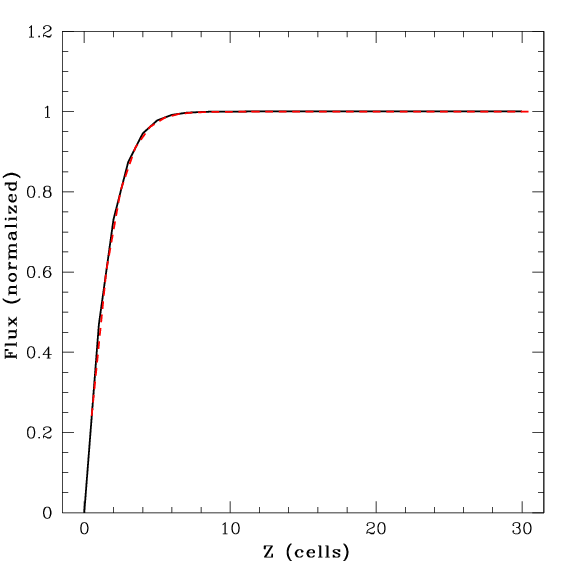}
\caption[The same as Figure 3.2, but with $\tau_m=5$]
{The same as Figure 3.2, but with $\tau_m=5$. Once again, the departure
of the BDNL temperature solution is mainly due to the breakdown of the
approximations used to calculate the analytic curve.}
\label{fig3}
\end{figure}

\begin{figure}
\centering
\includegraphics[width=3.95in] {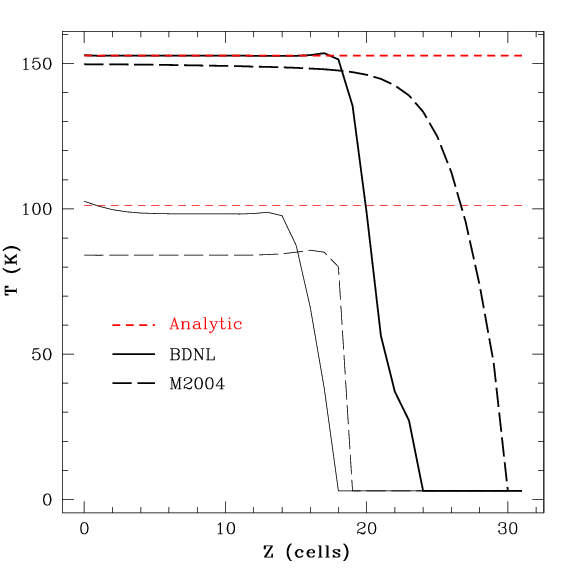} 
\caption[Temperature curves for the
low optical depth limit]
{Temperature curves for the
low optical depth limit. The red, dashed curve indicates estimated isothermal
temperatures from equation (3.5), the solid, dark curve indicates the results of
the BDNL routine, and the dashed, dark curve indicates the results of the M2004
routine.  The curves at the lower temperature correspond to $\tau_m=0.5$, and
the curves at the higher temperature correspond to $\tau_m=0.05$.  The lower
optical depth corresponds to the higher temperature because cooling is less
efficient.  The M2004 routine yields temperatures that are too cold, because it
always uses the free-streaming approximation when $\tau<2/3$.  However, the
M2004 routine converges to the correct solution as $\tau_m\rightarrow 0$.  The
BDNL routine is consistent with the estimated temperature for the $\tau_m=0.05$
case, and it is roughly consistent with the $\tau_m=0.5$ case.  The
inconsistency seen in the $\tau_m=0.5$ case is a result of the small $\tau_m$
assumption, which is used to derive equation (3.5).}
\label{fig4}
\end{figure}

\section{Contraction Test}

In order to study how accurately the radiation algorithms work with the
hydrodynamics routines and to study the effects of resolution, I allow the slab
to cool and follow the contraction.  If one assumes a constant opacity law and a
large $\tau_m$, then the contraction becomes homologous, and a relationship
between the midplane temperature and the cooling time is easily attainable.
Consider the cooling time
\begin{equation}t_{\rm cool}=\frac{U}{\sigma T_e^4}\sim \frac{p_m h
\tau_m}{T_m^4},\end{equation}
where $h\sim\Sigma/\rho_m$ is the scale height of the atmosphere and $U$ is the
internal energy per unit area.  If one assumes an ideal gas law, then one
expects that
\begin{equation}t_{\rm cool} \sim \frac{1}{T_m^3}.\end{equation}
Finally, because $U\sim p_m h\sim T_m$,
\begin{equation}t_{\rm cool}\sim \frac{1}{U^3}.\end{equation}
\begin{figure}
\centering
\includegraphics[width=3.95in]{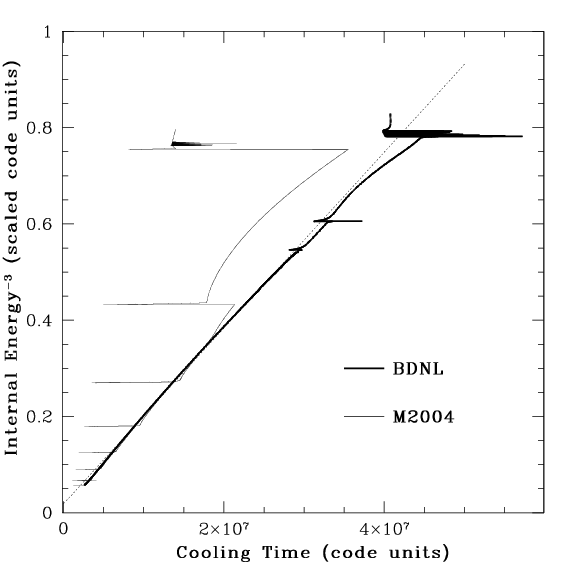}
\caption[Contracting slabs as shown in the $t_{\rm cool}$-$U^{-3}$ plane
for the same cases as shown in Figure 3.1]
{Contracting slabs as shown in the $t_{\rm cool}$-$U^{-3}$ plane
for the same cases as shown in Figure 3.1; all curves should be linear if the
slabs contract as expected.  Both schemes break down when the high optical depth
regime is contained within 5 cells ($U^{-3}\approx0.75$), but the M2004 routine
(light curve) starts to deviate from the expected solution once the slab is
resolved by 6 cells ($U^{-3}\approx0.43$).  The sudden decreases in cooling
times for the M2004 routine occur when the optically thin/thick boundary
transitions into another cell.  The dotted line shows the
expected behavior. }
\label{fig5}
\end{figure}
For this test, I take the relaxed slab shown in Figure 3.1 with $\tau_m=100$ and
turn the heating term off.  The contraction is followed until the scheme breaks
down.  Figure 3.5 demonstrates this contraction 
sequence in the $t_{\rm cool}$-$U^{-3}$ plane when the slab is evolved with the
BNDL scheme (heavy, dark curve) and  when it is evolved with the M2004 scheme
(lighter, gray curve).  Both cases follow the expected contraction closely,
which is indicated by the dotted line, until the optically thick atmosphere is
resolved by five  (BDNL) or six (M2004) cells. The sudden decreases in cooling
times for the M2004 routine occur when the optically thin/thick boundary
transitions into another cell.  

\section{Convection Test}

Another test I describe demonstrates whether the radiation scheme permits
convection when it should occur.  Lin \& Papaloizou (1980) and Ruden \& Pollack
(1991) show that convection is expected in a disk-like atmosphere when the
Rosseland mean optical depths are large and when $\beta > \beta_{\rm crit}$ for
$\kappa\sim T^{\beta}$.  For a $\gamma=5/3$ gas, the vertical entropy gradient
is driven to a negative value when the critical exponent $\beta_{\rm crit}
\approx1.5$. Thus, I present two cases for which almost identical atmospheres are
allowed to relax to an equilibrium.  In one case, $\beta=1$, which should make
the atmosphere convectively stable, and the other has $\beta=2$, which should
make it unstable.  As shown in the top panels of Figure 3.6, I find that the BDNL
routine produces convection when it should and does not when it should not.
Likewise, the bottom panels of Figure 3.6 demonstrate that the M2004 routine also
permits or inhibits convection correctly.  However, M2004 does seed an
artificial superadiabatic gradient at the boundary between the optically thin
and thick regions due to the temperature drop at the photosphere.  Nevertheless, convection does not occur for $\beta=1$ even with this seed, and
the superadiabatic gradients are an order of magnitude smaller for $\beta=1$
than for $\beta=2$ at that boundary. 

Finally, I measure the flux that is carried by convection in the $\beta=2$ case
for the BDNL scheme.  As indicated in Figure 3.7, convection carries about 30\% of
the flux.  Figure 3.7 also acts as a reminder that the energy must ultimately be
radiated away.  Boley et al.~(2006) and Rafikov (2007) argue that because the convective
flux is controlled by the radiative flux leaving the photosphere of the disk,
convection should not be expected to lead to rapid cooling and fragmentation in
protoplanetary disks, as claimed by Boss (2004a) and Mayer et al.~(2007) (see
Chapter 5).

\begin{figure}
\centering
\includegraphics[width=7cm]{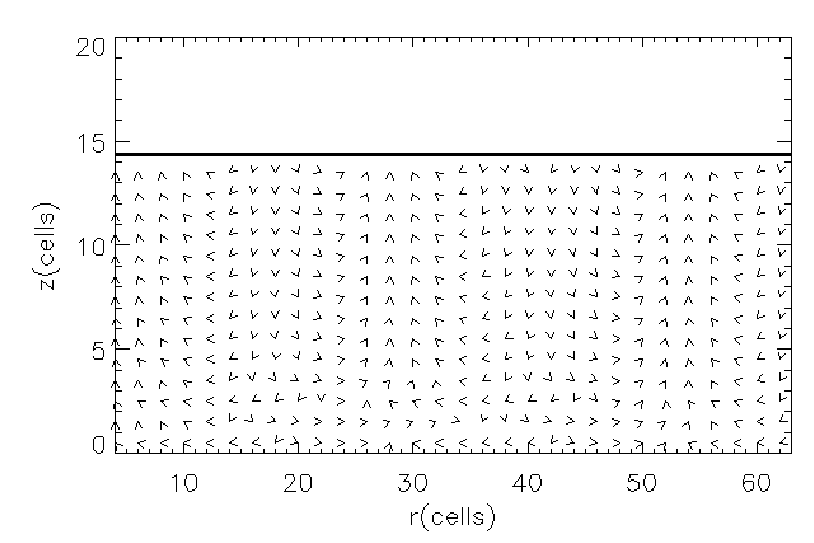}\includegraphics[width=7cm]{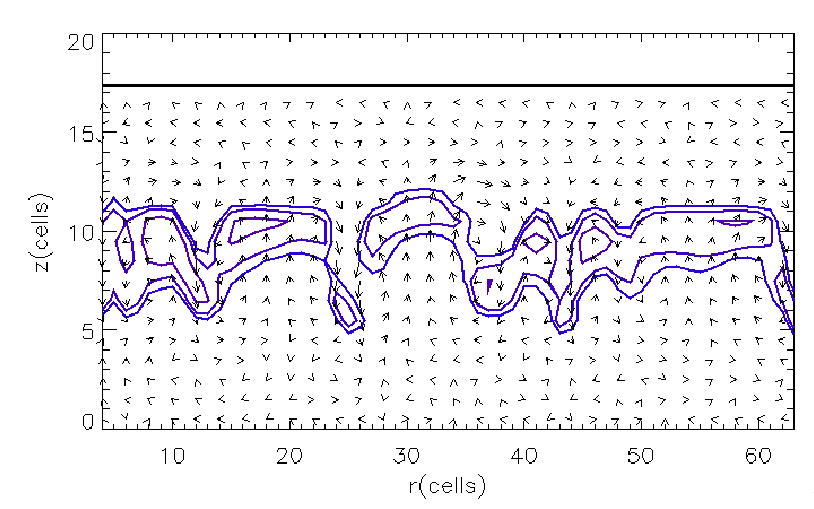} 
\includegraphics[width=7cm]{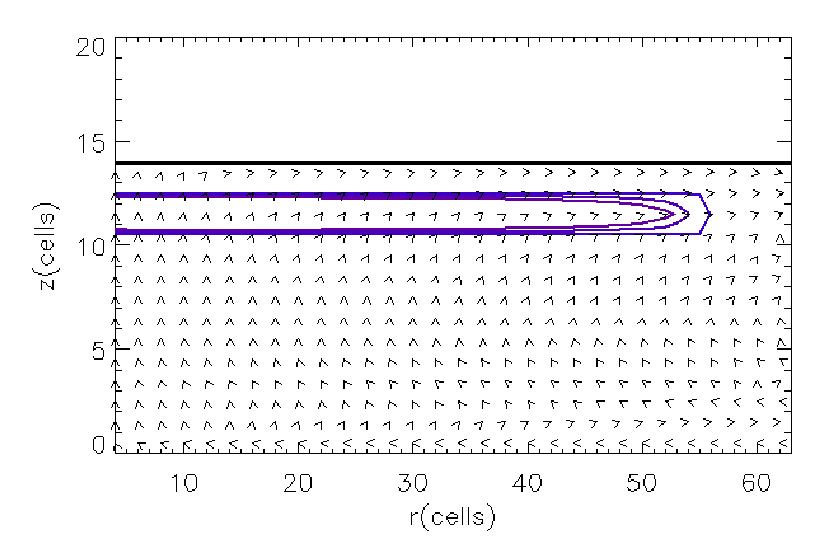}\includegraphics[width=7cm]{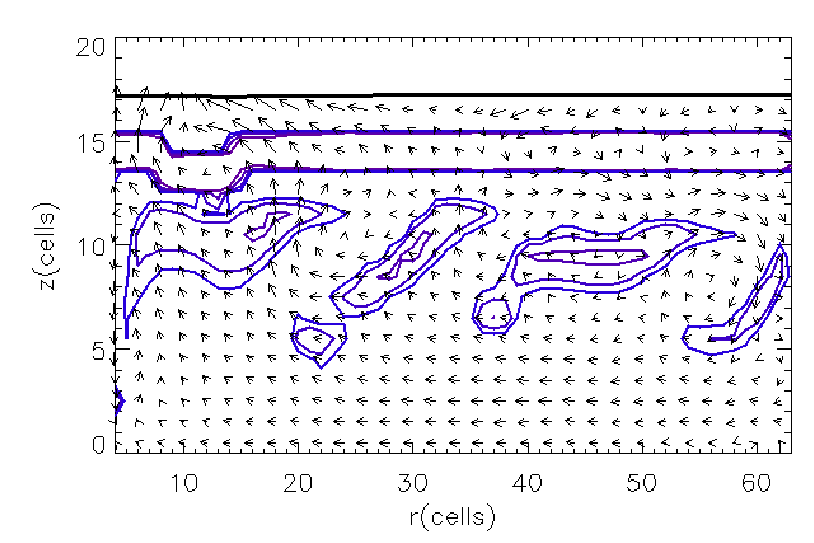} 
\caption[Convection test]
{Convection test.  The heavy black contour indicates the same density
contour for each panel.  Arrows show relative velocities in the $r$ and $z$
directions.  The blue contours indicate superadiabatic regions.  The motions in
the left panels are a few orders of magnitude smaller than the motions in the
right panels. Top: BDNL. The left panel shows the case $\beta=1$; convection and
superadiabatic gradients are absent, and the velocities represent low level
noise.  The right panel shows the case $\beta=2$; convective cells and
superadiabatic gradients are present. Bottom: Same as the top panel but for the
M2004 scheme.  The superadiabatic regions near the top density contour are due
to the artificial, sudden drop in temperature at the optically thin/thick
interface.  The superadiabatic gradients in the left panel are about an order of
magnitude smaller than those in the right panel.  }
   \label{fig6}
\end{figure}

\begin{figure}
   \centering
   \includegraphics[width=3.95in]{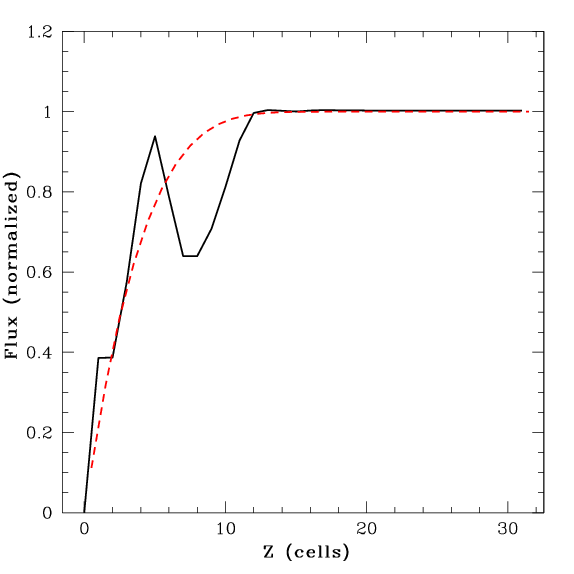}
   \caption[Average flux through the atmosphere with $\beta=2$ and with the BDNL
scheme]
{Average flux through the atmosphere with $\beta=2$ and with the BDNL
scheme. Convection carries about 30\% of the flux at maximum, but almost all the
flux is radiative in the photospheric region near cell 15.  }
   \label{fig7}
\end{figure}

\section{2D Test}

The M2004, C2006, and BDNL radiation routines include more cell-to-cell coupling in the
vertical direction than the radial or azimuthal directions. Vertically integrated optical depths are also
used to determine the radiation boundary conditions.  Generally, this
vertical bias is justified in the highly flattened disk systems that are evolved in the SV and in CHYMERA. 
However, the degree to which temperature structures are skewed by this bias should be investigated 
using 2D and 3D tests.  Although the 2D test is still being developed, I present a few results here for the BDNL routine.

A spherical, constant density mass distribution is loaded onto the grid and held fixed. This isolates
the radiation algorithm in the code.  A constant energy input source is included in just the center cells, i.e., the first computational cell closest to the $r$, $z$ origin for all $\phi$.  The opacity is set to some constant, and the magnitude of the constant is adjusted to vary the optical depth regimes for different tests.  Figure 3.8 shows the results.  The temperature contours are roughly spherical for the high optical depth case, which indicates that the optically thick/thin boundary conditions are reasonable. 

The low optical depth case does not perform nearly as well.  For this regime, the temperature contours are highly skewed along the $r$ axis, and there is an odd temperature inversion at the photosphere near the $z$ axis (Fig.~3.8).  At optical depths of about 5, the radiation transport is not quite in the diffusion limit (see \S 3.1), and so fluxes along $\hat{e}_r$ and $\hat{e}_z$ are disparate.  This creates the temperature skew along the $r$ axis because radiation from the heat source in the center cell is coupled over several cells in the $z$ direction by the ray solution, but is only coupled between adjacent cells in the $r$ direction by the diffusion equation.

Near the $z$ axis, the temperature inversion in the photosphere is also due to the difference in handling radial and vertical fluxes.  The combination between the flux-limited diffusion approximation in low-optical depth regimes and the radiation boundary condition described in \S 2.4.2 introduces too much flux along $r$, and these boundary cells cool more than expected.  When the flux along the $r$ direction
is set to zero for the boundary cells, the temperature inversion is corrected.

In addition to temperature contours, the fluxes along the $r$ and $z$ axes are shown in Figure 3.9 for the high and low optical depth cases.  The fluxes for the high optical depth case follow the expected $x^{-2}$ profile.  There is some deviation near $r=z=0$, but this is likely due to the cylindrical energy input source near the center of the sphere.  The low optical depth case shows much more deviation from the expected profile.  As discussed in Chapter 6, routines that include complete cell-to-cell coupling in all directions is a critical next step for radiation algorithms in disk codes.  Although it should be noted that for both optical depths, the total luminosity of the sphere reaches the expected value.

\begin{figure}
   \centering
   \includegraphics[width=7cm]{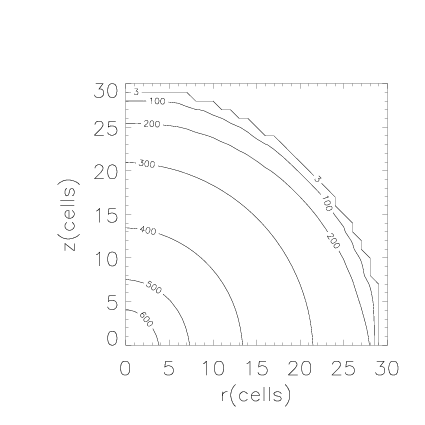}\includegraphics[width=7cm]{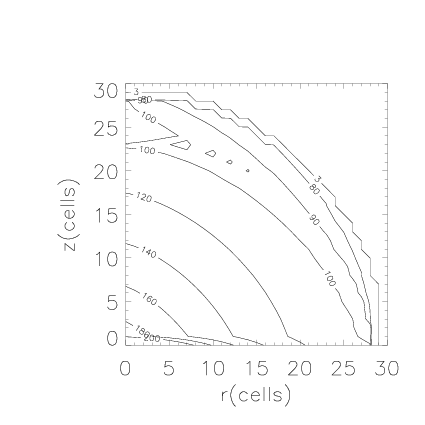}
   \includegraphics[width=7cm]{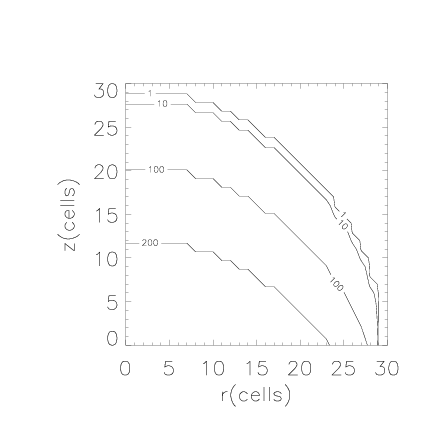}\includegraphics[width=7cm]{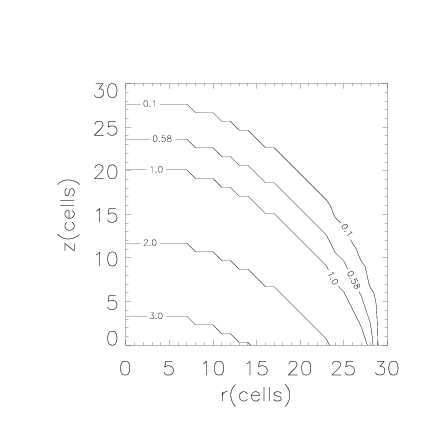}
   \caption[2D test temperature and $\tau$ contours]
   {Temperature (left panels) and vertically integrated $\tau$ (right panels) plots for the 2D test of the BDNL algorithm.  
   The high optical depth test (top) shows a slight flattening of the
   temperature contours, but is satisfactory overall.  The low optical depth test (bottom) shows much more flattening
   of the temperature contours and odd behavior along the $r$ axis and near the $z$ axis at the photosphere (see text). } 
\end{figure}

\begin{figure}
   \centering
   \includegraphics[width=11cm]{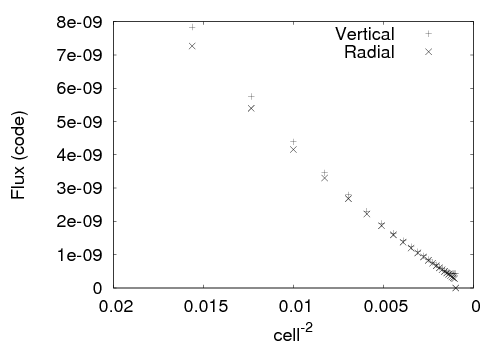}
   \includegraphics[width=11cm]{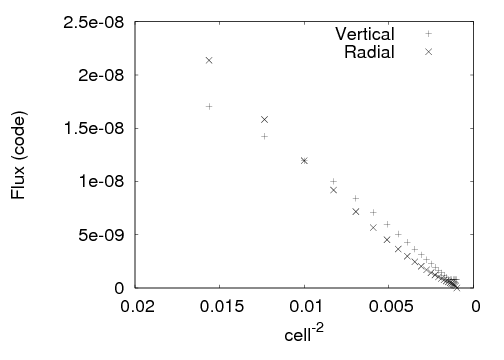}
   \caption[Flux profiles for the 2D  high and low optical depth tests]
   {Vertical and radial flux profiles for the 2D  high (top) and low optical depth (bottom) tests.  The high optical depth profiles converge to 
   an $x^{-2}$ profile, as expected.  There is a deviation (only shown in part) as one approaches $r=z=0$, but this may be due to the nonspherical energy  source.  The low optical depth case shows odd behavior.  For the flux along the $z$ axis, the flux strongly deviates (only shown in part) from the expected profile near $r=z=0$, but converges as one moves outward.  In contrast, the flux along the $r$ axis follows
   an $x^{-2}$ profile near the center, but deviates as one moves outward.  The kinks at the very end of each flux profile are due to the edge of the sphere.  Although not understood at the moment, the 
   magnitudes of the flux profiles are also different.} 
\end{figure}

\chapter{INITIAL MODEL GENERATION}

In this Chapter I describe the various initial models that are used in
these studies. 

\section{The Mej\'ia, Light, and Moderate Disks}

As discussed in Mej\'ia (2004), Mej\'ia et al.~(2005), and Cai (2006),
initial conditions can be generated as star+disk equilibrium models using a
modified Haschisu (1986) self-consitent field (SCF) relaxation method.  In this
method, the density and angular momentum distributions, assuming a polytropic
equation of state, are iteratively solved until convergence to the desired model
is reached. A star+disk model is created by specifying $M_{\rm star}/M_{\rm total}$,
$r_{\rm pol}/r_0$, and $p$, where $r_0$ is the outer, initial edge of the disk, $r_{\rm pol}$ is
the polar radius of the protostar, and $\Sigma(r)\sim r^{-p}$ 
is the desired surface density. During the iterations, the angular momentum
distribution is adjusted to create the desired law for the disk, while the
protostar is specified to be either a uniformly rotating or non-rotating
polytrope.  The goodness of the equilibrium model is evaluated by measuring the
virial test of the model: $ VT = \big | E_{\rm grav} + 2 E_{\rm rot} + 3 \mathcal{P}\big | /
\big | E_{\rm grav}\big|$, where $\mathcal{P}=\int p dV$ and $VT = 0$ for perfect
equilibrium.  

Because the star+disk model is created in dimensionless code units, the model is
scalable.  However, scaling the disk beyond a few AU results in an unrealistic
 size for the central star.  In addition, the hydrodynamics time step is restricted by the
Courant condition, and so the hot star severely limits the time step size,
which is problematic for long disk simulations.  To obviate these difficulties,
the central star is removed from the grid, but its gravitational potential is
stored (see Mej\'ia 2004) or replaced by a point mass potential with the origin
at the center of the grid.  Replacing the potential with a point mass only
creates small disturbances in the equilibrium model, and it allows for easy
expansion of the grid.  For implications regarding keeping the central mass'
potential fixed, see Chapter 6.

For the studies presented here, three of the equilibrium models were
generated by the SCF method: the Mej\'ia disk, the
moderate disk, and the light disk.  Their model parameters are given in Table
3.1.  The Mej\'ia disk has been used for the initial conditions in multiple
studies, including Mej\'ia (2004), Mej\'ia et al.~(2005), Boley et al.~(2005),
Cai et al.~(2006, 2007), Cai (2006), Boley \& Durisen (2006), Boley et
al.~(2006, 2007a, 2007c). The minimum $Q$ for the Mej\'ia disk is about 2.  The
light disk and moderate disks were used as the initial conditions for the shock
bore studies (see Chapter 5) discussed in Boley et al.~(2005) and Boley \& Durisen (2006).  Both
disks are stable against GIs, with the minimum $Q\approx 10$ for the light disk and 5 for
the moderate disk. For Boley \& Durisen (2006), the moderate disk was cooled in a 2D
version of the SV (Pickett 1995; Mej\'ia 2004) so that the minimum $Q\approx 2$.
Figure 4.1 shows density contour plots scaled to the central density of the now
removed protostar, $\rho_c$, for each model.

\section{Flat-Q Disk} The flat-Q disk is created for evolution 
with a variable first adiabatic index with an ortho/para = 3:1 (Chapter 2). 
To create an equilibrium model
with the SCF method, significant modifications would be required. Moreover,
even with a simple adiabatic index, it was very difficult for
the SCF method to converge to an equilibrium solution for the desired model
parameters (below).  
Because there was little time for extensive code modifications
and because computer cycles were readily available on department rendering
machines, a far less eloquent method was used.  

Consider a disk that is vertically polytropic, is axisymmetric, has a constant
$\gamma$, and is roughly Keplerian.  Also assume that self-gravity is negligible for the
vertical structure. 
For this case, I find that
\begin{eqnarray}\Sigma(r)&=&\pi^{-(3\gamma+1)/4}\left(\frac{2}{\gamma-1}
\left(\frac{\Gamma[\gamma/(\gamma-1)]}{\Gamma[(3\gamma-1)/(2\gamma-1)]}\right)^2
\right)^{(1-\gamma)/4}\\\nonumber &\times&\left(\gamma
K(r)\right)^{1/2}\left(GQ(r)\right)
^{-(\gamma+1)/2}\Omega(r)^\gamma,\end{eqnarray}
where $K$ is the polytropic coefficient for any given $r$, $G$ is the
graviational constant, and $\Omega$ is the Keplerian angular speed.  To create a
model, one needs to specify power laws for $\Sigma$ and $Q$, $\Sigma$ and $K$, or
$K$ and $Q$.  In practice, I specify $\Sigma$ in order to control the mass of
the disk and then choose either a constant $K$ profile or a constant $Q$
profile. Once the profiles have been determined, the disk structure can be
derived by using the following equations.
\begin{eqnarray}\rho(z) &=& \rho_0\left(1-z^2/h^2\right)^{1/(\gamma-1)},\\ 
h^2 & = & \frac{2\gamma K}{\Omega^2(\gamma-1)}\rho_0^{\gamma-1},{\rm~and}\\ 
\rho_0 &=&\left(\Sigma \Omega\left(\frac{\gamma-1}{2\gamma \pi K}\right)^{1/2}
\frac{\Gamma[(3\gamma-1)/(2\gamma-2)]}{\Gamma[\gamma/(\gamma-1)]}\right)^
{2/(\gamma+1)},
\end{eqnarray}
where $h$ is the disk scale height. Note that when the $K$ is
constant and $Q$ is constant, the surface density profile follows a
$r^{-1.5\gamma}$ profile.

Equations (4.1-4.4) allow one to solve for a model disk.  However, the disk will
not be in equilibrium because a Keplerian rotation was assumed and self-gravity
was neglected.  Moreover, a constant $\gamma$ was still assumed.
Why use this analytic method over the SCF method?  (1) Both
methods are guaranteed to produce a model that is out of equilibrium when loaded
onto the grid if a variable $\gamma$ is introduced. (2) The SCF method has
difficulty converging models for disk parameters desired for the flat-Q case. (3) The model
is intended for a radiation hydrodynamics simulation, and although both the SCF
method and the analytic model produce a polytropic model in rough equilibrium,
the models are out of radiative equilbrium. Because both methods require
some relaxation in the hydrodynamics code, the easiest and most straightforward
method is the analytic model.  Unfortunately, the relaxation of the analytic
model to an equilibrium structure proves to be somewhat difficult, and requires
employing numerical tricks, e.g., alternating between dampening the momentum in
the radial and vertical directions. Some oscillations of the outer disk could
not be dampened, and remain in the initial model. 

By using the analytic method described above, the flat-Q model is produced with
$\gamma=5/3$, $\Sigma\sim r^{-2}$, $Q=\rm constant$, and $K\sim r$.  The
total disk mass is approximately $0.09 M_{\odot}$, and in accordance with the
$\Sigma$ and $K$ profiles, the midplane temperature profile $T\sim r^{-1}$.
The initial disk has a radius of 10 AU, and the surface density at 5 AU is 3300
$\rm g~cm^{-2}$.  The actual $Q$ for the model after relaxation is about 2 everywhere. 

In order to study spiral waves, shocks, and mass
transport when a dead zone becomes highly unstable,  the flat-Q model is
modified to drive a region of the disk toward instability.  Mass is added to
the disk, linearly in time, in a Gaussian profile, with the peak centered on 5
AU.  The internal energy density is adjusted so that the specific internal
energy remained the same. The radial
FWHM of the Gaussian is chosen to be 3 AU, which is roughly the size of the
most unstable wavelength $\lambda_u\approx 2\pi h\approx 0.2\pi r$ (Durisen et al.~2007b). 
Mass is added until nonaxisymmetry is visible in midplane density images, which
took approximately 190 yr, i.e., 6 orp of evolution (1 orp is defined as 1 outer
rotation period at 10 AU for this model). The total mass added is
about 0.08 $M_{\odot}$, and so if one were to imagine a corresponding accretion
rate, it would be about $4\times 10^{-4}~M_{\odot}$ yr$^{-1}$. It is probably
unphysical to add so much mass to the disk so quickly without increasing the temperature of
the disk.  However, I remind the reader that the study is 
a numerical experiment, as described in Chapter 7.  

The mass build-up drops $Q$ below unity in a narrow region around 5 AU just
before the disk bursts. Because the ring is grown over 6 orp, 
the instabilily is probably not overshot. The mass-weighted average $Q_{\rm av}$ over
the FWHM of the Gaussian centered at 5 AU is approximately unity when the ring
becomes unstable.  This implies that $Q$ must approach unity for a
spatial range of at least an unstable wavelength before GIs will activate in a
disk.    
 
Figure 4.2 shows density contours for the flat-Q model and the flat-Q model with
the mass concentration.  As in Figure 4.1, the density is normalized to a
$\rho_c$, which is based on the typical magnitude difference between
the maximum disk density and maximum protostar central density in the SCF models.

\begin{table}
\center
\begin{center} 
\caption[Model parameters for initial disks generated by the SCF method]
{Model parameters for initial disks generated by the SCF method.} 
\begin{tabular}{l l l l l l}\\\hline
 Model &
$M_{\rm star}/M_{\rm total}$ & $r_{\rm pol}/r_0$ & $p$ & $VT$ &$Q_{\rm min}$\\\hline 
Mej\'ia & 0.875 & 20 & 0.5 & 2.6e-3 & 2\\ 
Light  & 0.982 & 10 & 0.5 & 6.5e-4 & 10 \\ 
Moderate & 0.964 & 10 & 0.5 & 6.3e-4& 5\\
Cooled Moderate & 0.964 & 10 & 0.5 & 6.3e-4 & 2\\ \hline 
\end{tabular} 
\end{center} 
\end{table}

\begin{figure} \begin{center}
\includegraphics[width=3in]{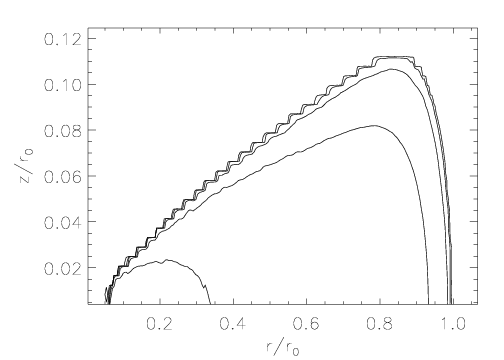}\includegraphics[width=3in]{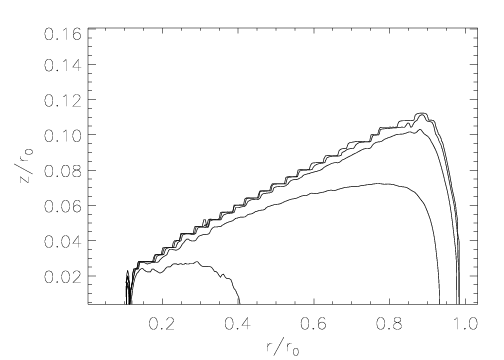}
\includegraphics[width=3in]{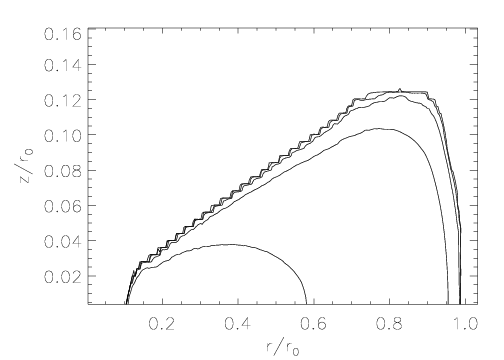}\includegraphics[width=3in]{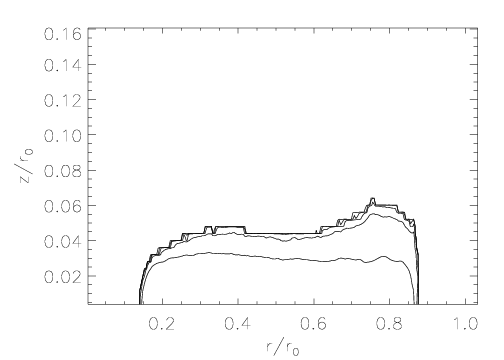}
\caption[Density contours for the Mej\'ia disk, the light disk, the moderate disk, and the
cooled moderate disk]
{Density contours, with $\rho/\rho_c =$ 1.e-4,~1.e-5,~1.e-6,~1.e-7, and
1.e-8, for the Mej\'ia disk, the light disk, the moderate disk, and the cooled
moderate disk, from the top left to bottom left respsectively. The odd density
enhancement at the inner edge of the light disk is an artifact of removing the
protostar; it is partly accreted to the protostar and partly subsumed into the
disk during the beginning of the simulation.} \end{center} \end{figure}

\begin{figure} \begin{center}
\includegraphics[width=3in]{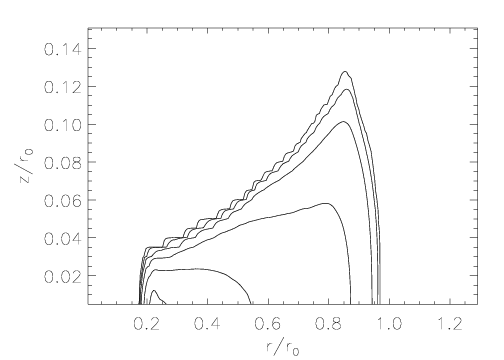}\includegraphics[width=3in]{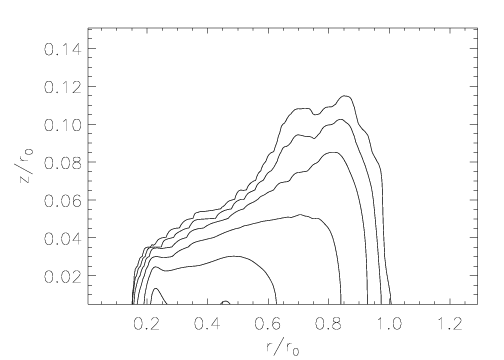}
\caption[Similar to Figure 4.1, but for the flat-Q model]
{Similar to Figure 4.1, but with $\rho/\rho_c =$
1.e-3,~1.e-4,~1.e-5,~1.e-6,~1.e-7, and 1.e-8 for the flat-Q model. The left panel is the initial
flat-Q model and the right panel is the same model, but with the density
enhancement.  For these models $r_0=10$ AU.} \end{center} \end{figure}

\chapter{SHOCK BORES}

\section{Introduction}

To address the dynamic evolution of protoplanetary disks properly, the
three-dimensionality of spiral waves and shocks must be understood.  Fully
three-dimensional waves in accretion disks have been studied by several authors
(e.g., Lubow 1981; Lin et al.~1990; Korycansky \& Pringle 1995; Lubow \& Ogilvie
1998; Ogilvie 2002a,b; Bate 2002), typically in the context of tidal forcing.
It has been noted (see Lubow \& Ogilvie 1998, hereafter LO98) that these waves
act like fundamental modes ($f$-modes), which correspond to large surface
distortions in the disk.  These waves can affect the disk's evolution through
wave dissipation at the disk's surface and through gap formation (Bryden et
al.~1999; Bate et al.~2003).  The ability of radially propagating waves to
transport angular momentum, and also influence gap formation, is dependent on
the thermal stratification of the disk (e.g., Lin et al.~1990; LO98). Finally,
consider the following time scale argument.  The dynamical time $t_d$ at a given
radius in a disk is about the orbital period, i.e, $t_d= P_{\rm rot}=2\pi\Omega$, 
where $\Omega$ is the orbital frequency.  The vertical time scale can be thought
of as the time it takes for a wave launched at the midplane to propagate to the
disk surface and back.  If $h$ is the disk scale height, and $c$ is the midplane
sound speed, then the vertical time scale $t_v = 2h/c \approx 2\Omega =
P_{\rm rot}/\pi$.  To within a factor of order unity, which depends on the thermal
stratification of the disk, the time scales are commensurate. Thus the
vertical direction, although often ignored, plays a very important role in disk
dynamics.

Independent work in the context of gravitational instabilities (GIs) has also
found that once GIs become well developed, e.g., in the asymptotic state
of Mej\'ia et al.~(2005), the spiral waves resulting from GIs in the disk also behave
like $f$-modes and involve large surface distortions (Pickett et al.~1998, 2000,
2003; Durisen et al.~2003).   Furthermore, it was noted in these studies that the
spiral wave activity is highly nonlinear and involves many modes of comparable
strengths (see also Gammie 2001; Lodato \& Rice 2004).  In the asymptotic disk,
sudden increases in disk scale height occur. These splashes of disk material
seem to be related to shocks in the disk and could have important consequences
for disk evolution through the generation of turbulence, breaking waves, and
chondrule formation (Boley et al.~2005, 2006).  In order to understand this
shock-related splashing, which is evidently highly nonlinear, an approach other
than $f$-mode analysis is required.

Martos \& Cox (1998, hereafter MC98) investigated shocks in the Galactic disk,
where the otherwise isothermal equation of state (EOS) is stiffened by magnetic
fields.  Their findings indicate that shocks occurring in semi-compressible
fluid disks have characteristics of hydraulic jumps and behave, in part, like
gravity modes ($g$-modes).   A classical hydraulic jump occurs in an
incompressible fluid, where the only way to reduce the kinetic energy of fluid
elements coming into the wave is to convert it to gravitational potential or
turbulent energy (e.g., Massey 1978).  Examples of hydraulic jumps can be found
in spillways wherever  there is an abrupt change in slope and the slowly moving
water has a greater height than the rapidly moving water.  The hydraulic jump is
the abrupt change in the height of the flowing water.  The same phenomenon can
be exhibited in a disk because, like the flowing water, the unperturbed disk is
usually hydrostatic in the $z$ direction and  work must be done against gravity
to expand the disk in the vertical direction.  Abrupt changes in the scale
height of a disk are thus possible for similar reasons.  MC98 present an
analytical theory, along with two-dimensional magnetohydrodynamics simulations,
in the context of hydraulic jump conditions, and note that the jumping gas can
lead to high-altitude shocks when the jumping material crashes back onto the
disk.  After MC98, Gom\'ez \& Cox (2002) simulated a portion of the galactic
disk in three dimensions, noting a behavior of the gas similar to that found by
MC98.  However, their analyses neglect self-gravity, which can
affect the morphology of a global shock.

I refer to these hybrids between hydraulic jumps and spiral shocks as {\it 
shock bores}. Following Boley \& Durisen (2006), I lay out in this 
Chapter a theory that combines the Rankine-Hugoniot 
jump shock conditions for both the adiabatic and isothermal
cases with the classical hydraulic jump conditions. 
 Fully three-dimensional
hydrodynamics simulations are discussed in \S 5.3 to illustrate simple
cases of spiral shocks in protoplanetary disks. I discuss some implications of
these results in \S 5.4 and summarize the main conclusions of this
study in \S 5.5.

\section{Shock Bores\label{hstheory}}
\subsection{Plane-Parallel Approximation}
  In a disk, abrupt vertical expansions will
typically occur after a shock.  The reason for a shock bore can be
understood by considering hydrostatic equilibrium (HE) and the EOS.  For
simplicity, assume that (1) the shock is planar, (2) it is propagating in the
$x$ direction, (3) the disk is vertically stratified in a vertical direction $z$
perpendicular to the $x$ direction, with the pre-shock region in vertical HE,
and (4), except for the discontinuity at the shock front,  ignore variations
in the $x$ direction.  With conditions (3) and (4) one may write
\begin{equation} \frac{1}{\rho}\frac{dp}{dz}=-\frac{d\Phi}{dz}
\end{equation}
for the pre-shock flow, where $\Phi$ is the total gravitational potential, $p$
is the pressure, and $\rho$ is the density.  Consider first the case of an
adiabatic shock.  Then for the Rankine-Hugoniot shock conditions,
\begin{equation}
\frac{p_2}{p_1}=\frac{2\gamma\mathcal{M}^2}{\gamma+1}-\frac{\gamma-1}{\gamma+1}\rm~and
\label{pfull:5.2}\end{equation} 
\begin{equation}
\frac{u_2}{u_1}=\frac{\rho_1}{\rho_2}=\frac{\gamma-1}{\gamma+1}+\frac{2}{\gamma+1}\frac{1}{\mathcal{M}^2},\label{rhofull:5.3}\end{equation}
where subscripts 1 and 2 represent the pre- and post-shock regions,
respectively, and where the gas speed relative to the shock is $u$, the ratio of
specific heats is $\gamma$, and $\cal{M}$ is the Mach number given by
\begin{equation} \mathcal{M}^2=\frac{\rho_1 u_1^2}{\gamma
p_1}.\label{mach:5.4}\end{equation}
If the pre-shock gas is initially in vertical HE, what is the state of vertical
force balance behind the shock?  For the adiabatic case, using equations (5.2)
and (5.3), one can write the ratio of the post-shock and pre-shock pressure body
forces as
\begin{eqnarray}\frac{1}{\rho_2}\frac{dp_2}{dz}\left(\frac{1}{\rho_1}\frac{dp_1}{dz}\right)^{-1} & = & \left(\frac{2\gamma\mathcal{M}^2}{\gamma+1}-\frac{\gamma-1}{\gamma+1}\right)
\left[\frac{\gamma-1}{\gamma+1}+\frac{2}{\left(\gamma+1\right)\mathcal{M}^2}\right]\\
& =& \frac{2\gamma\mathcal{M}^4\left(\gamma-1\right)-\mathcal{M}^2\left(1-6\gamma+\gamma^2\right)-2\left(\gamma-1\right)}{\mathcal{M}^2\left(\gamma+1\right)^2}.\label{pressurefraction:5.4}\end{eqnarray}
For the gravitational body force, let $\Phi_{\ast}$ be the background potential,
presumably due to the primary, and let $\Phi_{g_1}$ and $\Phi_{g_2}$ be the
contribution to $\Phi$ from the gas self-gravity in the pre- and post-shock
regions.  The ratio of the potential gradients then becomes
\begin{eqnarray}\frac{d\Phi_2}{dz}\left(\frac{d\Phi_1}{dz}\right)^{-1}
& =&  \frac{d_z\Phi_{\ast}+d_z\Phi_{g_2}}{d_z\Phi_{\ast}+d_z\Phi_{g_1}}\\
& = & \frac{q+d_z\Phi_{g_2}\left(d_z\Phi_{g_1}\right)^{-1}}{q+1},\label{potalmost:5.5}
\end{eqnarray}
where
\begin{eqnarray} q &=& \frac{\mathrm{d}\Phi_*}{\mathrm{d}z}\left(\frac{\mathrm{d}\Phi_{g_1}}{\mathrm{d}z}\right)^{-1}\label{qratio:5.6}\end{eqnarray}
 is the ratio of the background  and the pre-shock gas potential gradients.
Condition (4)  permits one to write
$\nabla^2\Phi\approx\nabla_z^2\Phi\propto\rho$; therefore, the ratio of the
self-gravity potential gradients is equivalent to the ratio of the densities.
This yields
\begin{eqnarray}\frac{d\Phi_2}{dz}\left(\frac{d\Phi_1}{dz}\right)^{-1}
 = \frac{q+a}{q+1},\label{potfraction:5.7}
\end{eqnarray}
where
\begin{eqnarray}a=\frac{\left(\gamma+1\right)\mathcal{M}^2}{2+\left(\gamma-1\right)\mathcal{M}^2}.\label{adef:5.8}\end{eqnarray}
Define the jump-factor $J_f$ to be the ratio of equations (5.5) and (5.7):
\begin{eqnarray}J_f & \equiv &\frac{1}{\rho_2}\frac{dp_2}{dz}\left(\frac{1}{\rho_1}\frac{dp_1}{dz}\right)^{-1}\left[\frac{d\Phi_1}{dz}\left(\frac{d\Phi_2}{dz}\right)^{-1}\right]\label{jf1}\\
& = & \frac{2\gamma\mathcal{M}^4\left(\gamma-1\right)-\mathcal{M}^2\left(1-6\gamma+\gamma^2\right)-2\left(\gamma-1\right)}{\mathcal{M}^2\left(\gamma+1\right)^2}\left[\frac{q+1}{q+a}\right]\label{jf2:5.9}.\end{eqnarray}
Note the limits of equation (5.13).  When self-gravity dominates ($q\rightarrow
0$) and $\mathcal{M}$ is large,
\begin{eqnarray}J_f \rightarrow \frac{2 \gamma \mathcal{M}^2\left(\gamma-1\right)^2 }
{\left(\gamma+1\right)^3},\label{jumplimit1:5.10}\end{eqnarray}
and, when the background potential dominates $(q\rightarrow\infty)$ and
$\mathcal{M}$ is large,
\begin{eqnarray}J_f \rightarrow \frac{2 \gamma \mathcal{M}^2\left(\gamma-1\right) }
{\left(\gamma+1\right)^2}.\label{jumplimit2:5.11}\end{eqnarray}
When self-gravity is negligible, $J_f$ becomes the post- and pre-shock
temperature ratio for a purely adiabatic shock. To understand the significance
of $J_f$, consider the vertical acceleration of the gas in the post-shock
region:
\begin{eqnarray}a_z=\frac{dv_z}{dt}&=&-\left(\frac{1}{\rho_2}\frac{dp_2}{dz}+\frac{d\Phi_2}{dz}\right).\label{accelpart1:5.12}
\end{eqnarray}
Using equation (5.12) in (5.16) yields
\begin{eqnarray}a_z=\frac{dv_z}{dt}&=&-
\frac{d\Phi_{2}}{dz} \left[J_f \frac{1}{\rho_1}\frac{dp_1}{dz} \left( \frac{d\Phi_{1}}{dz} \right)^{-1}+1 \right]\\
& = &\frac{d\Phi_{2}}{dz}\left(J_f-1\right).\label{accelpart2:5.13}\end{eqnarray}
In the limit of no self-gravity, where $q\rightarrow\infty$, $a_z\approx\Omega^2
z \left(J_f-1\right)$ for a thin disk, where $\Omega$ is the Keplerian circular
angular speed.  Equation (5.18) demonstrates the physical meaning of $J_f$.
When $J_f > 1 $, the gas is overpressured, and it will expand vertically, while
for $J_f < 1$, self-gravity causes the gas to compress.  If the shock is truly
strong, an expansion will always occur in the adiabatic case ($\gamma>1$).
However, self-gravity and $\gamma$ alter the strength of the adiabatic vertical
acceleration (see Fig.~5.1).   For low $\gamma$ and $q$, regimes with  $J_f < 1$
exist for low $\mathcal{M}$.

Now consider the case of an isothermal shock, which reduces the shock conditions to
\begin{equation} \frac{p_2}{p_1}=\frac{\rho_2}{\rho_1}=\mathcal{M}^2.\label{isofull:5.13}\end{equation}
For these conditions with the same assumptions as stated for the adiabatic case,
the resulting $J_f$ is
\begin{equation}J_f=\frac{q+1}{q+\mathcal{M}^2}.\label{jfiso:5.14}\end{equation}

\begin{figure}
\begin{center}
\includegraphics[width=3in]{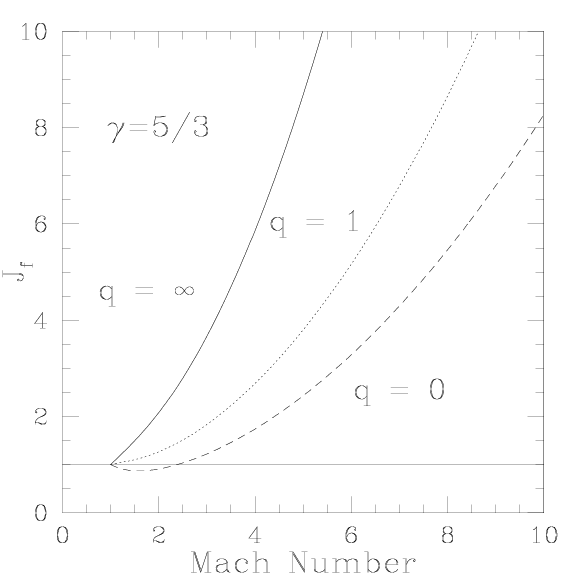}\includegraphics[width=3in]{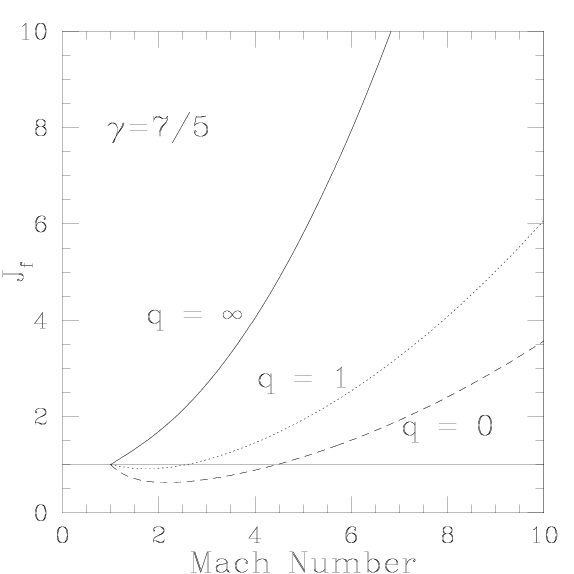}
\caption[Jump-factor curves]
{The three curves show the adiabatic $J_f$ with respect to $\mathcal{M}$
for $q\rightarrow\infty\rm\ (solid), \ 1\ (dotted),\ and\ 0\ (dashed)$,
corresponding to no self-gravity, equal background and self-gravity, and fully
self-gravitating, respectively.  Notice that for $\gamma=5/3$ (left), the disk
only collapses for the self-gravity dominated case and only at low Mach numbers.
However, when $\gamma=7/5 $ (right), the compressions are noticeable at low
$\mathcal{M}$ even when self-gravity is only just comparable to the background potential. }
\end{center}
\end{figure}

\noindent What one can immediately see from equation (5.20) is that if self-gravity is
important (typically $q\lesssim 100$), one
expects an isothermal gas to {\it compress} vertically in the post-shock region
because $\mathcal{M} \ge 1$. If $q\rightarrow\infty$, spiral shocks in an
isothermal disk will be essentially two-dimensional in the sense that vertical
HE is maintained and the waves propagate as spiral density waves without any
change in the disk scale height.   It should therefore be noted that spiral
waves in protoplanetary disks will only behave like pure density waves when the
self-gravity of the disk is negligible and the EOS is nearly isothermal.

To predict the height of a shock bore, consider a classical hydraulic jump.  The
height of a classical hydraulic jump can be predicted using the {\it Froude}
number $F$ of the pre-jump region, which measures whether the flow is {\it
rapid}, $F>1$, or {\it tranquil}, $F < 1$ (e.g., Massey 1970).  Assuming that
the jump is non-dissipative, the height of the post-jump flow $h_2$ can be found
by
\begin{equation} \frac{h_2}{h_1} = -\frac{1}{2}+\sqrt{\frac{1}{4}+2F^2}\label{classicjump:5.15},\end{equation}
where $h_1$ is the height of the fluid in the pre-jump region.  In the limit
that $F \gg1$, $h_2/h_1=\sqrt 2 F$.  This classical jump result can be used as a
model for understanding the maximum height a shock bore reaches during the
post-shock vertical expansion. First, consider the definition of the Froude
number,
\begin{equation}F \equiv \frac{u_1}{\sqrt{g h_1}},\label{froudedef:5.22}\end{equation}
where $g$ is the acceleration of gravity and $u_1$ represents the pre-jump flow
in the frame of the jump.  Second, consider a non-self-gravitating disk. In this
limit, one may write $g\approx-\Omega^2z$ and $h_1\approx c_s /\Omega$, where
$\Omega$ is the circular speed of the gas, $c_s$ is the midplane sound speed,
and $h_1$ is the scale height of the disk in the pre-shock region. Using these
relations in place of the corresponding terms in equation (5.22) reveals that
for a non-self-gravitating disk, $F\rightarrow\mathcal{M}$.   Any relation for the 
ratio of the scale heights before and after the expansion
should have a behavior similar to equation (5.21), since $F$ and $\mathcal{M}$
are closely related.

To relate the scale heights before and after a shock bore, consider equation
(5.13) to be a measure of the overpressurization in the post-shock gas.  From
this perspective, $a_z$ in equation (5.16) represents a
force field capable of doing work on the gas.  Furthermore, assume that the
difference in the gravitational potentials before the expansion and at the peak
of the expansion measures work done in the expansion.  Such an
approximation allows us to write
\begin{eqnarray} \int_0^{h_1} d z\ a_z  &\approx& \int_0^{h_2} d z\ \Omega^2 z -
\int_0^{h_1} d z\ \Omega^2 z \mathrm{,}\label{scalehapprox:5.17}\\
\frac{h_2}{h_1} &\approx &\sqrt{J_f},\label{scaleheights:5.18}\end{eqnarray}
where the RHS of equation (5.23) is the potential difference of the gas before
and after expansion and the LHS of equation (5.23) is a measure of the work that
can be done by the post-shock overpressure. Note that I neglect self-gravity in
this approximation because for many astrophysical situations $q\gtrsim100$.
This relation has a behavior similar to equation (5.21); when $\mathcal{M}=1$,
$h_2/h_1=1$, and, when $\mathcal{M}\gg1$, $h_2/h_1\sim\mathcal{M}$.  This does
not mean that every fluid element is expected to jump to the new height given by
equation (5.24).  Instead, the scale height, as a characteristic height of the
disk, will be changed.  For example, in equation (5.18), material near the
midplane, where $\partial \Phi_2/\partial z\rightarrow0$, will hardly be
affected by a single jump, while high altitude gas will have the strongest
response.

Equation (5.24) can also be derived using mass and momentum flux arguments
(e.g., MC98).  Mass conservation requires that $\Sigma_1 u_1 = \Sigma_2 u_2$,
and momentum conservation requires that $P_1-P_2$=$\Sigma_1 u_1 (u_2-u_1)$, for
$P_1=\int p_1 dz$.  Again, assume that self-gravity is negligible and that the
structure before and after the shock is homologous when in equilibrium; then
$p=A\rho_{0}h^2$, $P = AB \rho_0 h^3$, and $h=C\Sigma/\rho_0$, where $A$, $B$,
and $C$ are shape factors that depend on the details of the model and $\rho_0$
is the midplane density.  By using mass and momentum conservation, 
\begin{eqnarray}\left( \frac{h_2}{h_1} \right)^2 = \frac{u_2}{u_1}+\frac{\gamma\mathcal{M}^2}{BC}
\frac{u_2}{u_1}\left( 1-\frac{u_2}{u_1}\right).\label{scaleheights2:5.19}\end{eqnarray}

\begin{figure}
\begin{center}
\includegraphics[width=4in]{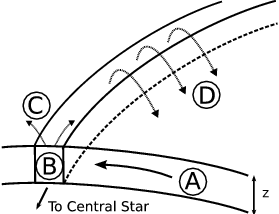}
\caption[Cartoon of a shock bore]
{A cartoon depicting the gas flow in a shock bore in the frame of the
spiral shock inside corotation.  The gas in the pre-shock region flows into the
spiral shock (A).  The shock (B) causes the material to be out of vertical force
balance and a rapid expansion results (C).  Due to spiral streaming and the loss
of pressure confinement, some of the gas will flow back over the spiral wave and
break onto the disk in the pre-shock region at a radius inward from where it
originated (D).  }
\end{center}
\end{figure}

\noindent Using equation (5.3) and assuming $BC\approx1$, equation (5.25) becomes equation
(5.24).  Depending on the EOS, the assumption that $BC\approx 1$ may a bit
inaccurate, but should introduce an error no larger than
about 10\% when using equation (5.24).  For example, in an isentropic gas $BC =\Gamma(1+m)\Gamma(1/2 +
m)/[\Gamma(3/2 + m)\Gamma(m)]$, where $m=\gamma/(\gamma-1)$, which yields
$BC=0.833$ for $\gamma=5/3$, and $BC\rightarrow 1$, as $\gamma\rightarrow1$.

\subsection{A Shock Bore in a Disk\label{realcase}}

 In a disk, shock bores turn into large waves that break onto the disk's
surface, resulting in flow that is considerably more complex than described in
\S 5.2.1.  Shock bores will not only create vertical undulations in the
disk, but will drive fluid elements to large radial excursions from their
circular orbits and result in stirring the nebula, possibly mixing it.  For
reasons I explain below, inside the corotation radius of the spiral shock,
these waves flow back over the spiral shock and break onto the pre-shock flow.
This behavior is confirmed in the numerical simulations in \S  5.3. 
I conjecture that shock bores will also result in breaking waves that
crash onto the pre-shock flow outside the corotation radius, but I do not
compute simulations outside corotation in this study.

For the following discussion, consider the point of view from inside the
corotation radius of the spiral shock.  The development of breaking waves can be
understood by recognizing two effects: When the gas crosses the shock front, the
shock-normal component of a fluid element's velocity will be diminished by a
factor given by the inverse of equation (5.3), while the tangential component
will be preserved, allowing the flow to be supersonic after the shock.  This
leads to streaming along the spiral arms, as demonstrated in streamline
simulations of fluid elements in spiral galaxies (Roberts et al.~1979).   In
addition, when the gas expands upward, the pressure confinement in the direction
normal to the shock front is lost, and the material expands horizontally,
causing some gas to flow back over the top of the shock.  As the jumping gas
moves inward and out over the pre-shock flow, it no longer has pressure support
from underneath and breaks back onto the disk.  The resulting morphology is a
spiral pattern moving through the disk with breaking surface waves propagating
along the disk's surface with the same pattern speed as the spiral wave (see
Fig.~5.2).   This morphology is only expected for a simple shock bore, i.e., there
are no additional waves and shocks except for what are produced by cleanly
defined spiral and breaking waves.  In a real disk with competing spiral waves,
the behavior can be much more complex.

In the simple case, there should also be a radius at which shock bores provide
the strongest corrugation in the disk's surface.  Consider the fluid elements in
the disk to be essentially on Keplerian orbits.  In a simple approximation, a
shock bore produces a vertical perturbation to the fluid element's orbit,
putting it on an inclined orbit leaving and then returning to the midplane in
about half a revolution.  In the inner disk, the orbit period of a fluid element
is much shorter than the pattern period of a spiral wave.  As a result, after a
fluid element encounters the first spiral shock, the pattern will have moved
only a small fraction of a pattern period by the time the fluid element
encounters another arm; all fluid elements end up elevated between shock
passages.   As one moves radially outward in the disk, the advancement of the
spiral shock becomes important and shock bores develop into breaking waves.
However, as one moves toward the corotation radius, the shocks become weak, and
shock bores are suppressed.


\subsection{Initial Models and Perturbations}

To study shock bores numerically, I generate spiral shocks by applying
nonaxisymmetric perturbations to two equilibrium axisymmetric disks. I use the
Mej\'ia and cooled moderate disks as initial conditions (see Chapter 4).  The
outer radius of each disk is scaled to 6 AU, and the central star is set to 1
M$_{\odot}$.  The resulting inner holes for the two models are about 0.3 AU and
0.6 AU in radius.  To relate these disks to Solar Nebula models, suppose that
$\Sigma(r)$ became $r^{-1.5}$ in the unmodeled disk outside 6 AU (see Lissauer
1987).  Then the total disk mass would be 0.82
M$_{\odot}$ out to 40 AU for the Mej\'ia disk and 0.21 M$_{\odot}$ for the
cooled moderate disk.  The reader should keep in mind that these disks are only
meant to demonstrate the dynamics of shock bores.   

Since shock bores can become very complex, especially when multiple jumps occur
near each other, I stimulate a single, two-armed spiral wave.  This is done using two
methods.  The first method forces spiral waves in the Mej\'ia disk by adding a
$\cos 2\varphi$ potential perturbation $\Phi_p$ centered near 5 AU with a radial
FWHM of about 0.5 AU.  This potential perturbation has the form
\begin{equation}\Phi_p\left(r,\varphi ; t\right)= A \cos\left(2\left(\varphi - \Omega_p t\right)\right)\cos^2\left(\pi | r-r_p | \Delta R^{-1}\right), \label{potpert:5.20}\end{equation}
where $r_p = 5$ AU defines the pattern rotational speed $\Omega_p$, which is
assumed to be Keplerian, $A$ is some scaling parameter, and $\Delta R=1$ AU. 
%
 %
This perturbation creates a well-defined, two-armed spiral that reaches all the way down to
the central hole (see Fig.~5.3) and does not change the center of mass.   Since the
perturbation is localized to 5 AU, the change in the potential far from 5 AU,
which stimulates the waves, can be thought of as  due to concentrations of mass at the
potential minima.  In this way, the $\cos 2\varphi$ potential perturbation
behaves like two stubby spiral arms that could be produced by gravitational
instabilities (see, e.g., Boss \& Durisen 2005a,b).

For the cooled moderate disk, a different $\cos 2\phi$ perturbation is applied.
Because the outer portion of the modeled disk has low mass, the stubby spiral
arms produced by equation (5.26) do not generate strong
shocks.  Instead, I place two 2.5 M$_{\mathrm{J}}$ corotating point masses at
$r = 5.2$ AU in the midplane of the disk, separated azimuthally by $\pi$
radians.  Again, this method keeps the center of mass of the system at the
center and creates two well-defined spiral arms, but it does allow for
additional, small arms to form.    Although I am treating the perturbers as
point masses, one should not interpret them strictly as protoplanets, because
any such perturber would open a gap within a few orbits (Bate et al.~2003).
Instead, envision these point masses to be transient clumps, formed by
gravitational instabilities in a surface density enhanced ring, perhaps a dead zone (see Chapter 7).

\subsection{Thermal Physics}

The behavior of shock bores depends on the EOS, as demonstrated in \S 5.2.1.
 To test the predictions of the shock bore theory, the Mej\'ia
disk is evolved two ways. (1) An energy equation that includes some cooling
plus heating by bulk viscosity is used with an EOS $p =
(\gamma-1)\epsilon$, where $\epsilon$ is the internal energy density and
$\gamma=5/3$. (2) An isothermal EOS is used, where
$p(r,\varphi,z)=\rho(r,\varphi,z)T_{0}(r)$ and $T_{0}$ represents the initial
axisymmetric midplane temperature.  The cooled moderate disk is evolved only with
an energy equation.  When the energy equation is used, a gentle volumetric
cooling rate with a constant cooling time is applied to the disk to partially
balance shock heating.  Without any cooling, the disk heats up in less than a
pattern period and surface waves are suppressed.  The cooling time for the
Mej\'ia disk is set to four pattern periods at the perturbation radius.  A
constant cooling time is also used for the cooled moderate disk to balance shock
heating, but with a cooling time of six pattern periods because the shocks are
weaker.

\begin{figure}
\begin{center}
\includegraphics[width=4in]{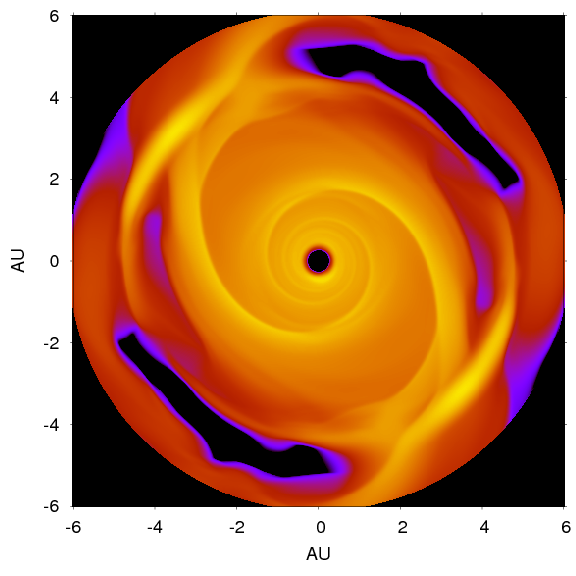}
\caption[Mej\'ia disk with perturbation]
{Logarithmic surface density colorscale spanning 4.5 orders of magnitude for the Mej\'ia disk.  Flow in this diagram is
counterclockwise. }
\end{center}
\end{figure}

\newpage
\begin{sidewaystable}
\begin{center}
\caption[Summary of parameters for disks in shock bore study]
{Summary of the parameters for the simulations presented in this study.
M$_d$ is the disk mass inside 6 AU, {\it Pert} is the type of perturbation used,
and $t_{\mathrm{cool}}$ is given in pattern periods.
The last column contains the number of cells used  for each coordinate.
Although two of the calculations have more $z$ cells, the spatial resolution is
the same.}
\vspace*{.25in}
\begin{tabular}{llllllll}\hline
Disk & Therm.\ Phys.\ & M$_d$/M$_{\odot}$ & Pert & $r_p$ (AU)& $Q_{\rm min}$&
$t_{\mathrm{cool}}$ & ($r$, $\varphi$, $z$) \\\hline
Mej\'ia & Energy Eq.\ & 0.143 & $\cos 2\varphi$ & 5.0 & 2 & 4 & (256, 512, 32)\\
&             Isothermal    & 0.143 & $\cos 2\varphi$& 5.0 & 2 & -  & (256, 128, 64)\\
Cooled moderate & Energy Eq.\ & 0.0370 & $2\times$2.5 M$_{\rm J}$& 5.2 &  2 & 6&
(256, 512, 64)\\\hline \end{tabular}
\label{models}
\end{center}
\end{sidewaystable}
\clearpage

\subsection{Fluid Element Tracer}

To investigate the shock parameters in both disks, the post-analysis
fluid element tracer (\S 2.5) is employed, with data roughly every 1/400th of
an orp.  Since the data storage demands are
extremely high to achieve this type of spatial and temporal resolution, the
disks are first evolved at a low resolution, 128 aziumthal zones, until spiral
shocks develop.  Once the shocks form a quasi-steady morphology in the midplane,
which takes about one or two pattern rotations, I switch to high resolution,
512 azimuthal zones, and continue the simulations for a little more than half a
pattern period.  The isothermal calculation is only done at low resolution,
because the low resolution is sufficient to demonstrate the general behavior of
shock bores under isothermal conditions. 

\section{Results}\label{results}

In the simulations with heating and cooling (Energy Eq.\ in Table 5.1), the disk
perturbations result in shock bores along the spiral arms.  The jumping material
creates very large waves that extend to nearly twice the height of the disk, and
the only steady features in the simulations are the presence of the shock fronts
and the shock bores.  The shock bores vary in strength, and the nonlinearity of
the waves creates transient features.

 \begin{figure}
\begin{center}
\includegraphics[width=6in]{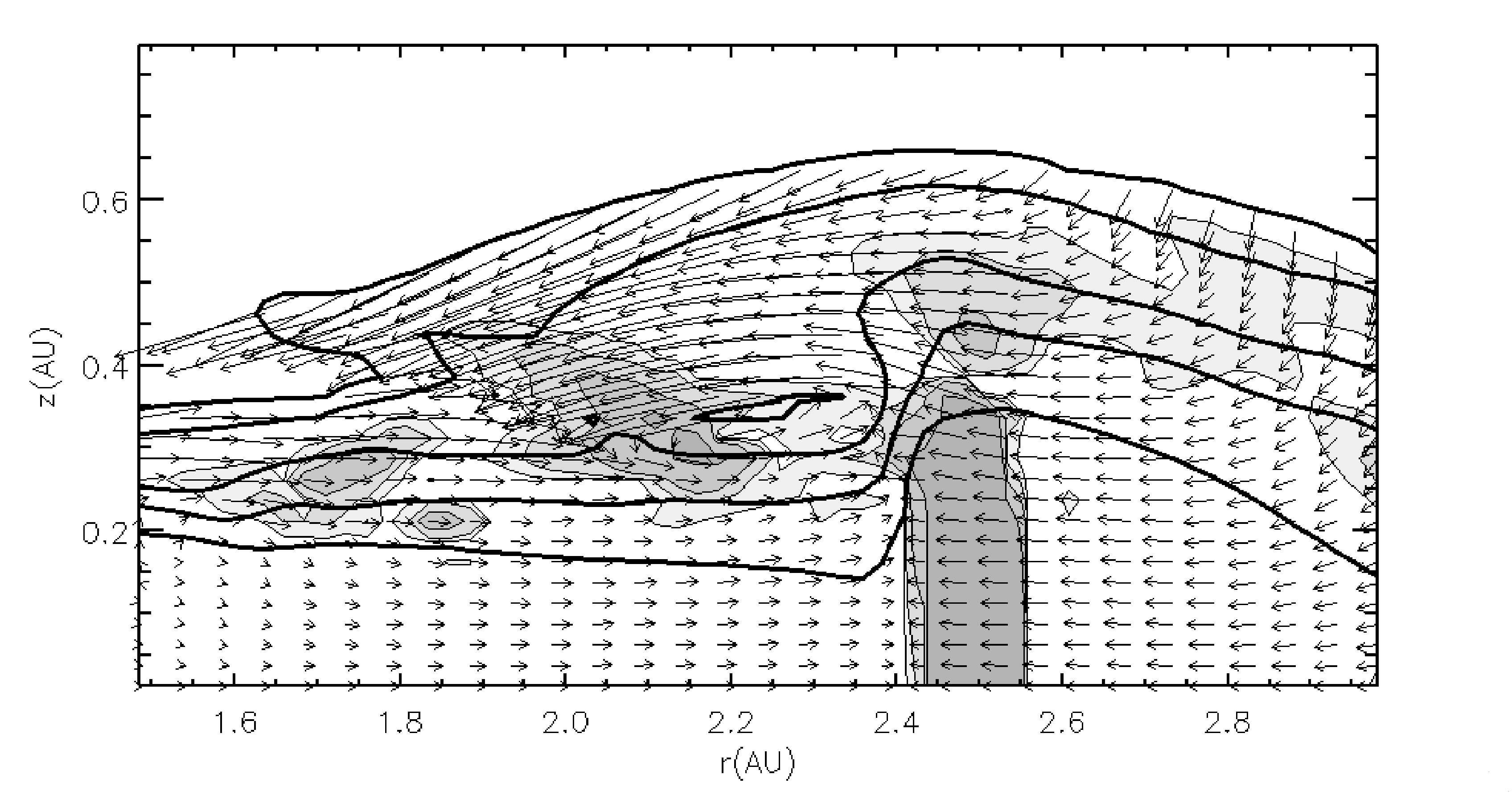}
\caption[Radial cross-section portraying the shock at $r=2.5$ AU for the Mej\'ia disk]
{A radial cross-section portraying the shock at $r=2.5$ AU for the Mej\'ia.  The thick
lines are density contours corresponding to 3.5(-12), 3.5(-11), 3.5(-10),\linebreak
7.1(-10), and 1.4(-9)$\rm~g~cc^{-1}$.  The gray, shaded regions indicate shock
heating corresponding to 5.6(-9), 2.2(-8 ), 9.0(-8), and 3.6(-7) erg cc$^{-1}$
s$^{-1}$.  The arrows show the velocity of the gas, with each component scaled to
its appropriate axis.}
\end{center}
\end{figure}
 
\begin{figure}
\begin{center}
\includegraphics[width=6in]{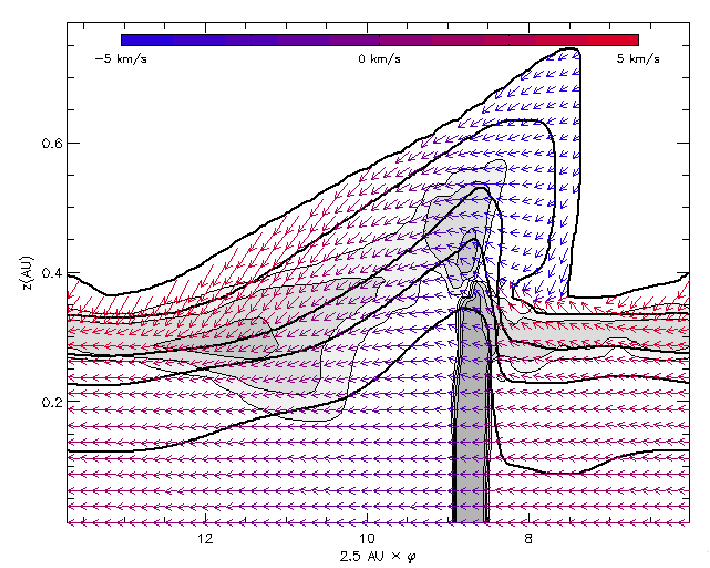}
\caption[Cylindrical cross-section at $r=2.5$ AU]
{Cylindrical cross-section at $r=2.5$ AU for the same time shown in
Figures 5.3 and 5.4.  The density and heating contours are the same as in
Figure 5.4. The arrows represent the flow of the gas in the frame of
 the potential perturbation and the color of the arrows represent the radial
flow of material through this cross-section (blue toward the reader or central
star and red away). This cross-section is shown from the perspective of an
observer at the central star, and $\varphi$ is measured counterclockwise from
the 3 o'clock position in Figure 5.3.}
\end{center}
\end{figure}

 These bores and breaking waves created by spiral shocks in disks are fully
three-dimensional; looking at only a single slice of a wave's morphology
and its corresponding gas flow can be  both confusing and misleading.  In the
following discussion, I present two 2D-cuts, $r$-$z$ and $\varphi$-$z$ at
different disk positions, as presented in Boley \& Durisen (2006).   When
considered together, they form a comprehensible picture of waves in the disk.
In the slices, contours delineate density and shock heating by artificial
viscosity,  and arrows represent gas velocity vectors.   It is important to
remember that one cannot trace fluid element trajectories by connecting gas
velocity vectors in any one plot.

\subsection{Mej\'ia Disk}
The cross-sections for the energy equation simulation show a very clean spiral
shock at $r=2.5$ AU, so I will describe it first.  As a point of reference,
Figure 5.3 is the midplane density grayscale of the disk at the same time that
the following cross-sections are shown.
Figure 5.4 shows the velocity vectors of the gas in an $r$-$z$ cut, which
corresponds to 8 o'clock in Figure 5.3.  The arrows represent the velocity
components of the gas, with each component scaled to its axis, the thick
contours represent the density structure, and the light contours with gray fill
represent shock heating due to artificial viscosity.  Near the midplane, 
where fluid elements are affected weakly
by the shock bores, fluid element trajectories are roughly arranged in
streamlines, creating the oval distortion seen in Figure 5.3, with the apastron
near the spiral shock.  This is indicated by the gas velocity vectors near 2.4
AU in Figure 5.4.  For $r>2.4$ AU, the gas at all altitudes has already passed
through a spiral shock at a larger radius and is flowing radially inward.  The
downward moving, high-altitude material that seems to appear from no origin is
passing through this cross-section and is part of a coherent flow.  For $r<2.4$
AU, the low altitude material ($z<0.3$ AU) represents the outwardly moving
pre-shock flow, while the mid- to high-altitude disk material has already passed
through a spiral shock and has now developed into a large breaking wave.   The
breaking wave produces high-altitude shocks over the pre-shock flow and a large
vortical flow, which appears to have some effect on the pre-shock flow even at
low-disk altitudes.  The disk height reaches a maximum near the spiral shock and
is slightly less than twice its unperturbed height. The height of the shock bore
will be discussed quantitatively in \S 5.4.

 Figure 5.5 is a $\varphi$-$z$  cross-section at $r=2.5$ AU, where $\varphi$ is
measured counterclockwise from the 3 o'clock position in Figure 5.3, with
velocities shown in the reference frame of the spiral wave, and complements
Figure 5.4.  In this figure, the arrows are colored to represent the radial flow
of the gas as seen by an observer positioned at the star.  Before the shock, the
gas is moving radially outward, and after the shock, it is moving radially
inward.  In addition, the mid- to high-altitude gas starts its vertical
expansion at that interface.   The sharp wall of gas at $2.5\mathrm{\
AU}\times\varphi=6$ AU is formed by gas  that jumped at a larger radius and is
now falling inward through this cross-section.  As this material crashes back
onto the disk, it creates high-altitude shocks that result in the expansion of
the high-altiude gas as the post-shock region transitions into the pre-shock
region of the next arm between 2.5 AU$\times \phi$ = 6 to 8 AU. The gas begins to move radially outward again due to the
roughly elliptical orbits of the gas.  The wall of material evolves with time, 
and can develop a tube similar to a breaking surface wave.  
The structure of shock bores is time-dependent and
transient\footnote{An animation showing the time-dependence of shock bores is
made available at http://hydro.astro.indiana.edu/westworld. Download the file {\it
Shock Front} under the {\it Movies} link.}.

Figure 5.6 shows the region 0.6 AU inward at about 6:30 o'clock in Figure 5.3.
The breaking wave drastically affects the high density gas flow.  A clear shock
front cannot be defined on the coarse 3D grid, but strong shocks are everywhere
in the wave.  A large vortical flow reaches down to the midplane at $r=1.8$ AU;
another vortical flow is present in the high-altitude gas at $r=2.0$ AU.  These
circulation cells are clearly related to the shock bores and waves.  In the
corresponding cylindrical cross-section, Figure 5.7, the complexity of the shock
structure is again highlighted, and it is difficult to distinguish the shock
bore, which is at about $1.9\times\varphi\rm\ AU=9\ AU$, due to the strong
disturbance by gas that jumped at a larger radius and is plunging into the disk
between  $1.9\times\varphi\rm\ AU=7$ and 8 AU.

 On the other hand, the shock bore  shown in  Figure 5.8 for $r=3.1$ AU at about
9:30 o'clock in Figure 5.3, which is near the Lindblad resonance of the imposed
$m=2$ perturbation, shows the development of a very large wave that has yet to
break onto the disk's surface.  The cylindrical cross-section for this location,
Figure 5.9, indicates that the wave is associated with a very sharp wall,
similar to the 2.5 AU morphology.
 
 \begin{figure}
\begin{center}
\includegraphics[width=6in]{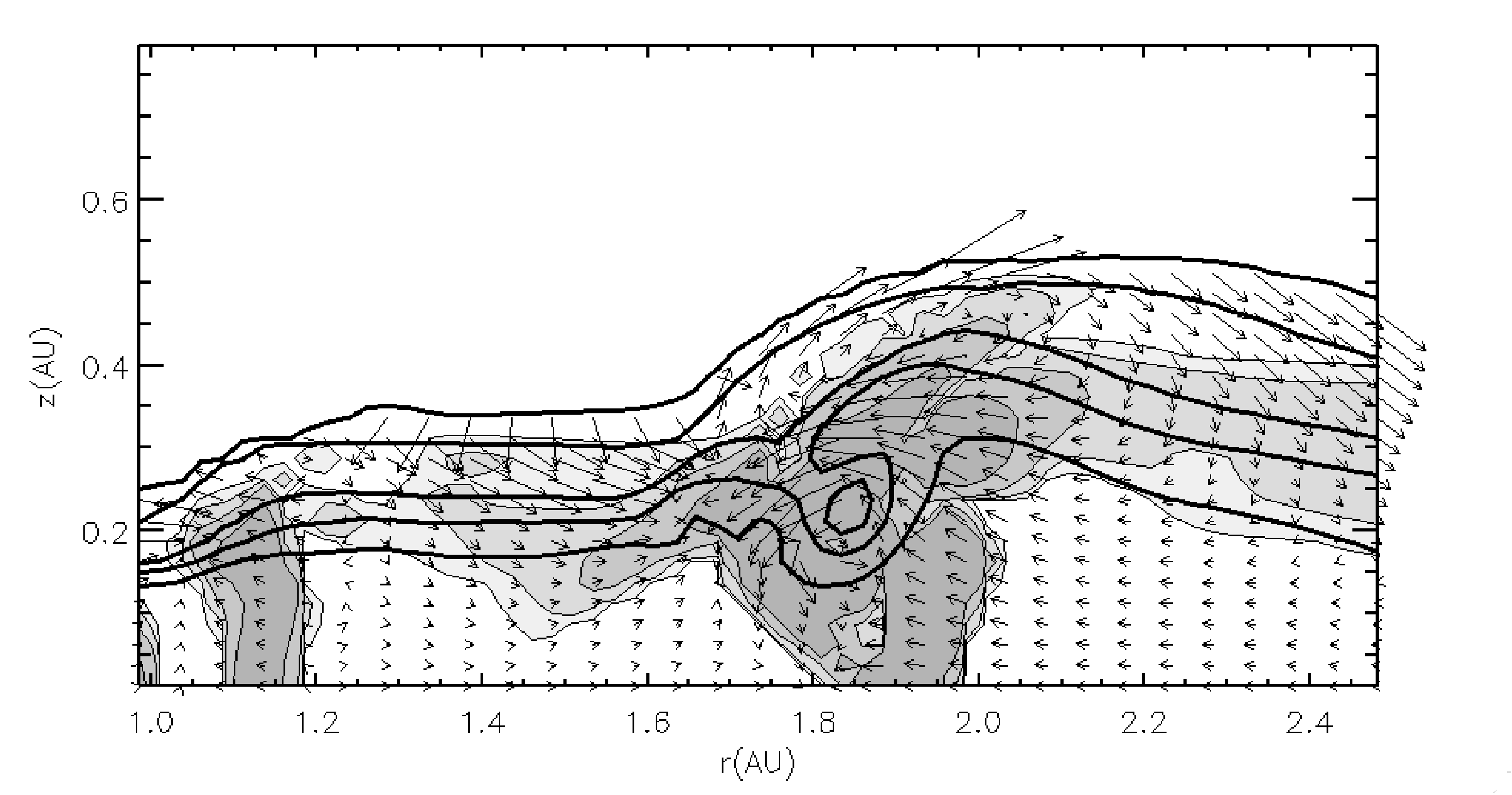}
\caption[The same as Figure 5.4, except cutting across the shock at $r=1.9$ AU]
{The same as Figure 5.4, except cutting across the shock at $r=1.9$ AU.}
\end{center}
\end{figure}
 
\begin{figure}
\begin{center}
\includegraphics[width=6in]{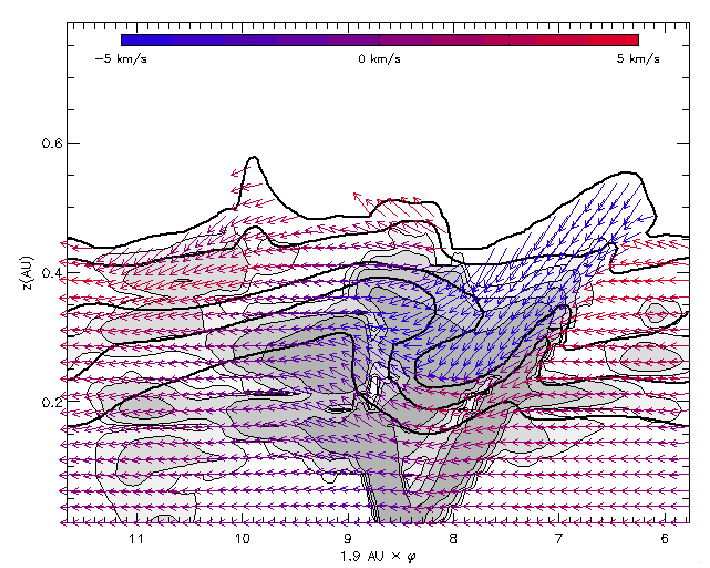}
\caption[Same as Figure 5.5, except for $r=1.9$ AU]
{Same as Figure 5.5, except for $r=1.9$ AU.  The shock morphology is
much more complex and the shock bore cannot easily be defined by a single shock.
Note the material, which jumped at a larger radius, plunging down into the disk
between 7 and 9 AU.} 
\end{center}
\end{figure}

These figures show that at all disk radii portrayed, the disk material at high
altitudes is moving differently from the material near the midplane, shocks are
present at all disk altitudes, and the shock bores lead to large breaking waves,
which create extensive vortical flows in the disk.  The effect that these bores
and waves have on the disk surface is emphasized by the isodensity surface
contour shown in Figure 5.10.

The high-mass disk was also evolved using an isothermal equation of state for
the same potential perturbation.  Figure 5.11 shows a $\varphi$-$z$
cross-section for a shock near $r=2.5$ AU.   Instead of expanding at the shock,
the gas compresses slightly and the height of the disk remains about the same.
I speculate that the small peak of downward-moving, low density material just
before the spiral shock is a remnant of transient features created by suddenly
switching the EOS of the disk and by forcing a strong disk perturbation, and it
is not part of the shock bore. The vertical compression of the gas at the shock
front is in agreement with equation (5.18) because $J_f \le 1$ for an isothermal
shock as given by equation (5.20).  In an isothermal gas disk, a spiral shock
will behave like a spiral density wave if self-gravity is unimportant, or it will
compress vertically along the shock front if self-gravity is important.

\begin{figure}
\begin{center}
\includegraphics[width=6in]{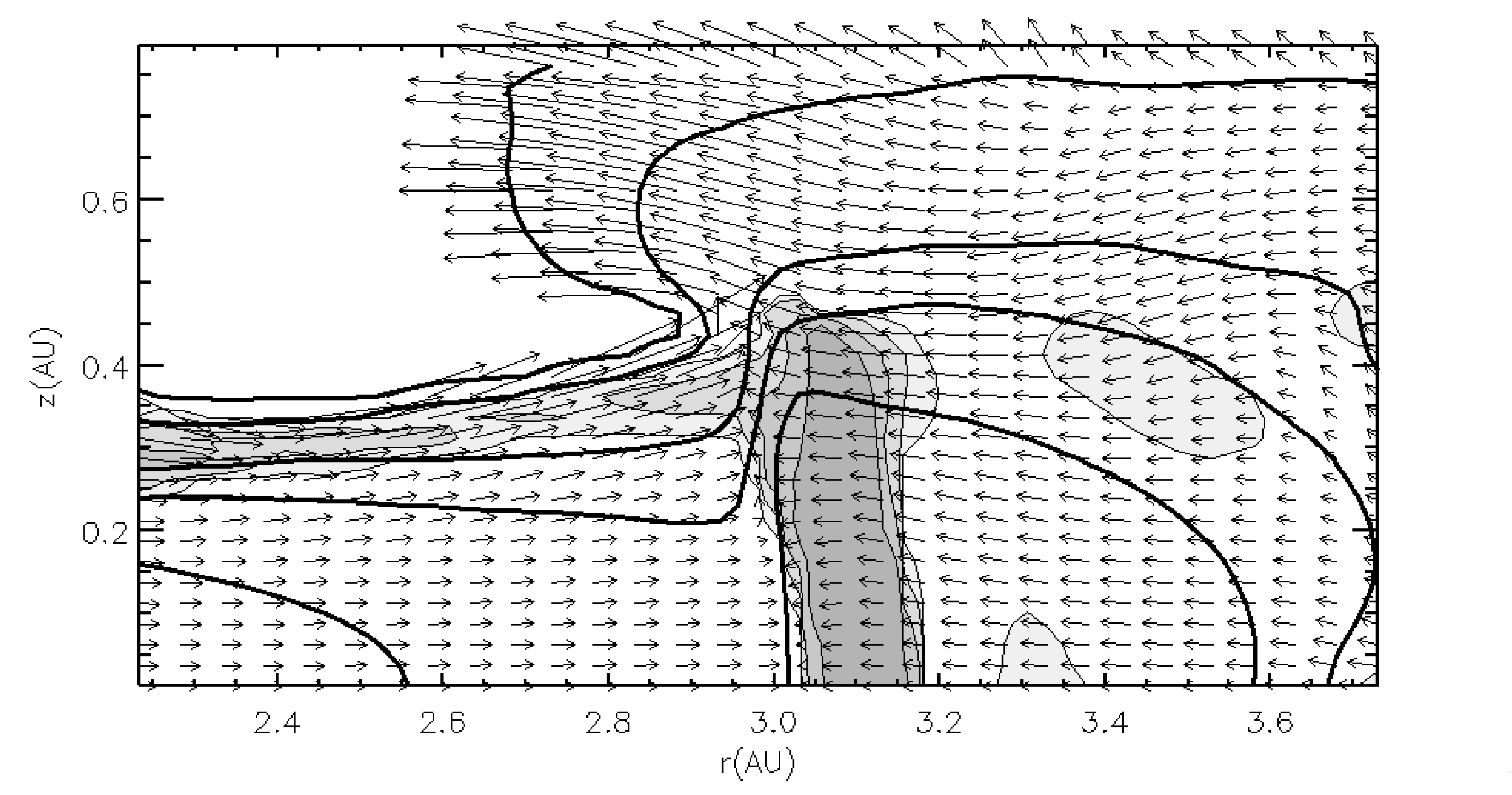}
\caption[The same as Figure 5.4, except cutting across the shock at $r=3.1$ AU]
{The same as Figure 5.4, except cutting across the shock at $r=3.1$ AU.}
\end{center}
\end{figure}

\begin{figure}
\begin{center}
\includegraphics[width=6in]{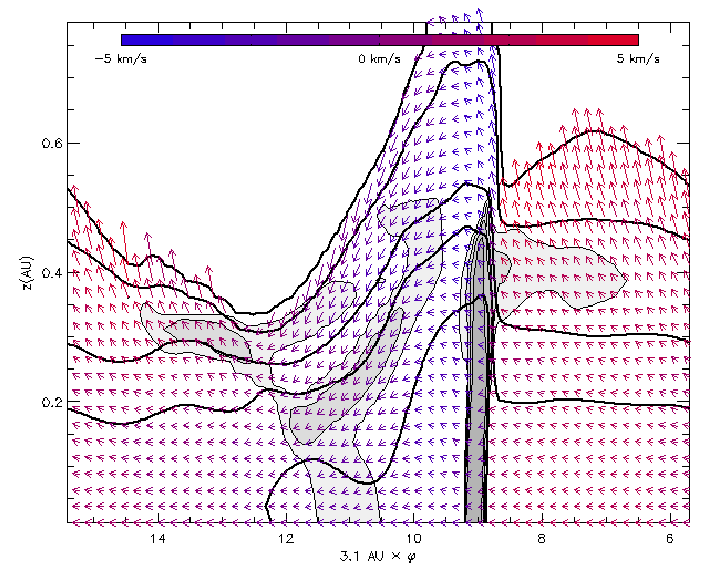}
\caption[Same as Figure 5.5, except for $r=3.1$ AU]
{Same as Figure 5.5, except for $r=3.1$ AU.  Even though this
cross-section is the closest to the corotation radius, a strong shock and
shock bores are still observed.  This is probably due to the large radial motions of
the fluid elements.} 
\end{center}
\end{figure}

\begin{figure}
\begin{center}
\includegraphics[width=6in]{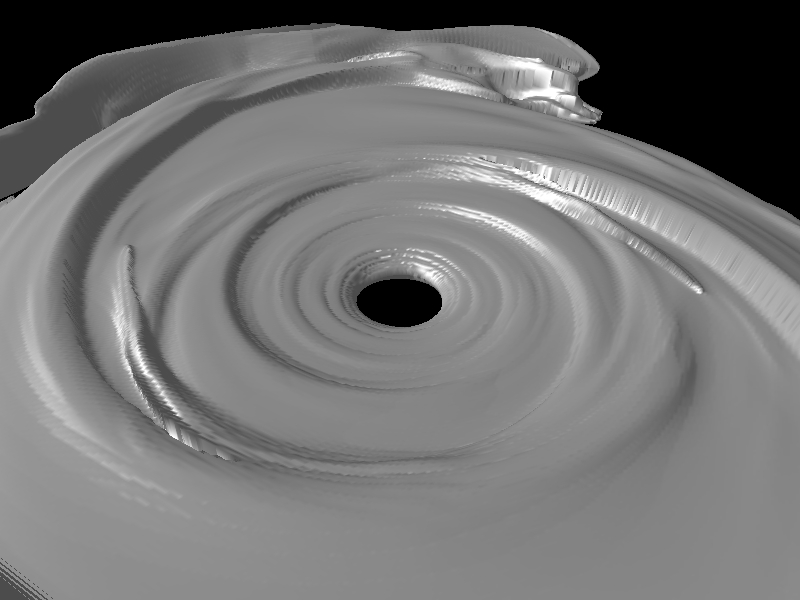}
\caption[Isodensity surface contour for $\rho=3.54(-10)$ g cc$^{-1}$ shown at
the same time as Figures 5.4-5.9]
{Isodensity surface contour for $\rho=3.54(-10)$ g cc$^{-1}$ shown at
the same time as Figures 5.4-5.9. The view is from above the disk looking from
about 5 o'clock to 11 o'clock in Fig.~5.3.  The surface contour routine is unable to show
breaking waves clearly.}
\end{center}
\end{figure}

\begin{figure}
\begin{center}
\includegraphics[width=6in]{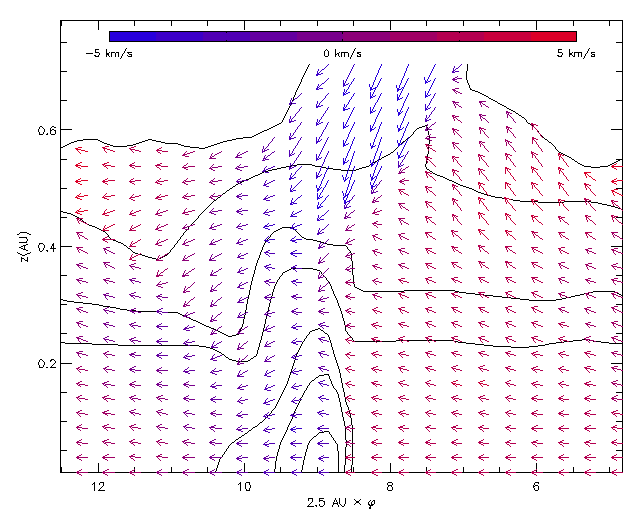}
\caption[A cylindrical cross-section at $r=2.5$ AU for the isothermal Mej\'ia
disk simulation]
{A cylindrical cross-section at $r=2.5$ AU for the isothermal Mej\'ia
disk simulation.  The density contours are the same as in Figure 5.5.  The shock
front does not cause rapid expansion in the post-shock region, but instead,
causes a compression.  I speculate that the small peak of downward-moving, low
density material just before the spiral shock is a remnant of transient features
created by suddenly switching the EOS of the disk and by forcing a strong disk
perturbation. Although this wave is not associated with the compression caused
by the shock bore, it is probably responsible for the slight undulatory
morphology in the post-shock region.}
\end{center}
\end{figure}

\begin{figure}
\begin{center}
\includegraphics[width=6in]{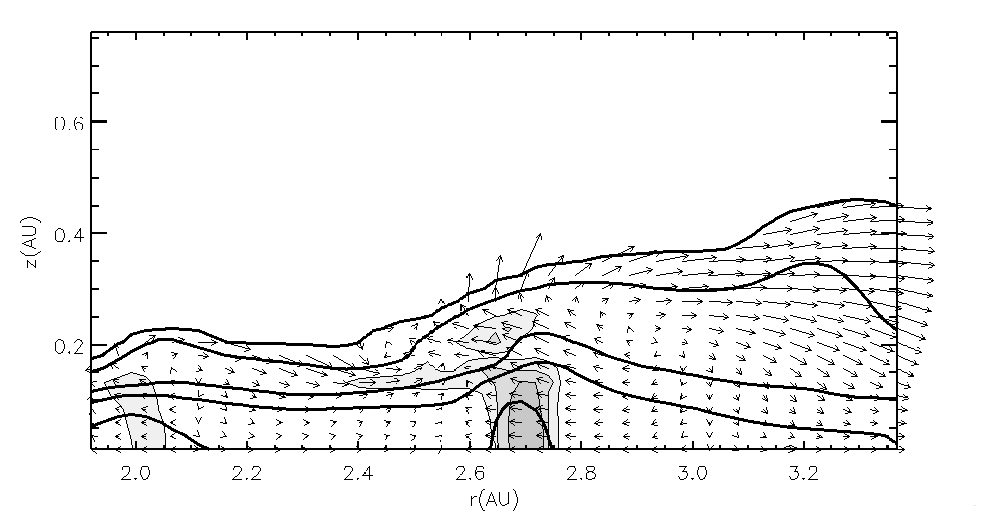}
\caption[The same as Figure 5.4, except for the moderate disk cutting
across the shock at $r=2.7$ AU]
{The same as Figure 5.4, except for the moderate disk cutting
across the shock at $r=2.7$ AU.}
\end{center} 
\end{figure}

\begin{figure}
\begin{center}
\includegraphics[width=6in]{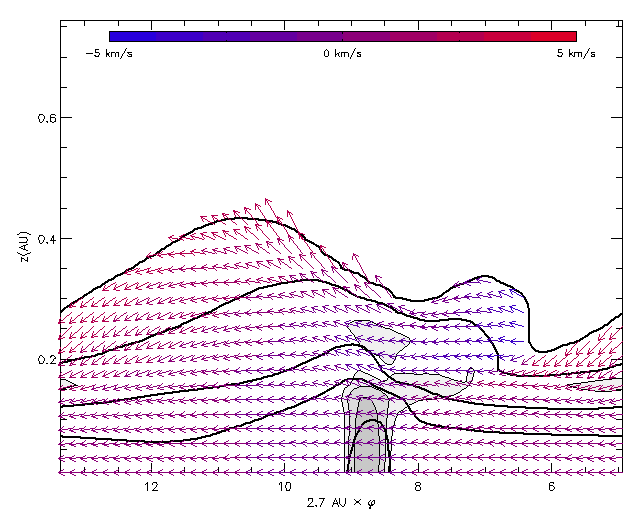}
\caption[The same as Figure 5.5, except for the moderate disk at $r=2.7$
AU]
{The same as Figure 5.5, except for the moderate disk at $r=2.7$
AU.  The shock is weaker but still strong enough to induce an shock bore.
In addition, the peak at 11 AU, which is moving radially outward, is not
obviously associated with the shock bore.  The nonlinear dynamics of waves in
these disks is complex; the shock bores are not necessarily the whole story. }
\end{center}
\end{figure}

\subsection{Cooled Moderate Disk}

The cooled moderate disk portrays a different shock morphology from the Mej\'ia
disk.  The shocks are weak, and therefore, the shock bores are weaker and are
localized.  The wave structure in the disk may be more closely related
to $f$-modes, as described by LO98.  Figure 5.12 shows the $r$-$z$ cross section
for the shock at $r=2.7$ AU, which is at the same relative position as the 2.5
AU cross-section of the Mej\'ia disk.  There is surface corrugation and a weak
shock bore at the spiral shock, but a breaking wave does not form. Figure 5.13,
the cylindrical cross-section, shows that a strong vertical wall formed by
inward falling material is, for the most part, absent. Nevertheless, the gas
does expand vertically after entering the shock.

\section{Discussion}

\subsection{Shock Bores}

Each of the three simulations shows different shock and wave morphologies.   In
the Mej\'ia disk, strong shocks are produced, which lead to the formation of
shock bores over most of the spiral wave.  The spiral waves in the isothermal
calculation propagate principally as spiral density waves.  The cooled moderate
disk only shows a bore-like morphology near $r = 2.7$ AU, and the disk has weak
spiral shocks.   It appears that shock bores are the extreme nonlinear outcome
when $f$-modes develop into strong spiral shocks.  To understand how well the
shock bore theory describes this nonlinear regime, I use the fluid element
tracer to quantify shock parameters.

Figure 5.14 shows a sample thermal history for a fluid element starting at $r =
2.5$ AU, $z$ = 0.176 AU, and $\varphi$ = 0$^{\circ}$ in the Mej\'ia disk.  The
fluid element is placed in the disk when the shocks first become strong, at the
same time that the grid is switched to high resolution. The fluid element is
followed for half a pattern period.  The shock is seen clearly in the density
and temperature/pressure plots.  It is also noticeable in the $u$ profile, where
$u$ is the shock normal speed in the frame of the pattern. To estimate $u$, it
is assumed that the pitch angle $i$ of the spiral arms is constant, nearly
20$^{\circ}$ for these calculations.  With this assumption
\begin{equation}u = v_{r}\cos i + (\Omega-\Omega_p)r \sin i\label{pitch:5.21}.\end{equation}
Although this will not precisely measure $u$, and therefore the Mach number, it
will  give a reasonable value for most of the shocks in the disk.  A
modification to this method is made for studying shock strengths in disks for
which the spiral waves are uncontrolled (Chapter 7).  Figure
5.15 indicates the Mach number as well as the trajectory of the fluid element
corresponding to Figure 5.14.  Even though the fluid element starts at $r=2.5$
AU, it is transported outward to 3.1 AU (Fig.~5.8) before it encounters the
shock.  The Mach number for the shock in Figure 5.8 is around 2.6, which roughly
agrees with the change in density. The resulting
change in the disk height should therefore be about 1.7, in the
non-self-gravitating limit, or 1.6 assuming a $q =10$; the actual vertical jump is
about 1.5.    Similar calculations indicate that the Mach
number for the spiral shock shown in Figure 5.4, the radius where the fluid
element is originally launched, is about 2.7.  The jump in disk height again
appears to be about 1.5-1.6.

Even though the jump morphology in the $r=1.9$ AU cylindrical cross-section of
the Mej\'ia disk is unclear, the shock bore description of spiral shocks
predicts the new height for the inner disk fairly accurately.  The Mach number
for the shocks at $r$ = 1.9 AU is also about 2.7.  According to equation (5.24),
the new disk height should also be about 1.6-1.7 times higher for
$q\rightarrow\infty$ than the original height. The initial height of the disk is
about 0.3 AU at $r$ = 1.9 AU and the height of the disk is about 0.5 AU in
Figure 5.7, which matches the prediction well.

The simple shock bore picture may be complicated by other wave dynamics near the
Lindblad resonance at $r=3.1$ AU (see Fig.~5.8 and 5.9), but shock bore theory
still seems to predict the behavior of the shocks well in this region.  Along
the spiral shock, a bore is clearly seen.  The fluid element tracer indicates
that the Mach number for the shock at mid-disk altitudes is near 2.6, as stated
above.  The Mach numbers probably remain high at these radii because the fluid
elements are on more highly elliptical orbits than in the inner radii, and the
radial component to $u$ becomes important, see equation (5.27), for these small
pitch-angle spirals.

\begin{figure}
\begin{center}
\includegraphics[width=6in]{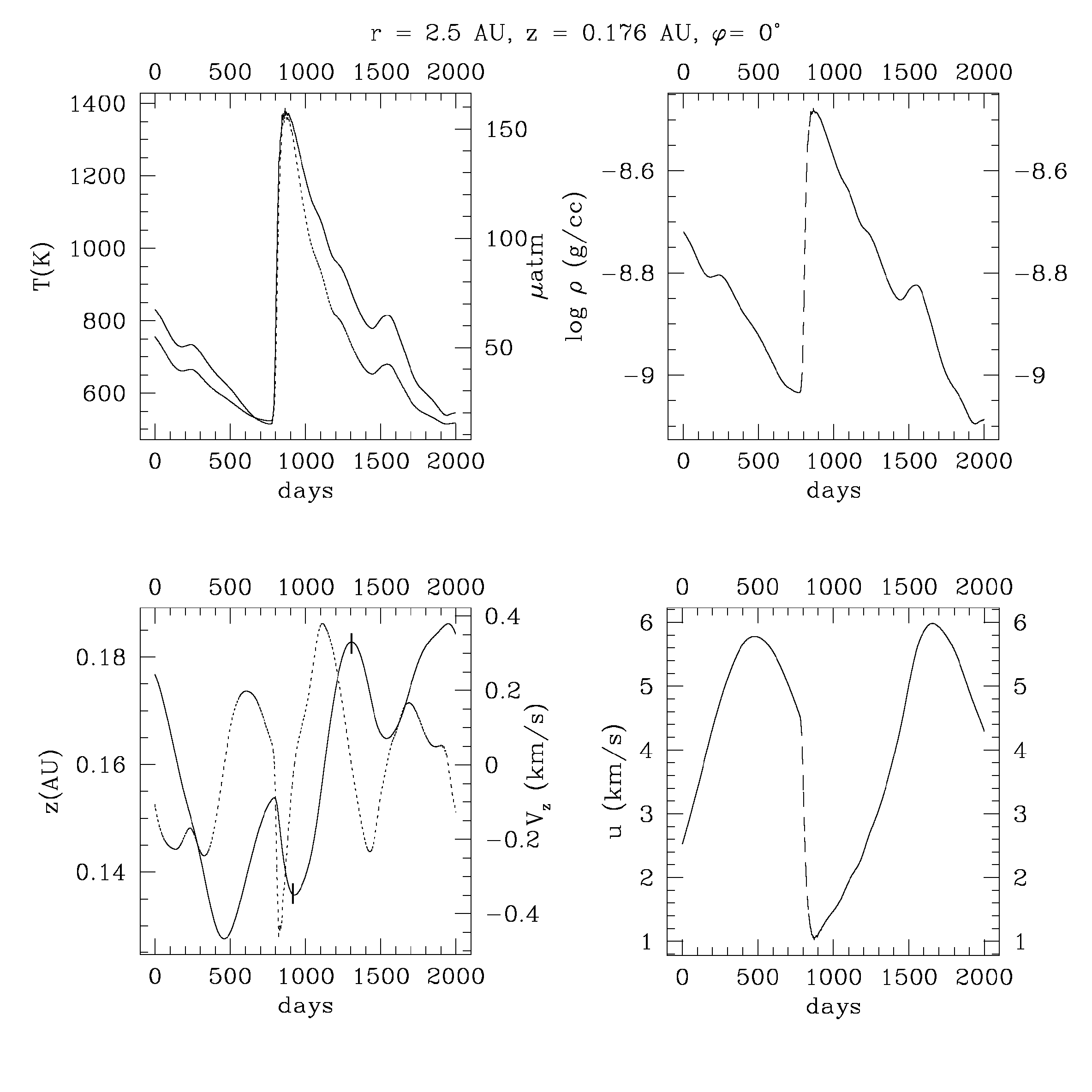}
\caption[An example of a thermal history for a fluid element starting at $r=2.5$
AU, $z = 0.176$ AU, and $\varphi = 0^{\circ}$]
{An example of a thermal history for a fluid element starting at $r=2.5$
AU, $z = 0.176$ AU, and $\varphi = 0^{\circ}$.  {\it Top left:} Temperature
(solid) and pressure (dotted) histories.  {\it Top right:} Density history. {\it
Bottom left:} Disk altitude trajectory (solid) and $V_z$ (dashed).  Although the
vertical velocity does become negative and the vertical altitude reaches a local
minimum while it is going through the shock, it quickly expands with a very
sharp change in the vertical velocity as soon as it is in the post-shock region.
Roughly, the shock bore is between the hash marks. {\it Bottom right:} The shock
normal velocity, with respect to the global spiral shock, in the frame of the
shock.} 
\end{center}
\end{figure}

\begin{figure}
\begin{center}
\includegraphics[width=6in]{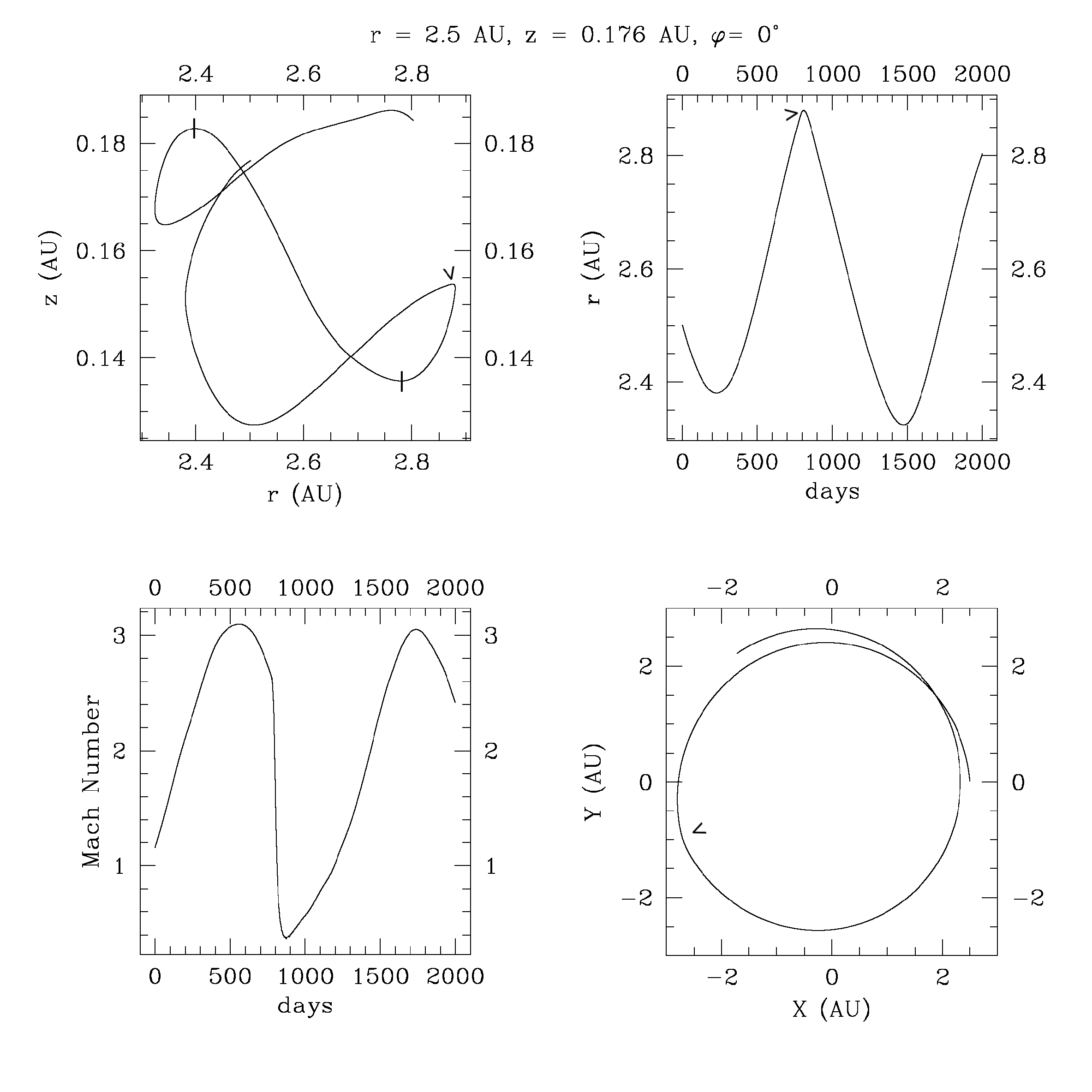}
\caption[To complement Figure 5.14, the fluid element trajectories are shown
above]{To complement Figure 5.14, the fluid element trajectories are shown
above. Clockwise, starting in the upper left: $r$ vs.\ $ z$, $r$ vs.\ time, the
projection of the motion onto the midplane in Cartesian coordinates, and  Mach number vs.\ time
based on the $P$, $\rho$ and $u$ information presented in Figure (5.14).
The open arrow heads show where the shock begins.  Roughly, the shock bore is
between the hash marks.}
\end{center}
\end{figure}

For the cooled moderate disk, the fluid element tracer indicates that the spiral
wave forms a distinct shock near $r=2.7$ AU.  Since the shock is weak, $u$ is
unreliable, so I put the measured density change into equation (5.3) to find a
Mach number of about 1.7.  Assuming self-gravity is negligible, I expect that
the maximum height of the jumping material to be about 1.3 times the initial
scale height, which is fairly consistent with Figure 5.12.

In discussing the changes of height in the disk, I have used the change in the
density contour morphology to estimate differences. However, this can be
misleading.  To give a better quantitative description, define a local disk 
scale height to be 
\begin{equation}h=\int_0^{\infty}dz\
\rho(r,\varphi,z)/\rho_{\mathrm{mid}}(r,\varphi).\end{equation}
Figure 5.16 plots this scale height for all three simulations at a similar
radius.  The scale heights in the pre-shock region for the energy and isothermal
calculations are similar, but the isothermal disk shows a decrease in the scale
height after the shock, as predicted, while the energy equation disk shows an increase.  The
change in disk scale height for the energy equation is about 1.5, compared to
1.6 or 1.7 predicted by equation (5.24) for the shock at this radius.  The
cooled moderate disk has a more gradual change in scale height than the other
calculations, because of its weak shocks.  The scale height changes by a factor
of 1.3, close to what is predicted by equation (5.24).

\begin{figure}
\begin{center}
\includegraphics[width=6in]{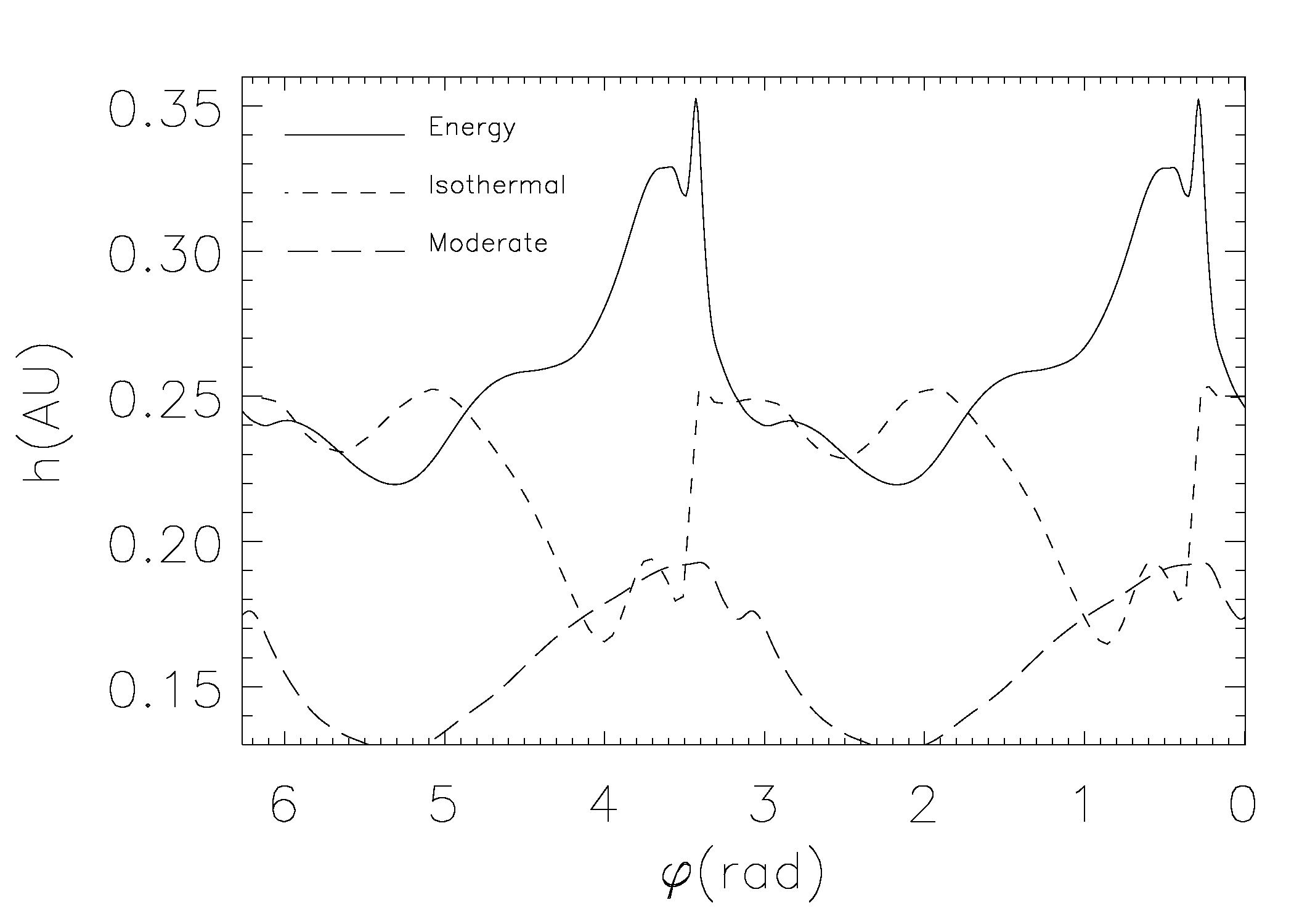}
\caption[The scale height of each disk vs.\ azimuth]
{The scale height of each disk vs.\ azimuth.  The Mej\'ia disk, both
energy and isothermal, are plotted for $r =2.5$ AU and the cooled moderate disk is
plotted for $r=2.7$ AU.  The shocks all occur between about 2 to 4 radians.}
\end{center}
\end{figure}

\begin{figure}
\begin{center}
\includegraphics[width=6in]{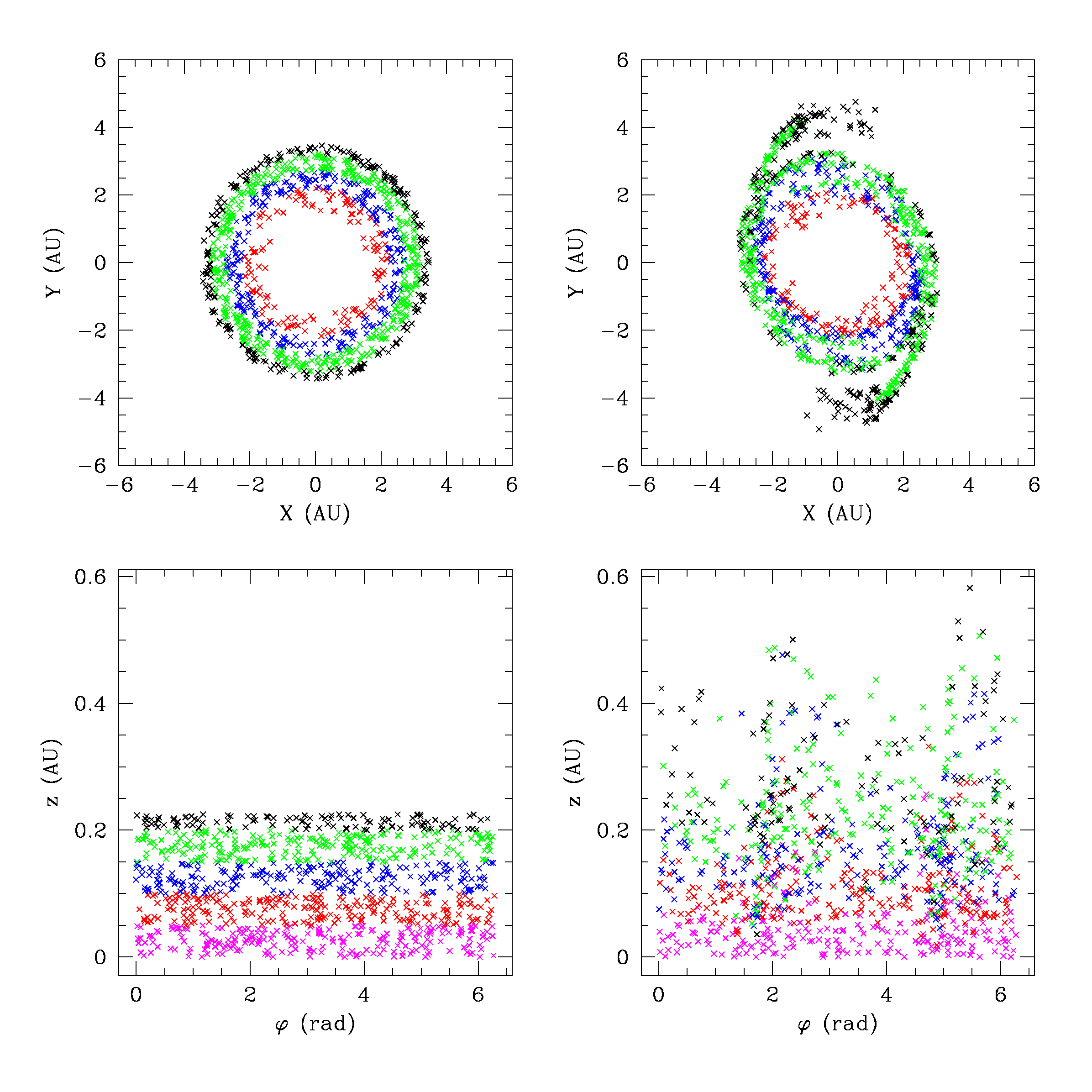}
\caption[The results of tracing 1000 randomly selected fluid elements within an
annulus for half a pattern period]
{The results of tracing 1000 randomly selected fluid elements within an
annulus for half a pattern period.  {\it Top left:} The initial radial and
azimuthal fluid element positions.  {\it Top right:} The position of the same
elements after half a pattern period. {\it Bottom left:} The initial vertical
distribution. {\it Bottom right:} The final vertical distribution.  Note that
even material near the midplane is starting to show signs of stirring.}
\end{center}
\end{figure}

\subsection{Processing and Mixing}

Large-scale vortical flows are present in both disks, which could result in the
mixing of disk material radially and vertically.   In addition, the
high-altitude disk material can be moving at a much greater speed, and even in a
different direction, than the mid- and low-altitude material.  This makes it is
possible to have a {\it slip surface} or {\it vortex sheet} at a Mach
intersection, i.e., an intersection of two parallel flowing streams with
different Mach numbers (e.g., Massey 1970).  This will provide additional
turbulence in the disk on a small scale and could influence the evolution of
solids in the disk, e.g., size-sorting of chondrules (e.g., Cuzzi et al.~2001).
Unfortunately, this effect cannot be modeled in these simulations since the typical cell
size is $\sim 10^{11}$ cm, which is too large to study turbulence on the
Kolmogorov scale where chondrules would be size-sorted (Cuzzi 2004).
Nevertheless, the energy contained in any shock bore, or even a strong wave, is
enough to provide the necessary chondrule size-sorting turbulence in a disk (see
Boley et al.~2005).

The large-scale vertical and radial mixing/stirring produced by
the spiral shocks and shock bores can be investigated by using the fluid element
tracer.  Figure 5.17  shows the results of tracking 1000 fluid elements for half
a pattern
period.  The fluid elements are randomly distributed in disk volume between $r =
1.44$ and 3.36 AU and between $z =0$ and 0.24 AU.  For this analysis
I cannot distinguish between {\it mixing} and {\it stirring}
of nebular material, mainly due to the short integration.  
Figure 5.17 demonstrates, however, that shock bores do at
least produce significant stirring over tenths of an AU in only about 5.5 years.

Episodic mixing and stirring of the nebula could be driven by spiral shocks that
form from repetitious clump or arm formation over several million years.   This
would also provide short-lived shocks that could process solids throughout the
nebula (Boss \& Durisen 2005a,b), and may generate disk
turbulence (Boss 2004b; Boley et al.~2005).  Waves and shock bores could represent the major
mechanism by which solids are processed and packaged into their parent bodies in
protoplanetary disks. I return to this idea in Chapter 7.

\subsection{Shock Bores and Convection}

Convection can activate in a disk when $\beta > (3\gamma-4)/(\gamma-1)$ for an
opacity law $\kappa=\kappa_0 T^{\beta}$, where $\gamma$ is the ratio of specific
heats (Lin \& Papaloizou 1980; Ruden \& Pollack 1991).  If this condition is
satisfied, the vertical entropy gradient is driven negative, and convective instabilities
can grow and convective cells can be established for a range of disk radii
(e.g., Lin \& Papaloizou 1980; Rafikov 2007).  However, these convective cells
can also be easily disrupted by additional dynamics.  
Boley et al.~(2006) found that for the M2004 simulation (see below), convective cells were
dispersed by the onset of spiral wave activity.  In contrast, other researchers
claim that convection not only is active in very dynamic disks, but it has stark
consequences for disk cooling.  In particular, Boss (2004a) claims that
convection leads to rapid cooling and disk fragmentation.  Moreover, Boss
(2002) finds that disk cooling is insensitive to metallicity due to convective
fluxes. In this section, I explore the possibility of convection in unstable
disks, and demonstrate that sudden vertical motions in dynamic disks are most
likely shock bores and not convection cells.

I search for
convection in several unstable disk simulations by looking for regions where
upward and downward motions are commensurate with regions of negative entropy
gradients and are not
associated with strong shocks, as strong shocks will result in a shock bore.
Figure 5.18 shows the Mej\'ia disk simulation discussed in this section.
Within some regions with a negative vertical entropy gradient,
vortical flows are seen. However, shocks are typically found in these regions as well.
Such an alignment suggests that the regions are dynamic, i.e., they are not
representative of regions where convection occurs in a stellar interior sense,
but are regions that are associated with shock bores and waves. However, one
does not expect for convection to be self-sustained in such a disk because the
cooling time is {\it ad hoc}. To provide further comment, I investigate 
simulations with realistic cooling.  

\begin{figure}
\begin{center}
\includegraphics[width=6in]{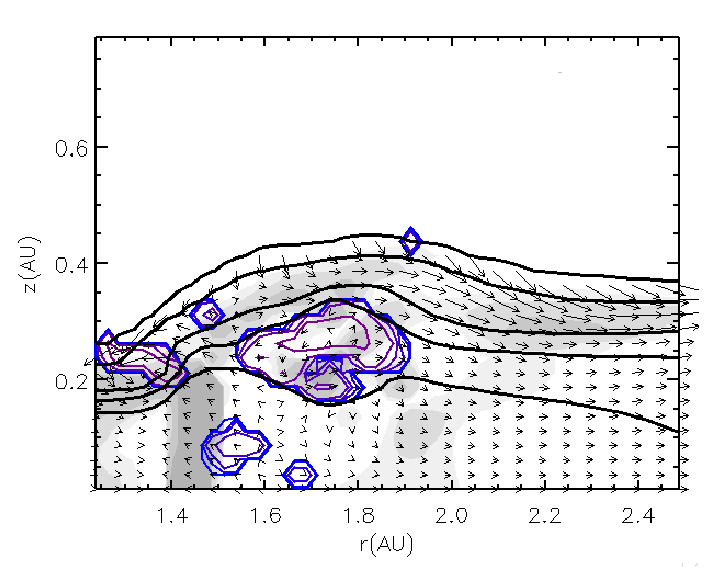}
\caption[Similar to Figure 5.4, but  at a different location in the high-mass
disk with additional contours denoting superadiabatic regions]
{Similar to Figure 5.4, but  at a different location in the Mej\'ia
disk with additional contours denoting superadiabatic regions. The thin contour
lines indicate where the vertical entropy gradient is negative.  }
\end{center}
\end{figure}

\begin{figure}
\centering
\includegraphics[width=6in]{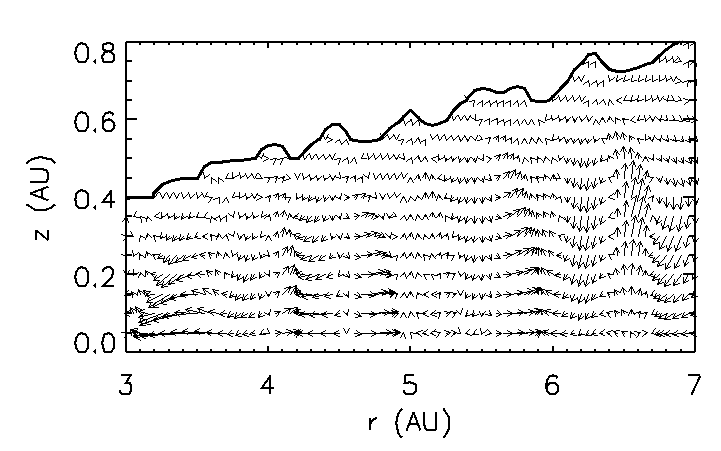}
\caption[Convection-like motions in an optically thick protoplanetary disk
model]
{Convection-like motions in an optically thick protoplanetary disk
model (see text in \S 5.6).  The heavy curve roughly indicates the disk's
photosphere, and the arrows, which are scaled to each axis and to the midplane
density for each column, indicate the momentum density.  Typical Mach numbers
for the gas range between a few hundredths to a few tenths. Convection-like
eddies are present throughout most of the disk.  However, cooling
times remain long in this disk, because ultimately, the energy must be radiated
away.} \label{fig20}
\end{figure} 

\begin{figure}
\begin{center}
\includegraphics[width=3in]{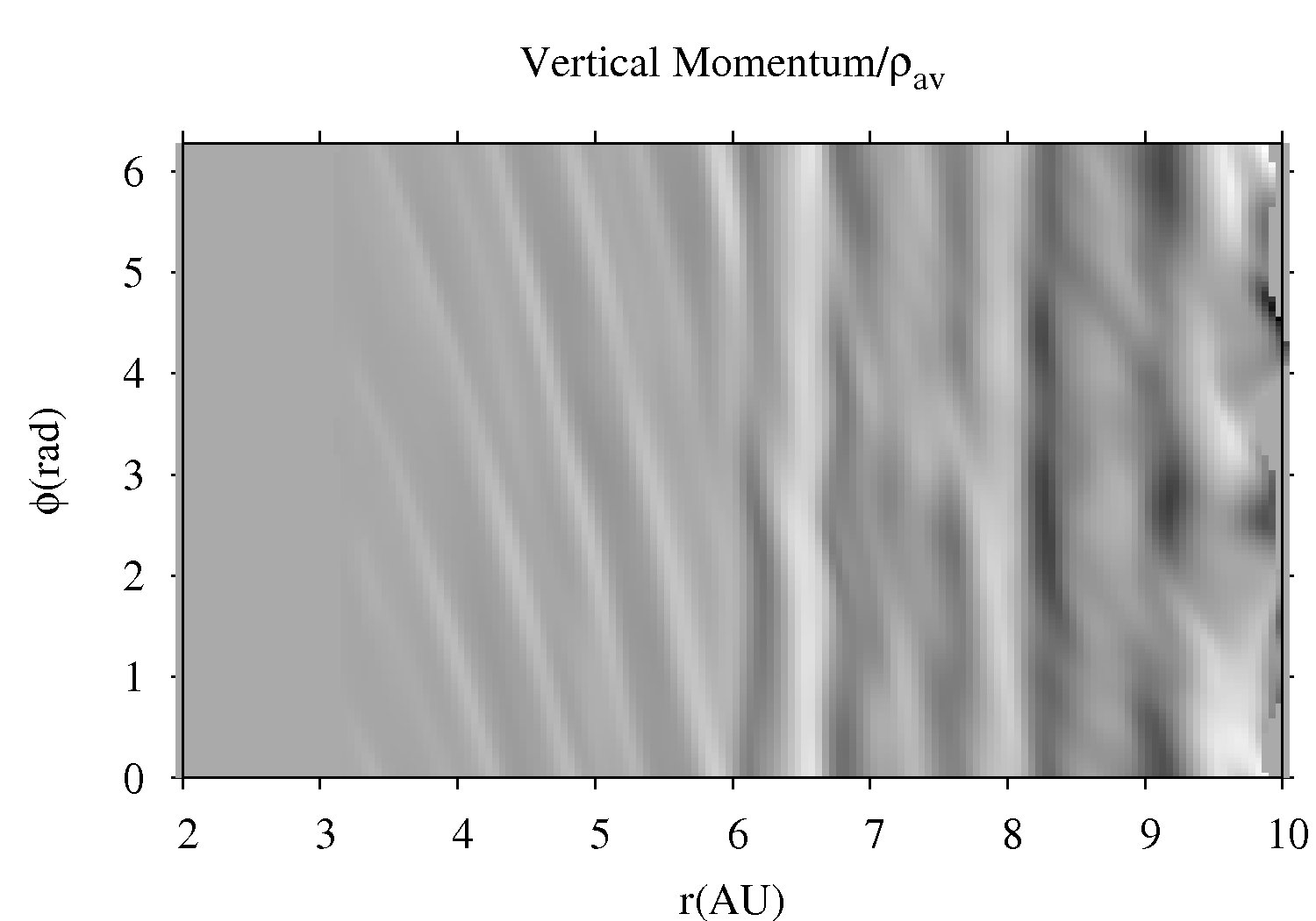}\includegraphics[width=3in]{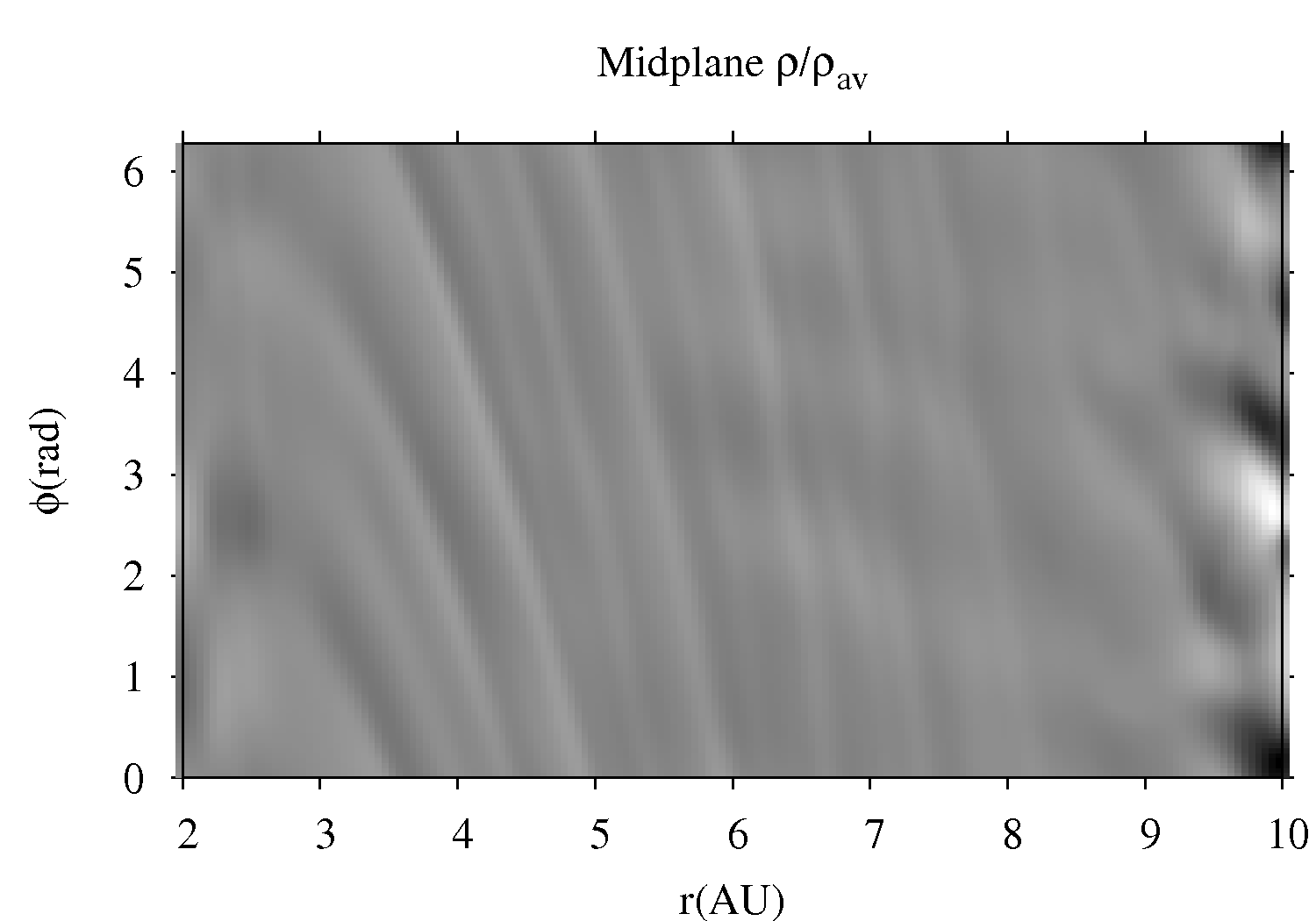}
\caption[Convection flow]
{The left panel shows the vertical momentum density through the $r-\phi$ plane at
$z=0.4$ AU, scaled to the azimuthally averaged midplane density $\rho_{av}$. For comparison,
the right panel shows the midplane density, scaled to $\rho_{av}$.  The vertical
motions inside 6 AU are associated with the development of a spiral wave, and so
it is difficult to determine whether such motions are related to a convective
flow or to wave dynamics, in particular, weak shock bores.  However, the strong
vertical motions centered on 6.5 AU and again at about 8 AU have no
corresponding density enhancement.  These are also the regions that have
large superadiabatic regions, and so those vertical motions appear to be {\it
bona fide} convective flows. }
\end{center}
\end{figure}

Consider the simulation presented by Boley et al.~(2006; see Chapter 6).  The initial disk is
the Mej\'ia disk scaled to 40 AU with a primary mass of 0.5 M$_{\odot}$, and the
disk was evolved with the M2004 radiation scheme (see \S 2.2) under an ideal gas
law with $\gamma = 5/3$.  Extended superadiabatic regions develop during the symmetric phase of the disk's evolution.  In these areas, the average Mach
speed in the convective eddies $\left< v_z/c_s \right> \approx 0.06$, where
$\left< v_z/c_s \right> = \int  \rho~\left| v_z\right| /c_s~{\rm d}V/\int
\rho~{\rm d}V$ and the integrals are evaluated over the volume spanning between
15 and 25 AU in $r$.  The value of $\left<v_z/c_s\right>$ fluctuates between
0.01 and 0.15 when each annulus of grid cells is evaluated separately ($\Delta
r= 1/6$ AU), and  some of the convective eddies result in nontrivial
compressional heating through artificial viscosity.   To evaluate energy
transport by convection, I estimate the convective flux $F_c = -1/2 c_p
\rho T v_z \ell \left({\rm d }\ln T/{\rm dz} -\nabla_{\rm ad} {\rm d}\ln P/{\rm
d}z\right)$ (Lin \& Papaloizou 1980) through each cell in the volume
described above. Here, the mixing length $\ell={\rm min}\left(z,P/\Omega_k^2
z\rho\right)$ and $c_p$ is the specific heat at constant pressure. By dividing
the total internal energy within the half disk between 15 and 25 AU in $r$ by
the total convective energy loss rate for that same region, I find that the
convective cooling time is about 1 ORP and that it is comparable to the
radiative cooling time.  However, this method measures the convective flux
based on a superadiabaticity estimate according to the mixing length formalism of
Lin \& Papaloizou (1980).
For a second measurement, I calculate the energy
carried by convective motions through a plane that cuts through the volume at
$z\approx1$ AU. The convective flux through the $i$th cell is $F_{c}|_i =
\rho v_z c_p \Delta T |_i$, where $\Delta T$ is the difference between the
actual temperature at the cell center and the azimuthally averaged temperature
for that $r$ and $z$.  I find a convective cooling time of about 2 ORPs with
this method.  According to either estimate  of the convective flux, both of
which are crude and uncertain, convection is efficient at redistributing energy
in the disk's interior. However, these high convective fluxes may be due to a
combination of the random perturbation to the initial constant vertical entropy
profile, to the seeding of superadiabatic regions at the interior/atmosphere
interface, which is typical of the M2004 and C2006 schemes (see Chapter 3), 
and to the exclusion of irradiation, which tends to
produce stabilizing temperature stratification.  Moreover, as discussed below,
the energy ultimately must be {\it radiated} away at the photosphere.

The convection cells that are established in the axisymmetric phase are
disrupted once nonaxisymmetry sets in and spiral waves dominate the dynamics.
Vertical motions in disks with strong spiral shocks are related to shock bores, which
create high-altitude shocks and help stabilize the disk against convection.

Because shock bores deposit energy into the upper layers of the disk, it is
reasonable to ask whether shock bores themselves provide convection-like
cooling.  This does not seem likely.  Shock bores
effectively reduce the temperature in the post-shock region as the jumping gas
expands vertically.  This removes thermal energy in the post-shock region and
deposits that energy in the upper layers via waves.   Therefore, I do expect
shock bores and possibly other wave effects (e.g., see Lubow \& Pringle 1993,
Lubow \& Ogilvie 1998, and Ogilvie 1998) to enhance disk cooling.  Although energy may be effectively
removed from the post-shock region and allowed to radiate away quickly, 
the interior of the disk can still only
cool by radiative diffusion because of the positive vertical entropy gradient
established by the shock bores. From this point of view, shock bores and other
wave effects limit the efficacy of spiral waves in heating the disk, but do not
enhance midplane cooling.
Distinguishing between shock bores and convection is not a point of semantics.
Shock bores are born of large-scale shocks in a disk, while thermal convection
and the criterion for convective instability are usually described in the
context of a disk in vertical hydrostatic equilibrium.

One objection to the above analysis of the Boley et al.~(2006) disk, and also
the similar simulation with the BDNL scheme (Boley et al.~2007c), is that the
disk is not highly optically thick, with the midplane Rosseland vertically integrated
optical depth $\tau_m \lesssim 100$. To address the role of convection in a highly
optically thick disk, with $\tau_m > 1000$ for a large region of the disk, I
use the flat-Q model without the dense ring for a numerical experiment.  

As a reminder, the disk is 10 AU in radius,
approximately 0.1 M$_{\odot}$ disk mass, and surrounds 1 M$_{\odot}$ star. I
reset $\gamma=1.4$ everywhere and force the opacity law $\kappa = (T/150~{\rm K})^3\rm
~cm^2~g^{-1}$.  
This opacity power
law, along with the chosen $\gamma$, ensures that the disk should be convective
as discussed above.  The model is moderately
stable to GIs, with $Q\approx 2$ for most radii.  Figure 5.19 shows that
convection appears to be active in this model. There are large pockets of
negative entropy gradients at mid-disk altitudes. However, are these motions
truly convection, or are they related to other wave phenomena?  Figure 5.20
compares the vertical mass flux through a plane at $z=0.4$ AU with the
the midplane density structure, each normalized by the azimuthally averaged 
density for every radius. The midplane density plot indicates that some of the
vertical motions might be explained, in part, by wave dynamics driven by the
 weak spiral structure
in the disk. However, near 6 AU where the convection-like motions are the
strongest, there is no indication that spiral wave activity is correlated with
the vertical motions.  

Crude measurements of energy
transport by vertical 
gas motions, with $F_c=\rho c_p v_z \Delta T$, indicate that vertical motions
are as important as radiation in transporting energy vertically.
Even though a large fraction of the energy in the low to middle regions of the disk can be
transported by vertical gas motion, I find $t_{\rm cool}\Omega\approx1000$ near 5 AU. For
comparison, fragmentation occurs when $t_{\rm cool}\Omega\lesssim12$ for a
$\gamma=7/5$ gas  (Rice et al.~2005). The
energy must ultimately be radiated away near the photosphere, and so the cooling
times remain long because the cooling time at the photosphere regulates
convection.  This result is consistent with Rafikov's (2007) analytic
predictions, with my numerical tests (see Chapter 3), the numerical simulations
of Boley et al.~(2006, 2007c), and the
numerical simulations of Nelson et al.~(2000) and Nelson (2000). I note again that
 Nelson et al.~assume a vertically isentropic density profile
when calculating the cooling times for their 2D SPH calculations, which is
similar to assuming efficient convection.

In the simulations by authors who claim to see fast cooling due to convection (see Boss
2004a; Mayer et al.~2007), the vertical motions are likely shock bores, and the
fast cooling due to ``convection,'' I argue, is likely a result of poor boundary
conditions for radiation physics or sudden changes in numerical parameters (see Chapter 7).  
 As discussed by Nelson (2006), Boley et al.~(2007c), and here, proper treatment
of radiation physics, especially near the photosphere, is crucial for estimating
proper cooling times.
In fact, in order to lower the cooling times for the flat-Q model decribed above
to those expected to lead to fragmentation, the effective temperature would need to be
approximately equal to the midplane temperature, which is about three times the
actual effective temperature at $r=5$ AU.

\section{Conclusions}\label{conclusions}

\subsection{Shock Bores And Waves}

When a strong spiral shock develops in a disk, the shock's highly nonlinear
behavior creates a shock bore, where the loss of vertical force balance in the
post-shock region results in a rapid vertical expansion of the gas.  The
resulting shock structure covers a large range of disk altitudes. The shocks are
not limited to the main spiral wave itself.  Jumping gas can fall back onto the
pre-spiral shock gas producing breaking waves, vortical flows, and mid- and
high-altitude shocks.  The analytic approximations in \S 5.3 do an accurate job
of predicting the height of the disk in the post-shock region as measured in 3D
hydrodynamics simulations.

Protoplanetary disks with clumps and spiral arms can have very complex wave
dynamics.  As described by several authors (e.g., Bate et al.~2003), waves can
transport angular momentum and lead to the formation of gaps.  In addition, as
suggested by these simulations, waves may provide turbulence in the
disk as well as large vortical flows, some of which may extend down to the
midplane. The combination of radial transport and vortical flows should work to
stir, if not mix,  the gas and the lighter solids over tenths of an AU in only an orbit period.
Moreover, the wave dynamics should lead to copious shocks in the disk.

I find that convection in disks is not a viable mechanism for significantly
reducing the cooling times to those required for fragmentation. 
 Moreover, vertical motions associated with
shocks are likely shock bores and not convective flows.

\subsection{Bearing On Reality}

As demonstrated here, in Boley et al.~(2005), and in Boley \& Durisen (2006),
 a perturbation must be
strong to produce dynamic waves and shock bores in protoplanetary disks.
However, such strong perturbations may be plausible.  The onset of gravitational
instabilities could create massive spiral arms and/or clumps that could drive
strong spiral shocks throughout the disk, including the inner regions (Boss \&
Durisen 2005a,b).  Even if the disk is gravitationally stable as a whole, there
still may be regions in the disk, e.g., a dense ring or annulus formed by a dead
zone (Gammie 1996), where gravitational instabilities could set in, produce a
burst of spiral wave activity, heat the region to stability, and possibly later
repeat the process (e.g., Armitage et al.~2001).  Such a transient mechanism
could provide strong enough perturbations to drive strong waves and shock bores
without opening a gap.  The simulations presented here are not self-consistent
inasmuch as a sudden perturbation is introduced and the simulations
are not evolved for a long stretch of time.  These
simulations do, however, demonstrate that shock bores occur in strong
spiral waves, and they provide a basis for comparison with much more complex
simulations of GI active disks.

\chapter{GRAVITATIONAL DISK INSTABILITY}

In this section, I describe general results of two disk instability studies,
namely the simulations presented in Boley et al.~(2006) and Boley et al.~(2007c).
I find that the GI activity in these disks relaxes to a thermally
self-regulating state, where heating balances radiative cooling. In this state, the effective 
Shakura \& Sunyaev (1973) $\alpha\sim10^{-2}$ due to GI-driven torques, the
transport is dominated by global modes, and the details of the mass
transport and GI strength-mode spectrum is sensitive to the treatment of
radiation physics.  The cooling rates are too long for disk fragmentation to occur, 
and none is observed.

The Boley et al.~(2006; hereafter B2006) calculation is the 
same as the Mej\'ia (2004; hereafter M2004) radiative transfer simulation with no irradiation, which was evolved with
the M2004 radiation scheme (\S 2.4.1; see
also Mej\'ia 2004, \S 4.1.2).  For the Boley et al.~(2007c; hereafter BDNL) simulation, the M2004 disk was restarted just after the burst, but evolved with the
BDNL radiation transfer scheme (\S 2.4.2) in the standard version (SV) of the
code (\S 2.1)  to study how sensitive the evolution is to the treatment of
radiation physics.

\section{Initial Conditions and Method}

The initial model is the Mej\'ia disk (see Chapter 4), with the disk radius
scaled so that it extends from 2.3 to 40.0 AU. The star is scaled to 0.5
$M_{\odot}$, and so the disk mass is 0.07 $M_{\odot}$. For these simulations, 1
orp (outer rotation period) $\approx 253$ yr, and represents the initial gas orbital period at $r=33$
AU. The M2004 simulation was run from the initial axisymmetric Mej\'ia disk,
while the BDNL simulation was restarted from the M2004 simulation at about
6.5 orp. Restarting just after the burst was a compromise between lowering the
numerical cost of the simulation and avoiding transient features introduced by
the switch in radiation physics routines as the disk approaches an asymptotic
state.  For both simulations, due to an error with the inclusion of helium in
the solar mix, the mean molecular weight used for the temperature and pressure
ranges in the simulation is around 2.7 when it should be close to 2.3.  This
introduces a systematic offset no larger than about 16\% into the temperature,
which directly affects the cooling rates. However, the opacity law is roughly
quadratic in temperature, which means the flux calculated from the diffusion
approximation ($F\sim T^4/\kappa$) is roughly quadratic in temperature, too. The
error in the mean molecular weight should be an error that typically enhances
the cooling by $1.16^2$ or about 4/3.  Because I expect the cooling to be
enhanced, fragmentation should be more likely in the M2004 and BDNL simulations
than it would be with the correct mean molecular weight. I also note that for
the BDNL simulation, only Rosseland opacities were used because the
interpolation described in \S 2.4.4 was still being developed at the start of
the BDNL simulation.

The temperatures in these simulations never reach the dust sublimation
temperature, near $T\sim 1400$ K (Muzerolle et al.~2003), so the opacity is mostly
due to dust. D'Alessio (2001) opacities are used, with minimum and maximum grain sizes
$a_{\rm min}= 0.005\mu$m and $a_{\rm max} = 1\mu$m, respectively (see Mej\'ia 2004 or Appendix A of B2006) and an
interstellar grain size power law $dn=a^{-3.5}da$. 
The minimum allowed temperature at all times is 3 K to simulate radiating
into empty space.  However, it should be noted that there is an inconsistency regarding the treatment of the hydrodynamics and the radiation physics.  In the
SV, wherever the temperature drops
below 3 K, the temperature is reset without correcting the specific internal
energy, leading to an inconsistent pressure.  This inconsistency is
corrected in CHYMERA by resetting the specific internal energy of the cell
as well as the temperature.  

Both simulations are evolved approximately up to 4000 yr,
i.e., about 16 orp.  The resolution used is the same as the constant cooling
time simulation presented in Mej\'ia et al.~(2005). For the M2004 simulation, the $z$
direction is fixed at 32 zones. As indicated by the radiation transfer tests (Chapter 3),
the BDNL scheme leads to a more extended atmosphere than the M2004 scheme,
and so the vertical direction was extended to 64 zones.  
The starting model fills the initial grid, 
namely $(r, \phi, z) = (256,128,32)$, and so when the disk rapidly expands
during the burst,
the grid is extended to
512 in $r$, while keeping the
same cell resolution of $\Delta r =\Delta z = 1/6$ AU.  M2004 and BDNL are
also run on a high azimuthal resolution grid, $(r, \phi, z)
=(512,~512,~32~\rm or~64)$ to test for fragmentation (see \S 5.2).   A random
cell-to-cell density perturbation of amplitude $\left|
\Delta\rho/\rho\right|=10^{-4}$ is applied at the very first step of the M2004 run,
which allows spiral modes to grow from the background noise as the disk cools.

\section{Disk Structures}

Figure 6.1 shows surface density images of the M2004 simulation.  The disk
undergoes the same four evolutionary phases described for the disks in Pickett
et al.~(2003), Mej\'ia (2004), Mej\'ia et al.~(2005), and Cai (2006), namely the
{\it axisymmetric}, {\it burst}, {\it adjustment}, and {\it asymptotic} phases.
The axisymmetric phase is the initial cooling phase, before the instabilties
activate. The burst phase describes the often violent onset of GIs.  The
adjustment or transition phase is the relaxation to the thermally
self-regulating GI-active phase during which heating is balanced by radiative
cooling, i.e., the asymptotic phase.  

During the axisymmetric phase, the disk begins to pulsate radially.
The pulsation is probably due to a combination of the loss of radial pressure support
due to cooling and of the absence of a mechanism for redistributing angular
momentum.  A dense, thin ring, in which $Q<1$, forms in the outer
disk just before the burst.  Once nonaxisymmetry develops, the ring becomes part
of the spiral structure. 
Between 2 and 3 orp, GIs fully activate, and the disk rapidly expands as open
spiral structure efficiently transports angular momentum outward.  During the
adjustment phase, the disk oscillates in size to a much greater extent than
noted in the axisymmetric phase.  In addition, the power in the now highly 
non-linear GI activity begins to dampen in response to shock heating and 
work done by gravity, and the disk transitions to a thermally self-regulating state.

One-armed structure does develop during the adjustment phase, but because the
star is held fixed at the origin, some of the dynamics are not accurately treated.
Methods like moving the star to a location that brings the center of mass back to the
center of the grid (Boss 1998) are not employed because this might incorrectly
treat the dynamics as well.  A preliminary leap-frog method has recently been
developed to release the star for the Wengen test 4 simulations (Mayer et al.~2007, in preparation), and a    
more accurate method is being developed for standard incorporation into CHYMERA,
but the simulations presented in M2004, B2006, and BDNL were completed 
before such routines were developed.  

\begin{figure}
\begin{center}
\includegraphics[width=5.95in]{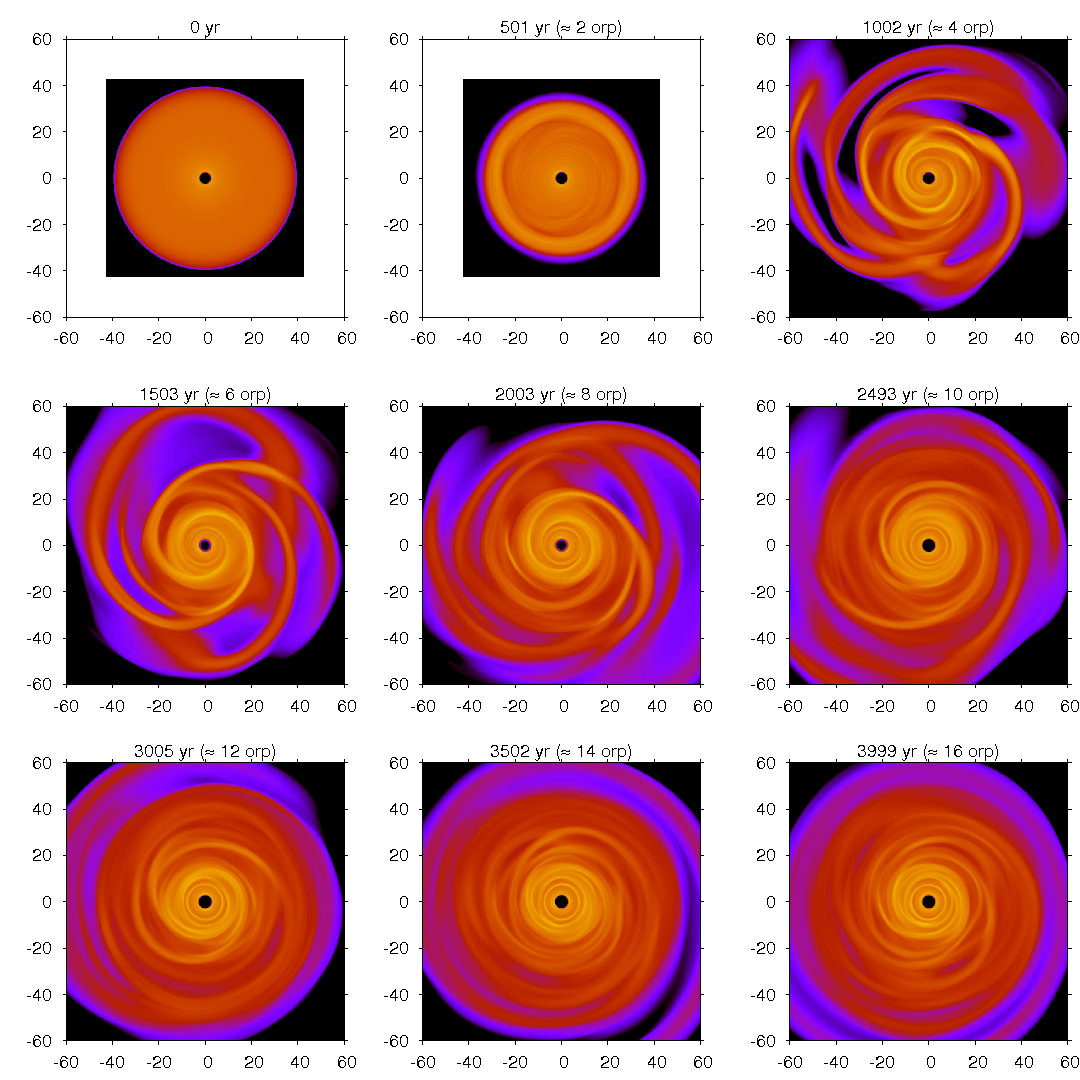} 
\caption[Surface density snapshots of the M2004]
{Evolution of the M2004 simulation. All
images show surface densities in logarithmic colorscale spanning 4.5 orders of magnitude, with the abscissa
and ordinate in AU. The computational grid is larger than what is shown in the
image (80 AU), and significant amounts of mass are not leaving the grid.
A movie of the simulation is available at {\it
http://hydro.astro.indiana.edu/westworld} under the {\it Movie s} link.}
\label{f1}
\end{center}
\end{figure}

\begin{figure}
\begin{center}
\includegraphics[width=5.95in]{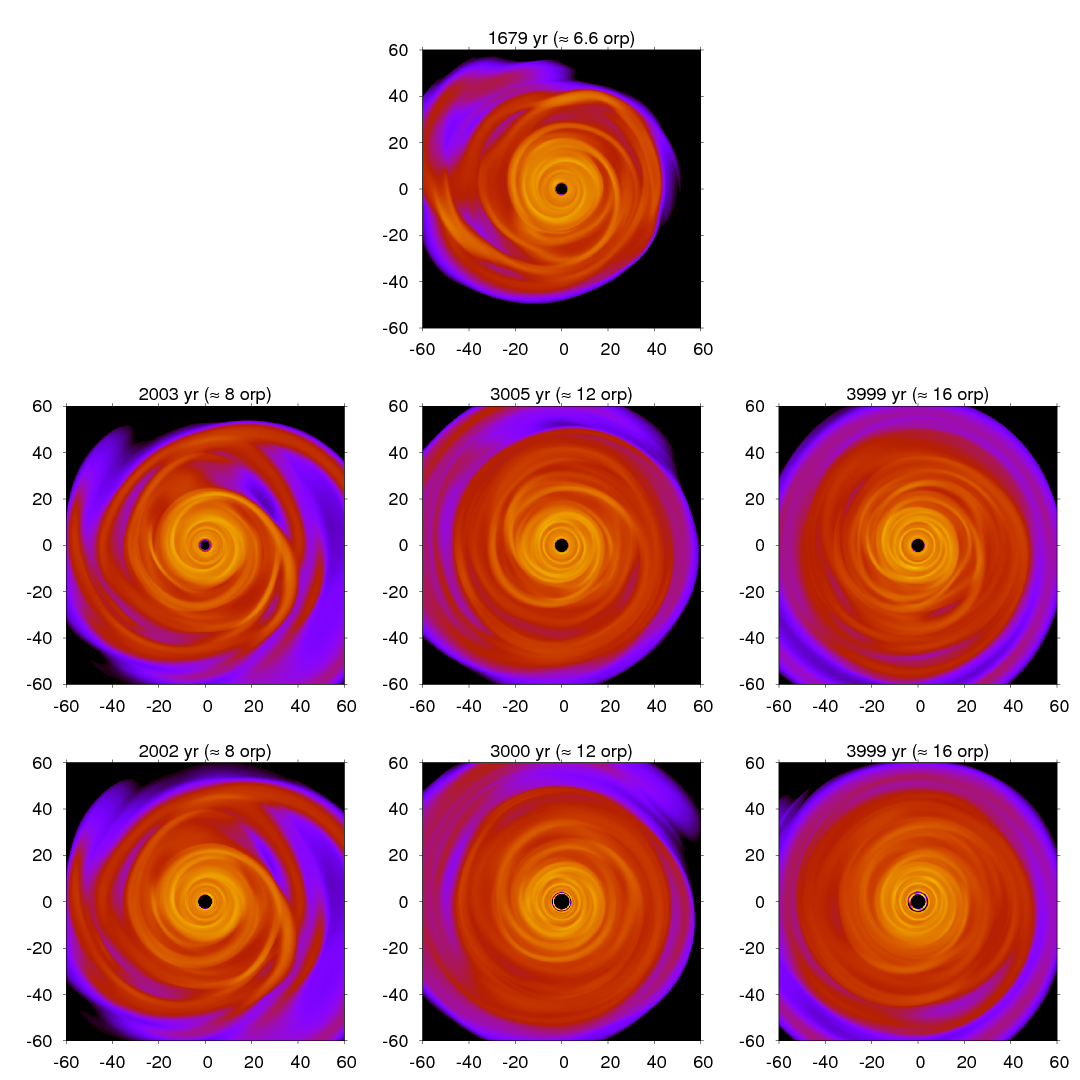} 
\caption[Comparison between surface densities for the M2004 and the BDNL
simulations: surface densities] {Comparison between the evolution of the M2004 (middle
row)
and BDNL (bottom row) simulations.  The scaling is the same as in Figure 6.1. Although fairly
modest, the differences between the simulations are discernible; the spiral
structure in the BDNL disk is more washed out than in the M2004 disk.} 
\label{f2}
\end{center}
\end{figure}

\begin{figure}
\begin{center}
\includegraphics[width=5.95in]{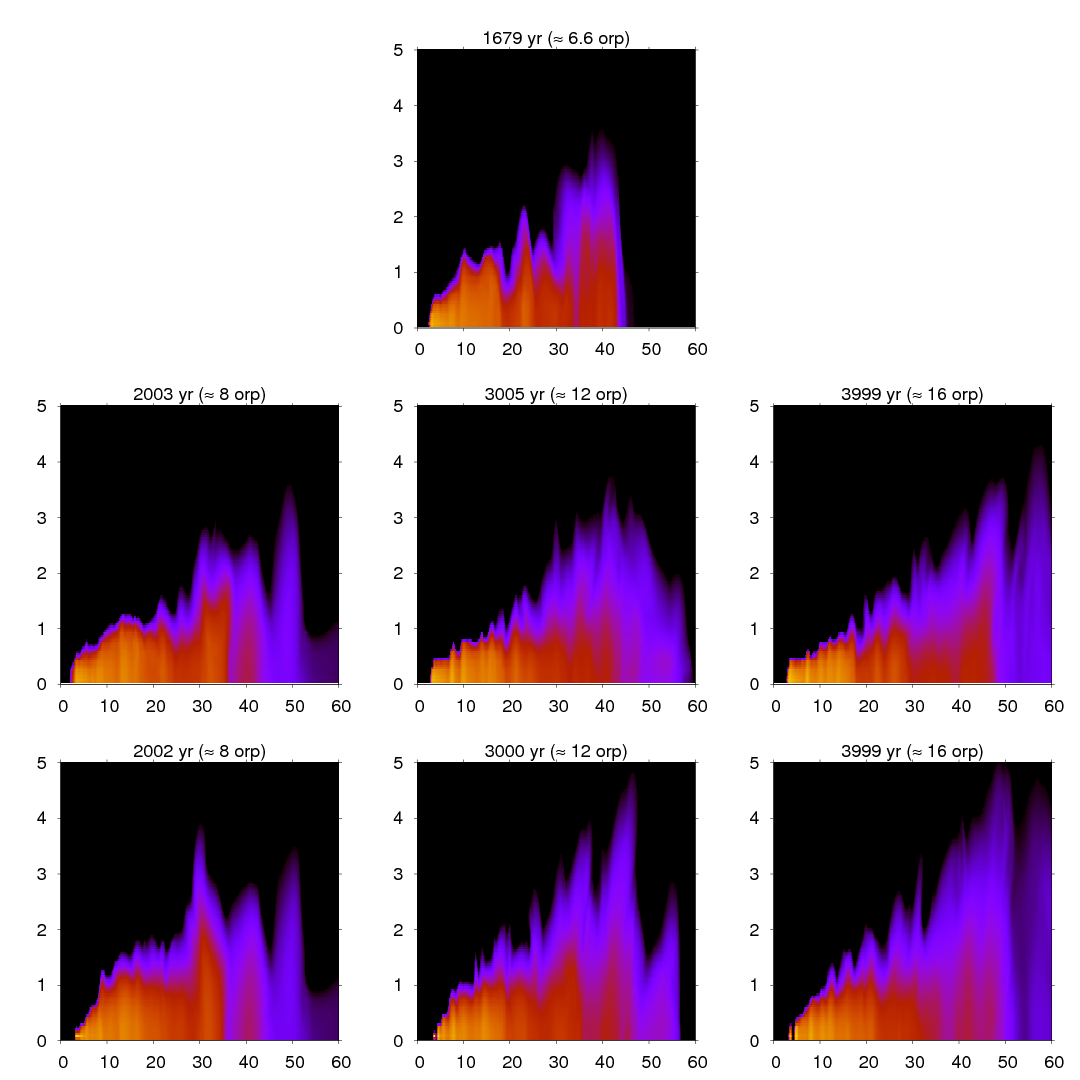} 
\caption[Comparison between meridional density slices for the M2004 and the BDNL
simulations]{Comparison between the evolution of the M2004 (middle row) and BDNL (bottom row)
simulations: merdional volume density slices at 3 o'clock in Figure 6.2.  The images show logarithmic density on a colorscale spanning 4.5
orders of magnitude.  The major differences between the images are that the BDNL
disk is more collapsed in the inner disk than the M2004 disk and is more flared
than the M2004 everywhere else.}  
\label{f3}
\end{center}
\end{figure}

\begin{figure}
\centering
\includegraphics[width=5.95in]{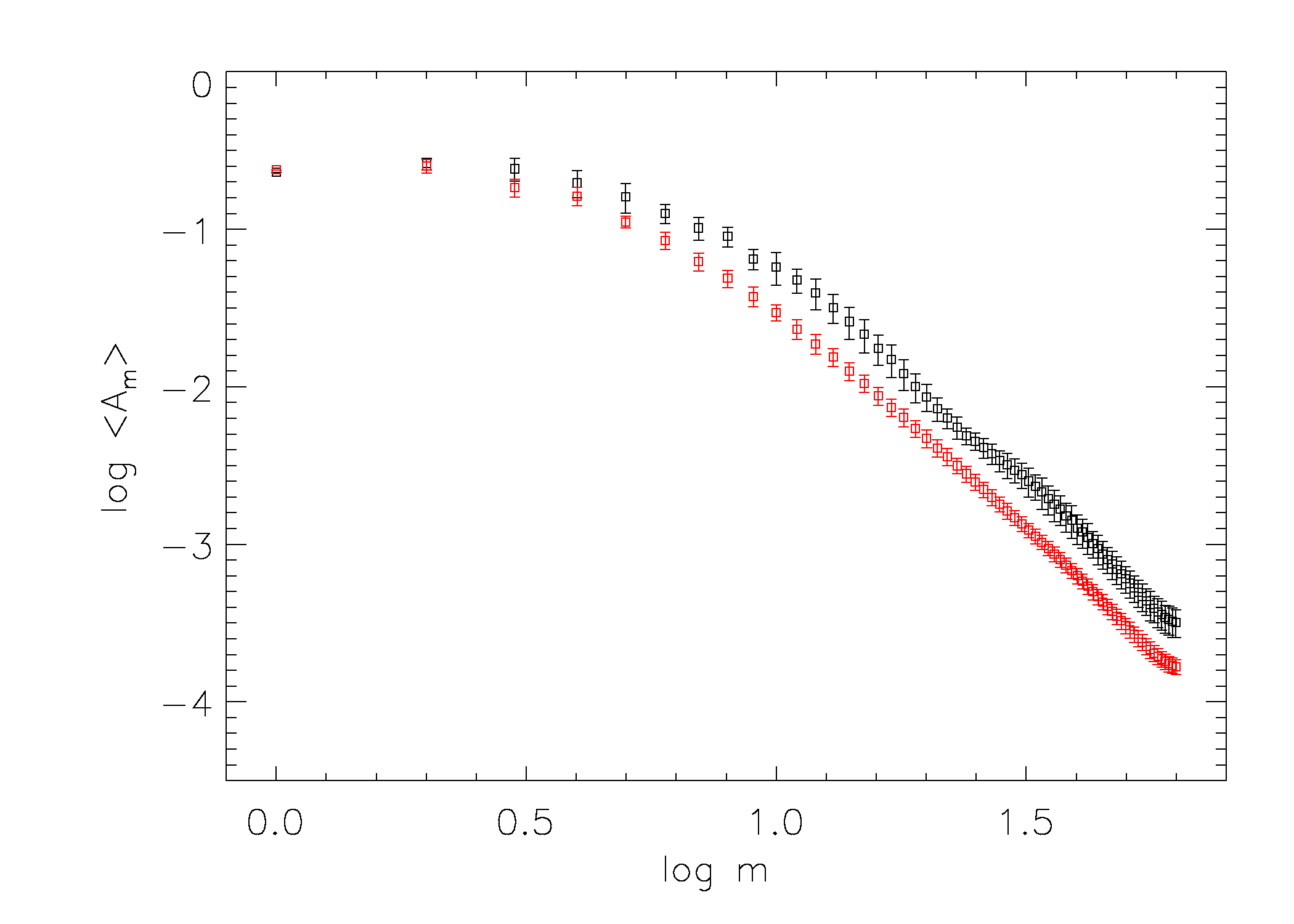}
\caption[Fourier amplitude spectrum for the M2004 and BDNL simulations]
{Fourier amplitude spectrum for the M2004 disk (black)  and the BDNL
disk (red), time averaged over the last two orp of each simulation.  The
bars represent typical fluctuations over the two-orp period.  The M2004
profile has
larger amplitudes everywhere except for $m=2$, which is consistent with the
M2004 disk being more nonaxisymmetric.  Both spectra can be fit with a
functional form of $A_m\sim(m^2+m_0^2)^{-n}$, where $n\approx 1.5$. 
This behavior at large $m$ may be indicative of gravitoturbulence,
nonlinear mode coupling, or both.}
\label{fig6.4}
\end{figure}

Although not extreme, qualitative differences between the M2004 and BDNL
simulations are
discernible in the surface density plots (Fig.~6.2) and meridional slices
$\phi\approx0$ (Fig.~6.3). The M2004 simulation has more pronounced spiral
structure than the BDNL simulation throughout most of the disk.  Also, the BDNL
simulation is more extended in the vertical direction because the disk's
atmosphere is hotter, as expected from the tests. The ring that forms at the
innermost edge of the BDNL simulation is due to poor numerical resolution; the
mass at the edge cooled until contained within one cell.  The M2004 simulation
does not exhibit similar behavior because the scheme is slighly more dependent on
resolution (Chapter 3).

In order to quantify the structural differences, I compute the global Fourier
amplitude spectrum for each simulation, where the sum
over the amplitudes is a measure of the nonaxisymmetric structure in the disk
and the spectrum is indicative of the dominate modes in the disk.  I compute
the time-averaged Fourier component $ \left<A_m\right>$ for $m$-arm structure by
\begin{equation}
A_m = \frac{\int\rho_m rdrdz}{\int \rho_0 rdrdz},
\end{equation}
where $\rho_0$ is the axisymmetic density component and $\rho_m$ is
the total Fourier amplitude of the $\cos(m\phi)$ and $\sin(m\phi)$ density
component.  The time-average is calculated by finding $A_m$ for a large number
of snapshots over the last two orps. The summed global Fourier amplitude
$\left<A_+\right>=\sum_{m=2}^{63}\left<A_m\right>=1.4$ for the M2004 disk, while
$\left<A_+\right>=1.1$ for the BDNL disk, which indicates that the M2004 disk
is more nonaxisymmetric.  I exclude $m=1$ from the summation because the
star is kept fixed.  The difference between the sums is also depicted by the Fourier
spectrum (Fig.~6.4).  The M2004 disk has larger amplitudes everywhere except for
$m=2$, which is consistent with the qualitative differences portrayed in Figures
6.2 and 6.3.  Including long-range transport of radiation in the BDNL scheme has resulted in weaker GIs.

 As
described in B2006, a curve with the  functional form
$A_m\sim\left(m^2+m_0^2\right)^{-n}$ can be fit to the data, where
$n\approx1.6$ and $m_0\approx 7.5$.  The BDNL disk is also consistent with this functional form, and
both disks roughly follow $n\approx 1.5$.  The
similar slopes at large $m$ may be indicative of gravitoturbulence (Gammie
2001) or nonlinear mode coupling (e.g., Laughlin et al.~1998). The meaning of the power spectrum remains unsatisfactorily understood.  

\section{Disk Energetics}

Figure 6.5 shows the evolution of the internal energy for each disk.  There is a
precipitous drop in energy over about an orp (253 yr) after switching the radiation
transport schemes.  However, the schemes roughly follow each other after the
drop, with the BDNL profile having a more shallow slope than the M2004 profile.  
The effective temperature profiles (Fig.~6.6) are also similar; 
each profile can be fit by an exponential. As discussed in B2006, this
profile is very steep when compared with observationally derived profiles that
are based on spectral energy distribution measurements. Observational studies typically
find that for $T_{eff}\sim r^{-q}$, $q\approx$ 0.4-0.8 (e.g., Beckwith et
al.~1990, Kitamura et al.~2002, and Dullemond et al.~2007; see also theoretical
models by Miyake \& Nakagawa 1995).  The exponential profile is likely a result
of the exclusion of stellar irradiation in these simulations.  

For each simulation, I calculate the time-averaged cooling time \linebreak $t_{cool} =
\int \epsilon dV/\int \mid\nabla \cdot {\bf F} \mid dV$ for each annulus on the grid, where
$\epsilon$ is the internal energy density of the gas and $\nabla \cdot {\bf F}$ is
the radiative cooling. The temporal average is taken to be about the last
six orp of evolution for the M2004 simulation and about the last 5 orp for the BDNL
simulation.  In Figure
6.7, I compare $t_{\rm cool}\Omega$ curves for each disk, where $\Omega$ is the
angular speed of the gas.  The cooling time is much longer for $r \lesssim 35$
AU in the BDNL disk than it is for the M2004 disk. This is likely due to a
combination of the different
opacities used by the two routines and of the free-streaming approximation, employed by the M2004
algorithm, in regions where long-range radiation coupling matters (see Chapter 3).  Outside $r\sim 35$ AU, the curves
converge.   The longer cooling times are consistent with the
washed out structure in the BDNL simulation.  For both disks, the cooling times
are well above the fragmentation criterion $t_{\rm cool}\Omega \lesssim$ 3 to 6 for a
$\gamma=5/3$ gas (Gammie 2001; Rice et al.~2003), so I expect neither disk to
fragment.  Regardless, I ran the BDNL simulation at 512 azimuthal divisions 
between 10 and 11 orp (2530 and 2783 yr) to test for fragmentation, when the total
internal energy reaches its minimum, and found no signs of
fragmentation, as one expects from the long cooling times. 

\begin{figure}
\centering
\includegraphics[width=5.95in]{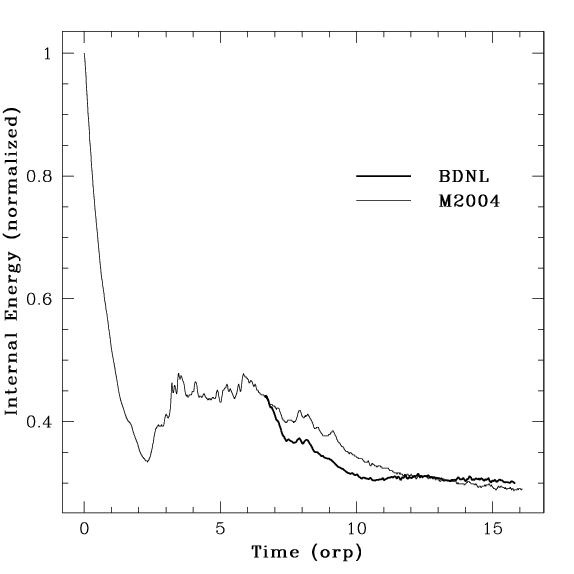} 
\caption[Intenral energy vs.~time for the M2004 and BDNL simulation]
{Internal energy for BDNL (heavy curve) and M2004 (light curve), normalized by the
M2004 initial value.  The
precipitous drop at 7 orp is a result of suddenly switching radiation schemes.  Between
about 7 and 10.5 orp (1671 and 2657 yr)  the curves approximately track each other.}
\label{fig10}
\end{figure}

\begin{figure}
\centering
\includegraphics[width=5.95in]{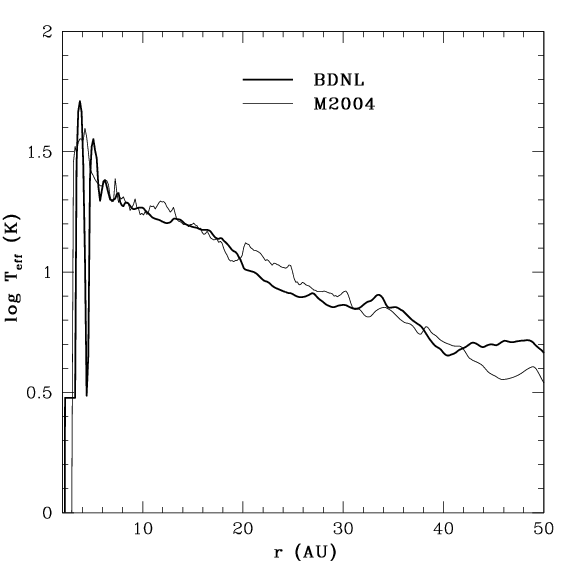}
\caption[Effective temperature profiles for the M2004 and BDNL simulations]
{Effective temperature profiles for the BDNL (heavy curve, time-averaged
over about the last 5 orp or 1165 yr and M2004 simulations (light curve, time-averaged over the last
6 orp or 1418 yr).  Both follow an exponential profile, and are reasonably consistent.
Their departure from a rough $r^{-1/2}$ effective temperature profile is
likely due to exclusion of stellar irradiation.} 
\label{fig6}
\end{figure} 

\begin{figure}
\centering
\includegraphics[width=5.95in]{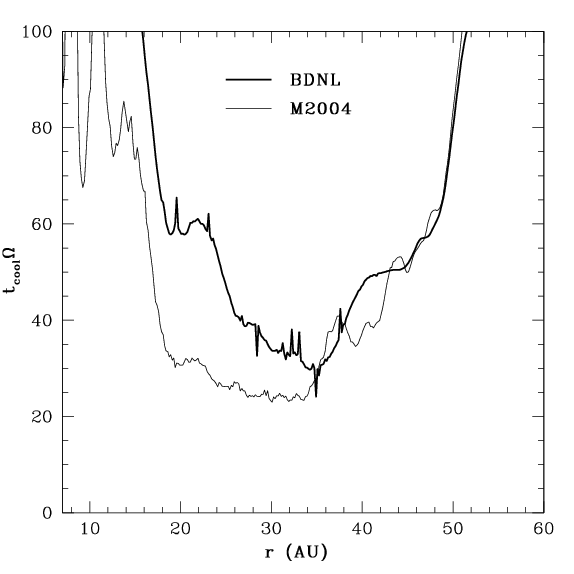}
\caption[Cooling time curves for the M2004 and BDNL simulations]
{Cooling time curves scaled by the local angular speed for the BDNL
(heavy curve) and M2004 (light curve) simulations for the time-averaged periods over about
the last 5 and 6 orp, respectively.  Both curves are relatively consistent for
$r\gtrsim 35$ AU, but they depart for inner radii.  This is likely due to a
combination of the different
opacities used by the two routines and of the free-streaming approximation, employed by the M2004
algorithm, in regions where long-range radiation coupling matters.} 
\label{fig7}
\end{figure} 

\section{Angular Momentum Transport}

In this section, I compare the angular momentum transport in each disk by
analyzing the gravitational torque on the inner disk due to the outer disk and
by measuring the effective Shakura \& Sunyaev (1973) $\alpha$.  As discussed in
\S 2.5.2, the gravitational torque is calculated by 
\begin{equation} C = \int_{V(R)} \rho {\bf x} \times {\bf \nabla}\Phi~dV,
\end{equation}
where $\Phi$ is the gravitational potential, ${\bf x}$ is the position vector,
and the integral is over the volume contained inside $r=R$.  For this analysis, I am concerned
with the vertical component $C_z$.  The time-averaged torque, averaged over the
last six orp for the M2004 disk and about the last 5 orp for BDNL disk, is shown for each
simulation in Figure 6.8.  The solid curves represent the torque profiles, with
the heavy curve indicating the BDNL disk.  The dashed curves show the mass flux
for each disk with arbitrary but consistent scaling; the peak mass flux
$\dot{M}=$ few$\times10^{-7}~M_{\odot}~\rm yr^{-1}$ for each disk. The torques
are of the same magnitude, but the torque profiles are noticeably different.
Based on the $\left<A_m\right>$ plots and the visual differences in disk
structures, the M2004 disk has a more complex morphology and stronger modes.
The multiple peaks in the M2004 torque profile are another indication of this
complex morphology and competing global, dominant modes.  The BDNL torque
profile also has multiple extrema, but the variations are not as extreme.  The hydrodynamical
stress (Chapter 2) may play an important role in the BDNL disk.

\begin{figure}[ht]
   \centering
   \includegraphics[width=5.95in]{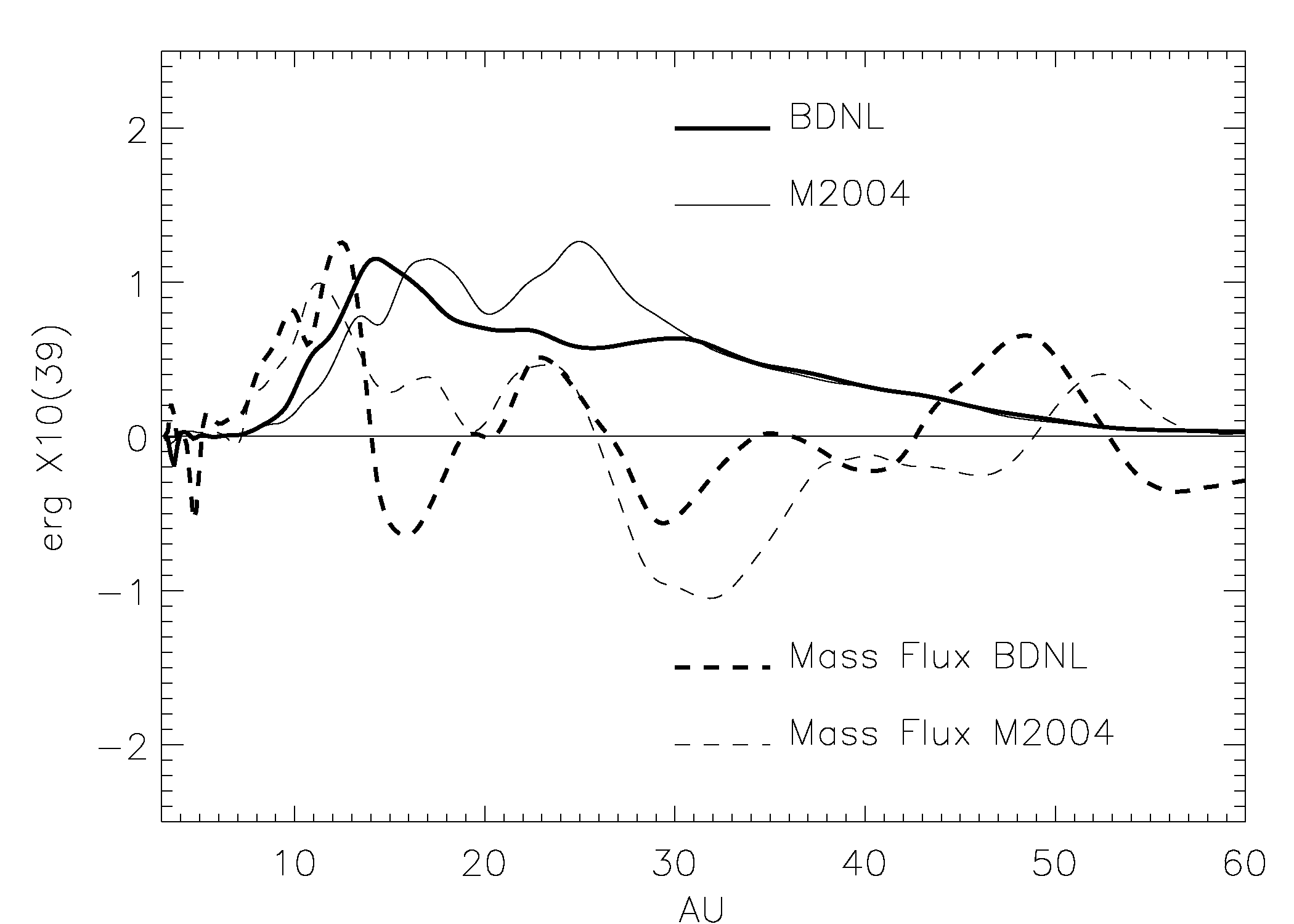}
   \caption[Gravitational torque profiles]
   {The negative of the gravitational torque (solid curves) and mass flux (dashed curves)
profiles for BDNL (heavy curves), time-averaged over the last 5 orp, and M2004
(light curves), time-averaged over the last 6 orp.  BDNL's torque profile
shows one strong peak and several minor peaks, while the M2004 torque profile
has two very strong peaks.   The mass fluxes for each disk (arbitrary but
consistent scaling) are consistent in magnitude.  }
   \label{fig8}
\end{figure}
 
\begin{figure}[ht]
   \centering
   \includegraphics[width=5.95in]{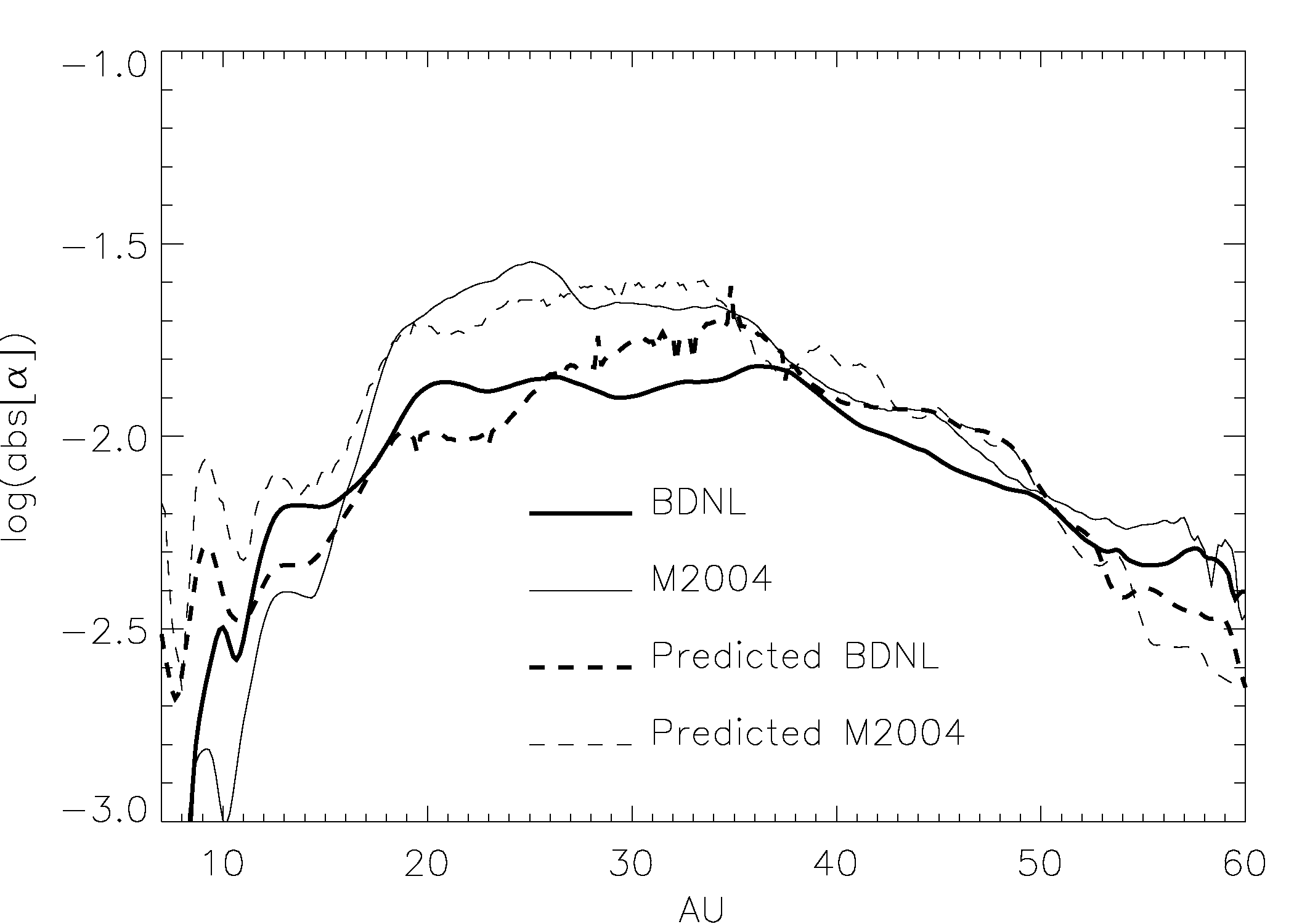}
   \caption[Effective $\alpha$ profiles for BDNL]
    {Effective $\alpha$ profiles for BDNL (heavy curves) and M2004
(light curves). The solid curves indicate the effective $\alpha$ derived from
the torque profiles, and the dashed curves indicate the predicted $\alpha$
based on an $\alpha$ disk prescription,
derived from the $t_{\rm cool} \Omega$ profiles (Gammie 2001) in Figure 6.7 with the
assumption of negligible self-gravity (see text).  Both disks roughly follow
the predicted $\alpha$ over a large range of radii. }
   \label{fig9}
\end{figure} 

\begin{figure}[ht]
   \centering
   \includegraphics[width=5.95in]{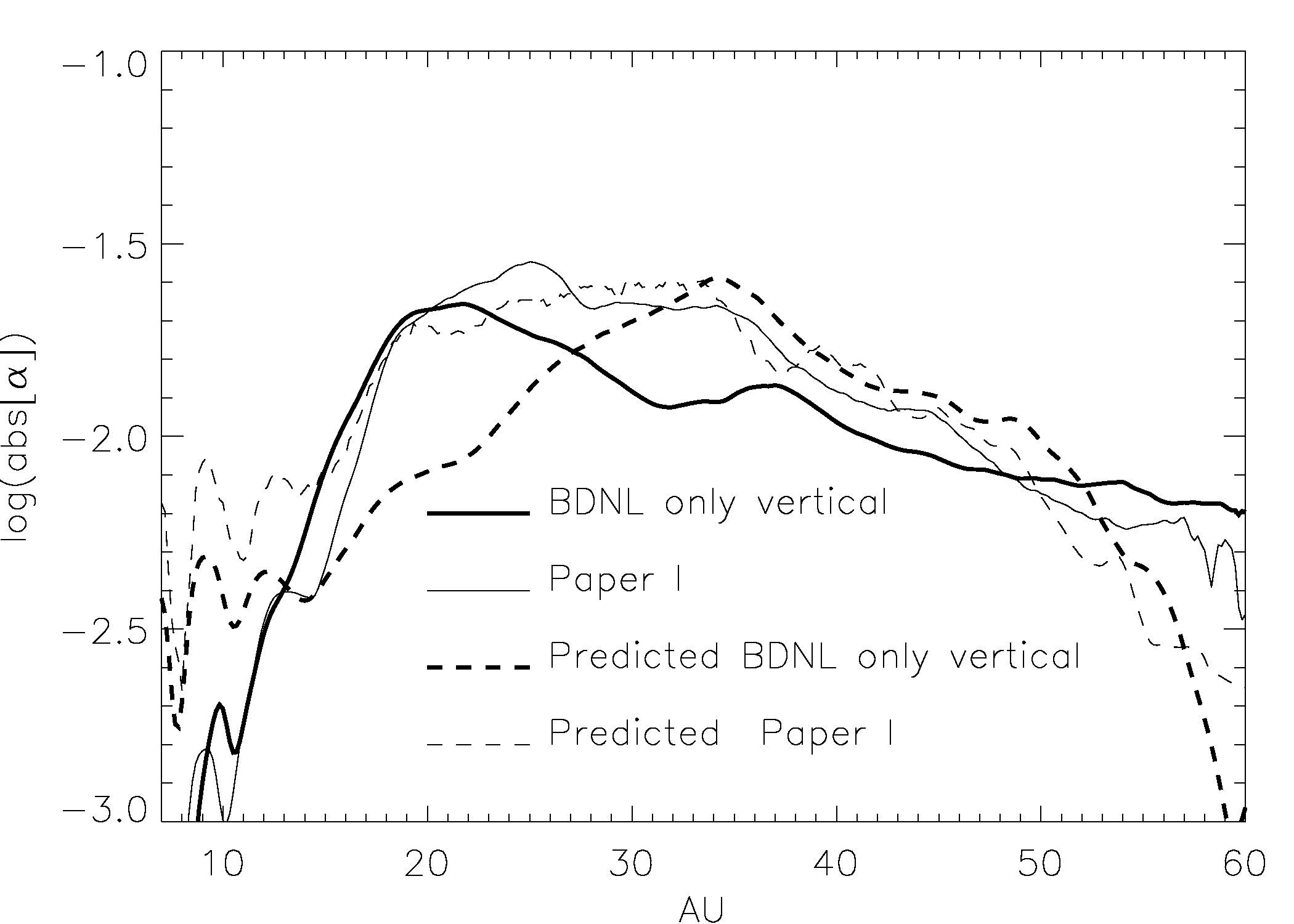}
   \caption[BDNL $\alpha$ profile for vertical radiation transfer only]
   {Same as Figure 6.9, but the BDNL simulation has only vertical
radiation transport. }
   \label{fig10}
\end{figure} 

Both disks have complicated mass flux profiles.  The principal inflow/outflow
boundary in the M2004 simulation is at $r\sim 26$ AU.  The BDNL disk,
by contrast, has two main inflow/outflow boundaries.  The $r\sim15$ AU
boundary corresponds to the peak torque in the BDNL disk, and I refer to this
as the principal inflow/outflow boundary because the mass fluxes are the highest
near it.  Roughly, the peak in each torque profile aligns with the principal
inflow/outflow boundary.  The agreement is imprecise, because the mass flux
average is based on differencing mass cylinders at different times, which
yields a time average based on the second-order mass flux integration.  The 
torques are derived in post-analysis calculations, and so the temporal sampling is
much sparser.  Moreover, the mass fluxes are highly variable with time, and
averages over slightly different time periods can result in different mass flux
profiles.  However, major inflow/outflow transitions are usually near torque
profile extrema. The mass fluxes for $r\gtrsim 40$ AU are complicated by pulsations
that begin just before the disk bursts and continue throughout the evolution.  

I calculate an effective $\alpha$ by relating the vertically integrated gravitational stress $\mathcal{T}_z$ 
to the gravitational torque through $\mathcal{T}_z=-C_z/2\pi r^2$ (equation [2.42]).  An effective $\alpha$ is then given by
 (Gammie 2001)
\begin{equation} \alpha =   \Big | \frac{d\ln \Omega}{d\ln r}\Big |^{-1}
\frac{\mathcal{T}_z}{\left< c^2 \Sigma \right>},\end{equation}
where the brackets indicate an azimuthally averaged quantity, $c$ is the
adiabatic midplane sound speed, and $ \Sigma$ is the surface density.
 The comparison between the effective $\alpha$ for the BDNL and M2004
disks is shown in Figure 6.9.  The profiles are of similar magnitude everywhere, but with
BDNL being significantly lower between about 20 and 36 AU. 

I also show in Figure 6.9 the $\alpha$ one would expect for an $\alpha$ disk
(see Gammie 2001; equation [2.41])
\begin{equation}
\alpha =  \left(\Big |\frac{d\ln\Omega}{d\ln r}\Big |^2
\gamma_{\rm 2D}\left(\gamma_{\rm 2D}-1\right)t_{\rm cool}\Omega\right)^{-1},
\end{equation}
where $\gamma_{\rm 2D}$ is the two-dimensional adiabatic index.  For $\gamma=5/3$,
$\gamma_{\rm 2D}\approx 1.8$ in a strongly self-gravitating disk and 1.5 in a
non-self-gravitating disk (Gammie 2001).  I use the $t _{\rm cool}\Omega$ profiles
(Fig.~6.7)  in equation (6.4) to plot the anticipated $\alpha$ for a local model
in the non-self-gravitating limit.  For most radii, both disks are roughly
consistent with this $ \alpha$ prescription, and $\alpha$ is roughly constant
between 20 and 35 AU.  The main difference between the two simulations is the
lower $\alpha$ in the BDNL simulation, which is consistent with the longer
cooling times.  Recall that the free-streaming approximation is used in the M2004 scheme
wherever $\tau<2/3$.  This can lead to additional cooling if $\tau\gtrsim$ a few
hundredths (see Fig.~3.4); the use of only the Rosseland mean opacity in the BDNL simulation likely plays a role as well.

Although the overall evolutions are in rough agreement, they demonstrate
sensitivity to the details of radiation transport.  By including the long-range
effects of radiative transfer and by using different opacities for different
optical depth regimes, the BDNL disk shows less structure, is more flared, and
has a lower effective $\alpha$ that deviates slightly more from equation (6.4).
These differences demonstrate the need for a radiation algorithm that includes
the long-range effects of radiative transfer in all three-dimensions, which
will be missed by diffusion approximations and is missed in the BDNL scheme in the
$r$ and $\phi$ directions.  To illustrate the importance of radiative transport
in all three directions, I show in Figure 6.10 the effective $\alpha$ profile for
a disk that was evolved with the BDNL radiative routine, but with the radial
and azimuthal diffusive radiation transport turned off.  According to this
plot, without any $r$ and $\phi$ transport, I would surmise that the effective $\alpha$  
deviates strongly
from the predicted $\alpha$.

\section{Mass Redistribution}
Once GIs active, mass in the M2004 disk is quickly reordered.  During the burst, roughly between 2 and 4 orp (500-2000
yr), accretion rates are as high as a \linebreak few$\times 10^{-5}~M_{\odot}~\rm yr^{-1}$.
During the adjustment phase (6-10 orp or 1500-2500 yr), 
the accretion rates decrease by an order of magnitude, and
for the asymptotic phase (10-16 orp or 2500-4000 yr), the typical accretion rate
decreases by another order of magnitude.  The mass inflow/outflow boundary
varies with time, and multiple inflow/outflow
boundaries develop.

Figure 6.11 (bottom) also illustrates how the $r^{-0.5}$ surface density profile of the
initial disk is lost, and approximately relaxes to a Gaussian of the form $\Sigma =
\Sigma_{0} 10^{-\left( r/r_e\right)^2}$ for $r\gtrsim20$ AU (dashed curve). The least-squares
fit in log-linear space with the dummy variable $x=r^2$ yields $r_e=46.7$ AU
when fit between $r= [20,60]$ AU.  However,  on the intervals $r=[20,43]$ AU and
$r=[43,60]$ AU the surface density profile seems to follow two different power
laws $\Sigma\sim r^{-\nu}$.  For the inner interval, $\nu=-1.93$, and for the
outer interval, $\nu=-5.97$.  The Spearman correlation coefficients are
$R=-0.992,- 0.986,-0.985$ for the exponential, the inner power law, and the
outer power law, respectively.

The disk inside 20 AU is not well-described by any simple function.
There is a ring at 7.5 AU, and another seems to be forming at 10.5 AU, both
resembling those that appeared in the constant cooling simulation with the
Mej\'ia disk (Mej\'ia 2004; Mej\'ia et al.~2005). Although the exact mechanism
is unclear, these rings
appear to be a numerical artifact that results from a combination between
poor vertical resolution and keeping the star fixed.  However, there is 
also a broad, radial mass concentration at about 15 AU due to tightly 
wrapped spiral arms, which is likely a real feature. I therefore speculate that 
if there are regions in disks that are kept hot for physical reasons, ring-like mass
concentrations could grow. 

\begin{figure}
\begin{center}
\includegraphics[width=10cm]{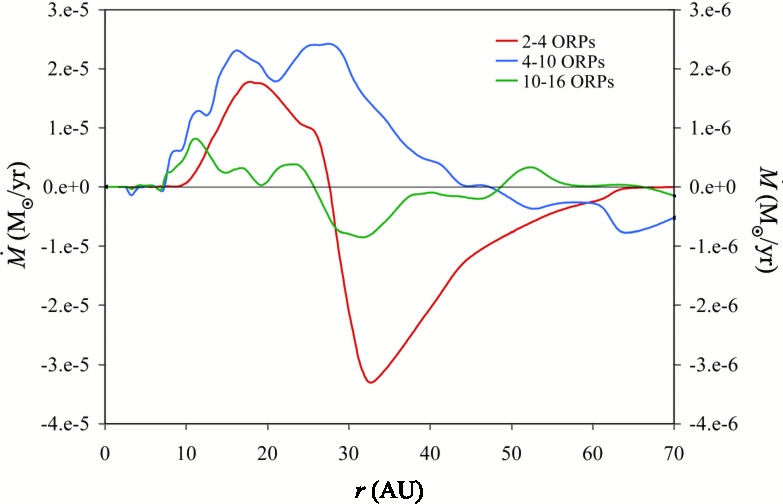}
\includegraphics[width=8cm]{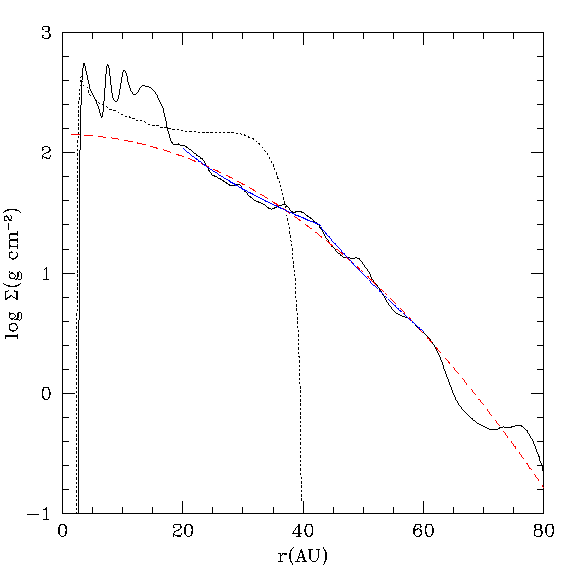}
\caption[Mass redistribution and surface density profile for the M2004
simulation]
{Global mass redistribution for the M2004 disk.  The top panel shows the average mass transport
rate calculated by differences in the total mass fraction as a function of
radius for three different times: between 2 and 4 orp, 4 and 10 orp, and 10 and
16 orp.  The red curve is scaled to the left ordinate while the blue and green
curves are scaled to the right ordinate.  The bottom panel shows the surface
density as a function of radius for the initial disk (dotted) and the final
state (solid), which reflects the density profile during the asymptotic state.
The dashed curve shows that the density profile for $r\gtrsim 20$ AU follows a
Gaussian distribution as described in the text.  However, the surface density
profile can also be broken down into two power laws (blue curves). (Same as
in B2006.)} \label{f3}
\end{center}
\end{figure}

\begin{figure}
\begin{center}
\includegraphics[width=5.95in]{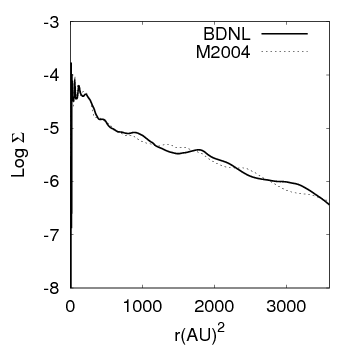}
\caption[Mass distributions for the M2004 and BDNL simulations]
{Comparison between the surface density profiles of the M2004 and BDNL disks.
Each profile is consistent with a Gaussian profile for $r\gtrsim 20$ AU.}
\label{f12}
\end{center}
\end{figure}

\begin{figure}
\begin{center}
\includegraphics[width=5.95in]{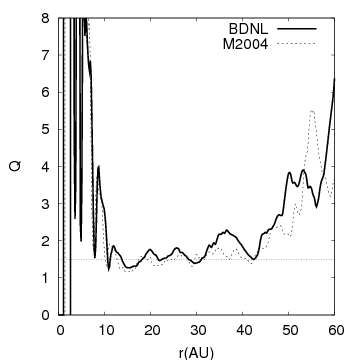}
\caption[Toomre $Q$ for the M2004 and BDNL simulations]
{Toomre $Q(r)$ for the M2004 and BDNL simulations. Both profiles are for snapshots 
at the end of the simulation.  However, each profile evolves little
between about 10 orp and 16 orp.  The line indicates $Q=1.5$.}
\label{f13}
\end{center}
\end{figure}

The BDNL simulation's mass distribution is very similar to that of the M2004
simulation, despite the difference in the strength of GIs in each disk.  Because the BDNL
disk was restarted after the burst and the burst is responsible for reordering mass quickly, 
it is expected for the mass profiles to be fairly similar.  
As displayed in Figure 6.12, both surface density profiles follow Gaussians over a large radial extent.

\section{Thermal Balance}

The Toomre $Q$ evolves little between
about 10 orp (2530 yr) and the end of the simulations (Fig.~6.13). 
 As noted by M2004 and B2006, the
average $Q$ in the M2004 disk between 14 and 43 AU is 1.5, while the average
$Q$ between the same region is 1.7 for the BDNL disk.  The BDNL disk having a
higher $Q$ than the M2004 disk is consistent with the previously discussed
assessments.  In addition, the steady $Q$ indicates that both disks are in an
asymptotic-like state, as was noted for the Mej\'ia et al.~(2005) constant
cooling time simulation, and so the GIs in these disks have led to a thermally
self-regulating state for which radiative cooling is balanced by work and
shock dissipation. The distinction asymptotic-{\it like} is made because the
cooling times in both disks are continually adjusting, and so the behavior of GIs will
evolve with the disk structure. To estimate the lifetime of the asymptotic
phases for each disk, I note that the energy loss for the last few orp of
evolution in both disks is roughly constant. Based on a linear fit to the energy
curves for $t\ge 14$ orp (3540 yr; Fig.~6.5), the M2004 disk can sustain the asymptotic
phase for about 18000 yr, while the BDNL disk can sustain the same phase for
about 25300 yr. For both disks, this asymptotic phase is relatively short-lived
compared with typical total disk lifetimes (e.g., Hartmann 2005). However,
effects like grain growth and irradiation would likely alter these calculations
significantly.  The simulations presented by Cai et al.~(2006, 2007) indicate
that grain growth and irradiation weaken GIs, which could result in slower
evolution.  Moreover, a relatively small disk (radially) is simulated in these
calcuations, and much more extended disks may have different evolution
timescales. 

\section{Coherent Structure}

A periodogram analysis (Horne \& Baliunas 1986) can be used to look for
periodicities in data sets.  When applied to the cosine of the phase angles of the radially
dependent Fourier components of the azimuthal density distribution, the
periodogram is a powerful tool for finding persistent, coherent structure in
disks.  In this case, the structure corresponds to $m$-arm spirals with steady
pattern speeds. Structure that maintains the same phase for an extended period
of time will have more power than noise or short-lived phase coherent
structures. I should note that power in a mode on the periodogram does not
necessarily mean that the mode is particularly strong, in the $\left<A_m\right>$ 
sense, although the two are often correlated. It means rather that the pattern is 
highly coherent.

Figure 6.14 shows periodograms for Fourier components $m=1$, 2, 3, and 4 in the M2004 disk. Overplotted are
the corotation speed, derived from the mass-weighted orbital motion of the gas,
$\Omega_w(r)$, and the inner and outer Lindblad resonances (ILR and OLR), which
are calculated from the epicyclic frequency $\kappa=(2\Omega_w
d\Omega_w/dr+4\Omega_w^2)^{1/2}$. 
There is significant power in mode $m=1$ at various points at corotation, and
at the OLR at about 2 orp$^{-1}$.  Because the power is fairly
localized at the resonances, the $m=1$ power is likely a
result of a combination between a superposition of modes (see below) and structure that
arises from keeping the primary fixed.  
For $m=2$, there are three strong swathes of power, with noise,
corresponding to 0.8, 1.5, and 2 orp$^{-1}$.  The corotation radius for the 1.5-orp$^{-1}$ swath
is commensurate with the ILR of the 0.8-orp$^{-1}$ swath and with the OLR of the 2-orp$^{-1}$
swath.  In addition, these swathes of power are coincident with mass flux and
torque profile features in Figure 6.8.  The inflow/outflow boundary at $r\sim 26$ AU
aligns with corotation of the 1.5-orp$^{-1}$ swath, which also aligns with the outer
peak in the torque profile.  The 2-orp$^{-1}$ swath aligns with the dip in the mass
flux profile and with the inner peak in the torque profile.

In the $m=3$
periodogram, the coherent structure is distinct. There are three strong signals at 0.8, 1.5, and 2 orp$^{-1}$, as in the  $m=2$ periodogram, and there is additional structure at about
4.4 orp$^{-1}$.  For $m=4$, almost all coherent structure is washed out.
Finally, I note that the $m=1$ power along corotation is commensurate with the
major $m=2$ and 3 swathes of power.  This is likely an indication
of mode coupling between the low-order modes in the disk (e.g., Laughlin \& Korchigan 1996; Laughlin
et al.~1998).

Figure 6.15 is similar to Figure 6.14, but for the BDNL disk.  Again, there is power
in the $m=1$ Fourier component along corotation. However, there is fairly coherent structure at 2 orp$^{-1}$
between 15 and 40 AU, with a break of about 5 AU. This will be
discussed more in a moment.
For $m=2$, there is very persistent structure between 10 and
30 AU, with a well-defined pattern at about 2  orp$^{-1}$. There are also
less coherent swathes of power at about 0.8 and 4.4 orp$^{-1}$.  The 4.4-orp$^{-1}$
swath is of some interest because it shows that the pattern period of the
inner-disk structure is still evolving. For $m=3$, the same swathes of power
that are in the $m=2$ periodogram are present.  However, power for the innermost
swath is much more well-defined, and the pattern period is closer to 4 orp$^{-1}$;
again, there is an indication of structure evolution. In addition, a swath of power at the 1.5-orp$^{-1}$ corotation is present in the $m=3$ periodogram, but is largely absent in the $m=2$ periodogram.   By
$m=4$, most of the structure is relatively muted.  As in the M2004 disk, the commensurate
power for modes $m=1$, 2, and 3 are likely indicative of some form of mode
coupling among low-order $m$, and ILR-corotation-OLR
alignments like those present in the M2004 periodograms, can be found for $m=2$ and 3.

Even though there are several persistent spiral structures in the
disk, the mass flux profile is confusing.  Nonetheless, there are some alignments of
interest.  The principal inflow/outflow boundary is at about 15 AU, which is
very close to the corotation of the 4.4- and 4-orp$^{-1}$ swathes for $m=2$ and 3,
 respectively.  The second mass inflow/outflow boundary at about 20 AU is
commensurate with the well-defined $m=2$ Fourier mode (2 orp$^{-1}$), and the third inflow/outflow
boundary at about 27 AU is commensurate with the corotation of 1.5-orp$^{-1}$ swath in the
$m=3$ periodogram.  

To evaluate further the importance of these $m$-arm structures for mass transport, I decompose
the BDNL disk into its respective density Fourier components and reconstruct the
disk with only certain modes. Figure 6.16 shows the results. The $m=1$
component has strong power in the periodogram, but it is a minor component of the torque
profile until $r\gtrsim
30$ AU.  Most of the power inside $r\sim30$ AU is generated by spirals with arms $m=2$, 3, and
4, and so mass transport should be modeled fairly accurately.  Regardless, this does emphasize
that routines for freeing the star in the SV and in CHYMERA need to be developed.
 
 \begin{figure}
\begin{center}
\includegraphics[width=5.95in]{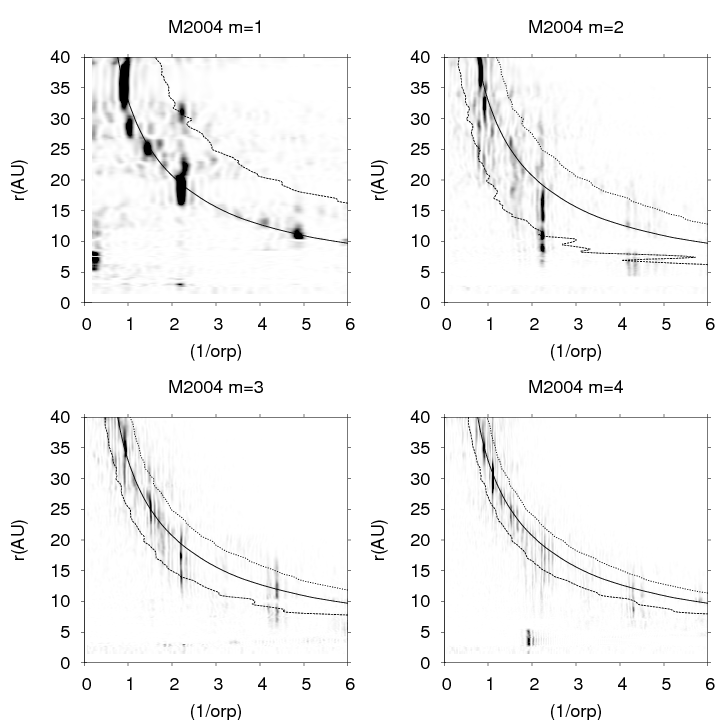}
\caption[M2004 disk periodograms]
{M2004 disk periodograms showing $m=1$, 2, 3, and 4.  
The curves indicate corotation and the ILR and OLR
(only  OLR and corotation for $m=1$).}
\end{center}
\end{figure}
 
 \begin{figure}
\begin{center}
\includegraphics[width=5.95in]{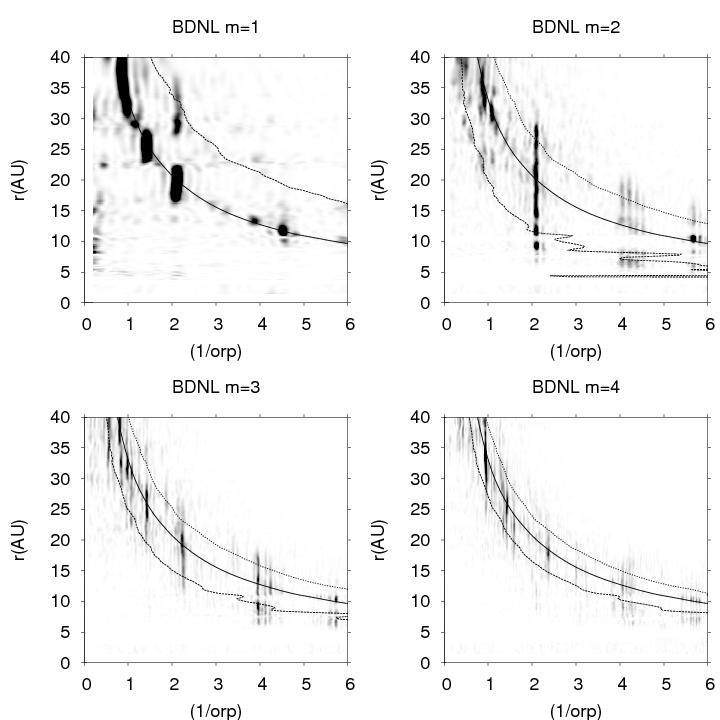}
\caption[BDNL disk periodograms]
{BDNL disk periodograms showing $m=1$, 2, 3, and 4.  
The curves indicate corotation and the ILR and OLR
(only  OLR and corotation for $m=1$).}
\end{center}
\end{figure}
 
\begin{figure}
\begin{center}
\includegraphics[width=5.95in]{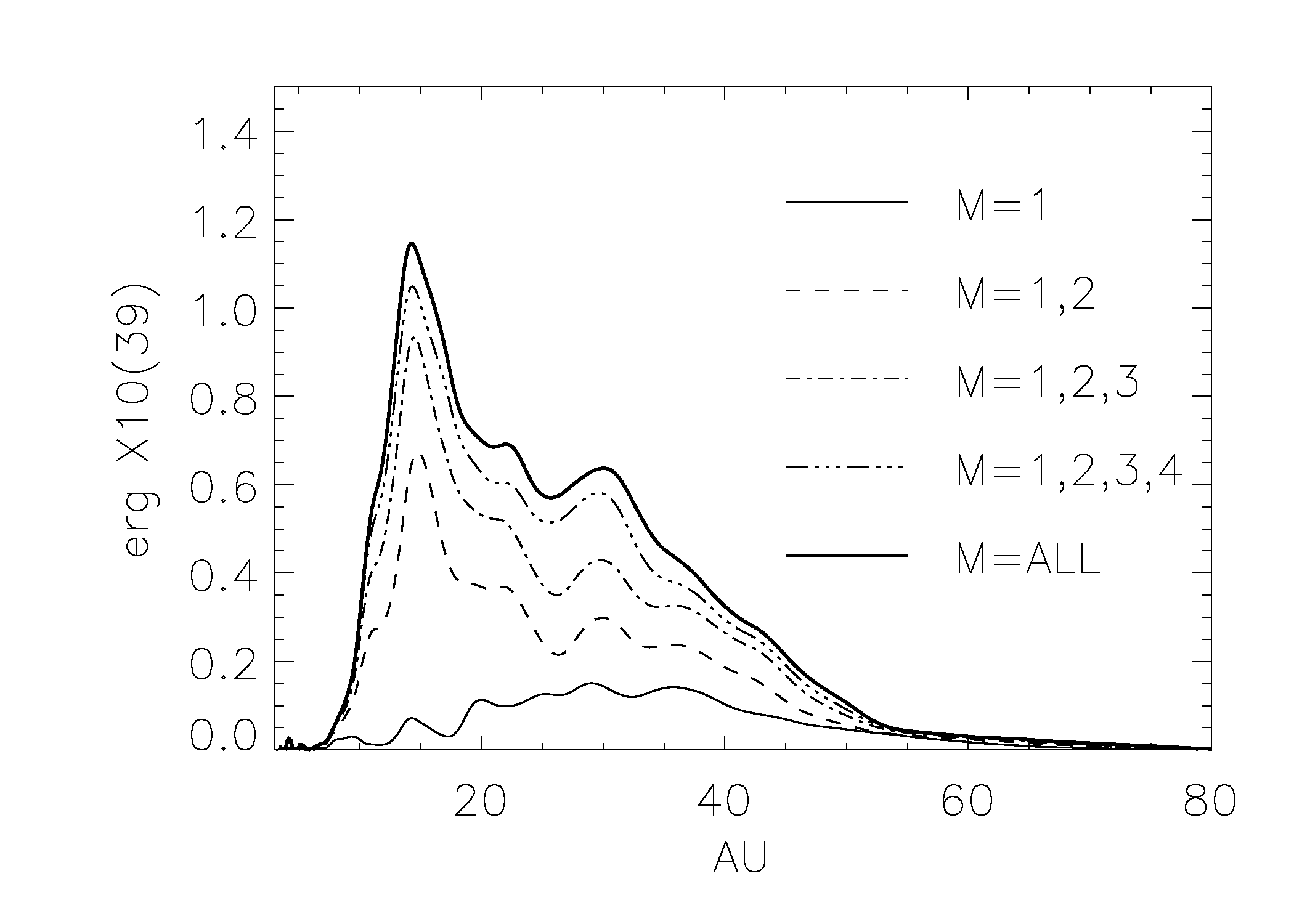}
\caption[BDNL disk torque reconstruction]
{BDNL disk torque reconstruction. The $m=1$ mode becomes important in the outer
disk, but the $m=2$, 3, and 4 modes dominate the torque profile for $r\lesssim
30$ AU.}
\end{center}
\end{figure}

\section{Local or Global Transport?}

The locality of mass and angular momentum transort in GI-active disks has been
addressed by multiple authors (e.g., Pringle 1981; Laughlin \& Rozyczka 1996; Balbus \&
Papaloizou 1999; Gammie 2001; Lodato \& Rice 2004; Mej\'ia 2005): Can mass
transport in  GI-active disks be modeled by a local $\alpha$ prescription, or do
the long-range torques and/or wave transport that GIs produce make such an approximation misleading?
If GIs have a global effect on mass transport in the disk, structure caused
by spatially and temporally variable accretion rates, such as rings or density
enhancements, can be missed by assuming that the disk evolves like an
$\alpha$ disk. 

The above analyses indicate that low-order spiral waves are the dominate
nonaxisymmetric structures, that the mass flux is correlated with resonances of
coherent, persistent low-order spiral waves, and that most of the time-averaged
torque is produced by these low-order structures. Even though the effective $\alpha$
during the asymptotic phase
for each disk is similar to what one would expect based on $\alpha$ disk theory, 
I caution anyone from interpreting these results as an indication that one
can evolve the disk as an $\alpha$ model. The mass fluxes are highly variable, with multiple
inflow/outflow regions due to the interplay between the large-scale, low-order
structure.   A local model does not seem likely to represent the mass fluxes in the
M2004 and BDNL
disks well. In particular, bursts of GIs would be incorrectly modeled. However, if the $t_{\rm cool}$ profile were known {\it a priori}, which is an unclear prospect, the $\alpha$ model would be successful in predicting gross disk
properties, such as the typical magnitude of the mass flux during the asymptotic phase.   

\section{Spectral Energy Distribution}

As emphasized by Nelson et al.~(2000), an important test of
physical relevance for a disk simulation is a comparison between SEDs of
observed systems and the derived SED from a simulation.  Because the M2004
simulation was followed through all four phases and because the effective
temperatures for both simulations are essentially the same, I only construct
SEDs for the M2004 disk as if it were at some large distance $d$ and viewed
face-on.
For simplicity, I assume that each $(r,\phi)$ column of the disk emits
radiation according to a blackbody law at its effective temperature over its
surface area.  Because I use a cylindrical grid, the $i$th area element has a
solid angle ${\rm d}\Omega_i=r_i{\rm d}r{\rm d}\phi/d^2$.  The specific flux,
or flux density, then can be tallied by
\begin{equation} { F}_{\nu}=\sum_i {\rm d}\Omega_i
B_{\nu}\left(T_i\right),\label{eq24}\end{equation}
where $B_{\nu}$ is the Planck function and $T_i$ is the effective temperature of
the $i$th cell. To avoid using distance, I choose to express the SED in terms
of $4\pi d^2\nu {\rm F}_{\nu}$, which is a typical observer's SED.  As a basis 
for comparison, I adopt the
approach of Nelson et al.~(2000) and define a fiducial SED to be one that is 
derived from a disk with a temperature law $r^{-0.6}$, which is the median
best fit law to the T Tauri disk sample presented in Beckwith et al.~(1990).

Figure 6.17 shows the SEDs derived from M2004 for three different stages in its
evolution.  The short dashed line delineates the SED for the disk near its
brightest period during the burst phase (2.4 orp), when the luminosity is
nearly 19 $L_{\odot}$ as integrated between $10^{10}$ and $10^{15}$ GHz.  The
long-dashed line delineates the SED for the disk as it enters the asymptotic
phase (10 orp), and the solid black line indicates the SED for the disk at the
end of the calculation (16 orp).  The star, which is assumed to have an
$R=2R_{\odot}$ and a $T_{\rm eff} = 4000$ K, is included in the SED profile.
The SED has dips in specific luminosity as it nears the star because it has a
2.3 AU hole and is missing a contribution from an inner and hotter portion of
the disk.  The red lines indicate fiducial SEDs based on assuming a $T_{\rm
eff}\sim r^{-0.6}$ temperature profile for a 0.0033, 0.01, and 0.033 $L_{\odot}$
disk and integrating between 0 and 60 AU.  The fiducial SEDs also have a slight
dip in their profile just before transitioning to the stellar portion of the
SED due to discretizing the temperature profile into grid cells 1/6 AU wide in
$r$.  The blue line is a fiducial SED calculated for a luminosity of 
$0.0024~L_{\odot}$, which is the disk luminosity at 16 orp, in the same way as the
other fiducials but with a 2.3 AU hole.  Even though the actual effective
temperature profile for M2004 is an exponential, the SED for M2004 is
observationally consistent with a profile $T_{\rm eff}\sim r^{-0.6}$ because
the inner 20 AU of the disk closely follows a $T_{\rm eff}\sim r^{-0.59}$. 

Although GIs appear to lead to an exponential $T_{\rm eff}$ profile for this
disk, stellar irradiation (D'Alessio et al.~1998, 2001) probably keeps the outer regions
of the disk warmer than what is modeled here, except for cases of strong shadowing.
The irradiation routine developed by Mej\'ia (2004) is not employed in these calculations
because, in her routine, the stellar irradiation is directly incident on the disk.
The optical photosphere for protoplanetary disks is typically three disk scale heights above the midplane (D'Alessio et al.~1998, 2001, 2006).   Directly modeling this low-density, high-temperature atmosphere, where the stellar irradiation is absorbed and scattered, along with the
relatively high-density, low-temperature disk is a computational challenge.

As will be discussed in more detail in Chapter 7, the SEDs of the M2004 disk
also suggest that a disk undergoing a burst due to GIs
can become very bright and lead to very high accretion rates.  The M2004 disk
reaches a peak luminosity of 19 $L_{\odot}$ at about 2.4 ORPs, which makes the
disk approximately 10,000 times brighter in the burst phase than in the
asymptotic phase.  Furthermore, the disk has a luminosity of only about 0.01
$L_{\odot}$ around 2.1 ORPs, so the disk luminosity increases by a factor of
about 1,000 in 80 yr.  The increase in luminosity is sudden and may be much
shorter than reported here due to coarse temporal resolution of the
files required to generate SEDs.  Bursting disks may be related to the FU Ori
phenomenon (see Chapter 1).

\begin{figure}
\begin{center} 
\includegraphics[width=5.95in]{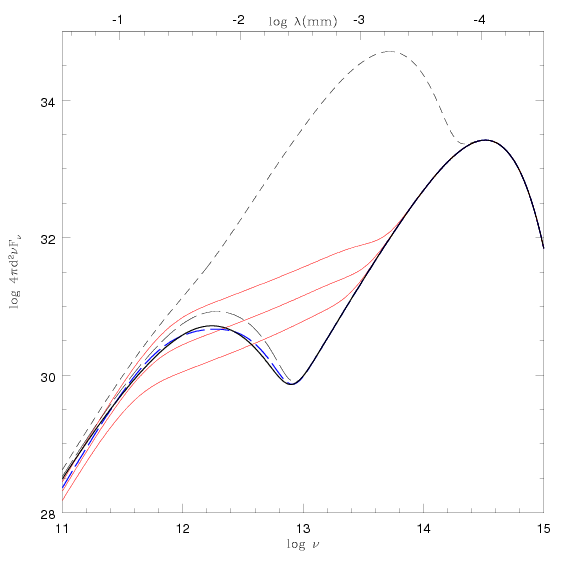}
\caption
[SEDs for the M2004 disk]
{The solid line indicates the SED for the M2004 disk at the end of the calculation
(16 orp or 4,050 yr).  The short-dashed and long-dashed curves delineate the SEDs at
2.5 orp and at 10 orp.  The red curves represent fiducial SEDs
for 0.0033, 0.01, 0.033 $L_{\odot}$ for a disk that extends from 1/6 AU to 60
AU and has a $T_{\rm eff}\sim r^{-q}$ for $q=0.6$.  The blue dashed curve
indicates the SED for a disk that is truncated at 2.3 AU with a $q=0.6$.  }
\label{f17} 
\end{center} 
\end{figure}

\section{Conclusions}

The main results of this comparative study are as follows.  The disks evolve
toward a thermally self-regulating, gravitationally unstable state, where work
and shock heating balance radiative losses.  The mass transport is highly
variable and dominated by global, low-order modes. The gross mass
transport is similar in magnitude to that expected by a local model.  However,
evolving the disk as an $\alpha$ model will miss the large mass flux
variability and will neglect the spiral shocks and shock bores in a disk, which
may be important for generating turbulence (Boley et al.~2005) and
for chondrule formation (Chapter 7). Moreover, the cooling times are too
long to lead to fragmentation, and as discussed in \S 5.4.3, convection eddies
are unable to lower the cooling times to those that result in disk fragmentation. 

What is the fate of these disks? 
As discussed in \S 6.6, the asymptotic phase can only be maintained
for a few$\times 10^4$ yr.  Because the models are incomplete, i.e., I exclude
the inner disk, irradiation, possible infall, dust settling, grain growth, and
other mass transport mechanisms, it is difficult to comment on the disk's
ultimate fate.  Regardless, I speculate about the big picture in Chapter 8 based
largely on the results of this comparative study.

\chapter{DEAD ZONE MODELS}

In Chapter 6, I discussed two disk simulations, where one was evolved with the M2004 radiation scheme
and the other with the BDNL scheme (Chapter 2). The results indicate that (1) disk evolution is sensitive to the details of the treatment of radiation physics, (2) cooling times are too
long for disk fragmentation, (3) the effective $\alpha\sim10^{-2}$ during the asymptotic phase, (4) high mass fluxes occur during GI bursts ($\sim 10^{-5}
M_{\odot}\rm~yr^{-1}$), and (5) the
shocks during the burst rapidly heat the disk such that
the disk can outshine the primary. 
The high mass fluxes during bursts may indicate an internal, repeatable process for driving the
FU Ori phenomenon (Armitage et al.~2001). As discussed in Chapter 1, an FU Ori outburst is
characterized by a rapid (1-10s yr) increase in optical brightness of a young T Tauri
object, typically 5 magnitudes, with mass accretion rates of the order $10^{-4} M_{\odot}~\rm yr^{-1}$ from the disk
onto the star (Hartmann \& Kenyon 1996). The most promising explanation for the FU Ori phenomenon is a thermal instability (Bell \& Lin 1994).  But even with this model, the driving mechanism that initiates the thermal instability remains unclear.   In this Chapter, I consider the possibility that a burst of GI activity in a mass-enhanced region of a disk could produce a cascade of instabilities that ultimately leads to an FU Ori event.  I envision the mass enhancement to be a result of reduced accretion in a dead zone, where the MRI-active layer is only a small fraction of the total mass in a given column. 

If episodic GI bursts drive the FU Ori phenomenon, then might these
bursts also produce shocks that are strong enough to form chondrules? Iida et al.~(2001), Desch \&
Connolly (2002), Ciesla \& Hood (2002), and Miura \& Nakamoto (2006) demonstrated that chondrules can be formed by gas drag friction when a chondritic precursor enters a shock.
Certain post-shock conditions allow for melts to cool at rates consistent
with laboratory experiments. 
One should note that the shock model for chondrule formation is not a model for the mechanism driving
the shock, and any strong shock has the potential to form chondrules.  Global spiral waves were
identified as a potential shock-driving mechanism by Wood (1996).  However, it remains unclear
how these waves can be repeatedly generated.  

For this study, I adopt the hypothesis that bursts of GI activity in dead zones drive the FU Ori phenomenon and produce shocks with chondrule-forming pre-shock conditions. In order to investigate this scenario, I designed a numerical experiment to evolve a
massive, highly unstable disk with an initial radial extent between 2 and 10 AU.
The experiment is conducted with three principal objectives in mind: (1) To
characterize shocks in
a disk with a very strong burst of GI activity near 5 AU and to determine whether 
such shocks can drive chondrule formation, (2) to determine whether a strong
burst can drive accretion rates high enough to start a cascade of instabilities,
resulting in an FU Ori outburst, and (3) to investigate cooling times in the 2-10 AU region
for a massive disk.

\section{Expected Shock Strengths}

The shock speed $u_1$ for a fluid element entering a logarithmic spiral wave with pitch angle $i$ is described by
\begin{equation}
u_1=v_r\cos i + \left(G M/r\right)^{1/2} \bigg| 1-\left(\frac{r}{r_p}\right)^{3/2}\bigg| \sin i,
\end{equation}
where I have assumed that the gas azimuthal motion is Keplerian and where $r_p$ is the pattern radius 
for some global spiral wave.  Neglect $v_r$ for simplicity.  This leads to 
\begin{equation}
u_1\approx30 {\rm~km~s^{-1}} \left(\frac{M}{M_{\odot}}\right)^{1/2}\left(\frac{r}{\rm AU}\right)^{-1/2}\bigg| 1-\left(\frac{r}{r_p}\right)^{3/2}\bigg|\sin i
\end{equation}
Notice the limiting behavior of equation (7.2).  When $r_p\gg r$, the shock speed limits to the Keplerian speed times $\sin i$, and so whether spiral waves with large $r_p$ produce chondrules is mainly dependent on the pitch angle of the spiral wave.  In contrast, when $r\gg r_p$, the speed increases
as $r$, and even shallow pitch angles can produce chondrules, depending on the model. 

\begin{figure}
\begin{center}
\includegraphics[width=5.95in]{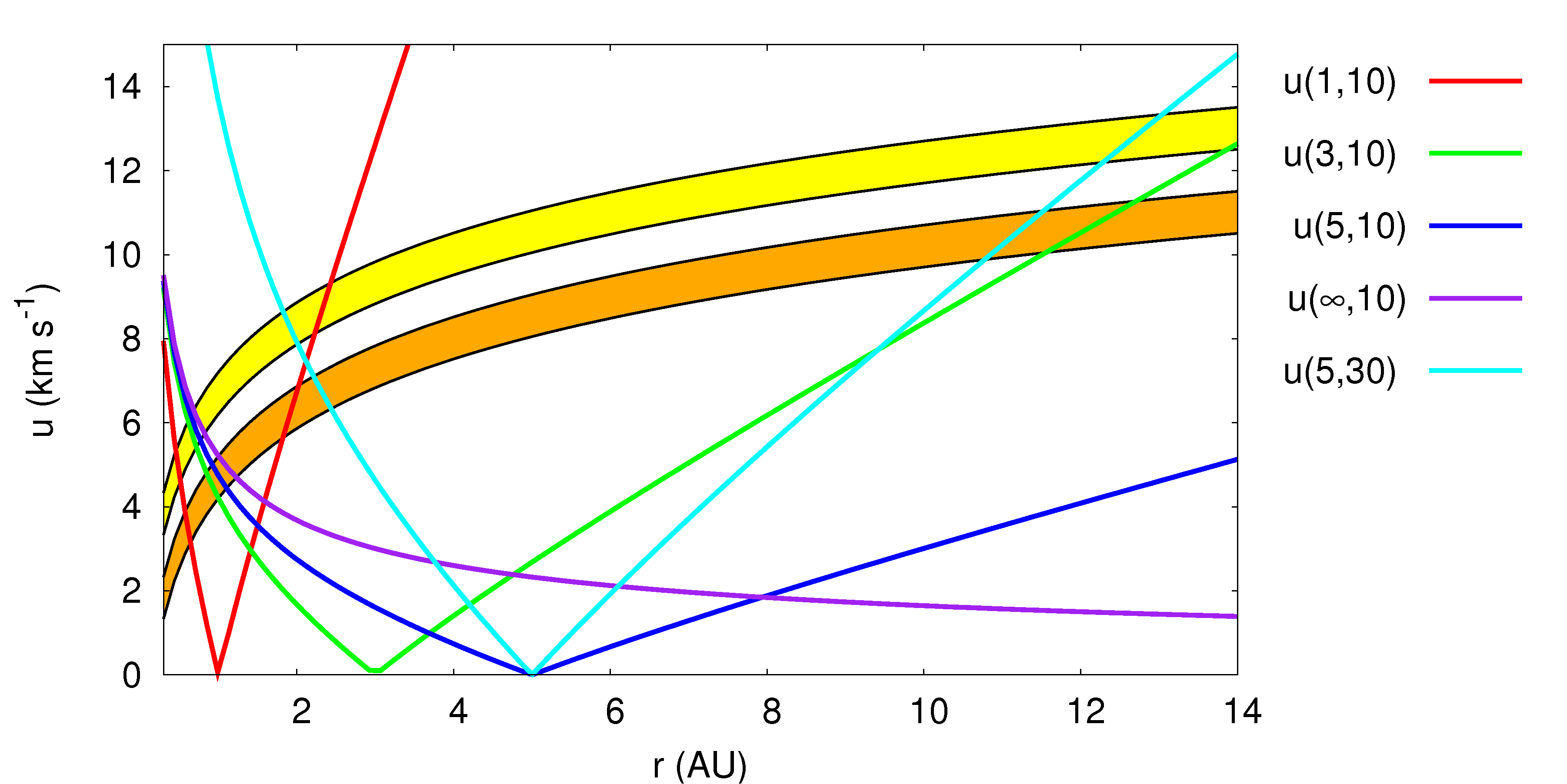}
\caption[Shock speed expectation]{Expected shock speeds $u$ based on equation (7.2). In the legend, $u(1,10) = u_1(r_p=1{~\rm AU,}~i=10^{\circ})$ for pattern radius $r_p$ and pitch angle $i$. The colored regions highlight where chondrule formation is expected based on the Desch \& Connolly (2002) shock calculations (see text), with the yellow region appropriate for a MMSN density distribution and the orange for the same density distribution but with 10$\times$ the mass.  Shocks
that occur inside corotaton for $i=10^{\circ}$ will not produce chondrules between 1 and 5 AU.  
However, chondrules can be produced by these low pitch angle spirals in shocks outside corotation.
If the pitch angle is fairly open, such as $i\approx30^{\circ}$, then a spiral wave with a corotation near 5 AU can produce chondrules in the asteroid belt and at comet distances.}
\end{center}
\end{figure}

For example, consider the Minimum Mass Solar Nebula (MMSN), where the midplane $\rho(r)=1.4\times10^{-9} \left(r/{\rm 1~AU}\right)^{-11/4}$ g cm$^{-3}$ (Hayashi et al.~1985). Figure 7.1 shows the $u$-$\rho$ plane with heavy, solid curves indicating shock speeds for
$r_p=1$, 3, 5,  and $10^6$ AU and with $i=10$ and 30$^{\circ}$.  The colored regions on the plot highlight where chondrule formation is expected in a MMSN (yellow) and a 10$\times$MMSM (orange).  The chondrule-forming curves are based on the results of Desch \& Connolly (2002), with $u_1\approx-\left(11+2\log \rho \left({\rm g~cm^{-3}}\right)\right) \pm0.5{\rm~km~s^{-1}}$.  The boundaries set by these curves are meant to be illustrative, not definitive.

Spiral shocks along shallow pitch angle spiral waves ($i\approx 10^{\circ}$) are mostly if not entirely out of the chondrule-forming range between $r=1$ and 5 AU for shocks inside corotation. However, spirals with corotation in the inner disk can produce chondrule-forming shocks for a wide range of radii, depending on the actual $r_p$.   If the spiral waves have very open pitch angles ($i\gtrsim 30^{\circ}$), then spiral waves with corotation near $r=5$ AU can effect chondrule-forming shocks near the asteroid belt and at cometary distances. The mass of the disk does not change these general behaviors. 

A major caveat for this simple-minded approach is that the fluid elements' motions may be poorly described by equation (7.2) if vertical and radial excursions induced by shock bores cannot be neglected.   As demonstrated in Chapter 5, strong spiral shocks can significantly alter the fluid flow.   Regardless, this analysis serves as a base description for expected shock speeds in the disk for a given spiral wave.
Because I am interested in exploring the production of chondrules near the asteroid belt and the annealing of solids in the outer disk, I have tried to design a numerical experiment that is biased toward producing strong shocks with a corotation near 5 AU and with open pitch angles.

\section{Methodology}

The numerical experiment is set up as follows. The
initial model is the flat-Q disk with the surface density enhanced ring (\S 4.2).  The
disk is initially between about 2 and 10 AU in radius, and the 
total disk mass is $0.167~M_{\odot}$.  The surface density ring enhancement is distributed as a Gaussian centered on 5 AU with
a FWHM of 3 AU.  The FWHM was chosen to be comparable to the most unstable wavelength at 5 AU (\S 4.2). 
The mass-weighted $Q$
and average surface density profiles for the disk are shown in Figure
7.2.  For comparison, the initial $\Sigma\sim r^{-2}$ is also shown. 
The instabilties activate near 4.5 AU once the region between 3 and 6 AU reaches a
mass-weighted average $Q\approx 1$, even though the minimum $Q$ at 4.5 AU is
well below unity.

For the treatment of H$_2$, I consider the arguments in \S 2.3.3.  Recall that energetic particles (EP) are attenuated with a
surface density of 100 $\rm g~cm^{-2}$. Figure 7.3 compares the height above the midplane where
$m(z)=\int^\infty_z \rho dz = 100~{\rm g~cm^{-2}}=m_c$ with the scale height of the
disk,
$h=m(0)/\rho$.  In
addition, it also shows the fraction of the half disk column mass 
that is
contained above one vertical attenuation length.  One should note that EPs will likely be
on random or oblique
trajectories, and so EPs may affect less of the disk than assumed here.  Cosmic
ray ionization will be negligible for most of the disk inside about 8 AU, and 
the disk is too cold for thermal ionization of alkalis. 
Production of H$^+$ and H$_3^+$ will be limited to
radioactive decay, with $^{26}$Al dominating the EP production at a rate of
$\zeta\sim10^{-19}~\rm s^{-1}$ (Stepinski 1992; Chapter 2). This low rate will result in a
frozen ortho/para 
ratio for most of the disk.  
However, because there will be some
protonated species in the disk, the ratio can evolve slowly to its thermal
equilibrium ratio while acting out of thermal equilibrium on the dynamic timescale.
Finally, even for the areas where cosmic rays can penetrate, the
depletion timescale due to dust is likely shorter than the dynamic timescale in
this disk
everywhere.  I use a
frozen ortho/para ratio of 3:1 in this study for these reasons.

The disk is evolved with the BDNL radiation physics algorithm.  The grain size
distribution is assumed to follow the ISM power law, $dn \sim a^{-3.5} da$
(D'Alessio et al.~2001), where the maximum grain size $a_{\rm max}$ is chosen to be
1 mm, and a weighted combination between Rosseland and Planck mean opacities
is used (see Chapter 2).  The particles are assumed to be well-mixed with the
gas.  However, the effects of dust settling are heuristically considered by
evolving the disk in several ways: standard opacity and 1/100, 1/1,000, and 1/10,000 of the
standard opacity (Table 7.1). This choice in opacities varies the midplane Rosseland mean optical depths from about $\tau=10^4$ to unity.  It should be noted that the standard opacity simulation has $\tau v/c\approx 1$ near the inner edge, and so it is close to the limit $(\tau v/c>1)$ where advection of photons becomes important (Krumholz et al.~2006), but this effect is not modeled in these simulations.  For a base simulation, the disk is also evolved adiabatically
for about 1 orp, where 1 orp for these simulations is one outer
rotation period at 10 AU, or about 32 yr.  In addition, no external radiation
is assumed to be shining onto the disk, except for a 3 K background temperature.
Although such a naked disk is likely to be unrealistic, except for a case of
strong self-shadowing, the assumption is made to bias the disk toward strong shocks
inasmuch as irradiation weakens GI strength (Cai et al.~2007).  If chondrule-producing
shocks cannot be created in simulations biased in their favor, then it would present a serious
problem for GIs as the source of chondrule processing.

Due to inefficient cooling in the standard simulation, high enough temperatures in
the disk are reached after 80 yr to create radiative timescales that are too short to resolve;  the
simulation is stopped.  As described below, the reduced
opacity simulations are able to cool much more efficiently, and the disk
temperatures remain manageable with the time-explicit algorithms used in CHYMERA.
However, the reduced opacity simulations are stopped after about 110 yr due to
the development of a strong, one-arm spiral, which is treated incorrectly with the
fixed star assumption. 

\begin{figure}
\begin{center}
\includegraphics[width=5in]{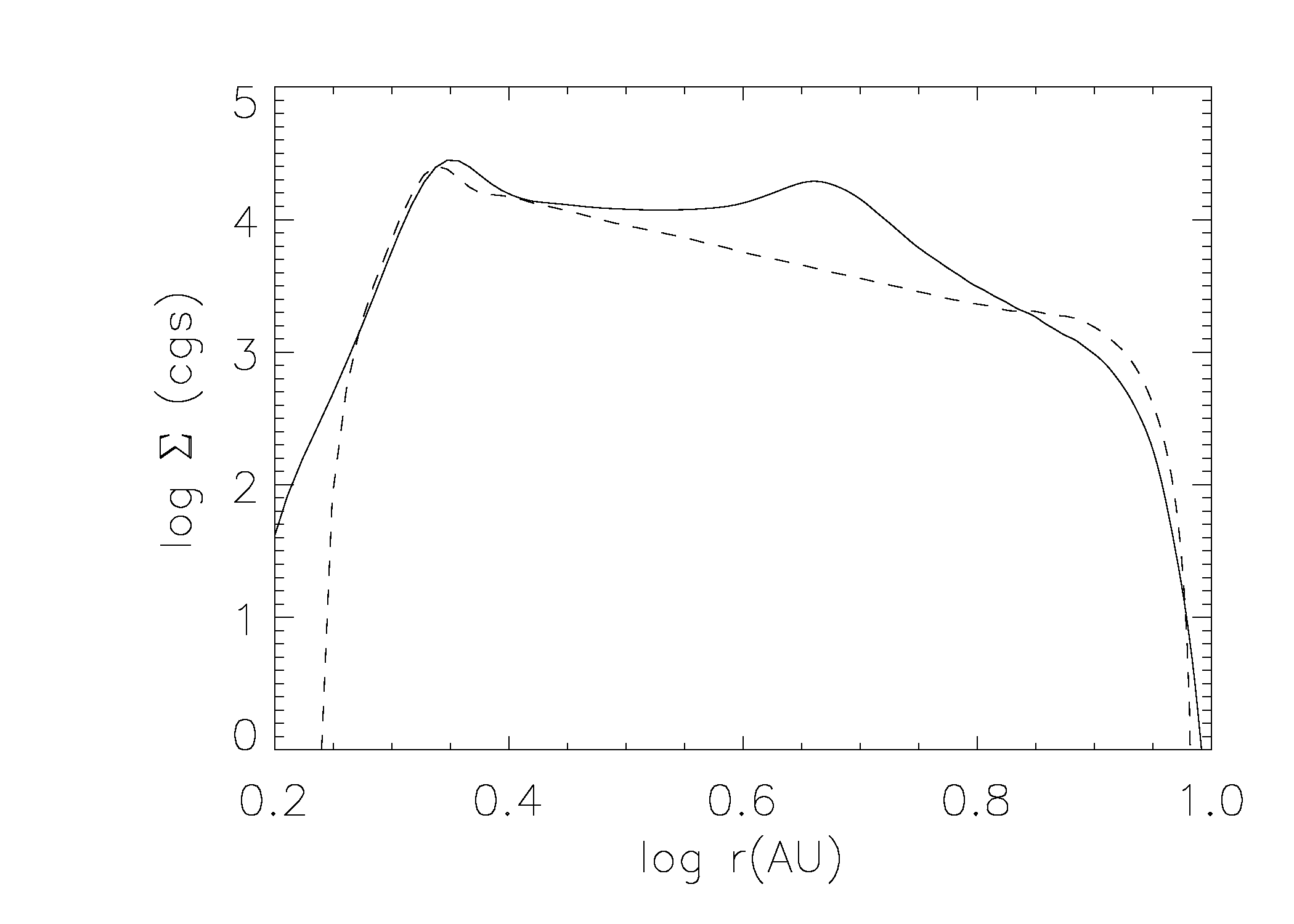}
 \includegraphics[width=5in]{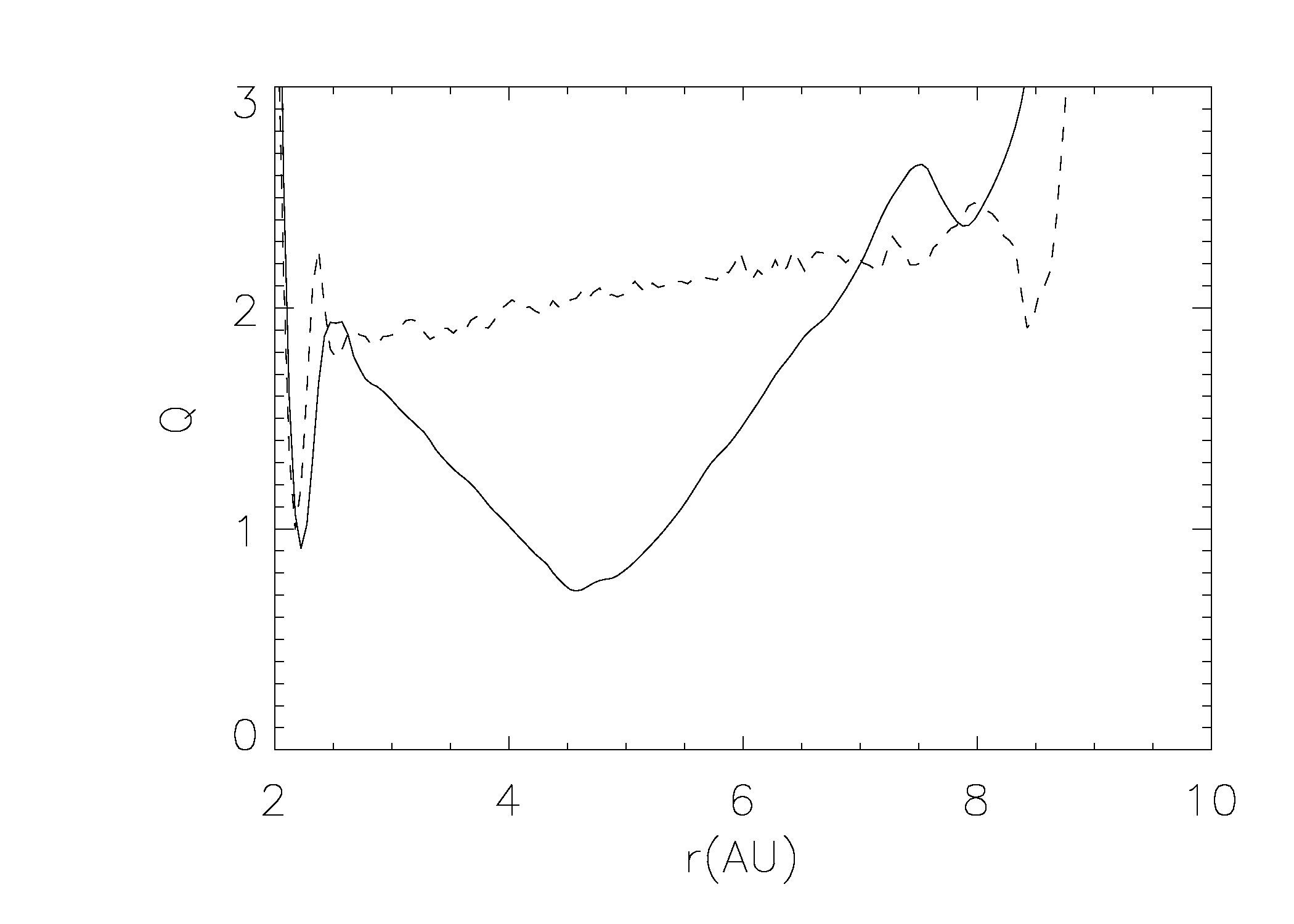}
\caption[Initial surface density profiles]{Top: Surface density profile for the flat-Q disk with (solid curve) and without (dashed curve) the surface density enhanced ring.  The surface density has a maximum
at about 4.5 AU instead of at the target 5 AU due to the steep initial surface density profile. 
Bottom: The initial $Q$
profile with (solid curve) and without (dashed curve) the surface density enhancement.  Even though the minimum $Q$ drops below unity, the GI burst only activates
when the mass-weighted average $Q$ approaches unity over a 3-AU annulus centered on $r=4.5$ AU.}
\end{center}
\end{figure}

\begin{figure}
\begin{center}
\includegraphics[width=5.95 in]{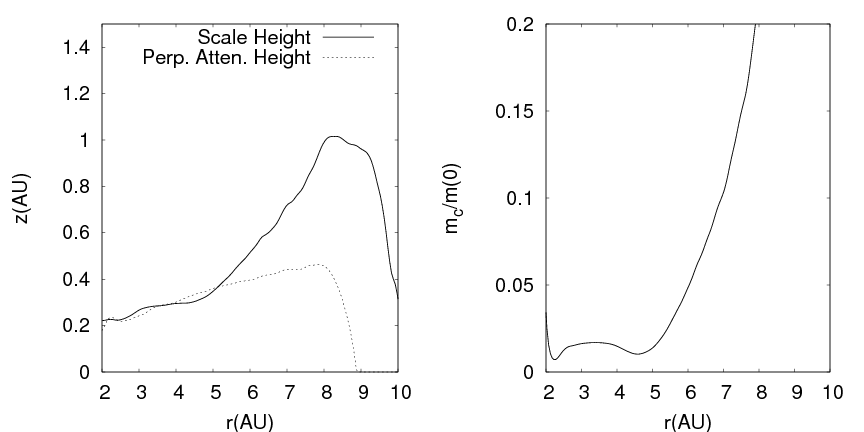}
\caption[Scale height and EP penetration depth]{Left: The scale height of the
disk (solid curve) is shown with the height above the disk where EPs are expected to pass
through one attenuation length (dotted curve). Right: The fraction of the half surface density
that is contained within the first attenuation length.}
\end{center}
\end{figure}

\begin{table}
\begin{center}
\caption[Simulation information]{Simulation information.  The 1024 1/1,000 and 1/10,000 $\kappa$
simulations are still being evolved at the time of this writing, and all 1024 simulations will be extended
to 110 yr if possible for future work. The name of the simulation indicates the fraction of the standard opacity that is used during the evolution of the disk.}
\begin{tabular}{c c c }\\ \hline
Sim.~Name & Resolution $r,~\phi,~z$ & Duration\\ \hline\hline
Adiabatic & 256, 512, 64 & 33 yr \\
512 Standard &  256, 512, 64 & 80 yr \\
512 1/100 $\kappa$&  256, 512, 64  & 110 yr  \\
512 1/1,000 $\kappa$&  256, 512, 64 & 110 yr \\ 
512 1/10,000 $\kappa$&  256, 512, 64  & 110 yr \\ 
1024 Standard & 512, 1024, 128 & 55 yr \\
1024 1/100 $\kappa$& 512, 1024, 128 & 55 yr  \\
1024 1/1,000 $\kappa$& 512, 1024, 128 & 33 yr  \\ 
1024 1/10,000 $\kappa$& 512, 1024, 128 & 33 yr \\ \hline
\end{tabular}
\end{center}
\end{table}

Each simulation is evolved, at least through the first 30 yr, at two resolutions:
$(r,~\phi,~z)=(256,512,64)$ and $(r,~\phi,~z)=(512,1024,128)$.  The lower resolution simulation (512 sim) has a grid spacing of 0.05 AU per cell in $r$
and $z$ and the higher 0.025 AU (1024 sim).  For both resolutions, $r\Delta \phi\approx \Delta r$ at about $r=4$ AU.
As described below, shock heating is more
intense in the 1024 simulations, and so the 1024 standard is
stopped after only about 55 yr for the same reason that the 512 standard is stopped.  The 1/1,000 and 1/10,000 $\kappa$ simulations are also being run at high resolution and have been evolved for about 33 yr.  For each simulation,
 $10^3$ fluid elements are randomly
distributed in $r$ between about 3 and 7 AU, in $\phi$ over $2\pi$, and in $z$
roughly within the
scale height of the disk. 
The fluid elements are integrated as the simulation is evolved.

\section{Evolution}

Surface density plots for the 512 simulations are shown in Figures 7.4 and 7.5 at
$t\approx 33$ and 77 yr. 
 The images at $t\approx33$ yr indicate that the burst
has a well-defined, three-arm spiral and that the spiral arms become denser
as the opacity is lowered.  By 77 yr, the disk structures are noticeably different.
Each disk has a visually distinct two-arm spiral except for the 1/10,000 $\kappa$ simulation. 
The lower opacity
simulations appear to have stronger amplitudes, with denser spiral arms and stronger mid-order Fourier components (e.g., $m=4$ through 10 in Figure 7.7).  As will be discussed in more detail below, the 1/10,000 $\kappa$ simulation is on the verge of fragmentation.   

For comparison with Figures 7.4 and 7.5, the 1024 simulations are shown in Figure 7.6 at $t\approx33$ yr, which is the time when significant deviations in spiral morphology become clearly discernible.
The higher resolution simulations have additional fine structure, but they do not develop clumps or dense knots.  In fact, some of the knotty structure that develops in the 512 1/10,000 $\kappa$ simulation is absent at higher resolution.

As discussed in Chapter 6, visual differences can be quantified by a Fourier mode
analysis. The $\left<A_+\right>$ for each disk, averaged over 55-77 yr, is 1.19, 1.43, 2.30, 
and 2.40 for the 512
standard, 1/100, 1/1,000, and 1/10,000 $\kappa$ simulations, respectively.  Figure 7.7 demonstrates
the corresponding $A_m$ spectra.  The bars denote typical
fluctuations, and they should not be mistaken for error
bars. The low-order structure clearly dominates, with the high-order structure falling
off steeply. For each simulation, the $m=2$ mode dominates during this time
interval, but $m=1$, 3, and 4 are strong in all simulations. The 512 standard and 1/100 $\kappa$ data track each other, 
with some variation in the power in the low- to mid-order Fourier compnents.  The 512 1/1,000 and 1/10,000 $\kappa$ data also track each other, but diverge from the 512 standard and 1/100 $\kappa$ amplitudes.  In particular, the power in the Fourier modes falls off more gently than it does in the 512 standard and 1/100 $\kappa$ simulations, which as discussed above, is noticeable in the surface density plots.

 \begin{figure}
\begin{center}
\includegraphics[width=5.95in]{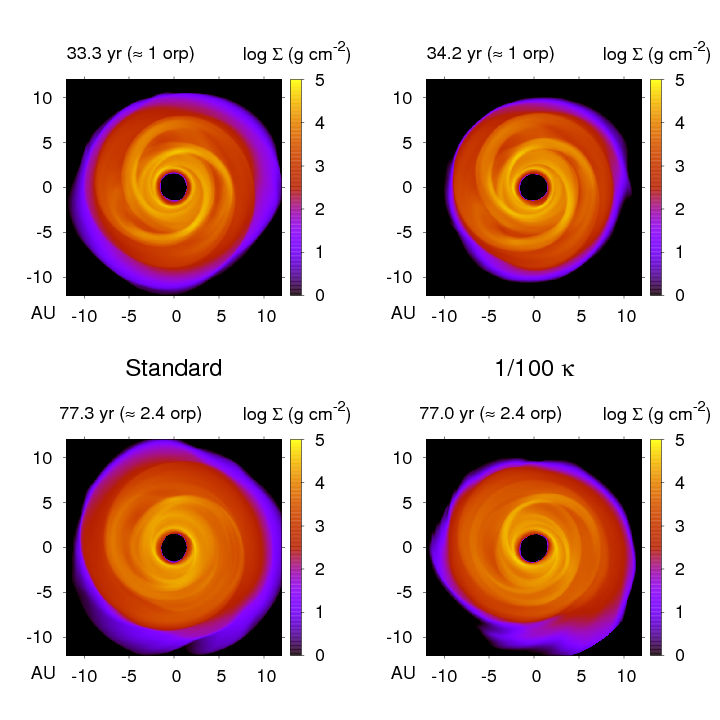}
\caption[Surface density maps for the 512 standard and $10^{-2}~\kappa$ simulations]{Surface density maps for the 512 standard and 1/100 $\kappa$ simulations during the burst and shortly after.  The strength of
the spiral waves increases as the opacity is lowered.}
\end{center}
\end{figure}

\begin{figure}
\begin{center}
\includegraphics[width=5.95in]{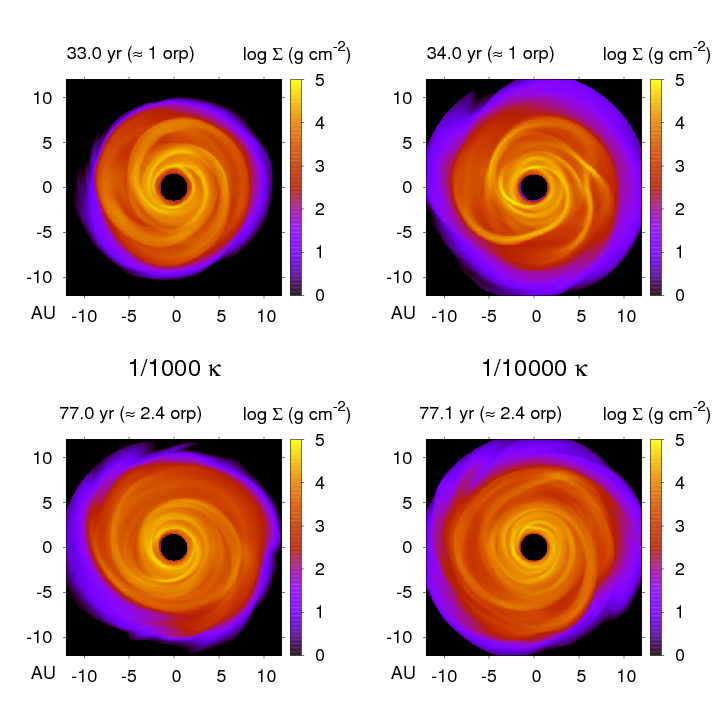}
\caption[Surface density maps for the 512 $10^{-3}$ and $10^{-4}$
$\kappa$ simulations]{Surface density maps for the 512 1/1,000 and 1/10,000
$\kappa$ simulations, which complement Figure 7.4.  }
\end{center}
\end{figure}

\begin{figure}
\begin{center}
\includegraphics[width=5.95in]{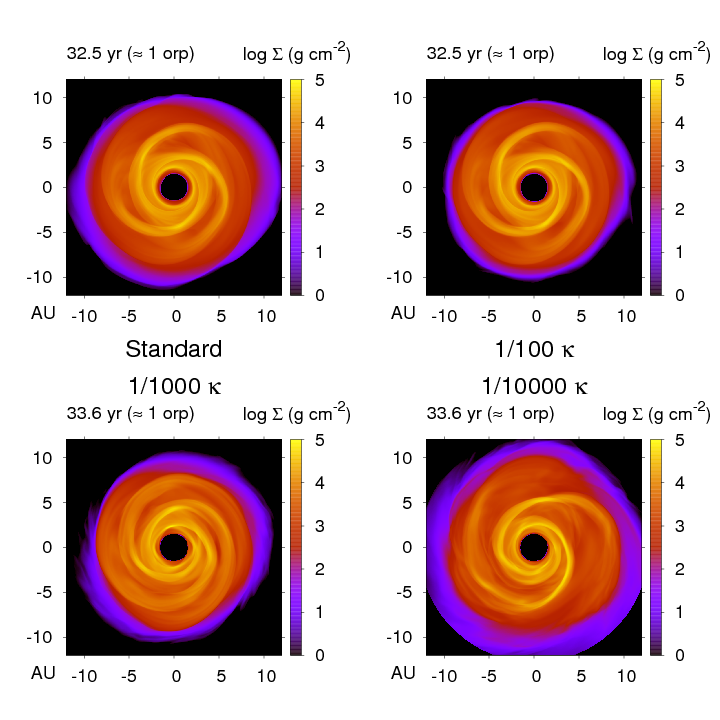}
\caption[Surface density maps for the 1024 simulations]{Surface density maps for the 1024 simulations at approximately the same time as in the top panels in Figures 7.4 and 7.5.}
\end{center}
\end{figure}

The energetics of the disk simulations are shown in Figures 7.8 and 7.9, where
cumulative energy loss by radiation (blue) and cumulative energy dissipation by
shocks (red) are plotted.  For the 512 standard case, the shock heating clearly
dominates over the cooling; the disk is heating up over the entire
evolution.  Energy is transported inefficiently for the highly optically thick
disk, and the disk evolves much like an adiabatic simulation.  When the opacity
is reduced by a factor of 100, the cooling becomes much more important than in
the standard simulation, with radiation energy losses becoming more similar to heating
by shocks. In addition, the shock heating rate becomes larger during the burst. 
 When the opacity is reduced by a factor of $10^3$, the
radiative cooling is even more efficient and surpasses shock heating. The $10^4$ opacity reduction shows the strongest
shock heating and the fastest disk cooling.
As one would expect from analytic arguments, the opacity has a profound effect on
disk cooling and, consequently, on spiral shocks. For the 1024 standard
and 1/100 $\kappa$ simulations, cooling and shock heating are similar to the 512 simulations for about the first 32 yr, but the 1024 runs exhibit more shock heating, while the cooling remains roughly the same.  This additional shock heating pushes the disks toward GI stability faster than what is seen in the 512 simulations.  The departure of the 1024 shock heating from the 512 curves is commensurate with the visual deviations of spiral structure in the surface density images.

\begin{figure}
\begin{center}
\includegraphics[width=5.95in]{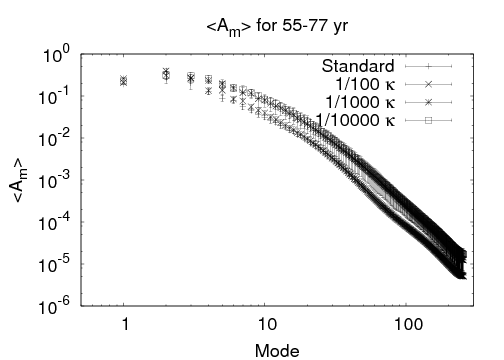}
\caption[$A_m$ spectra for the 512 standard, $10^{-2}$, $10^{-3}$, and $10^{-4}$ $\kappa$
simulations]{$A_m$ spectra for the standard, 1/100, 1/1,000, and 1/10,000 $\kappa$
simulations. The bars indicate typical fluctuations during the 55-77 yr time
period, and should not be mistaken for error bars. }
\end{center}
\end{figure}

\begin{figure}
\begin{center}
\includegraphics[width=4in]{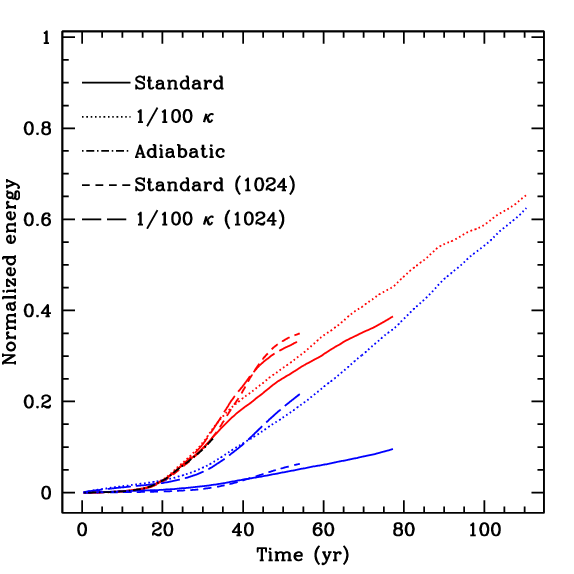}
\caption[Heating and radiation energy loss: high $\kappa$]{Radiation energy losses and shock
heating for the 512 and 1024 standard and 1/100 $\kappa$ simulations. 
The effect of opacity on
the energy loss and shock heating is clear: the lower the opacity, the faster
the cooling.  In addition, the 1024 simulations exhibit additional shock heating, but
comparable cooling to the 512 resolution simulations.  The short, adiabatic curve is difficult
to see because it tracks the standard curves closely.}
\end{center}
\end{figure}

\begin{figure}
\begin{center}
\includegraphics[width=4in]{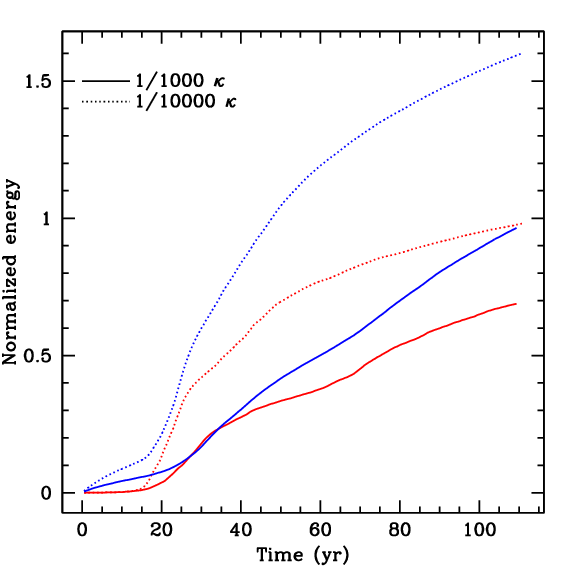}
\caption[Heating and radiation energy loss: low $\kappa$]{Complement to Figure 7.8: Radiation energy losses and shock
heating for the 512 1/1,000 and 1/10,000 $\kappa$ simulations are shown. }
\end{center}
\end{figure}

The effect of opacity on energy losses in the 512 disks is also demonstrated in
Figures 7.10 and 7.11.  In these figures, brightness temperature maps 
are shown for the same time as the top rows in Figures 7.4 and 7.5,  where $T_b=(\pi
I_+/\sigma)^{1/4}$ and $I_+$ is vertical outward intensity (see Chapter 2).  In addition, I
define a cooling time temperature $T_c=(\int_0^{\infty}\epsilon
dz/\sigma\int_0^{\infty}-\nabla\cdot{\bf F}dz)^{1/4}$, which is the temperature
that corresponds to a given column's effective flux if all of the energy were to
leave the column vertically.  If the column is being heated, $T_c$ is set to zero.

As the opacity is lowered, the spiral structure becomes more clearly outlined, and 
the disk becomes brighter because the photons can leave from hotter
regions.  The brightness maps also demonstrate that sustained fast cooling by convection 
is absent.  If hot gas near an optically thick midplane is quickly transported to altitudes where $\tau\sim1$, convection can, in principle, enhance disk cooling.  However, the efficacy of convective cooling is controlled by the rate that energy can be radiated from the photosphere of the disk.  If localized convective flows were responsible for fast cooling in the optically thick disks, one would expect to find strong, localized $T_b$ enhancements.  For example, the 512 standard and 1/100 $\kappa$ simulations would need to have regions as bright as the 512 1/10,000 $\kappa$ simulation.    
The corresponding energy flux that is needed for convection to be sustained and to
cool the disk does not occur (see Chapter 5 for additional discussion regarding
convection).  Finally, the $T_c$ maps evince that radiation can lead to net heating
in regions near strong shocks, e.g., the black outlines (areas experiencing net radiative heating) around the thin, red spiral waves.

\begin{figure} 
\begin{center}
\includegraphics[width=5.95in]{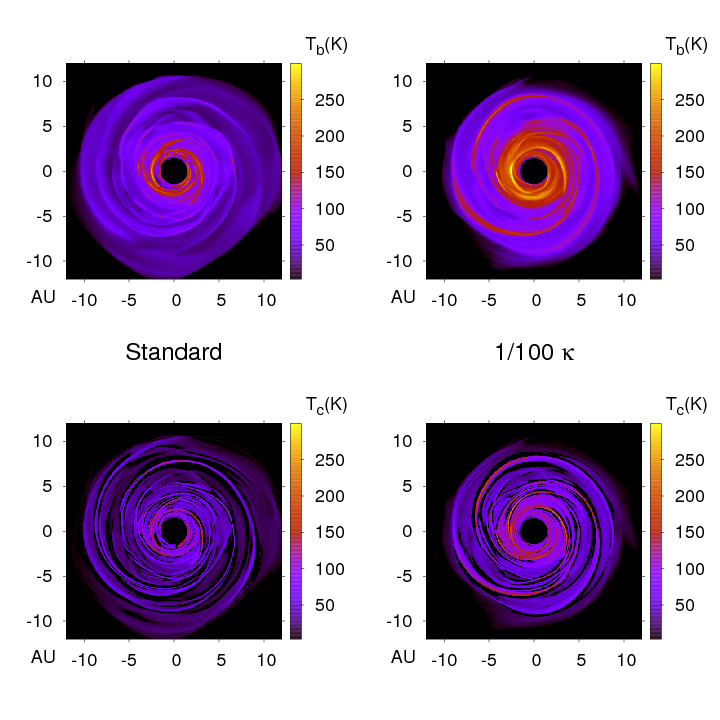}
\caption[$T_b$ and $T_c$ maps: high $\kappa$ simulations]{$T_b$ and $T_c$ maps for the
standard and 1/100 $\kappa$ simulations. As the opacity is lowered, the disks
become much more efficient at cooling.  See also Figure 7.11.}
\end{center}
\end{figure}

\begin{figure}
\begin{center}
\includegraphics[width=5.95in]{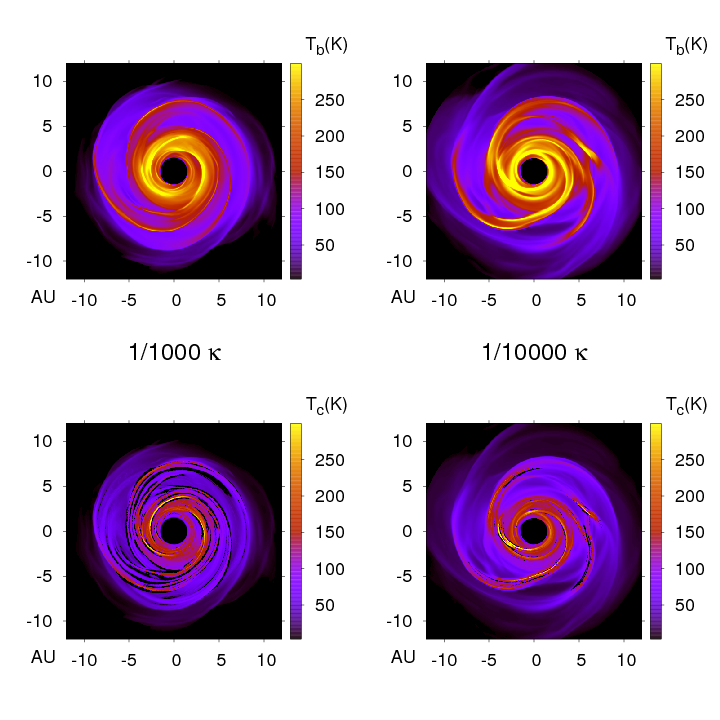}
\caption[Brightness and $T_c$ maps: low $\kappa$ simulations]{Complement to Figure 7.10: Brightness and $T_c$ maps for the
1/1,000 and 1/10,000 $\kappa$ simulations.}
\end{center}
\end{figure}

The azimuthally averaged brightness temperature profiles are shown in Figure 7.12.  As
shown in Figures 7.10 and 7.11, the brightness temperatures are higher for the lower opacity
simulations.  In addition, each profile can be roughly fit by a power law
$T_b\sim r^{-p}$, with $p=0.61$, 0.84, 0.91, and 0.85 for the 512 standard, 1/100, 1/1,000, and 1/10,000 $\kappa$ simulations, respectively.  These power laws are consistent with observed
effective temperature profiles for irradiated disks (e.g, Beckwith et al.~1990; Kitamura et al.~2002), but these disks have not reached an asymptotic phase, during which the effective profiles may be closer
to those found in the M2004 and BDNL simulations (Chapter 6).

The importance of the opacity for disk cooling is also shown in Figures 7.13-7.16.
On these plots, three quantites are shown: the Toomre $Q$, the mass-weighted
$\Gamma_1$, and the $t_{\rm cool}\Omega/f(\Gamma_1)$ profiles. For the $t_{\rm cool}$ curves, the
cooling times are calculated by dividing the azimuthally and vertically
integrated internal energy by the azimuthally and vertically integrated radiative cooling for
each annulus in the disk.
The $f(\Gamma_1)$ is the critical value of $t_{\rm cool}\Omega$, below which fragmentation is expected, for the corresponding $\Gamma_1$.  Rice et al.~(2005) demonstrated that
$f(7/5)\approx12$ and $f(5/3)\approx6$.  Based on these values, I assume $f(\Gamma_1)\approx-23\Gamma_1+
44$ for the stability analysis presented here. One should keep in mind that $f(\Gamma_1)$ is not a strict threshold and that the relation is approximate, especially for simulations that permit the cooling time to evolve with the disk (Johnson \& Gammie 2003).  Regardless, it serves as a general indicator for disk fragmentaiton.   

\begin{figure}
\begin{center}
\includegraphics[width=5.95in]{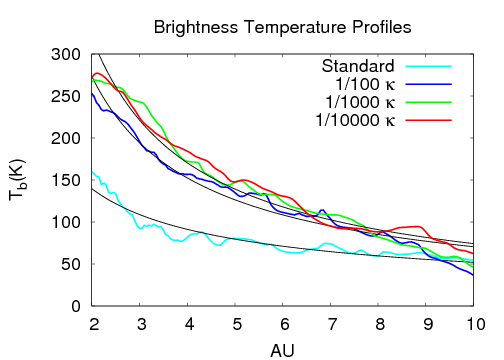}
\caption[Brightness temperature profiles for the 512 standard, $10^{-2}$, $10^{-3}$, and $10^{-4}$ $\kappa$ simulations]{Brightness temperature profiles for the standard, 1/100, 1/1,000, and 1/10,000 $\kappa$  simulations at about 77 yr.  Each profile can
be fit approximately by a power law $T_b\sim r^{-p}$ (smooth curves), with
$p=0.61$, 0.84, 0.91, and 0.85 for the standard, 1/100, 1/1,000, and 1/10,000 $\kappa$ simulations, respectively.  }
\end{center}
\end{figure}

\begin{figure}
\begin{center}
\includegraphics[width=5.95in]{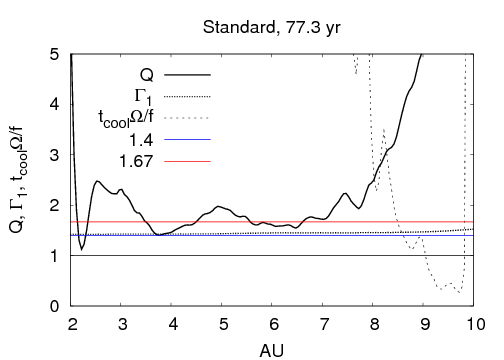}
\caption[Stability assessment for the 512 standard simulation]
{Stability assessment for the 512 standard simulation. The ordinate shows values for $Q$, $
\Gamma_1$, and $t_{cool}\Omega/f(\Gamma_1)$ as a function of $r$, where 
$f(\Gamma_1)\approx-23\Gamma_1+44$.  If
$t_{cool}\Omega/f(\Gamma_1) > 1$, then the disk is 
stable against fragmentation.}
\end{center}
\end{figure}

\begin{figure}
\begin{center}
\includegraphics[width=5.95in]{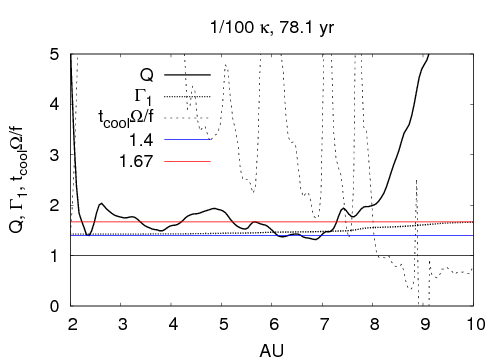}
\caption[Stability assessment for the 512 $10^{-2}$ $\kappa$ simulation]
{Same as Figure 7.13, but for the 512 1/100 $\kappa$ simulation.}
\end{center}
\end{figure}

\begin{figure}
\begin{center}
\includegraphics[width=5.95in]{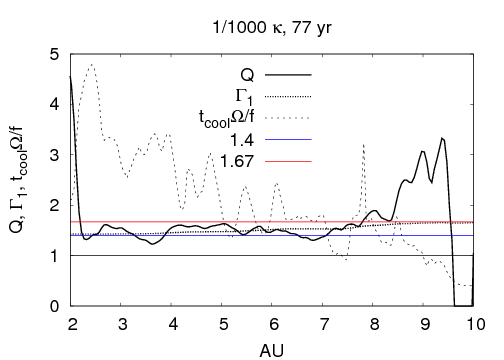}
\caption[Stability assessment for the 512 $10^{-3}$ $\kappa$ simulation]
{Same as Figure 7.13, but for the 512 1/1,000 $\kappa$ simulation.}
\end{center}
\end{figure}

\begin{figure}
\begin{center}
\includegraphics[width=5.95in]{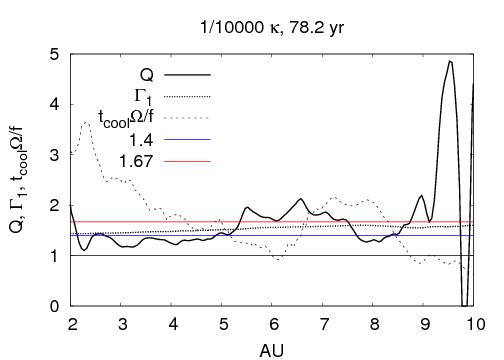}
\caption[Stability assessment for the 512 $10^{-4}$ $\kappa$ simulation]
{Same as Figure 7.13, but for the 512 1/10,000 $\kappa$ simulation.}
\end{center}
\end{figure}

\begin{figure}
\begin{center}
\includegraphics[width=5.95in]{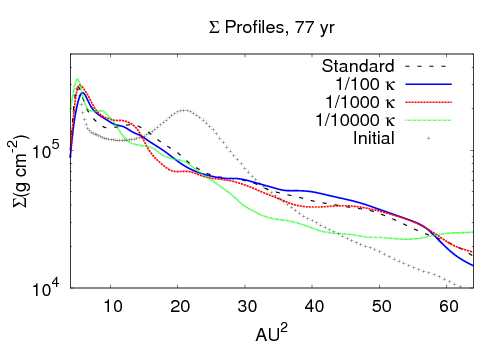}
\caption[Surface density profiles for the  512 standard, $10^{-2}$, $10^{-3}$, and $10^{-4}$ $\kappa$ simulations]{Surface density profiles for the 512 standard, 1/100, 1/1,000, and 1/10,000 $\kappa$ simulations at about 77 yr.  Each profile, except for the 1/10,000 $\kappa$ simulation, is consistent with a Gaussian between about 2 and 8 AU. The initial surface density profile with the density enhancement is shown with $+$s. The burst reorders each
disk efficiently.}
\end{center}
\end{figure}

\clearpage
\begin{figure}
\begin{center}
\includegraphics[width=3in]{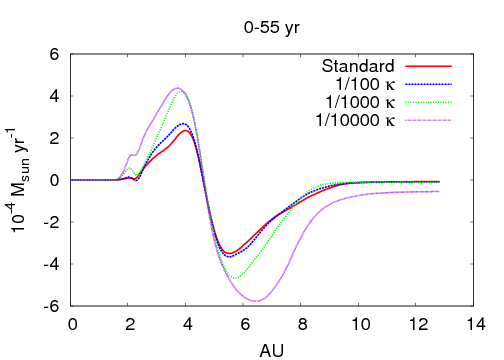}\includegraphics[width=3in]{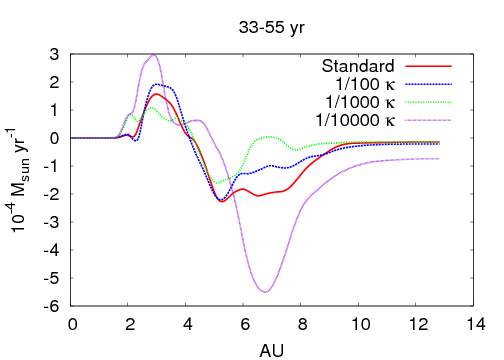}
\includegraphics[width=3in]{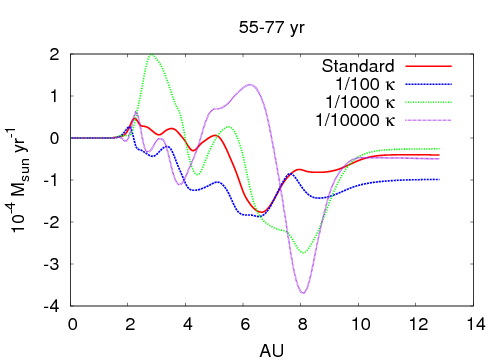}
\caption[Bursting dead zone mass fluxes]{Mass fluxes for the 512 standard, 1/100, 1/1,000, and 1/10,000 $\kappa$ simulations during the 0-55 yr, 33-55 yr, and 55-77 yr periods.
For each simulation, inward and outward mass fluxes are well above $10^{-4}~
M_{\odot}
~\rm yr^{-1}$.  Positive mass fluxes are inward.}
\end{center}
\end{figure}

As Figures 7.13-7.16 demonstrate, the
disk is approaching a state of constant $Q$ for a wide range of radii, as expected in an asymptotic phase (Chapter 6).  The
mass-weighted average $Q$s between 3 and 6 AU are 1.73, 1.69, 1.49, and 1.43 for the
standard, 1/100, 1/1,000, and 1/10,000 $\kappa$ simulations, respectively. The
variation in the average $Q$ is consistent with the $A_+$ measurements.

Because the cooling times fluctuate rapidly, I average the integrated cooling rates and
the internal energies for the
65- and 77-year snapshots.  The cooling time profiles for the appropriate $\Gamma_1$
only drop substantially below unity in regions of high $Q$; these
disks are stable against fragmentation.  As the opacity is lowered, the cooling times decrease
as well, with the 1/10,000 $\kappa$ simulation close to the fragmentation limit.
It should also be noted that the only disk that behaves similarly to a constant $\Gamma_1$ disk
is the standard opacity simulation, and for all other simulations, a constant $\Gamma_1$ is inappropriate.

In addition to the changes in the $Q$ profile, the surface density profiles
are altered during the burst, and roughly follow a Gaussian profile between
about 2 and 8 AU (Fig.~7.17), with the 1/10,000 $\kappa$ simulation showing the most
deviation.  This relaxation to a Gaussian was noted
by Boley et al.~(2006, 2007c) for the M2004 and BDNL simulations, and such a profile seems to be typical of
GI-driven accretion when low-order modes dominate. This rapid
reorganization of mass also indicates that the accretion rates are very high.
Figure 7.18 shows mass fluxes in the disk for several time intervals.  As
described in Chapter 6, mass fluxes are calculated by differencing the mass
inside a cylinder at two different times.  The inward and outward accretion
rates vary, and can be well above $10^{-4}~M_{\odot}$ yr$^{-1}$.

\section{Shock Strength}

Using the periodogram analysis (Chapter 6), I find that the
location of corotation for the $m=2$ and $m=3$ spiral waves is at  $r\sim 4$ AU, as measured between 0 and 77 yr.  The exception is the 512 1/10,000 $\kappa$ disk, in which weak signals indicate corotation near 4 AU for $m=3$ and 5 AU for $m=2$.  In addition, the mass flux inflow/outflow boundary in each disk for the 0-55 yr time period (Fig.~7.18) indicates that corotation is initially near 4.5 AU.  
Overall, corotation is about 1 AU inward from the target location, but still places
 chondrule-forming shocks within the asteroid belt and at cometary distances for large pitch angle spirals.  In all simulations except for the 512 1/10,000 $\kappa$ run, the bursts produced spirals with pitch angles of only $i\approx 10^{\circ}$.
Because the 512 1/10,000 $\kappa$ simulation is close to fragmentation (see above), the amplitudes
for the mid-order Fourier components are large, and the global spirals develop a flocculent morphology in areas. Based on the arguments in \S 7.1, one should not expect for many chondrule-producing shocks to be present in these simulations unless the vertical and radial motions are important.

\begin{figure}
\begin{center}
\includegraphics[width=5.95in]{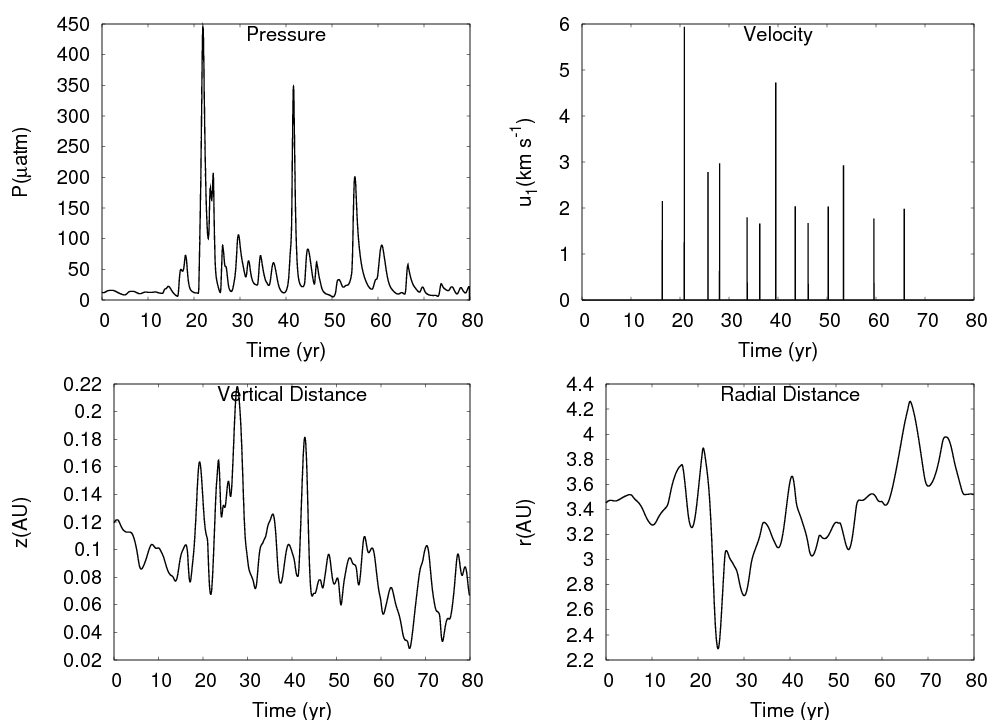}
\caption[Fluid element history]{Thermal and spatial histories for a fluid element in the 512
1/10,000 $\kappa$ simulation. Radial and vertical motions vary
considerably, as one would expect from shock bores.  The fluid element experiences many
shocks with multiple pre-shock conditions.  This fluid element's history includes one potential
chondrule-forming shock at $t\approx20$ yr.}
\end{center}
\end{figure} 

As discussed in \S 7.2, fluid elements are used to investigate shock
strengths and fluctuations in thermodynamic quantities, e.g., temperature, in the 512 simulations.
Figure 7.19 shows thermal and trajectory histories for a sample fluid element
in the 512 1/10,000 $\kappa$ simulation that experiences a potential chondrule-forming shock (see below).  Multiple shocks with a wide range
of pre-shock conditions are encountered. However, one should keep in mind that shocks in these simulations are unlike
simple, 1D shocks, where the length of the full shock occurs in roughly $10^{10}$
cm; these simulations can only resolve scales of about $10^{12}$ cm or larger.   Despite this potential difficulty, I use the
pressure difference before and after the shock to calculate the Mach number $\mathcal{M}$.
Once $\mathcal{M}$ is known, the velocity normal to the shock $u_1$, in the frame of the shock, can be inferred.   Using the pressure histories for each fluid element, a possible shock is identified whenever the $\partial p/\partial t$ changes from negative to positive then back to negative.  The pre-shock flow is then taken to be the first sign switch, and the post-shock flow is the second sign switch.  If $\mathcal{M}^2=\left(\gamma+1\right)\eta/\left( 2\gamma\right) + 1\ge 2$, the event is counted as a shock, where $\eta=\left(p_2-p_2\right)/p_1$ is the fractional pressure change and $\gamma$ is the average of the pre- and post-shock first adiabatic index.  Once $\mathcal{M}$ is determined, $u_1=\mathcal{M}c_{s1}$ is calculated, where $c_{s1}$ is the pre-shock adiabatic sound speed of the gas.  Other methods were tried, e.g., estimating the speed of the spiral wave, but local variations in pattern speed and three-dimensional shock orientations render such methods exceedingly difficult.    As can be seen in the pressure panel in Figure 7.19, using $\eta$ does a decent job of identifying shocks.  Some counting errors and poor estimations of shock strength are associated with this technique.  In particular, because shocks in these simulations are spread out over multiple cells by the AV algorithm, post-shock pressures could be reduced by radiation transport and three-dimensional wave effects, e.g., shock bores, which would cause underestimates for $\mathcal{M}$.

Table 7.2 indicates shock information as extracted from the fluid element information. Column two shows the
total number of detected shocks (TS) with $\mathcal{M}^2\ge2$, and column three displays
the average number of shocks per fluid element.  Columns four (TS Mass) and five (CS Mass) show the 
total dust mass (see below) that goes through a shock with $\mathcal{M}^2\ge 2$ and the total
dust mass that encounters a chondrule-forming shock, respectively.  Finally, column
six shows the total number (Total FE) of fluid elements that remain after the 70 yr time period, i.e., the elements that
were not accreted onto the star or fluxed into background density regions.

As described in \S 7.1, I consider a shock to be
chondrule-forming when $u_1$ lies in a 1 km s$^{-1}$ band
between 5 km s$^{-1}$ and 11 km s$^{-1}$ and the pre-shock density is between $\log
\rho(\rm~g~cm^{-3})$ = -8.5 and -10.5.  
Chondrule-forming shocks should be thought of as having the potential to form
chondrules, but may not yield chondritic material due to
incompatible dust to gas ratios, too high or low cooling rates, and fractionation. 
Likewise, by dust processing, I only mean that dust may be altered by the shock.

\begin{table}
\begin{center}
\caption[Shock information]{Shock information for the time period between 10 and 70 yr.  TS indicates the total shocks encountered by all fluid elements. TS/FE gives the average number of shocks for a fluid element. TS Mass is a rough estimate of the total dust mass pushed through shocks in each disk, and CS Mass is the total dust mass that encounters a chondrule-forming shock.  Finally, Total FE indicates the number of fluid elements that remain in the simulated disk at the end of the 60 yr time period.  The errors, when specified, are taken to be Poissonian, and they are only given for comparing the simulations to each other. }
\begin{tabular}{c c c c c c c}\\ \hline
512 Sim. & TS & $\frac{\rm TS}{\rm FE}$ & 
            TS Mass   (M$_{\oplus}$)& CS Mass (M$_{\oplus}$)   & Total FE\\ \hline\hline

Standard & $8714\pm93$ & $8.78\pm0.094$ & 4(3) & 0 & 992 \\
1/100 $\kappa$ & $8553\pm92$ & $8.64\pm0.093$ & 4(3) & 0 & 990 \\
1/1,000 $\kappa$ & $8343\pm91$ & $8.53\pm0.093$& 4(3) & 0 & 978 \\ 
1/10,000 $\kappa$ & $9287\pm96$ & $9.89\pm0.10$ & 4(3) & 3 & 939 \\ \hline

\end{tabular}
\end{center}
\end{table}

For the processed dust estimates, I assign each fluid element a mass by calculating the total
disk mass within some $\Delta r$ and distributing that material evenly among all fluid elements in that $\Delta r$.  
In addition, I assume a gas to solids ratio of 100 everywhere
for each simulation.  I note that this is inconsistent with the assumption that the
opacity is lowered in three of the simulations because of settling, which would
make the dust to gas ratio much higher in the midplane and lower everywhere else.  Because only
an order of magnitude estimate is sought, this detail is ignored  inasmuch
as the midplane will process more solids per shock and the high-altitude shocks
will process less.  

I check whether the total number of detected shocks  is reasonable by evaluating the expected number of shocks in a Keplerian disk for $m$-arm spiral waves.   If the pattern
radius for the spirals is $r_p$, the number of shocks 
during some time period $\Delta t$ for a fluid element orbiting at $r$ is 
\begin{equation}N_S=\frac{m (GM_{\rm star})^{1/2}}{2\pi r^{3/2}}\bigg | 1-
\left(r/r_p\right)^{3/2}\bigg | \Delta t.\end{equation}
Evaluating equation (7.3) for 1000 fluid elements
evenly distributed in annuli between 2 and 8 AU, with $r_p=4$ AU,  with $m=3$, and with $\Delta t=60$ yr yields approximately
8000 shocks.  So the number of detected shocks in these simulations is reasonable. 

To determine whether the differences in the number of shocks between each simulation are significant, I take
the uncertainty in these measurements to be Poissonian.  Within the uncertainty, 
the number of shocks per fluid element is roughly consistent for each disk except for the 512 1/10,000 $\kappa$ run.  The more flocculent spiral structure  in this simulation not only produces a larger number of shocks in the disk, but also effects chondrule-producing shocks.   Several thousand $M_{\oplus}$ of dust are pushed through shocks in each simulation, with a few $M_{\oplus}$ of dust experiencing chondrule-forming shocks in the 512 1/10,000 $\kappa$ run.  I remind the reader that these are only order of magnitude estimates.

All chondrule-forming shocks transpire within the first 40 yr of evolution, with three occurring within the first 21 yr (Table 7.3), and so the onset of the burst creates the strongest spiral shocks.  These shocks all occur between 3 and 5 AU.  Because the pattern radii for the global spiral waves are between 4 and 5 AU, radial and vertical motions and/or unusual wave geometry are likely responsible for the high shock speeds.  Figure 7.20 shows shock events with $\mathcal{M}^2\ge2$ on the $u_1$-$\rho$ plane.  The green strip indicates the chondrule-forming region.  Notice that there are a number of other shocks that fall just outside the green area.  The strip is only a rough estimate based primarily on one group's 1D shock simulations that are known to require refinements.

\begin{table}
\begin{center}
\caption[Chondrule-forming shocks information]{Information for chondrule-forming shocks in the 512 1/10,000 $\kappa$ simulation. Columns three and four indicate the $r$ and $z$ where the shock occurs, while columns five and six indicate the initial $r$ and $z$ for each fluid element.  All chondrule-forming shocks take place within the first three vertical cells ($\Delta z=0.05$ AU), which is also within roughly a third of the local disk scale height for these $r$.}
\begin{tabular}{c c c c c c c c}\\ \hline

$u_1(\rm km~s^{-1})$ & $\log \rho_1(\rm g~cm^{-3})$  
           & $r$(AU) & $z$(AU) & $r_i$(AU) & $z_i$(AU)& $t$(yr)\\ \hline
5.9 & -8.8 & 3.8 & 0.11 & 3.5 & 0.12 & 21\\
6.0 & -8.5 & 3.8  & 0.031 & 3.5 & 0.027 & 17\\
5.3 & -8.5 & 4.7 & 0.062 &  4.5 & 0.089 & 16 \\
5.8 & -8.7 & 3.2 & 0.079 & 3.6 & 0.071 & 40 \\ \hline
\end{tabular}
\end{center}
\end{table}

In addition to chondrule formation and dust processing, the relative importance of a given shock strength in heating the disk can be evaluated.   Figure 7.21 shows the number of shocks $N$ per $\Delta\mathcal{M}$ on a log-log plot.  Although there is variation among the profiles, each simulation roughly shows $N\Delta\mathcal{M}\sim\mathcal{M}^{-p}\Delta\mathcal{M}$ for $p\approx4$.  
  With this in mind, consider a
specific shock energy $e\sim\mathcal{M}^2$. The energy added to the disk per $\Delta\mathcal{M}$ is then $E\Delta\mathcal{M}\sim \mathcal{M}^2 N \Delta\mathcal{M}$.  
Based on this argument, dissipation is only evenly distributed among the shocks for $N\Delta\mathcal{M}\sim \mathcal{M}^{-2}\Delta\mathcal{M}$. 
Even though the shocks are mediated by global spiral waves during these bursts, the greatest
contribution to the heating by shocks comes from weak shocks over the $r$-range for which fluid trajectories have been followed.

\section{Discussion and Conclusions}

In this section, I review the implications of the results of this study for disk fragmentation, the effects of opacity on disk cooling, the FU Ori phenomenon, and chondrule formation.  I remind the reader that the simulations presented here are meant to be a numerical experiment that explores the possible connection between chondrules, FU Ori outbursts, and bursts of GI-activity.  Moreover, this experiment provides a systematic study of the effects of opacity on disk cooling, which complements the Cai et al.~(2006) metallicity study and provides a test bed for disk fragmentation criteria.

\begin{figure}
\begin{center}
\includegraphics[width=5.95in]{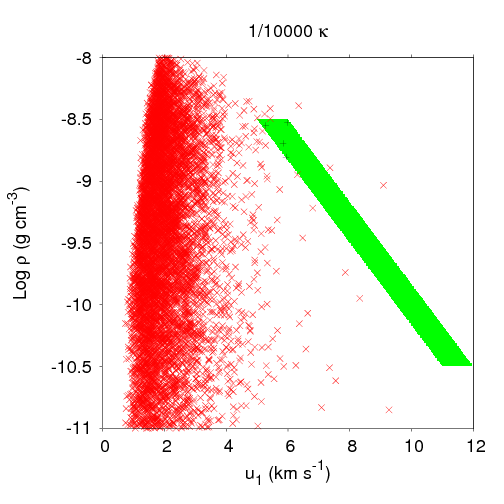}
\caption[Shocks on the $u_1$-$\rho$ plane]{Shocks on the $u_1$-$\rho$ plane for
the 512 1/10,000 $\kappa$ simulations.  The green area
shows the chondrule-forming region, based mainly on the results of Desch \& Connolly
(2002)  (see \S 7.1).  The green strip should not be thought of as definitive, and slight changes
in the strip's location could identify more chondrule-forming shocks, particularly at low $u_1$. }
\end{center}
\end{figure}

\begin{figure}
\begin{center}
\includegraphics[width=4in]{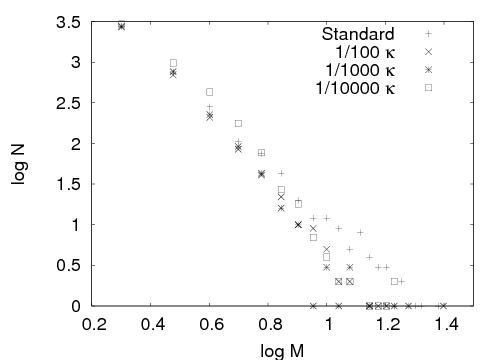}
\caption[Shock events vs.~Mach number.]{The number of shocks versus $\mathcal{M}$, binned in $\Delta\mathcal{M}=1$ intervals.  The profiles roughly follow a power law.}
\end{center}
\end{figure}

\subsection{Fragmentation}

 After the onset of the burst, none of the disks fragments, and only
the 512 1/10,000 $\kappa$ simulation shows dense knot formation during the peak of the burst (10-30 yr).  These knots do not break from the spiral wave even in the 1024 1/10,000 $\kappa$ simulation, and so clump formation does not seem to be missed due to poor resolution. One should also keep in mind that for this simulation, the opacity is suddenly dropped by a factor of 10$^{4}$, and such knot formation may not occur in a disk with more realistic settling timescales.  As discussed below, this is also a caveat for the chondrule formation results.  On the other hand, it does indicate that disk fragmentation may occur inside 10 AU under very extreme conditions.

The stability of these disks against fragmentation is supported by Figures 7.13-7.16.  These figures indicate that cooling rates for the 512 standard and 1/100 $\kappa$ simulations are too low to cause disk fragmentation.  The 512 1/1,000 and 1/10,000 $\kappa$ simulations do have areas where $\zeta = t_{\rm cool}\Omega/f(\Gamma_1) \lesssim 1$, but the disks approach stability against GIs in those regions ($Q\gtrsim1.7$; Durisen et al.~2007a).   As noted in \S 7.3, $\zeta \lesssim 1$ is not a strict instability criterion, but it does serve as an estimate for disk stability against fragmentation.  For no simulation is $\zeta$ well below unity where $Q$ is also $\lesssim1.7$.

\newpage

Lowering the opacity increases the cooling rates.  The 1/10,000 $\kappa$ simulation exhibits the most rapid cooling because the midplane optical depths are near unity, which results in the most efficient radiative cooling possible. {\it Changes in dust opacity have a profound effect on disk cooling}, contrary to what is claimed by Boss (2001, 2004).  Although not modeled, if the opacity were to continue to drop such that the midplane optical depth becomes well below unity, cooling would once again become inefficient.  In addition, supercooling of the high optical depth disks (standard and 1/100 $\kappa$) by convection does not occur.  These findings are consistent with analytic arguments by Rafikov (2005, 2007) and with numerical studies of disk fragmentation criteria by Gammie (2001), Johnson \& Gammie (2003), and Rice et al.~(2005). 

I have demonstrated that the BDNL radiation algorithm used for this study couples the high and low optical depth regimes well (Chapter 3).  My conclusions regarding fragmentation are based on mulitple analyses: surface density rendering (Figures 7.4-7.6), an energy budget analysis (Figures 7.8 and 7.9), cooling temperature and brightness maps (Figures 7.10 and 7.11), and a cooling time stability analysis (Figures 7.13-7.16).  Furthermore, the resolution in each direction for all simulations was doubled, and these 1024 simulations were evolved to at least the peak of the burst ($t\approx 20$-33 yr).  Radiative cooling and shock heating curves for the 1024 standard and 1/100 $\kappa$ simulations were compared with curves for the 512 simulations. The 512 and 1024 curves show similar radiative losses, but the 1024 simulations exhibit additional shock heating.  Fragmentation was not missed.  

\newpage

It is also pertinent to evince that the code can detect fragmentation when cooling rates are high and $Q$ is low.  Figure 7.22 shows snapshots for the 512 1/10,000 $\kappa$ simulation and a similar simulation but with the divergence of the fluxes artificially increased by a factor of two.  In the normal simulation, kinks in the spiral waves form during the onset of the burst.  
As discussed above, the disk is very close to the fragmentation limit, but the knots do not break from the spiral wave.  Figure 7.16 suggests, although for a later time, that increasing the cooling rates by a factor of two should drop $\zeta$ well below unity in low-$Q$ regions. As expected, the 512 1/10,000 $\kappa$ disk with enhanced cooling fragments.

Three clumps form, one for each spiral wave, between 4 and 5 AU.  The location of clump formation is consistent with the prediction by Durisen et al.~(2007b) that a spiral wave is most susceptible to fragmentation near corotation.  One of the clumps survives for several orbits and eventually passes through the inner disk boundary.  Resolution is always a concern for simulations.  Because these results are consistent with analytic fragmentation limits and numerical fragmentation experiments, it appears that CHYMERA can detect fragmentation at the resolutions employed for this study. 

\begin{figure}
\begin{center}
\includegraphics[width=3.5in]{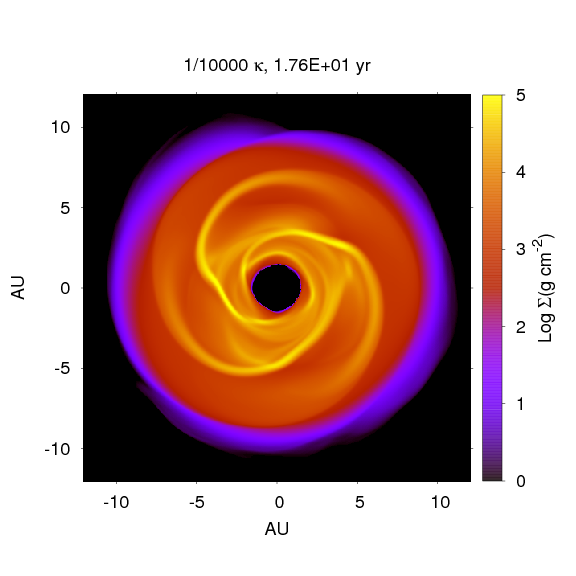}
\includegraphics[width=3.5in]{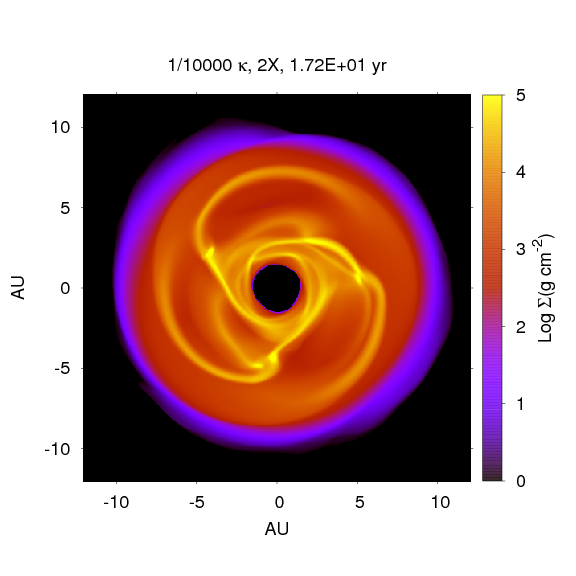}
\caption[Fragmentation]{The 512 1/10,000 $\kappa$ simulation and a similar calculation, but
with the cooling rates artificially increased by a factor of 2.  Based on Figure 7.16, one would
expect for the disk to fragment with this cooling enhancement, and it does. }
\end{center}
\end{figure}  

\subsection{FU Ori Outbursts}

In these simulations, the strong bursts of GI activity provide high mass fluxes  ($\dot{M}\gtrsim10^{-4}~M_{\odot}$ yr$^{-1}$) throughout each disk.  Even though corotation
is at $r\sim4$ AU for the major spiral arms, the 2 AU region of the disk is
strongly heated.  It is not difficult to speculate that if a larger extent of
the disk were modeled, the temperatures due to shocks would be large enough
to thermally ionize alkalis and drive a thermal MRI.
Figure 7.23 indicates maximum, average, and minimum temperatures of the fluid elements in the 512 disks. The
standard and 1/10,000 $\kappa$ simulations have the hottest fluid elements because the gas cannot
cool quickly or the shocks are extreme, respectively.  The other simulations show strong temperature variations
as well, but the peak temperatures of the disk are not as high. From these simulations it appears
to be plausible, but by no means proven, that a burst of GI activity as far out as 4 AU can
activate a thermal MRI inside 1 AU, which may then be responsible for a thermal instablility.  Although I am simulating a very
massive disk, I remind the reader that such density enhancements may be plausible for massive T Tauri systems (see Chapter 1). Additional studies need be conducted, preferably with a self-consistent build-up of a dead zone, to address the efficiency of this mechanism in low mass disks.

\newpage

There are at least three observable signatures for this mechanism.  First, if a
GI-bursting mass concentration at a few AU ultimately results in an FU Ori phenomenon, then
one would expect to see an infrared precursor, with a
rise time of approximately tens of years.  Second, one would also
expect for a large abundance of molecular species, which would normally be
frozen on dust grains, to be present in the gas phase
during the infrared burst due to shock heating
(T.~Hartquist 2007, private communication).  Third, approximately the first
ten AU of the disk should have large mass flows if the burst takes place near
4-5 AU.  I speculate based on these results that if the burst were to take place
at 1 AU, then high outward mass fluxes should be observable out to a few AU.

\subsection{Dust Processing} 
Each simulation shows that a large fraction of material
goes through shocks with $\mathcal{M}^2\ge2$. Although these shocks are weak,
their abundant numbers may result in the processing of dust to some
degree everywhere in the 2-10 AU region. In fact, such processing may be
necessary for prepping chondritic precursors for strong-shock survival  (Ciesla 2007, private communication). 

\begin{figure}
\begin{center}
\includegraphics[width=3in]{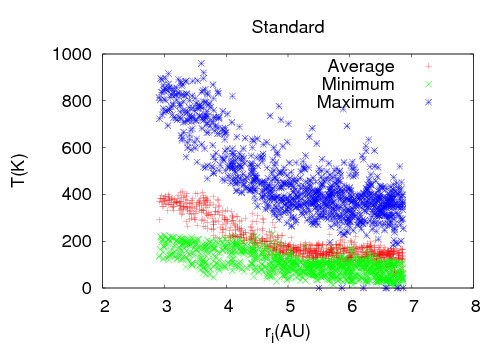}\includegraphics[width=3in]{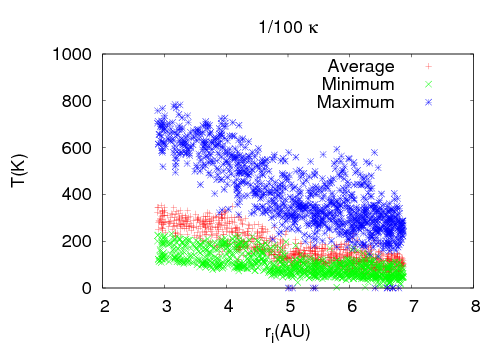}
\includegraphics[width=3in]{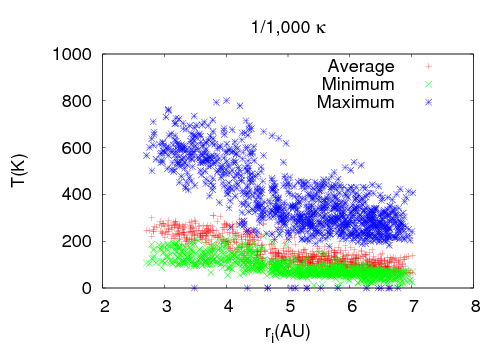}\includegraphics[width=3in]{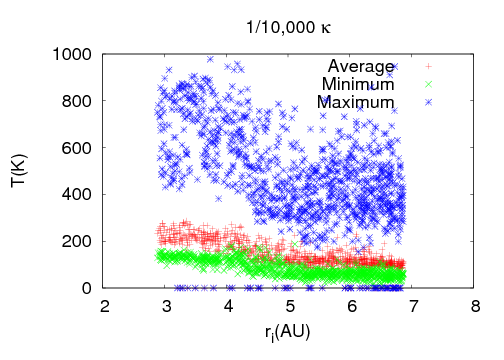}
\caption[Intial $r$ and minimum, average, and maximum temperature
plots]{Initial $r$ and minimum, average, and maximum temperature plots for fluid elements in each 512 case.  The temperature variations are quite large, and the standard and 1/10,000 $\kappa$ cases
approach maximum temperatures of 1000 K. The values on the abscissa are due to the fluid elements
that were lost from the simulated disk.}
\end{center}
\end{figure}  

Based on the arguments presented in \S 7.1, the intent was to produce spirals with large
pitch angles by constructing a disk biased toward a strong, sudden GI-activation near 5 AU.  Even with this bias, the pitch angles of the spirals remain small, with $i\approx10^{\circ}$.  
Why are the pitch angles so small?  According to the WKB approximation, $\cot i = \mid k_r r/m\mid$ (Binney \& Tremaine 1987), where $k_r$ is the radial wavenumber for some $m$-arm spiral.  As discussed in Chapter 4, the most unstable wavelength is roughly $\lambda_u\approx 2\pi c_s/Q\kappa\approx2\pi h/Q$ for disk scale height $h$.   For a disk unstable to nonaxisymmetric modes, $\lambda_u$ corresponds to some $m$-arm spiral (e.g., Durisen et al.~2007b).  By relating $k_r=2\pi\beta m/\lambda_u$, 
\begin{equation}
\cot i\approx \beta Qr/h
\end{equation} 
in the linear WKB limit ($\mid k_r r/m\mid \gg 1$), where $\beta$ is a factor of order unity.  Because $\beta Qr/h \sim10$  in gravitationally unstable disks, the linear WKB analysis may be marginally applicable.  Equation (7.4) predicts that linear spiral waves in these disks should have pitch angles $i\approx6$ to 11$^{\circ}$ for $\beta$ between 1 and 1/2, respectively.  It appears that this estimate for $i$ extends accurately to the nonlinear regime.

These spirals are efficient at heating the disk and transporting angular momentum, but not at producing chondrules.  Only for the 512 1/10,000 $\kappa$ simulation are chondrule-producing shocks detected; the 1024 simulations are still being analyzed.   This low opacity disk is on the verge of fragmentation, and so the presence of chondrule-forming shocks is likely due to the flocculent spiral morphologies, including kinks in spiral waves. 

To estimate the occurrence of a chondrule-forming shock, I employ a generous
$u_1$-$\rho$ criterion.  I do not take into account the optical depth criterion of
Miura \& Nakamoto (2006) on grounds that the large scale over which these shocks
take place can allow for chondrules to equilibrate with their surroundings (Cuzzi \& Alexander 2006; see Chapter 1).
However, one should be aware that the optical depth criterion of Miura \& Nakamoto
will likely exclude all chondrule-producing shocks inside 4 AU in the $u_1$-$\rho$ plane for
these simulations.  All chondrule-forming shocks occur between $r=3$ and 5 AU and at altitudes that are roughly less than a third of the gas scale height (Table 7.3).  Because a large degree of settling is assumed, these shocks are located in regions that may be consistent with the dusty environments in which chondrules formed (Wood 1963; Scott \& Krot.~2005). 

I estimate that  $\sim 1~M_{\oplus}$ of dust is processed 
through chondrule-forming shocks.  Because these shocks are limited to the onset of the GI burst,
a few $\times10~M_{\oplus}$ of dust would be processed in the 1/10,000 $\kappa$ disk if it went through about ten outbursts.  If more outbursts take place than those that lead to an FU Orionis event, then more chondritic material could be produced.  One should also remember that to produce these shocks, the disk was pushed toward fragmentation by suddenly dropping the opacity by a factor of 10,000.  In order for bursts near 4 or 5 AU to produce chondrules, the disk must be close to fragmentation.

An alternative to the fragmentation scenario is suggested in Figure 7.1.  In addition to driving the FU Ori phenomenon, bursts that occur near 1 AU can produce chondrules even with small spiral pitch angles.  Moreover, bursts that activate slightly further out, near 2-3 AU, could drive chondrule formation beyond the frost line and produce the annealed dust that is observed in comets (Harker \& Desch 2002;  Wooden et al.~2005; McKeegan et al.~2006).  There are several major advantages to this scenario, which I discuss in more detail in Chapter 8.

\chapter{CULMINATING CONCLUSIONS}

In this dissertation, I have addressed the behavior of spiral shocks in protoplanetary disks, and have attempted to make connections between these shocks and multiple disk phenomena and planet formation.  Spiral waves exhibit rich dynamics, and in general, they are by no means simple, two-dimensional beasts. They are likely responsible for some of the most energetic accretion events in disks that surround newly-forming, Sun-like stars.  I find that direct giant planet formation by disk instability requires demanding and possibly  unphysical conditions. Despite this, these waves are not inconsequential to building planets.   Spiral shocks can serve as the machinery for thermally processing dust throughout a planet-forming disk; therefore, they play a weighty role in the formation of planets.  In this final Chapter, I summarize the work and the principal conclusions for the studies in this dissertation. I also outline the hypothesis that gravitational instabilities in dead zones are responsible for chondrule formation and the FU Orionis phenomenon.

\section{Three-dimensionality of Spiral Waves}

As a spiral wave develops into a shock, the post-shock region is typically thrown out of vertical force balance and rapidly expands. One should only expect a spiral shock to behave like a density wave (two-dimensionally) when the gas is isothermal and when self-gravity can be neglected.  If self-gravity is important in an isothermal shock, then the gas will compress.  When the gas is stiffer than isothermal, the gas will almost always expand after the shock, except for cases of low Mach number and strong self-gravity.  This rapid expansion of the gas is called a shock bore, and these bores can create breaking waves and induce sudden radial as well as vertical  motions in the gas. Shock bores may also be an important source of disk turbulence.  As demonstrated in Chapter 5, large vortical flows are effected by spiral shocks, but with the analyses used in these studies, no distinction between large-scale stirring or mixing can be made.   

As cooling times decrease, shock heating rates increase and mid-order spiral structure becomes stronger.  Unless the cooling times drop below the fragmentation limit (see below),  radiation losses balance shock dissipation and work done by gravity on the gas.  Shocks are not limited to the midplane, but extend to all altitudes along the spiral wave front and elsewhere at mid- to high-disk altitudes as a result of waves generated by shock bores.   

 The pitch angles of spiral waves in GI-active disks are relatively small, even for large bursts.  Only when the disk approaches fragmentation are pitch angles with $i\gtrsim10^{\circ}$ present.  Based on fluid element histories, the frequency of shocks as a function of Mach number decreases approximately as a power law during bursts of GI activity, and the low Mach number shocks contribute the most to heating the disk. Shock frequencies during an asymptotic state have yet to be explored.

As shown in Chapters 6 and 7, low-order structure is primarily responsible for mass transport in GI-active disks, but as the cooling times approach the fragmentation limit, mid-order structure adds a significant  component to the gravitational torque. During the asymptotic phase of disk evolution, the effective $\alpha$
approaches the predicted $\alpha$ for a local model, despite the mass transport being dominated by global modes. Even when the $\alpha$ model is reasonably consistent with GI mass transport during the asymptotic phase, the picture of mass slowly diffusing through the disk is inaccurate; local mass fluxes in these disks are highly variable.  Moreover, using a local model to describe mass transport and disk heating by GIs during the asymptotic phase requires knowledge of the cooling rates {\it a priori}, which presents a challenge.   Finally, a local formalism for mass transport fails to capture GI bursts well, even though the pitch angles are small, $\mid k_r r/m\mid\sim$10, for the spirals in these simulations (\S 7.5.3).  In Chapter 7, it was shown that bursts of GIs centered on $r\sim4$-5 AU cause high mass fluxes over the entire simulated disk.  This is a concern for implementing a local model because bursts are a real possibility in disk evolution, e.g., in producing the FU Orionis phenomenon.  

\section{Disk Fragmentation}

The radiation hydrodynamics simulations presented in Chapters 6 and 7 have been analyzed using a suite of analyses that test multiple aspects of the physics that could affect disk fragmentation.   In particular, the results address the importance of convection, opacity, and radiation boundary conditions  (see also Nelson 2006) on disk cooling.

{\it Convection:}  The convection test results show that both the SV and CHYMERA allow for convection under appropriate conditions (\S 3.3).  As discussed in Chapters 5 and 6, convection is active in the M2004 disk during the axisymmetric phase. While active, the convective cooling rate is comparable to the radiative cooling rate, as is expected because energy must ultimately be radiated away at the disk's photosphere.  Once GIs activate,  convection is disrupted, and no fragmentation occurs when either the M2004, C2006, or BDNL algorithm is used for cooling.  

Mayer et al.~(2007) suggest that the M2004 simulation does not portray efficient cooling by convection because the optical depths are not large.   However, convection-like motions are also noted in the flat-Q disk without the density enhancement when the opacity law is tuned to induce convection.  The cooling times in this massive, highly optically thick disk are much longer than those expected to lead to fragmentation.  Moreover, in the dead zone models presented in Chapter 7, brightness temperature maps were shown to look for local regions where fast cooling by convection might be taking place.  No anomalous bright spots were noticeable. In the series of dead zone calculations, Rosseland mean midplane optical depths spanned between $\tau=10^4$ and 1.  Supercooling by convection was absent.  This is consistent with analytic arguments by Rafikov (2007), with our numerical tests, and with work by Nelson et al.~(2000), who assumed efficient convection in their estimations for the local disk effective temperature. Sudden vertical motions in disks along spiral shock fronts are shock bores.

{\it Opacity:} A systematic study of the effects of the opacity on the strength of spiral waves has been presented in Chapter 7.  The study demonstrates that {\it opacity can have a profound effect on disk cooling and on the structure of spiral waves.} This is evinced in the cumulative energy losses, the Fourier structure analysis, local cooling time plots, and the brightness temperature maps.  These results complement the study by Cai et al.~(2006), but are in disagreement with Boss (2002, 2004a), who posits that metallicity does not change the outcome of disk instability due to efficient cooling by convection.  The studies presented in this dissertation indicate that radiative cooling ultimately controls convection, which precludes cooling rates that lead to disk fragmentation in all disks I have studied ($r\lesssim 40$ AU).  I find, as expected from analytic arguments, cooling becomes more efficient as the Rosseland mean midplane optical depth approaches unity.  

{\it Cooling times:}  As the opacity is lowered, the cooling times approach the fragmentation limit.  Although the 1/10,000 $\kappa$ simulation is on the verge of fragmentation and knots do form in the spiral waves during the onset of the burst, none of the spirals or knots breaks into a clump.    When the resolution is doubled in each direction, which provides a very high resolution simulation with $(r,\phi,z)=(512,1024,128)$, the knots still do not turn into fragments.  The behavior of each simulation is consistent with the fragmentation criterion $t_{\rm cool}\Omega/f(\Gamma_1)=\zeta\lesssim 1$, and the 1/10,000 $\kappa$ disk only fragments when its cooling is artificially enhanced by a factor of two. 

As discussed above, when the opacity 
in the mass enhanced flat-Q disk is abruptly decreased by a factor of 10${^4}$, the disk shows fragmentation-like behavior.  A similar effect is reported by Mayer et al.~(2007), who find that their disk only fragments when the mean molecular weight is suddenly switched from $\mu=2.4$ to $\mu=2.7$.  It should be noted that the M2004 simulation was accidentally run with $\mu=2.7$, and Cai et al.~(2006) purposefully ran their simulations at the high $\mu$ for comparison with the M2004 results. Neither of the studies reported disk fragmentation.  This may indicate that when a disk fragments shortly after a sudden switch in a numerical parameter, e.g., opacity here and $\mu$ in Mayer et al., the fragmentation may be numerically driven rather than physically. Regardless, the results do indicate that disk fragmentation by GIs may be possible under extreme conditions. Demonstrating that such conditions can be realistically met remains to be seen.  Based on these results, disk fragmentation for $r\lesssim 40$ AU appears to be the exception rather than the norm.

It should also be noted that recent results by Stamatellos et al.~(2007), who find that disks can fragment at large radii ($r> 100$ AU), are not contradicted by this study or by analytic arguments (Rafikov 2005).

\section{Unified Theory}

A  hypothesis behind the work presented in Chapter 7 is that bursts of GI activity, dead zones, the FU Ori phenomenon, and chondrule formation are linked.  The general picture is that mass builds up in a dead zone due to layered accretion and that GIs erupt once $Q$ becomes low.  This activation of the instability causes a sudden rise in the mass accretion rate that heats up the disk inside 1 AU to temperatures that can sustain a thermal MRI.  The thermal MRI shortly thereafter activates the thermal instability (Armitage et al.~2001; Zhu et al.~2007).  In addition, the strong shocks process dust and form chondritic material in the asteroid belt and at comet distances.

The findings presented in Chapter 7 indicate that bursts as far out as about 4 AU can cause FU Ori mass accretion rates and significantly heat the inner disk.  Observational signatures of this scenario include high mass fluxes out to about 10 AU, an overabundance of typically ice phase molecular species in the gas phase, and a  mid to far infrared precursor to the optical FU Ori outburst.  For all opacities studied, multiple fluid elements reached peak temperatures near 800 to 1000 K between about 2 and 3 AU. The standard opacity simulation (midplane $\tau\sim 10^4$) and the 1/10,000 $\kappa$ simulation ($\tau\sim1$) show the highest maximum temperatures.

The spiral shocks for all simulations maintained relatively small pitch angles ($i\approx 10$) during the burst.  The exception is the 1/10,000 $\kappa$ simulation, which is also the only  simulation to contain chondrule-forming shocks.  This particular run is the closest to fragmentation, and has the most flocculent spiral structure.  From these results, I conclude that in order for bursts near 4 or 5 AU to bear chondrule-forming shocks, the disk must be on the verge of fragmentation (see also Boss \& Durisen 2005a,b).  Recall that the opacity in 1/10,000 $\kappa$ disk was suddenly lowered by a factor of 10$^{4}$ to model heuristically an extreme case of rapid growth and settling of solids.  As noted above, pushing a disk toward fragmentation may only be possible in extreme, and possibly unphysical, situations.  

As a result, I return to Figure 7.1.  Based on the simple, analytic argument, chondrule formation can take place at cometary distances and in the asteroid belt for GI-bursting dead zones between about 1 and 3 AU.  As suggested by Wood (2005), chondritic parent bodies may be representative of temporal as well as spatial formation differences.  In particular, he suggests that ordinary chondrites formed near the asteroid belt, while carbonaceous chondrules formed beyond the frost line based on petrological arguments.  Bursts in the inner disk seem to be able to accommodate these observations, which is congruent with driving the FU Ori phenomenon (Armitage et al.~2001).  As the disk evolves, the location of the dead zone can vary, with multiple bursts occurring at a few AU.  Accretion rates between $10^{-6}$ and $10^{-7}~M_{\odot}~\rm yr^{-1}$ could build up a 0.01 $M_{\odot}$ dead zone every $10^4$ to $10^5$ years, respectively.  Some of these bursts may drive the FU Ori phenomenon and produce chondrules, but others may only be responsible for chondrule-formation events.  In this scenario, the disk need not fragment, and there should be between 10 and 100 separate chondrule-formation events. 

\section{Numerical Improvements}

I have described the most up-to-date implementations of CHYMERA and the Standard Version (SV) of the Indiana University Radiation Hydrodynamics Code, including various code improvements that were necessary to accomplish the scientific objectives of this dissertation work.  CHYMERA is 
different from the SV in one significant way: the rotational states of molecular hydrogen
are taken into account for equilibrium statistics or for a given ortho/para mixture.  This modification
requires that $\epsilon$, not $\epsilon^{1/\gamma}$, be fluxed in the energy equation, which
in turn forces the $-p\nabla\cdot{\bf v}$ to be explicitly calculated during sourcing.  The ideal gas equation of state is cast in the form $p=k\rho T(\epsilon,\rho)/\mu m_p$, where $T$
is calculated from a table based on the cell's specific internal energy $e=\epsilon/\rho$.  Dissociation
and ionization of gas phase species are ignored for the current implementation, but could be added at a later date by changing the look-up table to account for a non-constant $\mu$.  Until this is done, the code can only model gas phase hydrogen behavior up to temperatures of approximately 1400 K.

Explicit modeling of the rotation behavior of molecular hydrogen is necessary due to the strong dependence of gas dynamics on the first adiabatic index $\Gamma_1$.  For example, $\Gamma_1$ changes the behavior of shock bores in disks (Chapter 5) and sets the critical cooling time $t_{\rm cool}\Omega=f(\Gamma_1)\approx-23\Gamma_1+44$ for fragmentation.
  Moreover, seemingly innocent approximations for the first adiabatic index, e.g.,  $\Gamma_1$ = constant or quadratic, may cause erroneous dynamics (\S\S  2.3.2 and 7.3).  

In Chapter 2, I reviewed the M2004 and C2006 radiation algorithms, and I introduced a new routine (the BDNL algorithm) that uses vertical rays to couple the optically thin and thick regions of the disk.  The accuracy of these algorithms has been evaluated (Chapter 3) and shown to follow the expected behavior in a steady state test, a contraction test, and a convection test.  The limits of each routine have also been demonstrated. Each algorithm needs to resolve the disk vertically with about five cells when the midplane optical depths are large.  Although this number is not absolute, it provides a rule of thumb for SV and  CHYMERA users.  

The BDNL routine is an improvement over the M2004 and C2006 schemes because it allows for complete cell-to-cell coupling in the vertical direction.  However, significant improvements can be made by including rays to account for radiation transport in all directions.  Such an approach would also help to correct the odd temperature contours that can arise if diffusion is used in low opacity regimes (\S 3.4).

\section{Future Code Development}

I present here several suggestions for algorithms that may improve CHYMERA's efficiency and extend its repertoire of physics.   The suggestions are only meant to sketch the possible modifications. These items will be listed on {\it CHYMERA's Den} (http://hydro.astro.indiana.edu/chymera), a wiki that provides code updates, documentation, and bug information for CHYMERA.  Some of the items listed below have already been added to the wiki, and some of the suggestions are being pursued by myself or other principal CHYMERA users.  

\subsection{Physics Modifications}

The star should be freed and allowed to respond to disk so that $m=1$ spirals and subsequent dynamics can be treated well.  It is my opinion that this algorithm should be implemented before any other improvements are made.  

Using vertical rays to solve for part of the divergence of the flux demonstrated that complete cell-to-cell radiative coupling has a noticeable effect on the evolution of the disk (Chapter 6).  Indirectly, this suggests that such coupling in the other directions can have just as important of an effect.  The next step toward improving the BDNL algorithm could be to include rays along the $x$ and $y$ directions by interpolating between cylindrical and Cartesian grids.  The $z$ direction can be treated as done in the BDNL algorithm.

The current opacity calculation in CHYMERA for the BDNL radiation algorithm uses the midplane Rosseland optical depth to set the interpolation between the Planck and Rosseland means.  A way to smoothly interpolate between the Planck and Rosseland mean opacities {\it along the ray} is desirable.  This will ensure that the Planck mean is used in regions of low optical depth regardless of the Rosseland midplane optical depth.  

As discussed in \S 2.4.3, limiters are employed to help amalgamate hydrodynamics, shock physics, and radiation physics in the SV and in CHYMERA.  The use of limiters might be significantly reduced by subcycling the radiation algorithm once the radiation timescale becomes smaller than, say, 1/100 the Courant time.  The radiation timescale can be evaluated at the beginning of each step by looking for the minimum over the grid of $\epsilon_i/\mid \nabla\cdot {\bf F}\mid_i$, which is the ratio of the internal energy density to the divergence of the flux for the $i$th cell.  The time step can then be set to some fraction of that minimum. 

Stellar irradiation can be employed using the technique described in D'Alessio et al.~(1999), where stellar irradiation's contribution to the mean intensity is included by modeling the optical surface.  This method requires knowing the structure of the optical photosphere of the disk, which is typically at altitudes three times the pressure scale height and which is not modeled in these simulations. However, several assumptions can be made to allow one to proceed.  For example, a direct correlation between the optical surface and the surface at one disk scale height can be assumed, or no correlation between the surfaces can be assumed.  Numerical experiments would determine the effects of these assumptions, and studies of shock bores in disks with extended, hot atmospheres could indicate relations between the disk midplane and the optical surface.

CHYMERA can be modified to evolve a two-fluid grain distribution.  As long as the dust grain sizes are in the Epstein drag limit (Weidenschilling 1977), a distribution of grains could be evolved as a pressureless fluid (e.g., Johansen \& Klahr 2005).  In combination with a well-mixed grain component that always remains entrained with the gas, the opacity for a given cell can be adjusted based on the evolving dust content.  This crude approach would allow for a first step investigation using CHYMERA on the effects of dust settling and dust concentration/depletion on the strength of GIs. 

Accretion onto the disk from a surrounding envelope can be simulated, for example, by assuming an Ulrich (1976) flow and by modifying the upper and outer boundary cells to permit inflow as well as outflow during fluxing (see Chapter 2).  However, preliminary work indicates that shocks due to this accretion flow may create very hot regions and cause highly disparate radiation and hydrodynamics timescales.  When radiation algorithms that include some form of cell-to-cell coupling are used (M2004, C2006, and the BDNL routines), a numerical catastrophe like the one discussed in \S 2.4.3 could occur.  The radiation timescale problem should be addressed before envelope accretion is implemented.  

In order to model other accretion mechanisms like the MRI, at least heuristically, an $\alpha$-viscosity should be included. Eventually, the equations of magnetohydrodynamics should be added as well.

\subsection{Efficiency}

In order to run CHYMERA on Columbia, a NASA Advanced Supercomputing (NAS) Division computer cluster, a new parallelization for CHYMERA using OpenMP libraries was necessary.  Under the old parallelization, which is the only version available at this time through the CVS repository (see CHYMERA's Den), the maximum speed-up is about 3-5$\times$ the serial version, depending on system architecture, with no gain beyond about eight processors.  Because this version of parallelization strongly inhibits running CHYMERA in parallel on Columbia, a team of specialists from Parallel Software Products, Computer Sciences Corporation, and the NAS Division of NASA Ames used a high-end optimization tools on CHYMERA to identify inefficiencies in the parallelization.  Using this tool, the team of specialists produced a parallel version of CHYMERA that can scale up to 128 processors with a speed-up of about 50$\times$ that of the serial version.  The new parallelization is currently being added to the repository.  

This underscores the need for periodic efficiency checks of the code.  A set of test problems and timing tools should be made available to principal CHYMERA users, and the results of the efficiency tests should be posted on CHYMERA's Den.   It should be noted that a message passing version of CHYMERA (distributed memory parallelization) may be released in the near future (see PSP Final Consultancy Report on CHYMERA's Den), but there is no guarantee.   If a major reorganization of CHYMERA occurs within the next several years, the new algorithms should be designed with message passing in mind to make a parallelized version of CHYMERA portable to distributed as well as shared memory machines. 

There are several ways that the CPU cost of running CHYMERA can be significantly reduced; I suggest two.  Roughly half of the cells at any time in the code contain background densities.  At the beginning of every or every few time steps, the location of the active grid boundary can be evaluated.  If some reliable criterion can be determined, then a floating grid boundary could be introduced.  Such an algorithm could potentially cut the CPU cost of a simulation in half.  

The second suggestion is to implement a multiple time step algorithm.  The Courant time is almost always determined near the inner disk boundary for these simulations, which can be several factors smaller than an appropriate Courant time for the middle of the disk.  The result is that the mid and outer disk is over-resolved temporarily.  One can take advantage of these disparate evolution timescales by introducing a multi-stepping algorithm that advances the inner region of the disk more often than the outer regions, similar to what is done for high and low resolution blocks in nested grid simulations.

\section{Future Work}

The results of these studies elicit additional questions, and provide a base for future work.  

\begin{itemize}
\item What is the effect of complete, 3D cell-to-cell radiative coupling on GIs? 
\item How does stellar irradiation affect the effective $\alpha$ when the incident irradiation depends in part on the geometry of the disk?  
\item What happens to spiral wave dynamics in unstable disks that have both self-shadowed regions and regions with normal incident irradiation? 
\item How does the effective $\alpha$ in disks vary with star and disk mass? 
\item Are the Fourier amplitude spectra and the effective $\alpha$s in these disks converged with the resolutions employed? 
 \item Could dust settling and grain growth or a sudden increase in the EP ionization rate due to a nearby source, e.g., a supernova, change the ortho/para statistics? What are the dynamic consequences for this change? 
 \end{itemize}

In addition to these questions, the hypothesis that the FU Ori phenomenon and chondrule formation are driven by GI activation in dead zones should be explored in greater detail.  In particular, the slow build up of mass near 1 AU should be followed through a burst, and accretion rates as well as shock strengths should be determined.  More generally, shock strengths and frequencies should be evaluated for disks in asymptotic phases as well as for less violent bursts than explored here. Finally, GIs in disks could have cosmochemical consequences, some of which may be observable.  Gathering fluid element histories for several stages of disk evolution would indicate long- and short-term thermodynamic variations that disk material may experience.

\section{Concluding Remarks}

Why study protoplanetary disks? One reason is curiosity, a desire to understand not just how planetary systems are built, but how the Solar System formed and how our civilization developed.     A picture of the formation of the Solar System will not come from describing any one process.  Our story can only be pieced together by understanding how multiple astrophysical phenomena are related. Such an investigation requires the combination of meteoritics, observations, theory, and numerical simulations.  Throughout history, astronomy has helped shape the way civilizations perceive their surroundings, and with the other sciences, has helped us understand our place in the universe.  It is my sincere hope that this dissertation aids in at least some small way to sharpen our cosmic perspective.

%
 
\renewcommand{\bibname}{BIBLIOGRAPHY}
\begin{spacing}{1.5}
\newcommand{\chondrites}{2005, in ASP Conf.~Ser.~341, Chondrites and the protoplanetary disk, ed 
A.~N.~Krot, E.~R.~D.~Scott, \& B.~Reipurth (San Francisco: ASP)}

\addcontentsline{toc}{chapter}{\bibname}

\end{spacing}
\newpage
\pagestyle{empty}
\renewcommand{\baselinestretch}{1}
\vspace*{-.5in}
\noindent \large Aaron Christopher Boley\hrule
\vspace*{.125in}

\normalsize
\begin{center} {\sc contact information}\end{center}

\small
\noindent \begin{minipage}[t]{2.5in}
	Swain Hall West 319\\           
	Department of Astronomy\\    
	Indiana University\\  
	Bloomington, IN  \\    
	47405-7105, USA  
\end{minipage}\begin{minipage}[t]{3.5in}
	{\it Phone:} 812.855.6911 \\  
	{\it Fax:}   812.855.8725 \\         
	{\it E-mail:}  acboley@astro.indiana.edu\\      
	{\it Web site:} http://orion.astro.indiana.edu/$\sim$aaron 
\end{minipage} \hfill

\begin{center}{\sc Education}\end{center}

\noindent \parbox[t]{7cm}{
	{\bf Indiana University}, \\Bloomington, IN, USA\\
	\vspace*{-.25in}
	\begin{itemize}{}
		\item[] Ph.D.~Astrophysics, September 2007
	\begin{itemize}{}
	\vspace*{-.125in}
		\item Advisor:  Richard H.\ Durisen
	\end{itemize}
	\vspace*{-.125in}
		\item[] M.A. Astronomy,  October 2004
	\end{itemize}
}
\parbox[t]{8cm}{
	{\bf Mount Union College},\\ Alliance, OH, USA\\
	\vspace*{-.25in}
	\begin{itemize}{}
	\item[] B.S. Physics \& Astronomy, German,  May 2002
	\begin{itemize}{}
		\vspace*{-.125in}
	\item Magna Cum Laude
			\vspace*{-.05in}
	\item Minor Mathematics
	\end{itemize}
	\end{itemize}
}

\begin{center}{\sc Research Interests}\end{center}

\noindent Computational/theoretical astrophysics: chondrule formation and dust processing, planet formation,  protoplanetary disk evolution, and radiative hydrodynamics.

\begin{center}{\sc Research Experience}\end{center}

\noindent{\bf Indiana University:} (Supervisor - Richard H.\ Durisen)  \\
\vspace*{-.2in}
\begin{itemize}
\item[] {Graduate student, Ph.D.~research} \hfill {\bf 2002-2007}
\end{itemize}

\noindent{\bf NASA GRC, Cleveland, OH:} (Supervisor -  Jeffrey Wilson)\\ 
\vspace*{-.2in}
\begin{itemize}
\item[]  {Computational Modeling of Left-Handed Metamaterials}   \hfill {\bf Summer 2002}
\end{itemize}

\noindent{\bf NRAO, Socorro, NM:} (Supervisor -  Mark Claussen)\\ 
\vspace*{-.2in}
\begin{itemize}
\item[] {Methanol Masers in Star Forming Regions}   \hfill {\bf Summer 2001}
\end{itemize}

\begin{center}{\sc Honors and Awards} \end{center}
\begin{itemize}
\item NASA Graduate Student Research Program Fellow, 2005-2007
\item Indiana University Astronomy Department's Research Recognition, 2006
\item Indiana Space Grant Fellow, 2005
\item Indiana University Astronomy Department's Teaching Assistant Recognition, 2004
\end{itemize}

\newpage

\begin{center}{\sc Teaching Experience}\end{center}
\noindent\begin{itemize}
\item NSF Sponsored Research Experience for Undergraduates (REU) Mentor, June-July 2005 \& 2006
\item Associate Instructor: A100 {\it Introduction to the Solar System } and A105 {\it Stars and Galaxies}, 6 Sessions between 2003 and 2005
\item Teaching Assistant: A100 and General Astronomy II, Sept.-Dec.~2002, Jan.-May~2003
\end{itemize}

\begin{center}{\sc Public Outreach Activities}\end{center}
\noindent\begin{itemize}
\item {\em Kirkwood Open House}, host for public observing sessions,  2002-present
\item {\em Physics and Astronomy Open House}, hosted astronomy activities for the public,  Fall 2002-2006
\item{\em Brownie Math and Science Day}, discuss various astronomy topics with children, Fall 2003 and 2004
\item {\em Science Olympiad}, test writer/grader, Spring 2002-2005
\end{itemize}

\begin{center}
{\sc Refereed Publications}
\end{center}
	\newcounter{Lcount}
	\newcounter{num}
	\setcounter{num}{8}
	\begin{list}{\arabic{num}}{}
		\usecounter{Lcount}

\item  ``Gravitational Instabilities, Chondrule Formation, and the FU Orionis Phenomenon,''\\
{\bf A.~C.~Boley} \& R.~H.~Durisen~2007, {\it ApJ Letters}, submitted.
\addtocounter{num}{-1}

\item  ``The Thermal Regulation of Gravitational Instabilities in Protoplanetary Disks. IV. 
Simulations with Envelope Irradiation,''\\
 K.~Cai, R.~H.~Durisen, {\bf A.~C.~Boley}, M.~K.~Pickett, \& A.~C.~Mej\'ia 2007,
{\it ApJ}, submitted.
\addtocounter{num}{-1}

\item   ``3D Radiative Hydrodynamics for Disk Stability Simulations: A Proposed Testing Standard,''\\
{\bf A.~C.~Boley}, R.~H.~Durisen, \AA.~Nordlund, \& J.~Lord 2007, {\it ApJ}, 665,
1254. 
\addtocounter{num}{-1}

\item ``The Internal Energy for Molecular Hydrogen in Gravitationally Unstable Protoplanetary Disks,''\\
{\bf A.~C.~Boley}, T.~W.~Hartquist, R.~H.~Durisen, \& S.~Michael 2006, {\it ApJ}, 656, L89. 
\addtocounter{num}{-1}

\item ``The Thermal Regulation of Gravitational Instabilities in Protoplanetary Disks III. Simulations with Radiative Cooling \& Realistic Opacities,''\\
{\bf A.~C.~Boley}, A.~C.~Mej\'ia, R.~H.~Durisen, M.~K.~Pickett, \& P.~D'Alessio 2006), {\it ApJ}, 651, 517.  
\addtocounter{num}{-1}

\item ``Hydraulic/Shock-Jumps in Protoplanetary Disks,''\\
{\bf A.~C.~Boley} \& R.~H.~Durisen 2006, {\it ApJ}, 641, 534. 
\addtocounter{num}{-1}

\item ``The Effects of Metallicity \& Grain Size on Gravitational Instabilities in Protoplanetary Disks,''\\
K.~Cai, R.~H.~Durisen, S.~Michael, {\bf A.~C.~Boley}, A.~C.~Mej\'ia, M.~K.~Pickett, \& P.~D'Alessio
 2006, {\it ApJL}, 636, 149.  
\addtocounter{num}{-1}

\item  ``The Three-Dimensionality of Spiral Shocks in Disks: Did Chondrules Catch a Breaking Wave?''\\
 {\bf A.\ C.\ Boley}, R.\ H.\ Durisen, \& M.\ K.\ Pickett 2005, In {\it Chondrites and the Protoplanetary 
 Disk}, ASPC Series, 341, 839.
\addtocounter{num}{-1}

	\end{list}

\begin{center}
{\sc Selected Conference Presentations}
\end{center}
	\setcounter{num}{7}
	\begin{list}{\arabic{num}}{}
	
		\item ``A Test Suite for 3D Radiative Hydrodynamics Simulations of Protoplanetary Disks,''\\
{\bf A.~C.~Boley}, R.~H.~Durisen, \AA.~Nordlund, \& J.~Lord 2006, {\it BAAS}, 38, 995. \addtocounter{num}{-1}

		\item ``3D Radiative Hydrodynamics Simulations of Protoplanetary Disks: A Comparison Between Two Radiative Cooling Algorithms,''\\
		J.~Lord, {\bf A.~C.~Boley}, \& R.~H.~Durisen 2006, {\it BAAS}, 38, 996 \addtocounter{num}{-1}

		\item ``The Effects of Varied Initial Conditions on Protoplanetary Disks,''\\
S.~Michael, {\bf A.~C.~Boley}, \& R.~H.~Durisen 2006, {\it BAAS}, 38, 1052. \addtocounter{num}{-1}

		\item ``Hydraulic/Shock-Jumps in Protoplanetary Disks,''\\
{\bf A.~C.~Boley} \& R.~H.~Durisen 2005, {\it BAAS}, 37, 1164. \addtocounter{num}{-1}

		\item ``Star Formation in the Outer Disk of NGC 5964,''\\
{C.~P. McKinney}, {\bf A.~C.~{Boley}}, L.~{van Zee}, D.~{Schade}, \& S.~{C{\^o}t{\'e}} 2005, {\it BAAS}, 37, 1392.  \addtocounter{num}{-1}

		\item ``Linking Chondrules to the Formation of Jupiter Through Nebular Shocks,''\\
{\bf A.~C.~Boley} \& R.~H.~Durisen 2005, LPI 1286, 8177. \addtocounter{num}{-1}

		\item ``The Three-Dimensionality of Spiral Shocks in Disks: Did Chondrules Catch a Breaking Wave?''\\
{\bf A.~C.~Boley}, R. H. Durisen, M. K, Pickett 2004, {\it Workshop on Chondrites \& the Protoplanetary Disk}, 9016.

	\end{list}

%

\begin{thebibliography}{}

\bibitem[Adams et al.(1987)]{adams_etal_1987}
Adams, F.~C., Lada, C.~J., \& Shu, F., H.~1987, 312, 788

\bibitem[Akima(1970)]{akima1970}
Akima, H.~1970, JACM, 17, 4

\bibitem[Amelin et al.(2002)]{amelin_etal_2002}
Amelin, Y., Krot, A.~N., Hutcheon, I.~D., \&
Ulyanov, A.~A.~2002, Science, 297, 1678

\bibitem[Andre et al.(1993)]{andre_etal_1993}
Andr\'e, P., Ward-Thompson, D., \& Barsony, M.~1993, ApJ, 406, 122

\bibitem[Andrews \& Williams(2005)]{andrews_williams_2006}
Andrews, S.~M., \& Williams, J.~P.~2005, 619, 175

\bibitem[Armitage et al.(2001)]{armitage_etal_2001}
Armitage, P.~J., Livio, M., \& Pringle, J.~E.~2001, MNRAS, 324, 705

\bibitem[Balbus \& Hawley(1991)]{balbus_hawley1991}
{Balbus}, S.~A. \& {Hawley}, J.~F.~1991, ApJ, 376, 214

\bibitem[Balbus \& Hawley(1998)]{balbus_hawley1998}
--.~1998, RvMP, 70, 1

\bibitem[Balbus \& Papaloizou(1999)]{balbus_papaloizou_1999}
Balbus, S.~A., \& Papaloizou, J.~C.~B.~1999, ApJ, 512, 650

\bibitem[Bate et al.(2003)]{bate_etal_2003}
Bate, M.~R., Lubow, S.~H., Ogilvie, G.~I., \& Miller, K.~A.~2003, MNRAS, 341, 213

\bibitem[Bate et al.(2002)]{bate_etal_2003}
Bate, M.~R., Ogilvie, G.~I., Lubow, S.~H., \& Pringle, J.~E.~2002, MNRAS, 341, 213

\bibitem[Beckwith et al.(1990)]{beckwith_etal_1990}
Beckwith, S.~V.~W., Sargent, A.~I., Chini, R.~S., \& Guesten, R.~1990, AJ, 99, 924

\bibitem[Bell \& Lin(1994)]{bell_lin_1994}
Bell, K.~R., \& Lin, D.~N.~C.~1994, ApJ, 427, 987

\bibitem[Binney \& Tremaine(1987)]{binney_tremaine}
Binney, J., \& Tremaine, S.~1987, Galactic dynamics (Princeton: Princeton Univ.~Press, 1987)

\bibitem[Bizzarro et al.(2004)]{bizzarro_etal_2004}
Bizzarro, M., Baker, J.~A., Haack, \& H.~2004, Nature, 431, 275

\bibitem[Black \& Bodenheimer(1975)]{black_bodenheimer_1975}
Black, D.~C., \& Bodenheimer, P.~1975, ApJ, 199, 619

\bibitem[Bodenheimer et al.(1990)]{bodenheimer_etal_1990}
{{Bodenheimer}, P., {Yorke}, H.~W., {Rozyczka}, M., \& {Tohline}, J.~E.}~1990,
ApJ, 355, 651

\bibitem[Boley \& Durisen(2005)]{boley_durisen_2005}
Boley, A.~C., \& Durisen, R.~H.~2005, LPI, 1286, 8177

\bibitem[Boley \& Durisen(2006)]{boley_durisen_2006}
--.~2006, ApJ, 641, 534

\bibitem[Boley et al.(2007c)]{boleyetal2007c}
Boley, A.~C., Durisen, R.~H., Nordlund, \AA, \& Lord, J.~2007c, ApJ, 665, 1254

\bibitem[Boley et al.(2005)]{boleyetal2005}
Boley, A.~C., Durisen, R.~H., \& Pickett, M.~K.~\chondrites, 839

\bibitem[Boley et al.(2007a)]{boleyetal2007a}
Boley, A.~C., Hartquist, T.~W., Durisen, R.~H., \& Michael, S.~2007a, ApJ, 656, L89

\bibitem[Boley et al.(2007b)]{boleyetal2007b}
--.~2007b, ApJ, 656, L89

\bibitem[Boley et al.(2006)]{boley_etal_2006}
{{Boley}, A.~C., {Mej{\'{\i}}a}, A.~C., {Durisen}, R.~H., {Cai}, K., {Pickett}, M.~K., \& {D'Alessio}, P.}~2006, 651, 517

\bibitem[Boss(1984a)]{boss_1984a}
Boss, A.~P.~1984a, ApJ, 277, 768

\bibitem[Boss(1984b)]{boss_1984b}
--.~1984b, MNRAS, 209, 543

\bibitem[Boss(1989)]{boss_1989}
--.~1989, ApJ, 346, 336

\bibitem[Boss(1997)]{boss_1997}
--.~1997, Science, 276, 1836

\bibitem[Boss(1998)]{boss_1998}
--.~1998, ApJ, 503, 923

\bibitem[Boss(2001)]{boss_2001}
--.~2001, ApJ, 562, 367

\bibitem[Boss(2002)]{boss_2002}
--.~2002, ApJ, 567, L149

\bibitem[Boss(2004a)]{boss_2004a}
--.~2004a, ApJ, 610, 456

\bibitem[Boss(2004b)]{boss_2004b}
--.~2004b, ApJ, 616, 1265

\bibitem[Boss(2005)]{boss_2005}
--.~2005, ApJ, 629, 535

\bibitem[Boss(2007)]{boss_2007}
--.~2007, ApJ, 661, L73

\bibitem[Boss \& Durisen(2005a)]{boss_durisen_2005a}
Boss, A.~P., \& Durisen, R.~H.~2005a, ApJ, 621, L137

\bibitem[Boss \& Durisen(2005b)]{boss_durisen_2005b}
--.~2005b, in ASP Conf.~Ser.~341, Chondrites and the protoplanetary disk, ed A.~N.~Krot, E.~R.~D.~Scott, \& B.~Reipurth (San Francisco: ASP), 821

\bibitem[Bryden et al.(1999)]{brydenetal1999}
Bryden, G., Chen, X., Lin, D.~N.~C., Nelson, R.~P., \& Papaloizou, J.~C.~B.~1999, ApJ, 514, 344

\bibitem[Cai(2006)]{cai_phd}
Cai, K.~2006, Ph.D.~Thesis, Indiana University


\bibitem[Cai et al.(2007)]{cai_etal_2007}
Cai, K., Durisen, R.~H., Boley, A.~C., Pickett, M.~P., Mej\'ia, A.~C.~2007, arXiv:0706.4046

\bibitem[Cai et al.(2006)]{cai_etal_2006}
{{Cai}, K., {Durisen}, R.~H.,  {Michael}, S.,  {Boley}, A.~C., 
{Mej{\'{\i}}a}, A.~C., {Pickett}, M.~K., \& {D'Alessio}, P.}~2006, ApJ, 636, L149


\bibitem[Calvet et al.(2005)]{calvetetal2005}
Calvet, N., Brice\'no, C., Hern\'andez, J., Hoyer, S., Hartmann, L., Sicilia-Aguilar, A., Megeath, S. T., \& D'Alessio, P.~2005, AJ, 129, 935

\bibitem[Cameron(1978)]{cameron_1978}
Cameron, A.~G.~.W.~1978, M\&P, 18, 5

\bibitem[Cassen(1981)]{cassen_moosman_1981}
Cassen, P., \& Moosman, A.~1981, Icarus, 48, 353

\bibitem[Chandrasekhar(1960)]{chandrasekhar_radtran}
Chandrasekhar, S.~1960, Radiative transfer (New York: Dover, 1960)

\bibitem[Ciesla \& Hood(2002)]{ciesla_hood_2002}
Cielsa, F.~J., \& Hood, L.~L.~2002, Icarus, 158, 281

\bibitem[Cohl \& Tohline(1999)]{cohl_tohline_1999}
Cohl, H.~S., \& Tohline, J.~E.~1999, 527, 86

\bibitem[Cox \& Giuli(1968)]{cox_guili_1968}
{{Cox}, J.~P., \& {Giuli}, R.~T.}~1968, Principles of Stellar Structure (New York:
Gordon and Breach)

\bibitem[Cuzzi(2004)]{cuzzi2004}
Cuzzi, J.~N.~2004, Icarus, 168, 484


\bibitem[Cuzzi \& Alexander(2006)]{cuzzialexander2006}
Cuzzi, J.~N., \& Alexander, C.~M.~O'D.~2006, Nature, 441, 483

\bibitem[Cuzzi et al.(2005)]{cuzzi_etal_2005}
Cuzzi, J.~N., Ciesla, F.~J., Petaev, M.~I., Krot, A.~N., Scott, E.~R.~D., \&
Weidenschilling, S.~J.~\chondrites, 732

\bibitem[Cuzzi et al.(2001)]{cuzzietal2001}
Cuzzi, J.~N., Hogan, R.~C., Paque, J.~M., \& Dobrovolskis, A.~R.~2001, ApJ, 546, 496

\bibitem[D'Alessio et al.(2001)]{dalessio_etal_2001}
{{D'Alessio}, P., {Calvet}, N., \& {Hartmann}, L.}~2001, ApJ, 553, 321

\bibitem[D'Alessio(2006)]{dalessio_etal_2006}
D'Alessio, P., Calvet, N., Hartmann, L., Franco-Hern\'andez, R., \& Serv\'in, H.~2006, ApJ, 638, 314

\bibitem[D'Alessio et al.(1999)]{dalessioetal1999}
D'Alessio, P., Calvet, N., Hartmann, L., Lizano, S., \& Cant\'o, J. 1999, ApJ, 527, 893

\bibitem[Dalgarno et al.(1973)]{dalgarno_etal_1973}
Dalgarno, A., Oppenheimer, M., \& Black, J.~H.~1973, Nature, 245, 100

\bibitem[Decampli et al.(1978)]{decampli_etal_1978}
{{Decampli}, W.~M., {Cameron}, A.~G.~W., {Bodenheimer}, P., \& {Black}, D.~C.}~1978,
ApJ, 223, 854

\bibitem[Desch(2004)]{desch_2004}
Desch, S.~J.~2004, ApJ, 608, 509

\bibitem[Desch \& Connolly(2002)]{deschconnolly2002}
Desch, S.~J., \& Connolly, H.~C., 
Jr.~2002, M\&PSA, 37, 183

\bibitem[Desch \& Cuzzi(2000)]{deschcuzzi2000}
Desch, S.~J., \& Cuzzi, J.~N.~2000, Icarus, 143, 87

\bibitem[Draine et al.(1983)]{draine_etal_1983}
{{Draine}, B.~T., {Roberge}, W.~G.,~\& {Dalgarno}, A.}~1983, ApJ, 264, 485 

\bibitem[Dullemond(2007)]{dullemond_etal_2007}
Dullemond, C.~P., Hollenbach, D., Kamp, I., \& D'Alessio, P.~2007, in Protostars and
Planets V, ed.~B.~Reipurth, D.~Jewitt, \& K.~Keil (Tucson: Univ. Arizona Press), 555

\bibitem[Durisen et al.(2007a)]{durisen_ppv}
Durisen, R.~H., Boss, A., Mayer, L., Nelson, A., Quinn, T., \& Rice, K.~2007a,
in Protostars and Planets V, ed.~B.~Reipurth, D.~Jewitt, \& K.~Keil (Tucson: Univ. Arizona Press), 607


\bibitem[Durisen et al.(1986)]{durisen_1986}
Durisen, R.~H., Gingold, R.~A., Tohline, J.~E., \& Boss, A.~P.~1986, ApJ, 305, 281

\bibitem[Durisen et al.(2007b)]{durisen_etal2007b}
Durisen, R.~H., Hartquist, T.~W., \& Pickett, M.~K.~2007b, arXiv:0709.1445


\bibitem[Durisen et al.(2003)]{durisenetal2003}
Durisen, R.~H., Mej\'ia, A.~C., \& Pickett, B.~K.~2003, Recent Research
Developments in Applied Physics, 1, 173

\bibitem[Durisen et al.(1998)]{durisen_1998}
Durisen, R.~H., Yang, S., Cassen, P., \& Stahler, S.~W.~1989, ApJ, 345, 959 

\bibitem[Eisner(2006)]{eisner_carpenter_2006}
Eisner, J.~A., \& Carpenter, J.~M.~2006, ApJ, 641, 1162

\bibitem[Fleming \& Stone(2003)]{flemming_stone2003}
Fleming, T.~P., \& Stone, J.~M.~2003, ApJ, 585, 908

\bibitem[Flower \& Watt(1984)]{flower_watt_1984}
Flower, D.~R., \& Watt, G.~D.~1984, MNRAS, 209, 25

\bibitem[Flower et al.(2006)]{flower_etal_2006}
Flower, D.~R., Pineau Des For\^ets, G., \& Walmsley, C.~M.~2006, A\&A, 449, 621

\bibitem[Fouchet et al.(2003)]{fouchet_etal_2003}
Fouchet, T., Lellouch, E., \& Feuchtgruber, H.~2004, Icarus, 161, 127

\bibitem[Fromang et al.(2004)]{frommang_etal_2004}
Fromang, S., Balbus, S.~A., Terquem, C., \& De Villiers, J.~2004, ApJ, 616, 364

\bibitem[Fuente et al.(1999)]{fuente_etal_1999}
Fuente, A., Martin-Pintado, J., Rodr\'iquez-Fern\'andez, N.~J., 
Rodr\'iguez-Franco, A., de Vicente, P., \& Kunze, D.~1999, ApJ, 518, 45

\bibitem[Furlan et al.(2006)]{furlan_etal_2006}
Furlan, E., Hartmann, L, Calvet, N., D'Alessio, P., Franco-Hern\'andez, R.,
Forrest, W.~J., Watson, D.~M., Uchida, K.~I., Sargent B., Green, J.~D., 
Keller, L.~D., \& Herter, T.~L.~2006, ApJS, 165, 568 

\bibitem[Gammie(1996)]{gammie_1996}
Gammie, C.~F.~1996, ApJ, 457, 355

\bibitem[Gammie(2001)]{gammie_2001}
--.~2001, ApJ, 553, 174

\bibitem[Gomez \& Cox(2002)]{gomez_cox2002}
G\'omez, G.~C., \& Cox, D.~P.~2002, ApJ, 580, 235


\bibitem[Green et al.(2006)]{green_etal_2006}
Green, J.~D., Hartmann, L., Calvet, N., Watson, D.~M., 
Ibrahimov, M., Furlan, E., Sargent, B., \& Forrest, W.~J.~2006, ApJ, 648, 1099

\bibitem[Greene et al.(1994)]{greene_etal_1994}
Greene, T.~P., Wilking, B.~A., Andre, P., Young, E.~T., \& Lada, C.~J.~1994, ApJ, 434, 614


\bibitem[Hachisu(1986)]{hachisu1986}
Hachisu, I.~1986, ApJS, 61, 479


\bibitem[Harker \& Desch(2002)]{harkerdesch2002}
Harker, D.~E., \& Desch, S.~J.~2002, ApJ, 565, 109

\bibitem[Hartmann(1998)]{hartmann_1998}
Hartmann, L.~1998, Accretion Processes in Star Formation 
(Cambridge: Cambridge Univ.~Press)

\bibitem[Hartmann(2005b)]{hartmann_2005b}
--.~\chondrites, 1003

\bibitem[Hartmann(2006)]{hartmann_etal_2006}
Hartmann, L., D'Alessio, P., Calvet, N., \& Muzerolle, J.~2006, ApJ, 648, 484

\bibitem[Hartmann \& Kenyon(1996)]{hartmann_kenyon_1996}
Hartmann, L., \& Kenyon, S.~J.~1996, ARA\&A, 34, 207

\bibitem[Hawley et al.(1984)]{hawley_etal1984}
Hawley, J.~F., Smarr, L.~L., \& Wilson, J.~R.~1984, ApJ, 277, 296

\bibitem[Hayashi(1985)]{hayashi_etal_1985}
Hayashi, C., Nakazawa, K., Nakagawa, Y.~1985, in Protostars and Planets II, 
ed.~D.~C.~Black, \& M.~S.~Matthews (Tucson: Univ. Arizona Press), 1100


\bibitem[Heinemann(2006)]{heinemann_etal_2006}
Heinemann, T., Dobler, W., Nordlund, \AA, \& Brandenburg, A.~2006, A\&A, 448, 731

\bibitem[Herbig(1960)]{herbig_1960}
Herbig, G.~H.~1960, ApJS, 4, 337

\bibitem[Herbig(1977)]{herbig_1977}
--.~Herbig, G.~H.~1977, ApJ, 217, 693

\bibitem[Hewins et al.(2005)]{hewins_etal_2005}
Hewins, R.~H., Connolly, H.~C., Lofgren, G.~E.~, Jr., \& Libourel, G.~2005, in ASP Conf.~Ser.~341, Chondrites and the protoplanetary disk, ed A.~N.~Krot, E.~R.~D.~Scott, \& B.~Reipurth (San Francisco: ASP), 286

\bibitem[Hewins et al.(1996)]{hewins_etal_1996}
Hewins, R., Jones, R., \& Scott, E.~1996, Chondrules and the Protoplanetary
Disk (Cambridge, UK: Cambridge University)



\bibitem[Horne & Baliunas(1986)]{horne_baliunas_1986}
Horne, J.~H., \& Baliunas, S.~L.~1986, ApJ, 302, 757

\bibitem[Hubeny(1990)]{hubeny_1990}
Hubeny, I.~1990, ApJ, 351, 632

\bibitem[Huss et al.(2005)]{huss_etal_2005}
Huss, G.~R., Alexander, C.~M.~O'D, Palme, H., Bland, P.~A., 
\& Wasson, T.~J.~\chondrites, 701

\bibitem[Iida et al.(2001)]{iida_etal_2001}
Iida, A., Nakamoto, T., Susa, H., \& Nakagawa, Y.~2001, Icarus, 153, 430

\bibitem[Itoh et al.(2002)]{itoh_etal_2002}
Itoh, S., Rubin, A.~E., Kojima, H., Wasson, J.~T., Yurimoto, H.~2002,
LPI, 33.1490

\bibitem[Jayawardhana et al.(2001)]{jayawardhana_etal_2001}
Jayawardhana, R., Hartmann, L., \& Calvet, N.~2001, ApJ, 548, 310

\bibitem[Johansen \& Klahr(2005)]{johansen_klahr2005}
Johansen, A., \& Klahr, H.~2005, 634, 1353

\bibitem[Johnson \& Gammie(2003)]{johnson_gammie_2003}
Johnson, B.~M., \&~Gammie, C.~F.~2003, ApJ, 597, 131

\bibitem[Jones et al.(2005)]{jones_etal_2005}
Jones, R.~H., Grossman, J.~N., \& Rubin, A.~E.~\chondrites, 251

\bibitem[Kenyon \& Hartmann(1987)]{kenyon_hartmann_1987}
Kenyon, S.~J., \& Hartmann, L.~1987, ApJ, 323, 714

\bibitem[Kenyon \& Hartmann(1991)]{kenyon_hartmann_1991}
--.~1991, ApJ, 383, 664

\bibitem[Kitamura et al.(2002)]{kitamura_etal_2007}
Kitamura, Y., Momose, M., Yokogawa, S., Kawabe, R., Tamura, M., 
\& Ida, S.~2002, ApJ, 581, 357

\bibitem[Korycansky \& Pringle(1995)]{korycansky_pringle1995}
Korycansky, D.~G., \& Pringle, J.~E.~1995, MNRAS, 272, 618


\bibitem[Krumholz et al.(2006)]{krumholz_etal_2006}
Krumholz, M.~R., Klein, R.~I., McKee, C.~F., \& Bolstad, J.~2006, arXiv:astro-ph/0611003

\bibitem[Krumholz(2007)]{krumholz_etal_2007}
Krumholz, M.~R., Klein, R.~I., \& McKee, C.~F.~2007, 656, 656



\bibitem[Kuiper(1951)]{kuiper1951}
Kuiper, G.~P.~1951, PNAS, 37, 1

\bibitem[Lada(1987)]{lada_1987}
Lada, C.~J.~1987, IAUS, 115, 1

\bibitem[Lada \& Wilking(1984)]{lada_wilking_1984}
Lada, C.~J., \&~Wilking, B.~A.~1984, 287, 610

\bibitem[Landau \& Lifshitz(1987)]{landau_lifshitz_1987}
Landau, L.~D., \& Lifshitz, E.~M.~1987, Fluid Mechanics (2nd edition: Butterworth-Heinemann)

\bibitem[Larson(1984)]{larson_1984}
Larson, R.~B.~1984, MNRAS, 206, 197

\bibitem[Laughlin \& Bodenheimer(1994)]{laughlin_bodenheimer1994}
Laughlin, G., \& Bodenheimer, P.~1994, 436, 335

\bibitem[Laughlin \& Korchagin(1996)]{laughlin_korchagin_1996}
Laughlin, G., \& Korchagin, V.~1996, 480, 855

\bibitem[Laughlin et al.(1998)]{laughlin_etal1998}
Laughlin, G., Korchagin, V., \& Adams, F.~C.~1998, 504, 945

\bibitem[Laughlin \& Rozyczka(1996)]{laughlin_rozyczka1996}
Laughlin, G., \& Rozyczka, M.~1996, ApJ, 456, 279

\bibitem[Lawson et al.(1996)]{lawson_etal_1996}
Lawson, W.~A., Feigelson, E.~D., \& Huenemoerder, D.~P.~1996, MNRAS, 280, 1071

\bibitem[Le Bourlot(2000)]{lebourlot_2000}
Le Bourlot, J.~2000, A\&A, 360, 656

\bibitem[Lin \& Papaloizou(1980)]{lin_papaloizou_1980}
Lin, D.~N.~C., \& Papaloizou, J.~1980, MNRAS, 191, 37

\bibitem[Lin et al.(1990)]{linetal1990}
Lin, D.~N.~C., Papaloizou, J.~C.~B., \& Savonije, G.~J.~1990, ApJ, 364, 326 


\bibitem[Lissauer(1987)]{lissauer1987}
Lissauer, J.~J.~1987, Icarus, 69, 249

\bibitem[Lodato \& Rice(2004)]{lodato_rice_2004}
Lodato, G., \& Rice, W.~K.~M.~2004, MNRAS, 351, 630

\bibitem[Lubow(1981)]{lubow1981}
Lubow, S.~H.~1981, ApJ, 245, 274

\bibitem[Lubow \& Ogilvie(1998)]{lubow_ogilvie1998}
Lubow, S.~H., \& Ogilvie, G.~I.~1998, ApJ, 504, 983

\bibitem[Lubow \& Pringle(1993)]{lubow_pringle1993}
Lubow, S.~H., \& Pringle, J.~E.~1993, ApJ, 409, 360 

\bibitem[Lynden-Bell \& Kalnajs(1972)]{Lynden-bell_kalnajs_1972}
Lynden-Bell, D., \& Kalnajs, A.~J.~1972, MNRAS, 157, 1

\bibitem[MacPherson et al.(2005)]{macpherson_etal_2005}
MacPherson, G.~J., Simon, S.~B., Davis, A.~M., 
Grossman, L., \& Krot, A.~N.~\chondrites, 225

\bibitem[Martos et al.(1998)]{martosetal1998}
Martos, M.~A., \& Cox, D.~P.~1998, ApJ, 509, 703

\bibitem[Massey(1970)]{massey1970}
Massey, B.~S.~1970, Mechanics of Fluids (2nd ed.; London: Van Norstrand Reinhold)

\bibitem[Mayer et al.(2007)]{mayer_etal_2007}
Mayer, L., Lufkin, G., Quinn, T., \& Wadsley, J.~2007, ApJ, 661, 77

\bibitem[Mayer et al.(2004)]{mayer_etal_2004}
Mayer, L., Quinn, T., Wadsley, J., \& Stadel, J.~2004, ApJ, 609, 1045

\bibitem[McKeegan et al.(2006)]{mckeegan2006}
{McKeegan}, K.~D. et al.~2006, Science, 314, 1724

	
\bibitem[Mej\'ia(2004)]{mejia_phd}
Mej\'ia, A.~C.~2004, Ph.D.~Thesis, Indiana University

\bibitem[Mej\'ia et al.(2005)]{mejia_etal_2005}
Mej\'ia, A.~C., Durisen, R.~H., Pickett, M.~K., \& Cai, K.~2005, ApJ, 619, 1098

\bibitem[Mihalas \& Weibel-Mihalas(1984)]{mihalas_weibelmihalas_1986}
Mihalas, D., \& Weibel-Mihalas, B.~1984, Foundations of radiation hydrodynamics (New
York: Oxford University Press, 1984)

\bibitem[Miura \& Nakamoto(2006)]{miuranakamoto2006}
Miura, H., \& Nakamoto, T.~2006, ApJ, 651, 1272

\bibitem[Miyake \& Nakagawa(1995)]{miyake_nakagawa}
Miyake, K., \& Nakagawa, Y.~1995, ApJ, 441, 361

\bibitem[Monaghan(1992)]{monaghan1992}
Monaghan, J.~J.~1992, ARA\&A, 30, 543

\bibitem[Morfill et al.(1998)]{morfill_etal_1998}
Morfill, G.~E., Durisen, R.~H., \& Turner, G.~W.~1998, Icarus, 134, 180

\bibitem[Muzerolle et al.(2003)]{muzerolleetal2003}
Muzerolle, J., Calvet, N., Hartmann, L., \& D'Alessio, P.~2003, ApJ, 597, L149

\bibitem[Natta(2007)]{natta_etal_2007}
Natta, A., Testi, L., Calvet, N., Henning, T., Waters, R., \& Wilner, D.~2007, in
Protostars and Planets V, ed.~B.~Reipurth, D.~Jewitt, and K.~Keil (Tucson: Univ. Arizona Press), 767

\bibitem[Nelson(2000)]{nelson_2000}
Nelson, A.~F.~2000, ApJ, 537, 65

\bibitem[Nelson(2006)]{nelson_2006}
--.~2006, MNRAS, 373, 1039

\bibitem[Nelson et al.(2000)]{nelson_etal_2000}
Nelson, A.~F., Benz, W., \& Ruzmaikina, T.~V.~2000, ApJ, 529, 357

\bibitem[Norman \& Winkler(1986)]{norman_winkler_1986}
Norman, M.~L., \& Winkler, K.~H.~A.~1986, in NATO Advanced Research Workshop
on Astrophysical Radiation Hydrodynamics

\bibitem[Ogilvie(1998)]{ogilvie_1998}
Ogilvie, G.~I.~1998, MNRAS, 297, 291

\bibitem[Ogilvie(2002a)]{ogilvie2002a}
--.~2002a, MNRAS, 330, 937

\bibitem[Ogilvie(2002b)]{ogilvie2002b}
--.~2002b, MNRAS, 331, 1053

\bibitem[Oishi et al.(2007)]{oishietal2007}
Oishi, J.~S., Mac Low, M., Menou, K.~2007, arXiv:astro-ph/0702549

\bibitem[Osorio et al.(2003)]{osorio_etal_2003}
Osorio, M., D'Alessio, P., Muzerolle, J., Calvet, N., \& Hartmann, L.~2003, ApJ, 586, 1148

\bibitem[Osterbrock(1962)]{osterbrock1962}
Osterbrock, D.~E.~1962, ApJ, 136, 359

\bibitem[Padgett(1999)]{padgett_etal_1999}
Padgett, D.~L., Brandner, W., Stapelfeldt, K.~R., Strom, S.~E., Tereby, S., \& Koerner, D.~1999, AJ, 117, 1490

\bibitem[Pathria(1996)]{pathria1996}
Pathria, R.~K.~1996, Statistical Mechanics (2nd ed.; Oxford: Butterworh-Heinemann)

\bibitem[Petaev \& Wood(2005)]{petaev_wood_2005}
Petaev, M.~I., \& Wood, J.~A.~\chondrites, 373

\bibitem[Palla \& Stahler(1990)]{palla_stahler_1990}
Palla, F., \& Stahler, S.~W.~1990, ApJ, 360, 47

\bibitem[Pickett(1995)]{pickett_phd}
Pickett, B.~K.~1995, Ph.D.~Thesis, Indiana University

\bibitem[Picket et al.(1998)]{pickett_etal_1998}
Pickett, B.~K., Cassen, P.~M., , Durisen, R.~H.,  \& Link, R.~P.~1998,
ApJ, 504, 468

\bibitem[Pickett et al.(2000a)]{pickett_etal_2000a}
--.~2000a, ApJ, 529, 1034


\bibitem[Pickett \& Durisen(2007)]{pickett_durisen_2007}
Pickett, M.~K., \& Durisen, R.~H.~2007, ApJ, 654, L155

\bibitem[Pickett et al.(2000b)]{pickett_etal_2000b}
Pickett, B.~K., Durisen, R.~H., Cassen, P.~M., \& Mej\'ia A.~C.~2000b, ApJ, 540, L95

\bibitem[Pickett et al.(1996)]{pickett_etal_1996}
Pickett, B.~K., Durisen, R.~H., \& Davis, G.~A.~1996, ApJ, 458, 714

\bibitem[Pickett et al.(1997)]{pickett_etal_1997}
Pickett, B.~K., Durisen, R.~H., \& Link, R.~1997, Icarus, 126, 243

\bibitem[Pickett et al.(2003)]{pickett_etal_2003}
Pickett, B.~K., Mej\'ia, A.~C., Durisen, R.~H., Cassen, P.~M., Berry, D.~K., \& Link, R.~P.~2003, ApJ,
590, 1060

\bibitem[Pilippetal(1998)]{pilippetal1998}
Pilipp, W., Hartquist, T.~W., Morfill, G.~E., \& Levy, E.~H.~1998, A\&A,
331, 121

\bibitem[Pollack et al.(1994)]{pollack1994}
Pollack, J.~B., Hollenbach, D., Beckwith, S., Simonelli, D.~P.,
Roush, T., \& Fong, W.~1994, 421, 615

\bibitem[Press et al.(1986)]{press_etal1986}
Press, W.~H., Flannery, B.~P., \& Teukolsky, S.~A.~1986, Numerical recipes.
The art of scientific computing (Cambridge: University Press)

\bibitem[Pringle(1981)]{pringle_1981}
Pringle, J.~E.~1981, ARA\&A, 19, 137

\bibitem[Rafikov(2005)]{rafikov_2005}
Rafikov, R.~R.~2005, ApJ, 621, L69

\bibitem[Rafikov(2007)]{rafikov_2007}
--.~2007, ApJ, 662, 642

\bibitem[Reipurth(2005)]{reipurth_2005}
Reipurth, B.~2005, in ASP Conf.~Ser.~341, Chondrites and the protoplanetary disk, ed A.~N.~Krot, E.~R.~D.~Scott, \& B.~Reipurth (San Francisco: ASP), 54

\bibitem[Rice et al.(2003)]{rice_etal_2003}
Rice, W.~K.~M, Armitage, P.~J., Bate, M.~R., \& Bonnell, I.~A.~2003, MNRAS, 339, 1025

\bibitem[Rice et al.(2005)]{rice_etal_2005}
Rice, W.~K.~M, Lodato, G., \& Armitage, P.~J.~2005, MNRAS, 364, L56

\bibitem[Roberts et al.(1979)]{robertsetal1979}
Roberts, W.~W., Huntley, J.~M., \& van Albada, G.~D.~1979, ApJ, 233, 67

\bibitem[Rodriguez-Fernandez et al.(2000)]{rodriquez_fernandez_2000}
Rodr\'iguez-Fern\'andez, N.~J., Mart\'in-Pintado, J., de Vicente, P., Fuente, A., 
H\"uttemeister, S., Wilson, T.~L., \& Kunze, D.~2000, A\&A, 356, 695

\bibitem[Ruden \& Pollack(1991)]{ruden_pollack_1991}
Ruden, S.~P., \&~Pollack, J.~B.~1991, ApJ, 375, 740

\bibitem[Russell et al.(2005)]{russell_etal_2005}
Russell, S.~S., Krot, A.~N., Huss, G.~R., Keil, K., 
Itoh, S., Yurimoto, H., \& Macpherson, G.~J.~\chondrites, 317

\bibitem[Sano et al.(2000)]{sano_etal_2000}
Sano, T., Miyama, S.~M., Umebayashi, T., \& Nakano, T.~2000, ApJ, 543, 486

\bibitem[Scott \& Krot(2005)]{scott_krot_2005}
Scott, E.~R.~D., \& Krot, A.~N.~\chondrites

\bibitem[Shakura \& Sunyaev(1973)]{shakura_sunyaev_1973}
Shakura, N.~I., \&~Sunyaev, R.~A.~1973, A\&A, 24, 337

\bibitem[Shuetal(2001)]{shuetal2001}
Shu, F.~H., Shang, H., Gounelle, M., Glassgold, A.~E., \& Lee, T.~2001,
ApJ, 548, 1029

\bibitem[Sod(1978)]{sod1978}
Sod, G.~A.~1978, JCoPh, 27, 1

\bibitem[Spitzer \& Tomasko(1968)]{spitzer_tomasko_1968}
Spitzer, L., \& Tomasko, M.~G.~1968, ApJ, 152, 971

\bibitem[Stahler(1983)]{stahler_1983}
Stahler, S.~W.~1983, ApJ, 274, 822

\bibitem[Stahler(1988)]{stahler_1988}
Stahler, S.~W.~1988, ApJ, 332, 804

\bibitem[Stamatellos et al.(2007)]{stamatellos_etal_2007}
Stamatellos, D., Whitworth, A.~P., \& Ward-Thompson, D.~2007, MNRAS, 379, 1390

\bibitem[Stepinski(1992)]{stepinski_1992}
Stepinski, T.~F.~1992, Icarus, 97, 130

\bibitem[Sternberg \& Neufeld(1999)]{sternberg_neufeld1999}
Sternberg, A., \& Neufeld, D.~A.~1999, ApJ, 516, 371

\bibitem[Stone \& Norman(1992)]{stone_norman_1992}
Stone, J.~M., \& Norman, M.~L.~1992, ApJS, 80, 753

\bibitem[Strom \& Suzan(1993)]{strom_suzan_1993}
Strom, S.~E., \& Edwards, S.~1993, ASPC, 36, 235


\bibitem[Tachibana\& Huss(2005)]{tachibanahuss2005}
Tachibana, S., \& Huss, G.~R.~2005, GeCoA, 69, 3075

\bibitem[The et al.(1994)]{the_1994}
Th\'e, P.~S., Perez, M.~R., \& van den Heuvel, E.~P.~J.~1994, ASPC, 62, 23

\bibitem[Tohline(1980)]{tohline_1980}
Tohline, J.~E.~1980, ApJ, 235, 866

\bibitem[Tomley et al.(1991)]{tomley1991}
Tomley, L., Cassen, P., \& Steiman-Cameron, T.~1991, ApJ, 382, 530

\bibitem[Tomley et al.(1994)]{tomley1994}
Tomley, L., Steiman-Cameron, T.~Y., Cassen, P.~1994, ApJ, 422, 850


\bibitem[Toomre(1964)]{toomre_1964}
Toomre, A.~1964, ApJ, 139, 1217

\bibitem[Ulrich(1976)]{ulrich_1976}
Ulrich, R.~K.~1976, ApJ, 210, 377

\bibitem[Umebayashi \& Nakano(1981)]{umebayashi_nakano_1981}
Umebayashi, T., \& Nakano, T.~1981, PASJ, 33, 617

\bibitem[van Albada et al.(1982)]{van_albada_etal_1982}
van Albada, G.~D., van Leer, B., \& Roberts, W.~W.~1982, A\&A, 108, 76

\bibitem[van den Ancker et al.(1997)]{vandenancker_etal_1997}
van den Ancker, M.~E., Th\'e, P.~S., Tjin A Djie, H.~R.~E., 
Catala, C., de Winter, D., Blondel, P.~F.~C., Waters, L.~B.~F.~M.~1997, A\&A, 324, 33

\bibitem[Vorobyov(2005)]{vorobyov_basu_2005}
Vorobyov, E.~I., \& Basu, S.~2005, ApJ, 633, 137

\bibitem[Vorobyov(2006)]{vorobyov_basu_2006}
--.~2006, ApJ, 650, 956

\bibitem[Wadsley et al.(2004)]{wadsley_etal_2004}
Wadsley, J.~W., Stadel, J., \& Quinn, T.~2004, New Astronomy, 9, 137

\bibitem[Walmsley et al.(2004)]{walmsley_etal_2004}
Walmsley, C.~M., Flower, D.~R., \& Pineau des For\^ets, G.~2004, A\&A, 418, 1035


\bibitem[Webber(1998)]{webber1998}
Webber, W.~R.~1998, ApJ, 506, 329

\bibitem[Weidenschilling(1977)]{weidenschilling_1977}
Weidenschilling, S.~J.~1977, Ap\&SS, 51, 153

\bibitem[Williams(1988)]{williams_1998}
Williams, H.~A.~1988, Ph.D.~Thesis, Louisiana State University

\bibitem[Whipple(1966)]{whipple1966}
Whipple, F.~L.~1966, Science, 153, 54

\bibitem[Whitehouse \& Bate(2006)]{whitehouse_bate_2006}
Whitehouse, S.~C., \& Bate, M.~R.~2006, MNRAS, 367, 32

\bibitem[Wood(1963)]{wood_1963}
Wood, J.~A.~1963, Icarus, 2, 152 

\bibitem[Wood(1996)]{wood_1996}
--.~1996, Meteoritics Planet.~Sci., 31, 641

\bibitem[Wood(2005)]{wood_2005}
--.~\chondrites, 953

\bibitem[Wooden(2005)]{woodenetal2005}
Wooden, D., Harker, D.~E., \& Brearley, A.~J.~\chondrites, 774

\bibitem[Yang(1992)]{yang_phd}
Yang, S.~X.~1992, Ph.D.~Thesis, Indiana University

\bibitem[Yorke et al.(1993)]{yorke_etal_1993}
Yorke, H.~W., Bodenheimer, P., \& Laughlin, G.~1993, ApJ, 411, 274

\bibitem[Zhu et al.(2007)]{zhu_etal_2007}
Zhu, Z., Hartmann, L., Calvet, N., Hernandez, J., Muzerolle, J., 
\& Tannirkulam, A.~2007, arXiv:0707.3429

\end{thebibliography}
\end{document}